\documentclass[12pt]{article}
\pdfoutput=1

\usepackage{xcolor}
\definecolor{darkblue}{rgb}{0.1,0.1,.7}
\usepackage[colorlinks, linkcolor=darkblue, citecolor=darkblue, urlcolor=darkblue, linktocpage,pagebackref]{hyperref} 

\usepackage[english]{babel}
\usepackage{amsmath,amssymb,graphicx,enumerate,bbm}
\usepackage{booktabs}
\usepackage{multirow}
\usepackage{geometry}
\usepackage[margin=10pt,font=small,labelfont=bf]{caption}

\usepackage{dsfont}

\usepackage{titlesec}
\titlespacing\section{0pt}{12pt plus 4pt minus 2pt}{0pt plus 2pt minus 2pt}
\titlespacing\subsection{0pt}{12pt plus 4pt minus 2pt}{0pt plus 2pt minus 2pt}
\titlespacing\subsubsection{0pt}{12pt plus 4pt minus 2pt}{0pt plus 2pt minus 2pt}
\titleformat*{\section}{\large\bfseries}
\titleformat*{\subsection}{\normalsize\bfseries}
\titleformat*{\subsubsection}{\normalsize\it}
\titleformat*{\paragraph}{\normalsize\bfseries}
\titleformat*{\subparagraph}{\normalsize\bfseries}

\usepackage[noadjust]{cite}

\usepackage[utf8]{inputenc}
\newcommand{\assign}{:=}
\newcommand{\backassign}{=:}
\newcommand{\comma}{{,}}
\newcommand{\nin}{\not\in}
\newcommand{\nobracket}{}
\newcommand{\nosymbol}{}
\newcommand{\rightarrowlim}{\mathop{\rightarrow}\limits}
\newcommand{\tmmathbf}[1]{\ensuremath{\boldsymbol{#1}}}
\newcommand{\tmop}[1]{\ensuremath{\operatorname{#1}}}

\newcommand{\tmtextbf}[1]{{\bfseries{#1}}}
\newcommand{\tmtextit}[1]{{\itshape{#1}}}
\newenvironment{enumeratenumeric}{\begin{enumerate}[1.] }{\end{enumerate}}
\newenvironment{itemizedot}{\begin{itemize} }{\end{itemize}}

\newcommand{\Lirr}{\Lcal_1}
\newcommand{\Lsup}{\Lcal_2}

\newcommand{\Lcal}{\mathcal{L}}

\newcommand{\Ocal}{\mathcal{O}}

\renewcommand{\hat}[1]{\widehat{#1}}
\newcommand{\la}{\langle}
\newcommand{\ra}{\rangle}
\newcommand{\psib}{\bar{\psi}}
\newcommand{\thetab}{\bar{\theta}}

\newcommand{\vareps}{\varepsilon}
\newcommand{\eps}{\varepsilon}
\newcommand{\del}{\partial}

\newcommand{\bR}{\mathbb{R}}

\newcommand{\calO}{\mathcal{O}}

\newcommand{\calF}{\ensuremath{\mathcal{F}}}

\newcommand{\calH}{\mathcal{H}}

\newcommand{\calB}{\ensuremath{\mathcal{B}}}

\newcommand{\calL}{\mathcal{L}}

\newcommand{\calS}{\mathcal{S}}

\renewcommand{\geq}{\geqslant}
\renewcommand{\leq}{\leqslant}
\newcommand{\NO}[1]{: {\hspace{-0.17em}}#1{\hspace{-0.17em}}:}
\newcommand{\tmH}{H}

\newcommand{\tr}{\mathrm{tr}}
\newcommand{\g}{\gamma}

\newcommand{\s}{\sigma}
\newcommand{\m}{\mu}
\newcommand{\n}{{\nu}}
\newcommand{\tma}{\alpha}
\newcommand{\e}{\varepsilon}
\newcommand{\tmb}{\beta}
\newcommand{\tmd}{\delta}

\newcommand{\vf}{\varphi}

\newcommand{\D}{\Delta}
\newcommand{\G}{\Gamma}

\numberwithin{equation}{section}

\usepackage{tocloft}
\usepackage{etoolbox}
\apptocmd{\thebibliography}{\setlength{\itemsep}{0em}}{}{}

\begin{document}
	
	\vspace*{-.6in} \thispagestyle{empty}
	\begin{flushright}
	\end{flushright}
	
	\vspace{.2in} {\large
		\begin{center}
			\bf  Random Field Ising Model and Parisi-Sourlas Supersymmetry. \\ Part II.
			Renormalization Group
		\end{center}
	}
	\vspace{.2in}
	\begin{center}
		{\bf 
			Apratim Kaviraj$^{a,b}$, \ Slava Rychkov$^{c,b}$,  \ Emilio Trevisani$^{b}$
		} 
		\\
		\vspace{.2in} 
		{\it $^{a}$Institut de Physique Th\'{e}orique Philippe Meyer, \\ $^{b}$
			Laboratoire de Physique de l’Ecole normale sup\'erieure, ENS, \\
			Universit\'e PSL, CNRS, Sorbonne Universit\'e, Universit\'e de Paris, F-75005 Paris, France}\\
		{\it $^{c}$ Institut des Hautes \'Etudes Scientifiques, Bures-sur-Yvette, France}\\
	\end{center}

	\vspace{.2in}

\begin{abstract}
  We revisit perturbative RG analysis in the replicated Landau-Ginzburg
  description of the Random Field Ising Model near the upper critical
  dimension 6. Working in a field basis with manifest vicinity to a
  weakly-coupled Parisi-Sourlas supersymmetric fixed point (Cardy, 1985), we
  look for interactions which may destabilize the SUSY RG flow and lead to the
  loss of dimensional reduction. This problem is reduced to studying the
  anomalous dimensions of ``leaders''---lowest dimension parts of
  $S_n$-invariant perturbations in the Cardy basis. Leader operators are
  classified as non-susy-writable, susy-writable or susy-null depending on
  their symmetry. Susy-writable leaders are additionally classified as
  belonging to superprimary multiplets transforming in particular $\textrm{OSp}
  (d | 2)$ representations. We enumerate all leaders up to 6d
  dimension $\Delta = 12$, and compute their perturbative anomalous
  dimensions (up to two loops). We thus identify two perturbations (with
  susy-null and non-susy-writable leaders) becoming relevant below a critical
  dimension {$d_c \approx 4.2$ - $4.7$}. This supports the scenario
  that the SUSY fixed point exists for all $3 < d \leq 6$, but becomes unstable
  for $d < d_c$.
\end{abstract}
\vspace{.3in}
\hspace{0.7cm} 
{September 2020}
\newpage
\tableofcontents

\newpage
\section{Introduction }\label{sec:intro}

This is the second part of our project dedicated to the puzzle of the Random
Field Ising Model. The first paper {\cite{paper1}} was about nonperturbative
CFT aspects, while here we will focus on the Renormalization Group (RG)
aspects and will propose a tentative resolution of the puzzle. The two papers
can be read largely independently.

As is well known, the usual ferromagnetic Ising model with the Hamiltonian
$\calH = - J \sum_{\langle i j \rangle} s_i s_j$, where $s_i = \pm 1$ are
spins on a regular $d$-dimensional lattice with nearest-neighbor interactions,
has a thermodynamic second-order phase transition in $d \geqslant 2$ which is
described by a non-Gaussian fixed point for $d < 4$. At the phase transition
the correlation length $\xi \rightarrow \infty$. This idealized Ising model
assumes no impurities, but real materials always have impurities. Sufficiently
near the critical temperature we will have $\xi > L$, the average distance
between impurities,\footnote{The cleanest electronics-grade silicon has $L
\sim 1000$ lattice spacings (about one impurity per billion atoms). Available
ferromagnetic materials have even more impurities.} and we should start
worrying about their effect. Will they change the universality class or not?

Specifically, in this work we are interested in impurities which have a random
and frozen magnetic moment (i.e.~some of the spins at assigned randomly chosen
values $+ 1$ or $- 1$, while others are allowed to fluctuate.\footnote{This
may be realizable in a ferromagnetic metal with randomly distributed magnetic
impurities forming a spin-glass state, due to the RKKY interaction whose sign
depends on the distance. The most common experimental realizations of the RFIM
is a randomly diluted antiferromagnet in a weak external magnetic field
{\cite{PhysRevB.29.505}}. See {\cite{Belanger}} for other experimental
realizations.} This is modeled by adding to the usual Ising Hamiltonian a
random magnetic field $h_i$ on each site:
\begin{equation}
  \calH = - J \sum_{\langle i j \rangle} s_i s_j + \sum_i h_i s_i
  \hspace{0.17em} . \label{RFIM}
\end{equation}
This equation defines our object of interest: the Random Field Ising Model
(RFIM). The real magnetic field $h = (h_i)$ is assumed to have a factorized
probability distribution
\begin{equation} \mathcal{P} (h) \mathcal{D} h = \prod P (h_i) d h_i, \end{equation}
so that $h_i$ are independent identically distributed random variables. It is
assumed that $h_i$ has zero mean and a finite variance: $\overline{h_i^2 } =
H$.

Observables are computed in two steps, first averaging over spin fluctuations
with a fixed magnetic field, and then over the magnetic field (this is called
quenched disorder average). E.g.~for the two-point function of spins:
\begin{equation} \overline{\langle s_i s_j \rangle_h} = \int \mathcal{D}h\,\mathcal{P} (h) \langle s_i
   s_j \rangle_h, \end{equation}
where $\langle s_i s_j \rangle_h$ is the thermodynamic average holding $h$
fixed, and the overbar will always denote a magnetic field average.

Near the phase transition, the lattice model {\eqref{RFIM}} may be replaced
by an effective Landau-Ginzburg Hamiltonian
\begin{eqnarray}
  &  & S [\phi, h] = \int d^d x \left[ \frac{1}{2} (\partial_{\mu} \phi)^2 +
  V (\phi) + h (x) \phi (x) \right],  \label{LGrfim}\\
  &  & V (\phi) = \frac{m^2}{2} \phi^2 + \frac{\lambda}{4!} \phi^4, \nonumber
\end{eqnarray}
where the random magnetic field has short-range correlations: $\overline{h (x)
h (y)} = H \delta^{(d)} (x - y)$. The relevance condition for the disordered
coupling is $\Delta_{\phi} < d / 2$ (``Harris criterion''). Since
$\Delta_{\phi} = d / 2 - 1 + \eta / 2$ and $\eta$ is small, the Harris
criterion is satisfied and the coupling $h (x) \phi (x)$ is strongly
relevant.\footnote{Alternatively, one could add the $h (x) \phi^2 (x)$
perturbation which is weakly relevant in 3d by the Harris criterion, using the
Ising fixed point dimension $\Delta(\phi^2)\approx 1.41$. This describes the phase
transition in a different lattice model: $\calH = - \sum_{\langle i j \rangle}
(J + \delta J_{i j}) s_i s_j$ where $\delta J_{i j}$ is a random perturbation
called bond disorder. Because the random $\phi^2$ perturbation is weakly
relevant, the bond-disorder phase transition is better understood than the
RFIM phase transition studied here, see e.g. {\cite{Komargodski:2016auf}} for
a recent discussion. Another related difference with bond disorder is
highlighted in footnote \ref{bond}.}

Thus, the phase transition in RFIM is different from the usual Ising model in
$d$ dimensions. In 1979, Parisi and Sourlas {\cite{Parisi:1979ka}} formulated
a conjecture relating it instead to the Ising model in $d - 2$ dimensions. It
is convenient to split the Parisi-Sourlas conjecture into two parts:
\begin{enumeratenumeric}
  \item {\it Emergence of SUSY:} The RFIM transition is described by a conformal
  field theory (CFT) in $d$ dimensions, possessing a non-unitary supersymmetry with scalar supercharges (Parisi-Sourlas SUSY);
  
  \item {\it Dimensional reduction:} A Parisi-Sourlas supersymmetric CFT in $d$
  dimensions (SCFT$_d$) has the same critical exponents as an ordinary,
  non-supersymmetric CFT in $d - 2$ dimensions.
\end{enumeratenumeric}
The dimensionally reduced $\tmop{CFT}_{d - 2}$ has the same global symmetry
$\mathbb{Z}_2$ as the parent $\tmop{SCFT}_d$ and is expected to be the
ordinary Ising fixed point in $d - 2$ dimensions. Hence, the RFIM transition
in $d$ dimensions should have the same exponents as the ordinary Ising
transition in $d - 2$ dimensions.

As subsequent work has shown, this conjecture is subtle: it is not quite
true, nor is it however totally false. In spite of much work, there seems to
be no agreement in the literature about why this happens (see Appendix
\ref{sec:history} for the review). Here are some relevant pro and contra
results:
\begin{itemizedot}
  \item It works in perturbative expansion near the upper critical dimension
  $d = 6 - \varepsilon$.
  
  \item It fails in $d = 3, 4$: numerical simulations show a non-SUSY
  continuous phase transition in the 3d and 4d RFIM {\cite{Fytas3,Picco1}}.
  Dimensional reduction also fails: the 1d Ising does not even have a phase
  transition, while the 2d Ising exponents do not agree with 4d RFIM
  {\cite{Picco1}}.
  
  \item It might be correct in $d = 5$ where there is numerical evidence for
  both SUSY and dimensional reduction {\cite{Picco2,Picco3}}.
  
  \item Both SUSY and dimensional reduction work perfectly in a parallel story
  for the random field $i \phi^3$ model relevant for the description of
  branched polymers, which maps on the Lee-Yang universality class in $d - 2$
  dimensions (see section \ref{sec:LY}). 
\end{itemizedot}
The first paper of our project {\cite{paper1}}\footnote{See also an online talk {\cite{zoom}} for an introduction.} performed new checks of
Part 2 of the Parisi-Sourlas conjecture, using nonperturbative CFT
techniques. We have not found any inconsistency from this point of
view.\footnote{First checks of dimensional reduction were based on
perturbation theory (see Appendix \ref{early}). Nonperturbative arguments for
dimensional reductions were advanced in
{\cite{CARDY1983470,KLEIN1983473,Klein:1984ff,Zaboronsky:1996qn}}. Our work
{\cite{paper1}} is different from these in that it does not rely on the use of
Lagrangians.} Since Part 2 held up to scrutiny, the problem must therefore lie
in Part 1. Here we will proceed to study Part 1, and try to understand why
sometimes it works and sometimes fails.

We will start in section \ref{PSrev} with a review of the Parisi-Sourlas
dimensional reduction. From many ways to the Parisi-Sourlas supersymmetry, we
choose to base our exposition on the method of replicas accompanied by the
``Cardy transform'': a judicious linear transformation of fields first
proposed by Cardy in 1985 {\cite{CARDY1985123}} but little used since. The
Cardy transform exhibits a Gaussian theory perturbed by various interactions,
some of which are weakly relevant and others are irrelevant in $d = 6 -
\varepsilon$ dimensions. The ``basic RG scenario'' (section \ref{basicRG})
consists in taking the $n \rightarrow 0$ limit and dropping the formally
irrelevant terms, which naively results in a supersymmetric theory (and hence
in dimensional reduction). This SUSY theory and its fixed point are discussed
in section \ref{RGbasic}, including a subtle point (section \ref{RGL0}) of how
SUSY emerges at long distances in the basic scenario, even though the
$S_n$-invariant regulator breaks it.

After a short recap in section \ref{Plan}, we plunge into the heart of our
study, which is to examine the validity of the basic RG scenario assumptions.
One of them (dropping the $n$-suppressed terms) is justified in section
\ref{sec:L2}, after having understood $S_n$ invariance in the Cardy basis
(section \ref{sec:Sn}). Section \ref{sec:Sn} also introduces the key concept
of the ``leader'' operator, which is the lowest-dimension part of an
$S_n$-singlet. Scaling dimension of the leader controls that of the full
perturbation, as we explain in section \ref{L0L1}. From here on, we work in
the strict $n = 0$ limit and examine if any perturbation irrelevant in $d = 6
- \varepsilon$ may become relevant in lower $d$, by looking at the leaders of
$S_n$-singlet perturbations. These leaders are classified (section
\ref{leaders}) into three classes: non-susy-writable, susy-writable and
susy-null, which have only triangular mixing among each other, simplifying the
anomalous dimension computations (section \ref{anomdim}). We list all leaders
up to dimension 12 in $d = 6$, which includes one or more leaders in each of
the three classes. Finally, using the computed one- or two-loop anomalous
dimensions, some of which are negative, we build a case for the loss of SUSY
via RG instability of the SUSY fixed point below some critical dimension value
$d_c$ (section \ref{Who}). Section \ref{sec:conclusions} is devoted to a
discussion of our results and to a list of open problems: developing our
method further, applying it in different situations (like for the branched
polymers), and checking our conclusions with alternative techniques.

Prior work on the RFIM phase transition being vast, we gather an extensive
review of the literature in Appendix \ref{sec:history}, which may be of
independent interest. Other appendices contain technical details referred to
from the main text.

\tmtextbf{Note on phi's.} This paper will have a
proliferation of phi's. Stroked $\phi$ is the original field in the random
field Landau-Ginzburg Lagrangian {\eqref{LGrfim}}. Stroked $\phi_i$ with an
index denotes replicated fields introduced in section \ref{replicas}. Loopy
$\varphi$ is a field from the Cardy transform basis, section \ref{sec:CT}. All
these live in $\mathbb{R}^d$. Big $\Phi$ is the superfield {\eqref{supefield}}
living in $\mathbb{R}^{d | 2 \nobracket}$. Finally, hatted $\widehat{\phi \,}$
is a scalar field in $\mathbb{R}^{d - 2}$ which appears in the dimensionally
reduced action {\eqref{Sdm2}}.

\subsection{Executive summary for RFIM experts}

The great length of this paper is justified by the complexity of the problem, and by our wish to make our work 
accessible to the readers without prior RFIM experience.
In this section we will provide a quick summary of our main ideas and results, which on the contrary will \emph{only} be understandable to the RFIM experts. All facts mentioned here are discussed in detail elsewhere in the paper (see the table of contents, the outline in the introduction, and a roadmap in section \ref{Plan}).

Via the method of replicas, RFIM phase transition is described by the Lagrangian ($n\to 0$)
\begin{equation}
\textstyle \sum\nolimits_{i = 1}^n \left[
\frac{1}{2} (\partial_{\mu} \phi_i)^2 + V(\phi_i) \right] - \frac{\tmH}{2}
\left( \sum\nolimits_{i = 1}^n \phi_i \right)^2 ,\quad V(\phi)=\frac {m^2}2 \phi^2+\frac{\lambda}{4!}\phi^4.
\end{equation}
Using the linear transformation of fields $\phi_1=\varphi+\omega/2$, $\phi_i=\varphi-\omega/2+\chi_i$ ($i=2\ldots n$), $\sum \chi_i=0$, the replicated Lagrangian is mapped to $\calL_0+\calL_1+\calL_2$, where
\begin{equation}
\calL_0 = \textstyle \del_\mu \varphi \del_\mu \omega - \frac{\tmH}{2} \omega^2
+ \frac{1}{2}  \sum\nolimits_{i=2}^n (\del_\mu \chi_i)^2 + \frac{m^2}{2}  \left( 2 \vf \omega +
\sum\nolimits_{i=2}^n \chi_i^2 \right) + \frac{\lambda}{4!} \left( 4 \omega \vf^3 + 6 \sum\nolimits_{i=2}^n
\chi_i^2  \vf^2 \right)  \hspace{0.17em} \label{L0preview},
\end{equation}
$\calL_1$ terms vanish for $n\to0$ and can be safely dropped, while $\calL_2$ terms are irrelevant and may be dropped in $d=6-\eps$, $\eps\ll 1$. The quartic interactions in $\calL_0$ are weakly relevant for $\eps\ll1$ and flow to an IR fixed point. Separating weakly relevant and irrelevant effects is the main point of this transformation. Replacing $-2$ linearly independent bosons $\chi_i$ by two scalar fermions $\psi,\bar{\psi}$, Lagrangian $\calL_0$ maps to an equivalent Parisi-Sourlas supersymmetric Lagrangian
\begin{equation}
\label{LSUSYpreview}
\textstyle \partial_\mu \varphi \del_\mu \omega - \frac{H}{2} \omega^2 +
\omega V' (\varphi) + \partial_\mu \psi \partial_\mu \psib + \psi \psib V''
(\varphi) \, .
\end{equation}
This derivation of Parisi-Sourlas supersymmetry (Cardy, 1985) works provided all the $\calL_2$ terms or any other $S_n$-invariant terms are irrelevant, as is the case for $\eps\ll 1$. Our work investigates for the first time whether this is still the case for $\eps=O(1)$.

To do so we build all possible $S_n$-invariant polynomial operators (not necessarily those included in the quartic replicated Lagrangian) and transform them into the field basis $\varphi,\omega, \chi_i$. We prove that the RG behavior of the transformed $S_n$-invariant operator is determined by the part of the lowest classical dimension (the leader). We compute one- or two-loop anomalous dimensions of many leaders, up to classical dimension $\Delta=12$, and classifying them into three types by their symmetry: 
\begin{itemize}
\item \emph{Susy-writable} leaders, which make sense in the SUSY field basis $\varphi,\omega, \psi,\bar{\psi}$. E.g. the leaders of $S_n$-invariant quadratic and quartic interactions $\sum_{i=1}^n \phi_i^2$ and $\sum_{i=1}^n \phi_i^4$ are, see \eqref{L0preview},  $ 2 \vf \omega +
\sum\nolimits_{i=2}^n \chi_i^2\to \vf \omega +
2\psi\bar{\psi} $ and $ 4 \omega \vf^3 + 6 \sum\nolimits_{i=2}^n
\chi_i^2  \vf^2 \to 4 \omega \vf^3 + 12 \psi\bar{\psi} \vf^2 $, so these are susy-writable, as are any leaders involving only $\chi^2_i$.

\item \emph{Susy-null} leaders, which can be transformed to the SUSY field basis, but which vanish after such a transformation because of $\psi^2={\bar\psi}^2=0$. The first such leader is $(\mathcal{F}_4)_L = (\sum_{i=2}^n \chi_i^2)^2\to (2\psi\bar{\psi})^2 = 0$ 
of the $S_n$-invariant operator $\mathcal{F}_4= \sum_{i,j=1}^n (\phi_i-\phi_j)^4$. 

\item \emph{Non-susy-writable} leaders, which cannot even be transformed to the SUSY field basis, because they involve $\chi_i$ raised to a power higher than 2. The first of them is the leader $(\mathcal{F}_6)_L = (\sum_{i=2}^n\chi_i^3)^2 - \frac{3}{2}
(\sum_{i=2}^n\chi_i^2) (\sum_{j=2}^n\chi_j^4)$ 
of the $S_n$-invariant operator $\mathcal{F}_6= \sum_{i,j=1}^n (\phi_i-\phi_j)^6$. 
\end{itemize}

Susy-writable and susy-null leaders have $O(n-2)$ symmetry in the field basis $(\varphi,\chi_i,\omega)$,  which is the same as symmetry of $\calL_0$, while non-susy-writable leaders break this symmetry to $S_{n-1}\subset O(n-2)$.

Susy-writable leader dimensions can be computed from the SUSY Lagrangian \eqref{LSUSYpreview} or, due to dimensional reduction, from the Wilson-Fisher Lagrangian in $d-2$ dimensions. On the contrary, susy-null and non-susy-writable leader dimensions are inaccessible from the SUSY Lagrangian or via dimensional reduction, and have to be computed from Lagrangian \eqref{L0preview}. 

We did not find any susy-writable leader which becomes relevant. On the other hand, the first susy-null leader $(\calF_4)_L$ and $(\calF_6)_L$ have sizable negative two-loop dimensions $8 - 2 \varepsilon - \frac{8}{27}
\varepsilon^2+\ldots$ and $12 - 3 \varepsilon- \frac{7}{9} \e^2+\ldots$, and appear to become relevant around $d_c\approx 4.5$. We thus predict that for $d<d_c$ the Parisi-Sourlas fixed point is destabilized, and the RFIM transition is described by another, non-supersymmetric, fixed point (about which we have nothing to say).

\section{Replicas and Cardy transform}\label{PSrev}

Our work begins from two ideas, one standard and one less so. The standard
idea is the method of replicas, used in essentially all known to us
renormalization group approaches to this problem.\footnote{One exception is
{\cite{Mukaida}}, which develops RG for the probability distribution of
magnetic impurities.} The less standard idea is the Cardy transform, a
linear transformation of replica fields first considered by Cardy
{\cite{CARDY1985123}} in 1985 but little used since. We use it because it
reveals the Gaussian fixed point and clarifies the renormalization group
picture.

\subsection{Method of replicas}\label{replicas}

We use the version of the method of replicas appropriate for the study of
correlation functions.\footnote{This is sometimes called ``second variant'',
see e.g. {\cite{Wegner:2016ahw}}, section 4.2.2.} We are interested in
quenched averaged correlators, defined first averaging over $\phi$ and then
over $h$:
\begin{equation}
  \label{formula:1copy} \overline{\langle A (\phi) \rangle} = \int
  \mathcal{D}h\,\mathcal{P} (h) \hspace{0.17em} \frac{1}{Z_h}  \int \mathcal{D}
  \phi\, A (\phi) e^{- \calS [\phi, h]} .
\end{equation}
Here $S [\phi, h]$ is given in {\eqref{LGrfim}}, $A (\phi)$ is any function of
the field $\phi$, e.g.~a product of $\phi$ at several points, $\mathcal{P}
(h)$ is the disorder distribution, and overbar denotes the disorder average.

We multiply the integrand in {\eqref{formula:1copy}} by $1 = Z_h^{n - 1} /
Z_h^{n - 1}$, rename $\phi \to \phi_1$, and represent $Z_h^{n - 1}$ in the
numerator as the product of partition functions of `replica' fields $\phi_2,
\ldots, \phi_n$, with the same action as $\phi_1$. We get:
\begin{equation}
  \label{formula:ncopies} \overline{\langle A (\phi) \rangle} = \int
  \mathcal{D}h\,\mathcal{P} (h) \frac{1}{Z^n_h}  \int \mathcal{D} \vec{\phi}\, A\,
  (\phi_1) e^{- \sum_{i = 1}^n \calS [\phi_i, h]} .
\end{equation}
This equation is independent of $n$. Particularly nice is the limit $n \to 0$,
since the denominator $Z_h^n \to 1$. With the usual provisos for going from
integer to real $n$ and commuting the limit and the integral, we get a simpler
formula:
\begin{equation}
  \label{replicatrick} \overline{\langle A (\phi) \rangle} = \lim_{n
  \rightarrow 0}  \int \mathcal{D}h\,\mathcal{P} (h)  \int \mathcal{D}
  \vec{\phi}\, A (\phi_1) e^{- \sum_{i = 1}^n \calS [\phi_i, h]} \hspace{0.17em}
  .
\end{equation}
As mentioned our disorder is mean zero and with short-range spatial
correlations:
\begin{equation}
  \overline{h (x)} = \int \mathcal{D}h\,\mathcal{P} (h) h (x) = 0
  \hspace{0.17em}, \qquad \overline{h (x) h (x')} = \int
  \mathcal{D}h\,\mathcal{P} (h) h (x) h (x') = \hspace{0.17em} \tmH
  \hspace{0.17em} \tmd (x - x') \hspace{0.17em} .
\end{equation}
The simplest distribution satisfying these properties is the Gaussian white
noise:
\begin{equation}
  \mathcal{P} (h) \propto e^{- \frac{1}{2 \tmH}  \int d^d x\,h (x)^2} .
\end{equation}
Assuming this distribution, the integral over $h$ in {\eqref{replicatrick}} is
Gaussian and can be performed. We obtain:
\begin{equation}
  \label{eq:Sr1} \overline{\langle A (\phi) \rangle} = \lim_{n \rightarrow 0} 
  \int \mathcal{D} \vec{\phi}\, A (\phi_1) e^{- \calS_n [\vec{\phi}]}
  \backassign \langle A (\phi_1) \rangle,
\end{equation}
\begin{equation}
  \label{Sr} \calS_n [\vec{\phi}] = \int d^d x \left\{ \sum_{i = 1}^n \left[
  \frac{1}{2} (\partial_{\mu} \phi_i)^2 + V (\phi_i) \right] - \frac{\tmH}{2}
  \left( \sum_{i = 1}^n \phi_i \right)^2 \right\} .
\end{equation}
This is a pleasing result: we can compute disorder-averaged correlation
functions from a theory where disorder is replaced by a coupling among $n \to
0$ replicas. This can be generalized to disorder-averaged products of several
correlation functions, e.g.
\begin{equation}
  \overline{\la A (\phi) \ra \la B (\phi) \ra \langle C (\phi) \rangle} =
  \lim_{n \to 0} \int \mathcal{D} \vec{\phi} \,A (\phi_1) B (\phi_i) C (\phi_j)
  e^{- \calS_n [\vec{\phi}]} = \langle A (\phi_1) B (\phi_i) C (\phi_j)
  \rangle, \label{form2}
\end{equation}
as long as all the three indices $1, i, j$ are all different. Note that the
replicated theory contains formally $\phi_i$ with any index, so there is no
contradiction in introducing 3 different fields as in {\eqref{form2}} which
will be compensated by $- 3$ fields when taking $n \rightarrow 0$ limit. Such
occurrences of a negative number of fields are a necessary feature of this
formalism; we will encounter it soon in section \ref{sec:CT}.

\subsection{Standard perturbation theory and the upper critical
dimension}\label{sec:uc}

From the quadratic part of the action $\mathcal{S}_n$ one derives the
propagator inverting the matrix
\begin{equation}
  \mathbf{G}^{- 1} = k^2  \mathds{1} - \tmH \mathbf{M} \hspace{0.17em},
\end{equation}
where $\mathbf{M}$ is an $n \times n$ matrix whose all elements are unity. An
easy computation gives
\begin{equation}
  \mathbf{G} = \frac{\mathds{1}}{k^2} + \frac{\tmH \mathbf{M}}{k^2  (k^2 - n \tmH)}
  \hspace{0.17em} . \label{eq:G}
\end{equation}
This propagator is employed in most perturbative studies of RFIM. Notice that
two terms have a different scaling with $k$, which renders perturbative
computations somewhat awkward.\footnote{Another displeasing feature is that
the second term in {\eqref{eq:G}} only acquires good scaling in the $n \to 0$
limit.} One usually has to go through the diagrams looking for terms most
singular in the limit $k \to 0$, hence most important at long distances, which
come precisely from the second term in {\eqref{eq:G}}. The effective expansion
parameter for these terms, deemed most important in IR, is therefore changed from $\lambda$
to $\lambda \tmH$. The $\tmH$ having mass dimension 2, $\lambda \tmH$ becomes
marginal at the upper critical dimension $d_{uc} = 6$.

This way of reasoning, while standard in much of RFIM work, seems like a
departure from the usual Wilsonian paradigm.\footnote{A related concept is
that of `zero-temperature fixed point' which we review in Appendix \ref{T=0}.}
Wilson taught us to think in terms of a Gaussian fixed point at which fields
have well-defined scaling dimensions. One then classifies perturbations into
strongly relevant, weakly relevant, and irrelevant. Strongly relevant
perturbations are tuned, irrelevant dropped, while the weakly relevant may
drive the RG flow to a non-Gaussian weakly-coupled fixed point nearby. This is
much more systematic and powerful than having to sift through diagrams. Cardy
{\cite{CARDY1985123}} showed that the disordered fixed point is not an
exception and can also be presented this way. We will now describe his
construction, which will form the basis for our work.

\subsection{Cardy transform}\label{sec:CT}

As mentioned, different components of the propagator {\eqref{eq:G}} have
different scaling dimensions. The idea of Cardy {\cite{CARDY1985123}} is to
make this manifest via a linear transformation in the field space. One then
drops the irrelevant terms in the resulting effective Lagrangian, and reaches
the disordered fixed point by RG flowing from a Gaussian fixed point perturbed
by a weakly relevant perturbation.

The Cardy transform can be guessed by the following argument. First one decides to
treat $\phi_1$ differently from $\phi_2, \ldots, \phi_n$ (perhaps motivated by
Eq.~{\eqref{replicatrick}} for the disordered correlated functions). One then
writes
\begin{equation}
  \label{eq:Cardy1} \phi_i = \rho + \chi_i \hspace{0.17em}, \quad (i = 2
  \ldots n) \hspace{0.17em}, \quad \text{with } \sum_{i = 2}^n \chi_i = 0
  \hspace{0.17em},
\end{equation}
i.e.~$\rho = \frac{1}{n - 1}  (\phi_2 + \ldots + \phi_n)$. The quadratic part
of {\eqref{Sr}} then separates nicely as ($\sum\nolimits' \equiv \sum_{i = 2}^n$):
\begin{equation}
  \frac{1}{2}  \left[ (\del \phi_1)^2 + (n - 1) (\del \rho)^2 - \tmH [\phi_1 +
  (n - 1) \rho]^2 \right] + \frac{1}{2}  \sum\nolimits' (\del \chi_i)^2 \hspace{0.17em}
  .
\end{equation}
In the $n \to 0$ limit this simplifies even further as
\begin{equation}
  \frac{1}{2}  \left[ (\del \phi_1)^2 - (\del \rho)^2 - \tmH (\phi_1 - \rho)^2
  \right] + \frac{1}{2}  \sum\nolimits' (\del \chi_i)^2 = \del \varphi \del \omega -
  \frac{\tmH}{2} \omega^2 + \frac{1}{2}  \sum\nolimits' (\del \chi_i)^2
  \hspace{0.17em}, \label{eq:n=0gauss}
\end{equation}
where we defined
\begin{equation}
  \label{eq:Cardy2} \varphi = \frac{1}{2}  (\phi_1 + \rho), \qquad \omega =
  \phi_1 - \rho \hspace{0.17em} .
\end{equation}
The Cardy transform is given, for any $n$, by Eqs.~{\eqref{eq:Cardy1}},
{\eqref{eq:Cardy2}}, which equivalently can be written as
\begin{eqnarray}
  \phi_1 & = & \varphi + \omega / 2, \nonumber\\
  \phi_i & = & \varphi - \omega / 2 + \chi_i \qquad (i = 2 \ldots n) . 
  \label{CT}
\end{eqnarray}
From the quadratic part of {\eqref{eq:n=0gauss}}, the transformed fields
$\omega$, $\varphi$, $\chi_i$ have well-defined scaling dimensions in the $n
\to 0$ limit:
\begin{equation}
  \Delta_\vf = \frac{d}{2} - 2 \hspace{0.17em}, \hspace{1cm} \Delta_\chi = \frac{d}{2} -
  1 \hspace{0.17em}, \hspace{1cm} \Delta_\omega = \frac{d}{2} \hspace{0.17em} .
  \label{dimensions}
\end{equation}
Note that it would be wrong to think of the $\omega^2$ term in
{\eqref{eq:n=0gauss}} as a mass term, because the kinetic term $(\del
\omega)^2$ is missing. In fact all propagators are scale
invariant:\footnote{\label{Piij}Here $\Pi_{i j}$ is an $(n - 1) \times (n -
1)$ matrix whose all elements are 1. Note that the $\chi \chi$ propagator is
consistent with the constraint $\sum_{i = 2}^n \chi_i = 0 .$}
\begin{equation}
  \langle \vf_k \vf_{- k} \rangle = \frac{H}{k^4}, \quad \langle \vf_k
  \omega_{- k} \rangle = \frac{1}{k^2}, \quad \langle \omega \omega \rangle =
  0, \quad \langle (\chi_i)_k (\chi_j)_{- k} \rangle = \frac{1}{k^2} \left(
  \delta_{i j} - \frac{1}{n - 1} \Pi_{i j} \right) . \label{propsL0}
\end{equation}
The $1 / k^2$ and $1 / k^4$ are the same powers as in {\eqref{eq:G}} but now
they are nicely separated. The dimension of $\vf$ is below the unitarity
bound---one sign that we are dealing with a non-unitary theory.

Applying the Cardy transform to the interaction term in {\eqref{Sr}}, we obtain
\begin{equation}
  V (\vf + \omega / 2) + \sum\nolimits' V (\vf - \omega / 2 + \chi_i) \hspace{0.17em} .
  \label{eq:cardyint}
\end{equation}
Taylor-expanding the quartic potential,\footnote{Only the quartic potential
will be treated in this work, while the cubic potential (branched polymers and
the Lee-Yang universality class) is dealt with in  \cite{papershort,paperphi3}. 
See section \ref{BP-LY}.} we organize the resulting terms by their scaling
dimension. Since $\vf$ has the smallest scaling dimension, the most relevant
term is obtained by keeping $\vf$ in the argument, which gives
\begin{equation}
  \label{eq:basicn} [1 + (n - 1)] V (\vf) = nV (\vf) \hspace{0.17em} .
\end{equation}
This is an example of an ``$n$-suppressed'' term, i.e.~term vanishing in the $n
\to 0$ limit. The naive expectation is that such terms should not matter.
Below we will discuss this in detail, analyze various subtleties, and confirm
the naive expectation. For the moment let us focus on the terms which survive
as $n \to 0$. The most relevant such terms appear when we expand either to
first order in $\omega$ or to second order in $\chi_i$ (the first order in
$\chi_i$ vanishes thanks to $\sum\nolimits' \chi_i = 0$):
\begin{eqnarray}
  &  & V' (\vf) \frac{\omega}{2} + (n - 1) V' (\vf)  \left( -
  \frac{\omega}{2} \right) = \omega V' (\vf) + \text{$n$-suppressed}
  \hspace{0.17em}, \\
  &  & \frac{1}{2} V'' (\vf)  \sum\nolimits' \chi_i^2 \hspace{0.17em} . \nonumber
\end{eqnarray}
These have the same scaling dimension $\Delta(V (\vf)) + 2$. We define the leading
Lagrangian $\calL_0$ including the quadratic terms and these most relevant
terms, in the $n \to 0$ limit:
\begin{equation}
  \label{L0} \calL_0 = \del \varphi \del \omega - \frac{\tmH}{2} \omega^2 +
  \omega V' (\vf) + \frac{1}{2}  \sum\nolimits' \left\{ (\del \chi_i)^2 + \chi_i^2 V''
  (\vf) \right\} \hspace{0.17em} .
\end{equation}
Explicitly, for the quartic potential (including the mass term) this is
\begin{equation}
  \label{L0phi4} \Lcal_0 = \del \varphi \del \omega - \frac{\tmH}{2} \omega^2
  + \frac{1}{2}  \sum\nolimits' (\del \chi_i)^2 + \frac{m^2}{2}  \left( 2 \vf \omega +
  \sum\nolimits' \chi_i^2 \right) + \frac{\lambda}{4!} \left( 4 \omega \vf^3 + 6 \sum\nolimits'
  \chi_i^2  \vf^2 \right) \hspace{0.17em} .
\end{equation}
We can now easily rederive the upper critical dimension $d_{uc}
= 6$ in this language: the quartic interactions have dimension $2 d - 6$ and become marginal at $d=d_{uc}$.\footnote{For the branched polymers (the cubic potential) analogous
considerations would give $d_{uc} = 8$.}

Expanding {\eqref{eq:cardyint}} to higher order, we get terms of higher
scaling dimensions. We include all such terms which survive in the $n \to 0$
limit into the subleading Lagrangian $\calL_1$. It is easy to see that the
lowest nontrivial terms in $\calL_1$ involve expanding to cubic order:
\begin{equation}
  \calL_1 \supset V''' (\varphi) \times \left\{ \sum\nolimits' \chi_i^3, \omega \sum\nolimits'
  \chi_i^2, \omega^3 \right\} \label{LIrr} .
\end{equation}
Comparing to the $V'' (\vf)  \sum \chi_i^2$ term present in $\calL_0$, we see
that these $\calL_1$ terms have dimension 1,2 and 4 units higher, so they are
irrelevant, at least in $d = d_{uc} - \eps$. The terms in $\calL_1$ proportional
to $V'''' (\varphi)$ (expanding to quartic order) would be even more
irrelevant.

Finally, we gather in $\calL_2$ all $n$-suppressed terms. They come from both
the quadratic part and the interactions, and some of them were already mentioned. E.g.
\begin{equation}
  \Lsup \supset n \left\{ (\del \varphi)^2, \varphi \omega, (\del \omega)^2, V
  (\varphi), \ldots \right\} \hspace{0.17em} . \label{Lsupp}
\end{equation}
We stress that the Cardy transform being just a linear
transformation of fields, it cannot introduce any mistake compared to the original replicated Lagrangian, unless one drops some terms. Any observable or correlation function which was computable
from the replicated Lagrangian can be equivalently computed in the Cardy basis $(\varphi,
\omega, \chi_i)$. E.g., applying the Cardy transform to $\phi_1, \phi_i,
\phi_j, \ldots$ in a general disordered correlator like {\eqref{form2}}, we
can express it as a linear combination of correlators of Cardy fields.

\subsection{Basic RG scenario}\label{basicRG}

Let us summarize the results so far. Starting from the random field action
{\eqref{LGrfim}} we used the method of replicas and the Cardy transform
{\eqref{CT}} to obtain a Lagrangian
\begin{equation}
  \label{lagmain} \Lcal_{\tmop{Cardy}} = \Lcal_0 + \Lirr + \Lsup
  \hspace{0.17em},
\end{equation}
where
\begin{itemize}
  \item $\Lcal_0$ contains the terms which are relevant and do not vanish as
  $n \to 0$,
  
  \item $\Lirr$ contains the terms which are irrelevant and do not vanish as
  $n \to 0$,
  
  \item $\Lsup$ contains all $n$-suppressed terms.
\end{itemize}
Classification relevant/irrelevant is for small $\e = d_{uc} - d$ and it is not a
priori clear what will be the fate of $\calL_1$ terms for larger $\e$. Let's
assume that (a) $\Lirr$ terms remain irrelevant and can be discarded, and (b)
that $\Lsup$ can be simply dropped in the $n \to 0$ limit. We will refer to
these two assumptions as ``basic RG scenario''. So we simply drop $\calL_1$ and
$\calL_2$ and assume that the IR physics of the disordered model is captured
by $\Lcal_0$ alone. Following this scenario we will draw some interesting
conclusions in sections \ref{PSSUSY} and \ref{sec:dimred}, concerning
supersymmetry and dimensional reduction. Starting from section \ref{RGbasic},
we will start carefully checking whether the two assumptions hold.

\subsection{Parisi-Sourlas SUSY}\label{PSSUSY}

Within the basic RG scenario, we need to understand the $n \rightarrow 0$
limit of the Lagrangian $\Lcal_0$. Here the dependence on $n$ appears only
through the $(n - 1)$ fields $\chi_i$ which sum to zero, so we have
effectively $(n - 2)$ linearly independent fields. Since $\Lcal_0$ is Gaussian
in $\chi_i$, integrating them out would give a result proportional to $\left(
\det [- \del^2 + V'' (\vf)] \right)^{- \frac{n - 2}{2}}$. When $n \to 0$ this
reduces to $\det [- \del^2 + V'' (\vf)]$, which is the usual result for a
fermionic Gaussian path integral (up to overall factors which cancel in the
computation of correlation functions). This motivates the substitution
\begin{equation}
  \label{fromchitopsi} \frac{1}{2}  \sum_{i = 2}^n \chi_i  [- \del^2 + V''
  (\vf)] \chi_i \overset{n \to 0}{\longrightarrow} \psi [- \del^2 + V'' (\vf)]
  \bar{\psi},
\end{equation}
where $\psi$ and $\bar{\psi}$ are two anticommuting real scalar fields. By
taking the limit $n \to 0$ of $\Lcal_0$ we thus obtain a Lagrangian of two
commuting real scalar fields $\vf$, $\omega$ and two anticommuting ones $\psi,
\psib$,\footnote{\label{onlyL0}Note that this formulation only works for
$\Lcal_0$. In fact the Lagrangians $\Lirr$ and $\Lsup$ may contain operators
proportional to $\sum\nolimits' \chi^k_i$ for $k > 2$, which are `non-susy-writable'
(cannot be written in terms of $\psi, \bar{\psi}$). This will be discussed in
detail below. In the following sections, to study the RG flow of $\Lirr$ and
$\Lsup$, we will therefore use the formulation in terms of the fields
$\chi_i$.}
\begin{equation}
  \Lcal_{\tmop{SUSY}} = \partial \varphi \del \omega - \frac{H}{2} \omega^2 +
  \omega V' (\varphi) + \partial \psi \partial \psib + \psi \psib V''
  (\varphi) \label{LSUSY1} .
\end{equation}
This is the Parisi-Sourlas Lagrangian, which is invariant under a
supersymmetry. This can be made manifest using an orthosymplectic superspace
with $x$ as a bosonic and $\theta$, $\thetab$ as two real Grassmann
coordinates. We can then combine all fields into one superfield
\begin{equation}
  \Phi (x, \theta, \thetab) = \varphi (x) + \theta \psib (x) + \thetab \psi
  (x) + \theta \thetab \omega (x) \hspace{0.17em} . \label{supefield}
\end{equation}
The action in superspace takes the form
\begin{equation}
  \label{SUSYPhi} \calS_{\tmop{superspace}} = \int d^d xd \thetab d \theta
  \left[ - \frac{1}{2} \Phi D^2 \Phi + V (\Phi) \right] \hspace{0.17em},
\end{equation}
where $D^2 \assign \del^2 - H \partial_{\theta} \partial_{\thetab}$ is the
super-Laplacian. It is straightforward to check that integrating over $\theta,
\thetab$ reduces {\eqref{SUSYPhi}} to $\int d^d x \hspace{0.17em}
\Lcal_{\tmop{SUSY}}$. Parisi-Sourlas supersymmetry transformations consist of
(super)translations $\mathbb{R}^{d | 2 \nobracket}$ and of $\tmop{OSp} (d| 2)$
(super)rotations which leave the superspace metric $d x^2 - \frac{4}{H} d
\theta d \thetab$ invariant.

The conclusion is that, if the basic RG scenario holds, the critical point of
a random field theory is in the same universality class as the IR fixed point
of the supersymmetric Parisi-Sourlas action {\eqref{SUSYPhi}}. This is Part 1
(Emergence of SUSY) of the Parisi-Sourlas conjecture.

This way to see the emergence of supersymmetry is different from the original
one {\cite{Parisi:1979ka}} based on classical solutions of
a stochastic partial differential equation. The original argument had some caveats (the solution may not
be unique, the fermionic determinant was missing the absolute sign, etc.). The
Cardy transform argument also has assumptions (can we drop $\mathcal{L}_{1}$ and $\mathcal{L}_{2}$?), but as we will see the validity of these assumptions may be easier to
check.

Supersymmetry leads to various nice consequences for correlation functions.
E.g.~$\langle \varphi (x) \varphi (0) \rangle$, \ $\la \varphi (x) \omega (0)
\ra$, $\la \omega (x) \omega (0) \ra$ correlators can be extracted from the
single correlator of superfields:
\begin{equation}
  \label{eq:pack} \langle \Phi (x, \theta_1, \thetab_1) \Phi (0, \theta_2,
  \thetab_2) \rangle = \langle \varphi (x) \varphi (0) \rangle + \theta_1
  \thetab_1 \la \omega (x) \varphi (0) \ra + \theta_1 \thetab_1 \theta_2
  \thetab_2 \la \omega (x) \omega (0) \ra + \ldots
\end{equation}
The l.h.s. being a function of $x^2 - \frac{4}{H} (\theta_1 - \theta_2)
(\nobracket \bar{\theta}_1 - \bar{\theta}_2) \nobracket,$this gives relations
\begin{equation}
  \langle \varphi (x) \omega (0) \ra = \langle \bar{\psi} (x) \psi (0) \ra = -
  \frac{4}{H} \frac{d}{d x^2} \langle \varphi (x) \varphi (0) \rangle, \qquad
  \langle \omega (x) \omega (0) \ra = 0. \label{Ward}
\end{equation}
While the IR scaling dimensions get corrections compared to the UV dimensions
{\eqref{dimensions}}, these supersymmetric relations imply that $\Delta_\omega =
\Delta_\varphi + 2,$ $\Delta_\psi = \Delta_{\bar{\psi}} = \Delta_\varphi+ 1$ remain true in the IR.

We can also trace what this implies for physical observables, which are
correlators of $\phi$'s. It is customary to consider connected and
disconnected 2-point functions:
\begin{equation}
\label{Gtmo}
  G_{\tmop{conn}} = \overline{\langle \phi (x) \phi (0) \rangle - \langle \phi
  (x) \rangle \langle \phi (0) \rangle}, \qquad G_{\tmop{disc}} =
  \overline{\la \phi (x) \ra \la \phi (0) \ra} .
\end{equation}
In the replica formalism these can be expressed as (see {\eqref{form2}})
\begin{equation}
  G_{\tmop{conn}} = \langle \phi_1 (x) \phi_1 (0) \rangle - \langle \phi_1 (x)
  \phi_i (0) \rangle, \qquad G_{\tmop{disc}} = \langle \phi_1 (x) \phi_i (0)
  \rangle .
\end{equation}
where $i \neq 1$ is arbitrary. Averaging over $i=2, \ldots, n$, Cardy-transforming, 
and using $\langle \omega \omega \rangle = 0$ as a
consequence of SUSY, we get
\begin{equation}
\label{conndisc}
  G_{\tmop{conn}} = \langle \varphi (x) \omega (0) \rangle, \qquad
  G_{\tmop{disc}} = \langle \varphi (x) \varphi  (0) \rangle .
\end{equation}
By {\eqref{Ward}}, this gives a relation between $G_{\tmop{conn}}$ and
$G_{\tmop{disc}}$.

We should warn the reader about various subtleties concerning the relations of
the Lagrangians $\mathcal{L}_0$ and $\mathcal{L}_{\tmop{SUSY}}$. First, while
the two are formally equivalent at the classical level, differences may appear
at the level of loop effects because the most natural $S_n$-invariant UV
regulator of $\mathcal{L}_0$ is not SUSY-invariant. We will resolve this
subtlety in section \ref{RGL0}.

Second, Lagrangians $\mathcal{L}_0$ and $\mathcal{L}_{\tmop{SUSY}}$ have
overlapping but not identical sets of correlation functions. Any
$\mathcal{L}_0$ correlator of operators made from $\varphi, \omega$ and
$\text{O} (n - 2)$-invariant objects quadratic in $\chi_i$ can be mapped to an
$\mathcal{L}_{\tmop{SUSY}}$ correlator via $\frac{1}{2} \sum\nolimits' \chi^2_i
\rightarrow \psi \bar{\psi}$, $\frac{1}{2} \sum\nolimits' (\partial \chi_i)^2
\rightarrow \partial \psi \partial \bar{\psi}$ etc. E.g.~we have the following
relation
\begin{equation}
  \left\langle \frac{1}{2} \sum\nolimits' \chi^2_i (x) \frac{1}{2} \sum\nolimits' \chi^2_i (0)
  \right\rangle = \langle \psi \bar{\psi} (x) \psi \bar{\psi} (0) \rangle,
\end{equation}
as is easy to check in the free theory ($\lambda = 0$). We extend this to other
bilinears and their products in Appendix \ref{chi-psi}. Some uncontracted
correlators can also be mapped, allowing for tensorial coefficients: e.g.
\begin{equation}
\langle \chi_i (x) \chi_j (0) \rangle = - (\delta_{ij} + \Pi_{ij}) \langle
\psi (x) \bar{\psi} (0) \rangle. \label{chi-chi}
\end{equation}
 However, this does not extend to general
correlators. E.g.~as we discuss in Appendix \ref{chi-psi}, it does not seem
possible to represent a general 4-point function $\langle \chi_i \chi_j \chi_k
\chi_l \rangle$ as a linear combination of $\langle \psi \psi \bar{\psi}
\bar{\psi} \rangle$ correlators (where $\psi$'s and are $\bar{\psi}$'s may be
inserted in arbitrary order at four points). So, while the Cardy basis still
contains an infinitude of different fields $\chi_i$, necessary to faithfully
represent general replicated observables {\eqref{form2}}, some of this
richness is gone in the SUSY theory which only has two fields $\psi,
\bar{\psi}$.\footnote{Going in the opposite direction, general SUSY
correlators of $\psi$, $\bar{\psi}$ at different points, like $\langle \psi
(x_1) \ldots \psi (x_n) \bar{\psi} (y_1) \ldots \bar{\psi} (y_n) \rangle$, do
not seem to have any particular meaning in the $\mathcal{L}_0$ theory.}

We will call ``susy-writable'' those $\mathcal{L}_0$ theory operators whose
correlators can be computed by SUSY theory $\mathcal{L}_{\tmop{SUSY}}$. Not
all $S_{n - 1}$-invariant operators belong to this class, the simplest
examples being $\sum\nolimits' \chi^k_i$ for $k > 2$, see footnote \ref{onlyL0}. These
operators are nontrivial, e.g. their 2-point functions are nonzero. Yet there
does not seem to be a way to compute them using the SUSY fields.

\subsection{Dimensional reduction}\label{sec:dimred}

Part 2 (Dimensional reduction) of the Parisi-Sourlas conjecture
{\cite{Parisi:1979ka}} states that the supersymmetric theory {\eqref{LSUSY1}},
{\eqref{SUSYPhi}} is related to a theory in two less dimensions with no
disorder nor supersymmetry. More concretely it says that correlation functions
of the SUSY theory can be mapped to correlation functions of a $(d -
2)$-dimensional model with the same interaction $V (\phi)$, by restricting the
coordinates to a codimension two hyperplane, and setting to zero the Grassmann
variables. In {\cite{paper1}} we tested the dimensional reduction for the
strongly coupled fixed point of the RG flow of the supersymmetric theory. We
argued that the map works at the level of axiomatic CFTs, due to particular
superconformal symmetry of the theory.

Let us illustrate how this works by considering the $2$-point functions of
$\Phi$ computed at the IR fixed point of the action {\eqref{SUSYPhi}} with a
given potential $V (\Phi)$ (e.g.~a quartic or a cubic). First by setting
$\theta = \thetab = 0$ in {\eqref{eq:pack}} we have a general SUSY relation:
\begin{equation}
  \langle \Phi (x_1, 0, 0) \Phi (x_2, 0, 0) \rangle = \langle \varphi (x_1)
  \varphi (x_2) \rangle .
\end{equation}
Next we pick a $d - 2$ hyperplane $\mathbb{R}^{d - 2} \subset \mathbb{R}^d$,
for definiteness spanned by the first $d - 2$ components. Dimensional
reduction means that by demanding $x$'s to lie in this hyperplane we get a
further equality:
\begin{equation}
  \label{dimredPhiPhi} \langle \varphi (x_1) \varphi (x_2) \rangle = \langle
  \widehat{\phi \,} (x_1) \widehat{\phi \,} (x_2) \rangle_{\text{CFT}_{d -
  2}}, \qquad (x_i \in \bR^{d - 2}) .
\end{equation}
The correlation function in the r.h.s. of {\eqref{dimredPhiPhi}} is computed
in a $(d - 2)$-dimensional conformal field theory, the RG fixed point of the
non-supersymmetric Landau-Ginzburg action
\begin{equation}
  \label{Sdm2} \calS = \frac{4 \pi}{H} \int d^{d - 2} x \left[ \frac{1}{2}
  (\partial \widehat{\phi \,})^2 + V (\widehat{\phi \,}) \right]
  \hspace{0.17em} .
\end{equation}
The potential is the same as the initial random field action, but this theory
lives in $2$ dimensions less and has no disorder fields. For simplicity we
stated {\eqref{dimredPhiPhi}} for 2-point functions, but it generalizes for
higher point functions and for composite operator insertions {\cite{paper1}}.

With prior studies
{\cite{CARDY1983470,KLEIN1983473,Klein:1984ff,Zaboronsky:1996qn}}\footnote{See
also recent rigorous work {\cite{Gubinelli1,Gubinelli2}}.} and our own tests
in {\cite{paper1}}, Part 2 of the Parisi-Sourlas conjecture appears to be on
rather solid ground, especially compared to Part 1. In this paper, we will
assume that Part 2 is true and we will use it as one of ingredients to
understand what may go wrong with Part 1. E.g.~we will need to understand the
spectrum of $S_n$-invariant perturbations of $\mathcal{L}_0$ theory, to see if
any of these become relevant. Those of these perturbations which are
susy-writable are captured by the SUSY theory. On the other hand, by
dimensional reduction, the spectrum of the SUSY fixed point can be understood
from the spectrum of the Wilson-Fisher fixed point, which is rather well known
(see sections \ref{sec:susywritableleaders} and \ref{susy-irr}). Of course,
dimensional reduction does not say anything about perturbations which are not
susy-writable, and those will have to be studied independently.

\section{RG flow in the basic scenario}\label{RGbasic}

In this section we will discuss in more detail the RG flow assuming the basic
RG scenario (i.e. dropping $\mathcal{L}_1$ and $\mathcal{L}_2$). We start in
section \ref{RGsusy} with some comments about the RG flow in the ``SUSY
theory'', i.e.~theory {\eqref{LSUSY1}}, {\eqref{SUSYPhi}} with field content
$\varphi, \psi, \bar{\psi}, \omega$ described by the Lagrangian
$\mathcal{L}_{\tmop{SUSY}}$ or, equivalently, the superspace action
$\mathcal{S}_{\tmop{superspace}}$. Then in section \ref{RGL0} we discuss the
RG flow in the theory $\mathcal{L}_0$ with field content $\varphi, \omega,
\chi_i$ and Lagrangian {\eqref{L0}}. We will see that the $\mathcal{L}_0$
theory is not quite equivalent to $\mathcal{L}_{\tmop{SUSY}}$ (even in the
fermion bilinear sector) because the $S_n$-invariant Wilsonian UV cutoff
partially breaks supersymmetry. Upon careful analysis we will see that these
SUSY breaking effects disappear at long distances.

\subsection{RG flow in $\mathcal{L}_{\tmop{SUSY}}$}\label{RGsusy}

In this section we will discuss the RG flow of the SUSY theory
{\eqref{LSUSY1}}, {\eqref{SUSYPhi}}. As already mentioned, this theory is
invariant under super-Poincar{\'e}, which is the semidirect product of
super-translations $\mathbb{R}^{d| 2}$ and super-rotations $\tmop{OSp} (d|
2)$:
\begin{equation}
  \text{super-Poincar{\'e}} =\mathbb{R}^{d| 2} \rtimes \tmop{OSp} (d| 2)
  \hspace{0.17em} . \label{eq:superp}
\end{equation}
All these transformations leave the superspace distance $x^2 - \frac{4}{H}
\theta \thetab$ invariant. Under super-translations $\delta \theta = -\vareps$,
$\delta \bar{\theta} = \overline{\vareps}$ the fields transform as
\begin{equation}
  \tmd  \vf = \overline{\vareps} \psi - \vareps  \psib, \hspace{1cm} \tmd \psi
  = \vareps \omega, \hspace{0.5cm} \tmd  \psib = \overline{\vareps} \omega,
  \hspace{1cm} \tmd \omega = 0 \hspace{0.17em} . \label{suptrans}
\end{equation}
Superrotations act in superspace as
\begin{equation}
  \delta x_{\mu} = \varepsilon_{\mu \theta} \theta + \varepsilon_{\mu \thetab}
  \thetab \hspace{0.17em}, \qquad \delta \theta = \frac{H}{2}
  \varepsilon_{\mu \thetab} x^{\m} \hspace{0.17em}, \qquad \delta \bar{\theta}
  = - \frac{H}{2} \varepsilon_{\mu \theta} x^{\m} \hspace{0.17em},
  \label{suprot}
\end{equation}
and the corresponding field transformations leaving the action invariant are
\begin{eqnarray}
  & \delta \varphi & = - \frac{H}{2} x^{\mu} \varepsilon_{\mu \theta} \psi -
  \frac{H}{2} x^{\mu} \varepsilon_{\mu \thetab}  \psib, \hspace{1cm} \tmd
  \omega = \varepsilon_{\mu \theta} \partial^{\mu} \psi + \varepsilon_{\mu
  \thetab} \partial^{\mu}  \psib \hspace{0.17em}, \nonumber \nonumber\\
  & \tmd  \psib & = - \frac{H}{2} x^{\mu} \varepsilon_{\mu \theta} \omega -
  \partial^{\mu}  \vf \varepsilon_{\mu \theta}, \hspace{1.4cm} \tmd \psi =
  \frac{H}{2} x^{\mu} \varepsilon_{\mu \thetab} \omega + \partial^{\mu}  \vf
  \varepsilon_{\mu \thetab} \hspace{0.17em} .  \label{suprot1}
\end{eqnarray}
There are also bosonic $\tmop{Sp} (2)$ transformations which rotate $\psi$,
$\bar{\psi}$ and leave $\varphi, \omega$ invariant; we do not write them
explicitly.

For the quartic potential and working in $d = 6 - \e$, we write the SUSY
Lagrangian as
\begin{equation}
  \Lcal_{\tmop{SUSY}} = \partial \varphi \del \omega - \frac{H}{2} \omega^2 +
  \partial \psi \partial \psib + m^2 \left( \omega \varphi + \psi \psib
  \right) + \frac{\lambda}{4!} \mu^{\eps}  \left( 4 \omega \varphi^3 + 12 \psi
  \psib \varphi^2 \right) \label{LSUSY2} .
\end{equation}
(where we introduced the RG scale $\mu$ and made the coupling $\lambda$
dimensionless). Standard techniques allow us to compute the RG flow
perturbatively. E.g.~the one-loop beta function for the dimensionless quartic
coupling $\lambda$ can be obtained in dimensional regularization as
\begin{equation}
  \beta_{\lambda} = - \varepsilon \lambda+ \frac{3 H \lambda^2}{64 \pi^3} + O (\lambda^3)
  \hspace{0.17em} . \label{betaSUSY}
\end{equation}
From this we obtain a fixed point at
\begin{equation}
  \lambda_{\ast} = \frac{64 \pi^3 \varepsilon}{3 H} + O (\e^2) \hspace{0.17em} .
  \label{eq:l*}
\end{equation}
(The fixed point lies at $m^2$=0 in dimensional regularization) One can check
that the renormalization of the fields $\vf, \omega, \psi, \psib$ turns out to
be equal, in agreement with supersymmetry. We elaborate on these computations
in Appendices \ref{ope} and \ref{2loop}. Another feature of the RG flow is
that the parameter $H$ does not get renormalized, since it enters in SUSY
transformations which cannot get deformed provided that the regulator
preserves SUSY, as turns out to be true for dimensional regularization (see
App. \ref{2loop}). Other regulators will be discussed below.

Finally, the SUSY RG flow is equivalent to the Wilson-Fisher flow in $\hat{d}
= 4 - \varepsilon$ dimensions with the Lagrangian:
\begin{equation}
  \Lcal_{\tmop{WF}} = \frac{1}{2} \left( \partial \widehat{\phi \,} \right)^2
  + \frac{m^2}{2} \widehat{\phi \,}^2 + \frac{\widehat{\lambda}}{4!} \widehat{\phi
  \,}^4
\end{equation}
upon identification of couplings
\begin{equation}
  \lambda = \frac{4 \pi}{H} \widehat{\lambda} . \label{lltilde}
\end{equation}
One can easily check that {\eqref{betaSUSY}} and {\eqref{eq:l*}} map under
these identification to the familiar Wilson-Fisher expressions, in particular
$\hat{\lambda}_{\ast} = (16 \pi^2 / 3) \varepsilon + O (\varepsilon^2)$. This,
of course, is a perturbative manifestation of dimensional reduction, and
{\eqref{lltilde}} follows from {\eqref{Sdm2}}.

As mentioned in section \ref{sec:dimred}, in this paper we assume dimensional
reduction (Part 2 of the Parisi-Sourlas conjecture) settled, so we assume full
equivalence between SUSY RG flow in $d$ dimensions and Wilson-Fisher RG flow
in $\widehat{d} = d - 2$ dimensions, both perturbatively and nonperturbatively.
For $\hat{d} \geqslant 2$, the Wilson-Fisher RG flow goes to a fixed point for
a particular value of the bare mass, and the corresponding $d = \hat{d} + 2$
SUSY RG flow will go to a SUSY fixed point for the same bare
mass.\footnote{The Wilson-Fisher fixed point is believed to exist also for $1
< \hat{d} < 2$, but we will not discuss this intermediate case in detail; see
{\cite{Golden:2014oqa}}.} On the other hand, for $\hat{d} = 1$, we get the 1d
Wilson-Fisher flow, which is just quantum mechanics with a quartic potential.
For whatever value of the mass, the quantum mechanical spectrum is discrete,
and the IR phase is massive. By the assumed exact correspondence, the 3d SUSY
RG flow thus should also flow to a massive phase, with exactly preserved
supersymmetry. We conclude that a nontrivial 3d SUSY RG fixed point does not
exist. Note that the absence of a SUSY IR fixed point does not imply
spontaneous breakdown of SUSY.\footnote{This point does not seem to be
universally appreciated in the literature. E.g.~Ref. {\cite{Tarjus1}} says
``even if the RG flow is started with initial conditions obeying
supersymmetry, a mechanism should be provided to describe a spontaneous
breakdown of supersymmetry.'' }

These simple observations show what exactly needs to be explained concerning
Part 1 of the Parisi-Sourlas conjecture, depending on $d$. Down to 4d, the
SUSY fixed point exists, so we need to understand if it is stable or not with
respect to the perturbations present in the $\mathcal{L}_0 +\mathcal{L}_1
+\mathcal{L}_2$. If the SUSY fixed point is unstable, then the flow will be
driven away from it, and the RFIM phase transition will be described by
another fixed point (about which we will have nothing to say in this paper).
The situation is different in 3d: the SUSY fixed point does not exist there,
so the RFIM phase transition must be \tmtextit{for sure} described by some
other fixed point. The only problem in 3d is to find this other fixed point,
not to explain the absence of SUSY.

Let us come back to the non-renormalization of $H$. It may look like we have a
one-parameter family of RG fixed points parametrized by the choice of $H$.
However all of these fixed points are trivially related to each other by
rescaling the fields, so in practice there is only one fixed point up to
equivalence. Rescaling
\begin{equation}
  \varphi \to r^{- 1} \varphi \hspace{0.17em}, \qquad \omega \to r \omega
  \hspace{0.17em} \qquad \psi, \bar{\psi} = \text{inv}, \label{rescFields}
  \hspace{0.17em}
\end{equation}
has the effect of rescaling $H \to r^2 H$, $\lambda \to r^{- 2} \lambda$.
Since $\lambda H$ is left invariant, the fixed point {\eqref{eq:l*}} is mapped
to an equivalent one characterized by another value of $H$.

We will see in section \ref{sec:susywritableleaders} that $\omega^2$ can be
seen as a member of superstress tensor multiplet,\footnote{More precisely
$\omega^2$ is a linear combination of a superstress tensor component and a
total derivative, see Eqs. {\eqref{stress}} and {\eqref{totdero2}}.} which
explains why it is exactly marginal, and why adding it to the action can be
undone by changing the superspace metric, which is what rescaling
{\eqref{rescFields}} secretly is. Usually, when a CFT is deformed by an
exactly marginal deformation, we get a different CFT with different scaling
dimensions and different OPE coefficients. This is clearly not the case when
deforming by $\omega^2$, since this leaves scaling dimensions invariant and
OPE coefficients change trivially due to rescaling, so we get a CFT equivalent
to the one we started with. In the renormalization group theory parlance, such
deformations which can be undone by a field redefinition are classified as
``redundant'' {\cite{Wegner}}. Usually redundant operators are those which are
proportional to the equations of motion, and they have zero correlation
functions at non-coincident points. Such operators and their scaling
dimensions do not even appear in CFT description. Operator $\omega^2$, although
``redundant'' in the sense described above, does have nonzero correlators at
non-coincident points, and is a bona fide CFT operator.

Finally let us discuss the SUSY RG flow in a Wilsonian scheme with a momentum
cutoff, as opposed to dimensional regularization. We have to regulate the
theory in a SUSY-preserving way, which requires some care in choosing momentum
cutoffs. Before cutoffs, the propagators are\footnote{Dimensional
regularization uses exactly these propagators and it is a SUSY-preserving
scheme.}
\begin{equation}
  \langle \vf_k \vf_{- k} \rangle = \frac{H}{k^4}, \quad \langle \vf_k
  \omega_{- k} \rangle = \frac{1}{k^2}, \quad \langle \omega \omega \rangle =
  0, \quad \langle \psi_k \bar{\psi}_{- k} \rangle = - \frac{1}{k^2} .
  \label{befCut}
\end{equation}
Momentum cutoff has to be imposed on the super-propagator. In position space,
the super-propagator must be a function of $x^2 - \frac{4}{H} \theta
\bar{\theta}$, while in supermomentum-space it is a function of $k^2 - H
\alpha \overline{\tma}$ where $\overline{\tma}, \tma$ are Grassmann
coordinates Fourier-conjugated to $\theta, \thetab$. This implies that
component propagators must be related by\footnote{It is also possible to
obtain these relations directly from position space {\eqref{Ward}} without
help of super-Fourier transform. For this, represent the radially symmetric
propagators as linear combinations of Gaussians $e^{- \alpha x^2}$ and do the
usual Fourier transform.}
\begin{equation}
  G_{\varphi \omega} (k) = G_{ \psib \psi} (k), \qquad G_{\varphi \varphi} (k)
  = - H \frac{d}{dk^2} G_{\varphi \omega} (k) .
\end{equation}
We see that Eqs. {\eqref{befCut}} satisfy these, and cutoffs must be
introduced in a way to preserve these relations. E.g.~we can choose
\begin{equation}
  G_{\varphi \omega} (k) = G_{\psib \psi} (k) = \frac{F_{\Lambda} (k^2)}{k^2},
  \qquad G_{\varphi \varphi} (k) = H \left( \frac{F_{\Lambda} (k^2)}{k^4} -
  \frac{F_{\Lambda}' (k^2)}{k^2} \right), \label{SUSYreg}
\end{equation}
where $F_{\Lambda} (k^2)$ is a function vanishing for $k^2 \geqslant
\Lambda^2$, $\Lambda$ the UV cutoff. Note the second term in $G_{\varphi
\varphi} (k)$, which is the price to pay for maintaining exact SUSY in a
Wilsonian scheme. E.g.~if $F_{\Lambda} (k^2) = \Theta (\Lambda^2 - k^2)$ we
see that we need to add a term proportional to $\delta (k^2 - \Lambda^2)$. If
the theory were regulated without this term, exact SUSY would be broken. E.g.
$H$ would be renormalized as a result. This will be discussed in the next
section.

\subsection{Emergence of SUSY from the $\mathcal{L}_0$ theory}\label{RGL0}

Let us now discuss RG flow in the $\mathcal{L}_0$ theory {\eqref{L0}} with the
quartic potential. As discussed in section \ref{PSSUSY}, this theory can be
mapped on $\mathcal{L}_{\tmop{SUSY}}$ via replacement $\frac{1}{2} \sum\nolimits'
\chi^2_i \rightarrow \psi \bar{\psi}$, $\frac{1}{2} \sum\nolimits' (\partial \chi_i)^2
\rightarrow \partial \psi \partial \bar{\psi}$. So at first glance this theory
has the same flow as the SUSY theory discussed in the previous section.
However there is a subtlety: the cutoff is not quite the same. The
$\mathcal{L}_0$ theory {\eqref{L0}} came from the replicated action
{\eqref{Sr}} possessing $S_n$ invariance. The replicated action had an
$S_n$-invariant regulator, and the $\mathcal{L}_0$ theory inherits this
regulator.

The kinetic part of the $\mathcal{L}_0$ theory had two pieces of different
origin: $\del \varphi \del \omega + \frac{1}{2}  \sum\nolimits' (\del \chi_i)^2$ which
came from the kinetic term of {\eqref{Sr}} and $-\frac{H}{2} \omega^2$ which
came from integrating out the magnetic field. In a regulated theory, these two
terms will have their own momentum cutoffs which do not have to coincide. We
can model this situation by writing the regulated kinetic term in momentum
space as
\begin{equation}
  F_{\Lambda} (k^2)^{- 1} k^2 (\varphi_k \omega_{- k} + \psi_k \bar{\psi}_{-
  k}) - \frac{H_{\Lambda} (k^2)}{2} \omega_k \omega_{- k}, \label{regkin}
\end{equation}
where $F_{\Lambda} (0) = 1$, $H_{\Lambda} (0) = H$, both $F_{\Lambda}$ and
$H_{\Lambda}$ go to zero at large momenta, and we already performed the map to
SUSY fields replacing $\frac{1}{2}  \sum\nolimits' \chi_{i, k} \chi_{i, - k}
\rightarrow \psi_k \bar{\psi}_{- k}$. We get the propagators:
\begin{equation}
  G_{\varphi \omega} (k) = G_{\psi \psib} (k) = \frac{F_{\Lambda} (k^2)}{k^2},
  \qquad G_{\varphi \varphi} (k) = H_{\Lambda} (k^2) \frac{[F_{\Lambda}
  (k^2)]^2}{k^4} . \label{SusyBr}
\end{equation}
Comparing these with {\eqref{SUSYreg}}, we see that SUSY is not in general
respected. In fact, while $G_{\varphi \omega} = G_{ \psib \psi}$ agree as they
should, the $G_{\varphi \varphi}$ propagators does not have the expected form.
Even if we choose $F_{\Lambda} (k^2) = H_{\Lambda} (k^2) = \Theta (\Lambda^2 -
k^2)$, $G_{\varphi \varphi}$ is missing the $\delta (k^2 - \Lambda^2)$ piece.

Thus, to understand the RG flow of $\mathcal{L}_0$, we have to understand the
RG flow of $\mathcal{L}_{\tmop{SUSY}}$ regulated in a non-SUSY invariant way.
One might worry that a regulator breaking SUSY can be very dangerous for its
fate, but fortunately all is not lost. The point is that the above regulator
breaks SUSY only partially, and what remains will be enough to have the full
SUSY emerge in the IR.

The complete preserved subgroup of super-Poincar{\'e} is the semidirect
product of the super-translations and of the bosonic $\tmop{OSp} (d| 2)$
subgroup $\tmop{SO} (d) \times \tmop{Sp} (2)$:
\begin{equation}
  \mathbb{R}^{d| 2} \rtimes [\tmop{SO} (d) \times \tmop{Sp} (2)]
  \hspace{0.17em}. \label{eq:subgroup}
\end{equation}
This will be referred to as ``partial SUSY''. That the usual bosonic
translations, rotations, and global fermionic $\tmop{Sp} (2)$ are preserved by
the propagators {\eqref{SusyBr}} is fairly obvious. A more subtle fact is that
the super-translations {\eqref{suptrans}} are also preserved. Indeed, for the
superfield two-point function, partial SUSY imposes the requirement
\begin{equation}
  \langle \Phi (x_1, \theta_1, \bar{\theta}_1) \Phi (x_2, \theta_2,
  \bar{\theta}_2) \rangle = A (x_{12}^2) + B (x_{12}^2)  (\theta_1 - \theta_2)
  (\bar{\theta}_1 - \bar{\theta}_2) \hspace{0.17em} .
\end{equation}
Expanding in components we get $G_{\varphi \varphi} (x) = A (x^2)$, $G_{\phi
\omega} (x) = G_{ \psib \psi} (x) = B (x^2)$, $G_{\omega \omega} = 0$. This is
precisely what Eq.~{\eqref{SusyBr}} says: that $G_{\phi \omega} = G_{ \psib
\psi}$ coincide while $G_{\varphi \varphi}$ may be unrelated. The functions
$A$ and $B$ are independent for the partial SUSY invariance, while the full
SUSY {\eqref{eq:superp}} requires the superfield two-point function to be a
function of $x_{12}^2 - \frac{4}{H}  (\theta_1 - \theta_2)  (\bar{\theta}_1 -
\bar{\theta}_2)$ and implies further relations {\eqref{Ward}}.

We can also write the regulated kinetic term {\eqref{regkin}} in superspace
as
\begin{equation}
  \int d^d x d \theta d \bar{\theta} \left[ - \frac{1}{2} \Phi F_{\Lambda}^{-
  1} (\partial^2) \partial^2 \Phi + \frac{1}{2} \Phi H_{\Lambda} (\partial^2)
  \del_{\theta} \del_{\bar{\theta}} \Phi \right], \label{kinpartSUSY}
\end{equation}
which makes it manifest that it preserves partial SUSY.

We are therefore led to consider RG flows which preserve only the partial
SUSY {\eqref{eq:subgroup}} (as well as the global Ising $\mathbb{Z}_2$
invariance which flips the sign of all fields). The most general term
invariant under {\eqref{eq:subgroup}} can be written as the superspace
integral of a local operator built from the superfield $\Phi$, allowing
contractions of the derivatives in $x$ and $\theta$ which preserve $\tmop{SO}
(d) \times \tmop{Sp} (2)$ and not necessarily the full $\tmop{OSp} (d| 2)$.
The two terms in {\eqref{kinpartSUSY}} are of such form. The structure of the
effective Lagrangian is thus less constrained than under the full SUSY.

However, and this is the key point which saves the day, at the relevant and
marginal level, we find only one new term which is invariant under partial
SUSY and not under full SUSY: this is the $\omega^2$ originating from $\Phi
\del_{\theta} \del_{\bar{\theta}} \Phi$. It is obviously invariant because
$\omega$ does not transform under supertranslations. All the other term
allowed by partial SUSY and breaking full SUSY are irrelevant.

Let us go through the list of candidates, starting from the SUSY mass term
$\varphi \omega + \psi \bar{\psi}$. It is fully super-rotation and
super-translation invariant, but in fact already partial SUSY
(supertranslations) fixes the relative coefficient, as is easy to check from
{\eqref{suptrans}}. Same for the quartic interaction $\omega \varphi^3 + 3
\psi \psib \varphi^2$. Terms $\varphi^2$ or $(\partial \varphi)^2$ are not
supertranslation invariant, very fortunately so since they would completely
ruin the structure of the quadratic Lagrangian if generated.

Due to this lucky circumstance, we expect that the following will happen. The
theory $\mathcal{L}_{\tmop{SUSY}}$ regulated in a partial-SUSY preserving way
will flow, for an appropriate bare mass value, to the fully SUSY fixed point
in the IR, and the only effect will be a renormalization of the coefficient of
$\omega^2$: $H_{\tmop{IR}} \neq H_{\tmop{UV}}$.\footnote{As discussed in
section \ref{RGsusy}, parameter $H$ is unphysical (redundant) when sitting at
a SUSY RG fixed point. But a change in this parameter is physical along an RG
flow which breaks SUSY.}

Let us see how this happens in detail in a toy model example. Let $\calS (H)$
denote the SUSY theory regulated in a fully SUSY-invariant way, $H$ being the
superspace parameter, while $\tilde{\calS}$ the same theory regulated in a way
which preserves only partial SUSY. We will model the cutoff by adding to the
action an irrelevant operator, a higher derivative term, which makes the
propagator decay faster in the UV (e.g.~$1 / k^2 \to 1 / (k^2 + k^4 /
\Lambda^2)$). So we take
\begin{equation}
  \tilde{\calS} = \calS (H) + \frac{g}{\Lambda^{\gamma}}  \int d^d x \hspace{0.17em}
  \widetilde{\calO}, \label{eq:Stilde}
\end{equation}
where $\Lambda$ is the UV cutoff, $\widetilde{\calO}$ an irrelevant operator
of dimension $d + \gamma$, $\gamma > 0$, which preserves partial SUSY but not
the full one, $g$ a dimensionless coupling. E.g.~we can choose a $\gamma = 2$
operator
\begin{equation}
  \widetilde{\calO} = \int d \theta d \bar{\theta}  \hspace{0.17em} \Phi
  (\del^2)^2 \Phi = \varphi (\del^2)^2 \omega + \psi (\del^2)^2  \psib
  \hspace{0.17em} .
\end{equation}
[Note that were we to choose
\begin{equation}
  \calO (H) = \int d \theta d \bar{\theta}\, \Phi (D^2)^2 \Phi = \vf (\del^2)^2
  \omega + \psi (\del^2)^2  \psib + H \omega \del^2 \omega, \label{eq:OSUSY}
\end{equation}
it would have been a fully SUSY-preserving regulator.]

\begin{figure}[h]
	\centering \includegraphics[width=300pt]{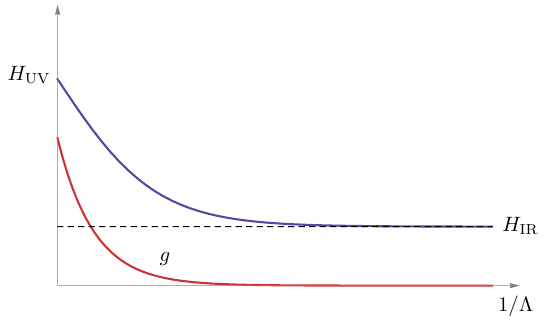}
  \
  \caption{Schematic RG flow of $H$ and $g$.\label{Hir} }
\end{figure}

Consider the structure of the RG flow within this toy model. After an RG step
$\Lambda \to \Lambda' = \Lambda / 2$ the irrelevant coupling decreases $g \to
g' = 2^{- \gamma} g$. The action $\calS (H)$ experiences the usual SUSY
renormalizations, on top of which we expect to generate a partial-SUSY
preserving (but full SUSY breaking) term $\omega^2$, with a coefficient
$\Delta H$ which should be interpreted as a change in $H$. This coefficient
vanishes in absence of interactions and in absence of $\calO$, thus $\Delta H
= O (\lambda g)$ where $\lambda$ is the quartic. Now we have the action $\calS
(H)$ which had a SUSY regulator adapted to $H$ but the new $H' = H + \Delta H$
has changed, so we have to change the SUSY regulator, e.g.~by moving a part of
$\widetilde{\calO}$ to $\calO (H)$ in {\eqref{eq:OSUSY}} which generates a
further change in $g$, $\Delta g = O (\Delta H)$. To summarize, after the RG
step, the full action at the scale $\Lambda'$ has the same form as
{\eqref{eq:Stilde}} with the couplings $H$ and $g$ replaced by
\begin{equation}
  H' = H + \Delta H, \quad \Delta H = O (\lambda g) \hspace{0.17em}, \qquad g'
  = 2^{- \gamma} g + O (\lambda g) \hspace{0.17em} .
\end{equation}
From here we draw the following conclusions. First, assuming that the quartic
$\lambda$ remains small, as it is the case at least for $\eps \ll 1$, the
irrelevant coupling $g$ approaches zero exponentially fast. Second, the series
made up of consecutive changes $\Delta H_i$ from infinitely many RG steps
needed to reach the IR fixed point converges. Therefore $H$ flows in the IR to
a finite value $H_{\mathrm{IR}}$. In particular we exclude the situation when
$H$ flows in IR to infinity.\footnote{The opposite situation when $H$ flows to
zero is also excluded as finetuned.} See Fig.~\ref{Hir}.

More abstractly, consider the RG flow of the SUSY theory perturbed by two
couplings breaking to partial SUSY, exactly marginal $\omega^2$ and irrelevant
$\widetilde{\calO} $:
\begin{equation}
  \calS (H) + \int g_0 \omega^2 + \frac{g}{\Lambda^{\gamma}}  \widetilde{\calO} .
\end{equation}
This time $ \widetilde{\calO}$ does not have to have the above quadratic form
and the discussion can be generalized easily to several $
\widetilde{\calO}$'s. On general grounds, the beta functions have the form
\begin{eqnarray}
  \beta_{g_0} & = & O (g), \nonumber\\
  \beta_g & = & (\gamma + O (\lambda, g_0) + O (g)) g . 
\end{eqnarray}
The small initial coupling $g$ will flow to zero in the IR if $\gamma + O
(\lambda, g_0)$ is positive. This quantity (up to $d +$) can be interpreted as
the scaling dimension of $\widetilde{\calO}$ at the SUSY fixed point corrected
by $g_0 \omega^2$ and since $\omega^2$ is exactly marginal, it should not
depend on $g_0$ at all: $\gamma + O (\lambda, g_0) \rightarrow \gamma + O
(\lambda_{})$. Operators $\widetilde{\calO}$ breaking full SUSY to partial
SUSY will reappear in section \ref{sec:susywritableleaders} as the
susy-writable leader operators, using the terminology to be introduced below.
We will see in sections \ref{susy-irr} and \ref{Who} that all such operators
remain irrelevant also in presence of $O (\lambda_{})$ corrections. Thus the
coupling $g$ flows to zero, and in the IR we recover the SUSY fixed point
perturbed by an exactly marginal deformation $\omega^2$, which as discussed in
the previous section amounts to a change in $H$.

\section{RG flow in the full Cardy theory: general plan}\label{Plan}

Let us recap. In section \ref{PSrev} we used the Cardy transform to rewrite the
replica action in terms of the variables $\vf, \chi_i, \omega$. This made
manifest the presence of a marginally relevant interaction close to the upper
critical dimension. We then dropped some terms in the action either because
they were irrelevant near $d = 6$ ($\Lirr$), or because they vanished in the
limit $n \to 0$ ($\Lsup$). This was  dubbed ``basic RG scenario'' in section
\ref{basicRG}. The remaining Lagrangian $\Lcal_0$ could be seen formally
equivalent to a supersymmetric one $\Lcal_{\tmop{SUSY}}$, replacing $\text{O}
(- 2)$-invariant bilinears made of fields $\chi_i$ by $\tmop{Sp}
(2)$-invariant bilinears made out of two Grassmann fields $\psi, \bar{\psi}$.
Then, in section \ref{RGsusy} we discussed RG flow in the SUSY theory,
concluding that it has a nontrivial RG fixed point down to $d = 4$ but not in
3d. This was based on dimensional reduction and the well-known Wilson-Fisher
fixed point properties. In section \ref{RGL0} we discussed the RG flow of
$\Lcal_0$ theory. Due to subtleties of the UV regulator the bare theory
preserves SUSY only partially (supertranslations but not superrotations), yet
at long distances full SUSY is recovered.

We will now come back to the full Cardy theory $\Lcal_0 + \Lirr + \Lsup$. We
will discuss the RG flow in this setup and examine the validity of the basic
RG scenario assumptions (a) and (b). We will work in the $\vf, \chi_i, \omega$
formulation rather then in the SUSY field basis $\vf, \psi, \psib, \omega$.
Indeed, the full Lagrangian contains some operators involving odd powers of
$\chi_i$ fields (e.g. $\sum\nolimits' \chi^3_i \vf$) which cannot be written in terms
of SUSY fermions. Such operators will play an important role in stability of
the RG flow.

Our plan is as follows. In section \ref{sec:Sn} we will describe a somewhat
peculiar form taken by the $S_n$ invariance in the Cardy basis. Here we will
introduce the notion of a ``leader''--the lowest-dimension part of an
$S_n$-singlet operator transformed to the Cardy basis--and of its
``followers'' which are the higher-dimension parts. In the short but important
section \ref{sec:L2} we will analyze the role of $n$-suppressed terms. Here we
will explain that the assumption (b) of the basic RG scenario holds, but in a
subtle sense: the theory with a small but finite $n$ is gapped, but as the $n
\rightarrow 0$ limit is taken, the approximately scale invariant region of the
RG flow becomes longer and longer. This is different from what happens in the
bond-(as opposed to field-) disordered Ising model, where the $n \neq 0$ fixed
point is believed to smoothly connect to $n = 0$, but it suffices for our
purposes: $\mathcal{L}_2$ can be dropped.

With $\mathcal{L}_2$ out of the way, in section \ref{L0L1} we will focus on
the $\Lcal_0 + \Lirr$ RG flow, working in the strict $n = 0$ limit. We will
examine if some $\Lirr$ perturbations (or other $S_n$-invariant perturbations
generated by RG) which are irrelevant in $d = 6 - \varepsilon$ dimensions, may
become relevant in lower $d$. The leader-follower distinction becomes very
handy here, because as we will see the relevance of an $S_n$-singlet
perturbation can be decided by computing the scaling dimension of the leader
alone.

Next, in section \ref{leaders}, we will classify the leader operators,
dividing them into three classes: non-susy-writable, susy-writable and
susy-null. These three classes RG-mix among each other only in a triangular
way, which will simplify the anomalous dimension computations (section
\ref{anomdim}). Exhaustive classification of leaders will be carried out up to
dimension 12 in $d = 6$, which includes one or more lowest-lying leaders in
each of the three classes. Anomalous dimensions will be computed at one or two
loops.

\begin{figure}[h]
	\centering \includegraphics[width=400pt]{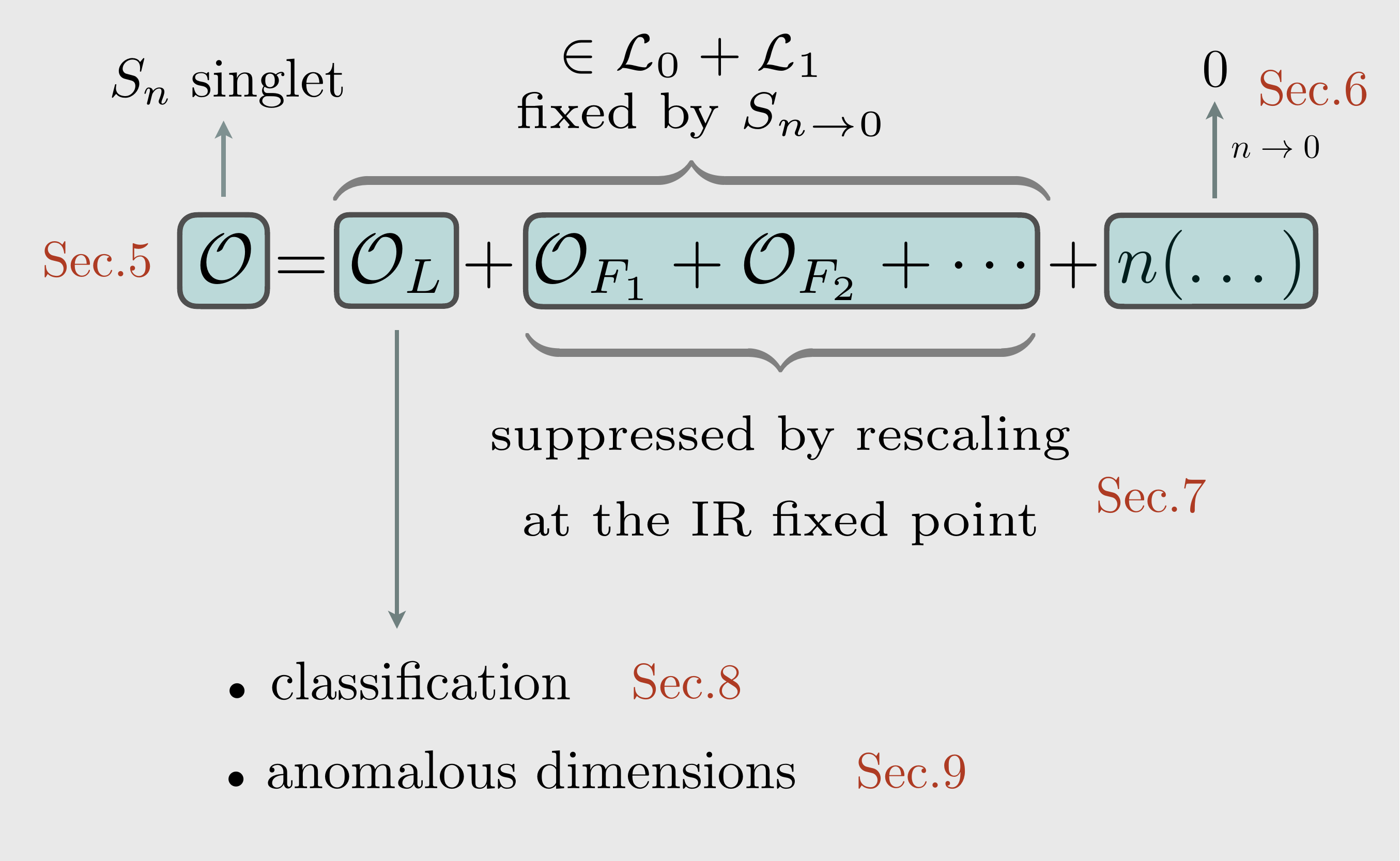}
  \
  \caption{\label{scheme}This figure illustrates how a generic $S_n$-singlet
  perturbation $\mathcal{O}$ is divided into various pieces (leader,
  followers, $n$-suppressed) and shows the sections of our paper where these
  pieces are discussed.}
\end{figure}

Fig.~\ref{scheme} is a roadmap for all these steps. Once completed, we will
see in section \ref{Who} what this implies for the loss of Parisi-Sourlas
SUSY.

\section{$S_n$ invariance in the Cardy basis}\label{sec:Sn}

The original replicated action {\eqref{Sr}} is invariant under permutations of
the $n$ replicas. While the Cardy transform clarifies many properties of our RG
flow, it somewhat obscures this $S_n$ symmetry, meriting a
discussion.\footnote{In this paper we will be content with using the standard
physics literature operational definition of what is meant by $S_n$ invariance
for $n \nin \mathbb{N}$: all algebraic manipulations are done with arbitrary
$n \in \mathbb{N}$ and the arising rational functions of $n$ are extrapolated
to $n$ non-integer or $n = 0$. Recently, Ref. {\cite{Binder:2019zqc}}
interpreted such manipulations in terms of Deligne categories, introducing a
notion of ``categorical symmetry''. Interestingly, for any group $G$, there is
also a Deligne category interpolating the replica symmetry $S_n \ltimes G^n$
{\cite{2006math5126K,2006math10552K}}, see {\cite{Binder:2019zqc}}, section
9.4. This may turn out useful in future rigorous mathematical justifications
of the method of replicas. In this paper we will not use categorical
language.} We have permutations $\phi_1 \leftrightarrow \phi_i$ and $\phi_i
\leftrightarrow \phi_j$ ($i, j \in \{ 2, \ldots, n \}$). After Cardy
transform, the latter give rise to permutations $\chi_i \leftrightarrow
\chi_j$ which generate $S_{n - 1}$ subgroup. This symmetry subgroup is
manifest in the Cardy basis: invariance of $\Lcal_{\tmop{Cardy}}$ under it
just means that $\chi_i$'s should enter in singlet combinations.

Consider now permutations of the former kind, $\phi_1 \leftrightarrow \phi_i$.
Without loss of generality we focus on $\phi_1 \leftrightarrow \phi_2$ since
together with $S_{n - 1}$ it generates the full $S_n$. Applying Cardy
transform with $\phi_2$ and $\phi_1$ interchanged, the relation between the
new and old Cardy fields is found from the equations
\begin{eqnarray}
  \phi_1 & = & \varphi + \frac{\omega}{2} = \varphi' - \frac{\omega'}{2} +
  \chi_2', \nonumber\\
  \phi_2 & = & \varphi - \frac{\omega}{2} + \chi_2 = \varphi' +
  \frac{\omega'}{2}, \nonumber\\
  \phi_i & = & \varphi - \frac{\omega}{2} + \chi_i = \varphi' -
  \frac{\omega'}{2} + \chi'_i \qquad (i = 3 \ldots n), 
\end{eqnarray}
where we renumbered the fields $\chi_i'$ so that their index always runs from
2 to $n$. We thus find
\begin{eqnarray}
  \varphi' & = & \varphi - \frac{2 - n}{2 (1 - n)} (\omega - \chi_2),
  \nonumber\\
  \omega' & = & \frac{\omega}{1 - n} - \frac{n}{1 - n} \chi_2, \nonumber\\
  \chi_2' & = & \frac{2 - n}{1 - n} \omega - \frac{\chi_2}{1 - n}, \nonumber\\
  \chi_i' & = & \chi_i + \frac{\omega - \chi_2}{1 - n} \qquad (i = 3 \ldots n)
  . 
\end{eqnarray}
We will mostly use the $n \rightarrow 0$ limit of this ``extra'' symmetry
transformation, which is
\begin{eqnarray}
  \varphi' & = & \varphi - 2 (\omega - \chi_2), \nonumber\\
  \omega' & = & \omega, \nonumber\\
  \chi_2' & = & 2 \omega - \chi_2, \nonumber\\
  \chi_i' & = & \chi_i + \omega - \chi_2 \qquad (i = 3 \ldots n) . 
  \label{n=0sym}
\end{eqnarray}
Let us see how $\mathcal{L}_{\tmop{Cardy}}$ behaves under this. The $\omega^2$
term in $\Lcal_0$, Eq.~{\eqref{L0phi4}}, is trivially invariant. It is more
interesting to check that the mass term is invariant:
\begin{equation}
  2 \vf \omega + \sum_{i = 2}^n \chi_i^2 \rightarrow 2 [\varphi - 2 (\omega -
  \chi_2)] \omega + (2 \omega - \chi_2)^2 + \sum_{i = 3}^n (\chi_i + \omega -
  \chi_2)^2,
\end{equation}
and using the constraint $\sum\nolimits' \chi_i = 0$ we see that the r.h.s. reduces to
the l.h.s. Analogously the kinetic part $\del \varphi \del \omega +
\frac{1}{2}  (\del \chi_i)^2$ is also invariant.

For the quartic term, $S_n$ invariance is realized in a still more interesting
way. The quartic term is the marginal (in $d = 6$) part of the $S_n$ invariant
term $\sigma_4 = \sum_{i = 1}^n \phi_i^4$ whose full $n \rightarrow 0$ limit
is
\begin{eqnarray}
  \sigma_4 & = & \left[ 4 \omega \vf^3 + 6 \sum\nolimits' \chi_i^2  \vf^2
  \right]_{\Delta = 6} \nonumber\\
  &  & + \left[ 4 \varphi \sum\nolimits' \chi_i^3 \right]_{\Delta = 7} + \left[ \sum\nolimits'
  \text{} \chi_i^4 - 6 \varphi \omega \sum\nolimits' \chi_i^2 \right]_{\Delta = 8}
  \nonumber\\
  &  & - \left[ 2 \omega \text{} \sum\nolimits' \chi_i^3 \right]_{\Delta = 9} + \left[
  \frac{3}{2} \omega^2 \text{} \sum\nolimits' \chi_i^2 + \varphi \omega^3
  \right]_{\Delta = 10},  \label{s4}
\end{eqnarray}
where we indicated the scaling dimensions in $d = 6$. The irrelevant terms in
the second and third lines have been assigned to the $\mathcal{L}_1$ part of
the Lagrangian. It is now possible to check that the sum of all terms is
invariant under the $n = 0$ symmetry {\eqref{n=0sym}}, although the first line
by itself is not.

We thus learn something very important: the sum $\mathcal{L}_0 +\mathcal{L}_1$
is invariant under the full $S_n$ symmetry in the $n \rightarrow 0$ limit
(denoted $S_{n \rightarrow 0}$), while individually the two parts are
invariant only under the $S_{n - 1}$ subgroup permuting $\chi_i$'s.

And what about $\mathcal{L}_2$? It was originally defined as consisting of the
$n$-suppressed terms, but now we can give an alternative description:
$\mathcal{L}_2$ consists of all terms which are not invariant under the extra
symmetry {\eqref{n=0sym}}, nor can be made invariant by adding further terms.
Consider e.g. $\varphi^2$, which is in $\mathcal{L}_2$ according to
{\eqref{Lsupp}}. Under the extra symmetry we have:
\begin{equation}
  \varphi^2 \rightarrow [\varphi - 2 (\omega - \chi_2)]^2 = \varphi^2 - 4
  (\omega - \chi_2) \varphi + 4 (\omega - \chi_2)^2 .
\end{equation}
This is obviously not invariant by itself, nor can it be made invariant by
adding other terms. E.g.~variation of $\varphi^2$ contains $- 4 \omega
\varphi$ and a moment's thought shows that this cannot be canceled by
variation of anything.

Since the terms in $\mathcal{L}_2$ are not invariant under the $S_{n
\rightarrow 0}$ symmetry, their coefficients must be proportional to $n$. This
explains why they are ``$n$-suppressed''. The advantage of this new
understanding is that it is not tied to the bare Lagrangian but can be used
along the RG flow. Since $\mathcal{L}_0 +\mathcal{L}_1$ are $S_{n \rightarrow
0}$ invariant, they can generate $\mathcal{L}_2$ terms only with
$n$-suppressed coefficients. This guarantees that a term $n$-suppressed in the
bare Lagrangian remains $n$-suppressed at a lower RG scale.

\subsection{$S_n$ singlets in the replicated basis}\label{Snsinglets}

As pointed out by Br{\'e}zin and De Dominicis {\cite{Brezin-1998}}, the
replicated Lagrangian {\eqref{Sr}}, in addition to the shown bare terms, will
generate infinitely many extra $S_n$ invariant terms upon RG
flow.\footnote{Depending on the circumstances, these extra terms may be
present already in the bare action, as was demonstrated explicitly in
{\cite{Brezin-1998}} via the Hubbard-Stratonovich transformation from the
lattice model.} Of course, these terms may or may not destabilize the RG flow
depending on their scaling dimensions. Leaving this more complicated question
for later, let us first learn to write general $S_n$ invariant terms (referred
to as ``singlets'' from now on). In the replicated basis, they can be
constructed as finite products:
\begin{equation}
  \left[ \sum_{i = 1}^n A (\phi_i) \right] \left[ \sum_{j = 1}^n B (\phi_j)
  \right] \left[ \sum_{k = 1}^n C (\phi_k) \right] \times \ldots 
  \label{genrepl}
\end{equation}
where $A, B, C \ldots$ are some polynomial\footnote{Limiting to polynomial
interactions is standard when dealing with perturbation of Gaussian fixed
points. Some literature on the RFIM (e.g. {\cite{Tarjus_2016}}) consider
interactions with non-polynomial field dependence, such as absolute value of
the fields (``cusps''). In App. \ref{cusp} we explain that cusp interactions
do not yield new perturbations of Gaussian fixed points, the full spectrum of
independent perturbations given by polynomial interactions.} functions of
$\phi_i$ and of its derivatives. For most part we will be interested in scalar
perturbations, which means that $A, B, C$ either do not contain derivatives,
or that all derivative indices are contracted.\footnote{Contractions of
derivative indices from different factors, e.g. from $A$ and $B$, are
allowed.}

We will use the notation ($\sum \equiv \sum_{i = 1}^n$)
\begin{eqnarray}
  &  & \sigma_k = \sum \phi_i^k, \nonumber\\
  &  & \sigma_{k (\mu)} = \sum \phi_i^{k - 1} \partial_{\mu} \phi_i,
  \nonumber\\
  &  & \sigma_{k (\mu \nu)} = \sum \phi_i^{k - 1} \partial_{\mu}
  \partial_{\nu} \phi_i, \nonumber\\
  &  & \sigma_{k (\mu) (\nu)} = \sum \phi_i^{k - 2} \partial_{\mu} \phi_i
  \partial_{\nu} \phi_i,\quad \tmop{etc} .  \label{skder}
\end{eqnarray}
The fields $\sigma_k$ were considered in {\cite{Brezin-1998}}, and the others
are natural generalizations. More singlets can be constructed by taking
products of these basic building blocks.

In this notation, e.g., the bare replicated Lagrangian (quartic potential) is
a linear combination of singlets
\begin{equation}
  \sigma_{2 (\mu) (\mu)}, \qquad \sigma_1^2, \qquad \sigma_2, \qquad \sigma_4
  .
\end{equation}
But we can clearly construct more singlets. E.g.~with four fields and no
derivatives the full list has five singlets {\cite{Brezin-1998}}:
\begin{equation}
  \sigma_4, \quad \sigma_1 \sigma_3, \quad \sigma_2^2, \quad \sigma_1^2
  \sigma_2, \quad \sigma_1^4. \label{gens4}
\end{equation}
Still more singlets are obtained by increasing the number of fields or
introducing derivatives. What are the scaling dimension of the corresponding
fixed point perturbations? Can they become relevant as the dimension is
lowered? We will study these questions systematically in the subsequent
sections.

Feldman {\cite{Feldman}}\footnote{Ref. {\cite{Feldman}} focuses on the Random
Field $O (N)$ Model, and the part starting from Eq.~(8) applies also to the
RFIM setting $N = 1$. In our work we will find support for some of Feldman's
results, but we will draw from them a different conclusion.} discussed a
family of singlet operators $\mathcal{F}_k$ $(k \in 2\mathbb{N})$ given by
\begin{eqnarray}
  \mathcal{F}_k & = & \sum_{i, j = 1}^n (\phi_i - \phi_j)^k = \sum_{l = 1}^{k
  - 1} (- 1)^l \binom{k}{l} \sigma_l \sigma_{k - l}   \label{Fk}
\end{eqnarray}
(the $l = 0, k$ terms vanish for $n \rightarrow 0$). Operators $\mathcal{F}_4$
and $\mathcal{F}_6$ will play an important role in our work.

\subsection{$S_n$ singlets in the Cardy basis}\label{singletsCardy}

Applying the Cardy transform to any singlet in the replicated basis, we get an
$S_{n \rightarrow 0}$ singlet in the Cardy basis. We have already seen such
expressions above, e.g.
\begin{equation}
  \sigma_2 = 2 \vf \omega + \sum\nolimits' \chi_i^2, \label{s2}
\end{equation}
while $\sigma_4$ is given in {\eqref{s4}}. We will use this procedure to
construct all singlets in the Cardy basis. Generality of this method follows
from the fact that the Cardy transform is an invertible linear transformation of
the field basis. We have the following master formula (here and below we drop
terms vanishing in the $n = 0$ limit):\footnote{Variational derivative
notation allows for the case when $A$ depends on the derivatives of $\phi$. In
this case these derivatives have to be distributed on the fields following
$\delta^k A / \delta \varphi^k$, in an obvious manner. Note that $(\delta^k A
/ \delta \varphi^k) \omega^k$ terms with $k$ even are $n$-suppressed ($A
(\varphi)$ and $(\delta^2 A / \delta \varphi^2) \omega^2$ being two
examples).}
\begin{eqnarray}
  \sum_{i = 1}^n A (\phi_i) & = & A \left( \varphi + \frac{\omega}{2} \right)
  + \sum\nolimits' A \left( \varphi - \frac{\omega}{2} + \chi_i \right) \\
  & = & \frac{\delta A}{\delta \varphi} (\varphi) \omega + \frac{1}{2}
  \frac{\delta^2 A}{\delta \varphi^2} (\varphi) \sum\nolimits' \chi_i^2 \nonumber\\
  &  & + \sum_{k = 3}^{\infty} \frac{1}{k!} \frac{\delta^k A}{\delta
  \varphi^k} (\varphi) \left[ \left( \frac{\omega}{2} \right)^k + \sum\nolimits' \left(
  - \frac{\omega}{2} + \chi_i \right)^k \right] .  \label{ACardy}
\end{eqnarray}
Let us introduce some useful terminology. By composite operators (composites,
for short) we will mean products of Cardy fields, their derivatives, and
linear combinations thereof. To each product composite we assign a classical
scaling dimension which is the sum of dimensions of its constituents, Eq.
{\eqref{dimensions}}. A linear combination of composites has a ``well-defined
classical dimension'' if all terms have the same dimension. Later on, we will
also discuss anomalous dimensions due to interactions. Due to mixing, only
some special linear combinations will have well-defined anomalous dimensions.

In this terminology, the terms in the first line of {\eqref{ACardy}} have the
same classical dimension, while those in the second line have a higher
dimension. For singlets involving at most two fields ($\sigma_1$, $\sigma_2$,
$\sigma_{1 (\mu)}$, $\sigma_{2 (\mu)}$, etc), the second line is absent
($\delta^k A / \delta \varphi^k \equiv 0$). Such fields, and products thereof,
are special: they have well-defined classical dimensions in the Cardy
basis.\footnote{This is only true in the $n = 0$ limit which is assumed here.}
One example is {\eqref{s2}} where both composites have dimension $d - 2$. Any
other singlet will becomes a linear combination of composites of different
dimensions in the Cardy basis. We have seen one example in {\eqref{s4}}.

Given any singlet $\mathcal{O}$, we can split it into parts with definite
classical dimension, which will come in unit steps:
\begin{equation}
  \mathcal{O}= [\mathcal{O}]_{\Delta} + [\mathcal{O}]_{\Delta + 1} + \ldots
\end{equation}
We will call the ``leader'' the lowest scaling dimension part of
$\mathcal{O}$, that is $[\mathcal{O}]_{\Delta}$, while $[\mathcal{O}]_{\Delta
+ k}$ with $k \geqslant 1$ will be called ``followers''. E.g.~the first line
of the r.h.s. of {\eqref{s4}} is the leader of $\sigma_4$, while the
subsequent lines contains the followers. The rationale for this terminology
will become clear in section \ref{L0L1}.

As an exercise which will turn out useful later on, let us transform Feldman
operators to the Cardy basis and extract the leader. Using definition
{\eqref{Fk}} we have:
\begin{eqnarray}
  \mathcal{F}_k & = & 2 \sum_{i = 2}^n (\omega - \chi_i)^k + \sum_{i, j = 2}^n
  (\chi_i - \chi_j)^k .  \label{Fk0}
\end{eqnarray}
In particular there is no dependence on $\varphi$ for this very special
operator. Expanding we have
\begin{eqnarray}
  \mathcal{F}_k & = & \sum_{l = 2}^{k - 2} (- 1)^l \binom{k}{l} \left( \sum\nolimits'
  \chi^l_i \right)  \left( \sum\nolimits' \chi^{k - l}_i \right) \label{FkCardy} \\
  &  & - 2 k \omega \left( \sum\nolimits' \chi^{k - 1}_i \right) + \ldots \nonumber
\end{eqnarray}
where we used $\sum\nolimits' \chi_i = 0$ and that $\sum\nolimits' \chi_i^k$ cancels between the
two terms for $n \rightarrow 0$. This shows the leader (first line) and the
first follower for $k \geqslant 4$. (For $k = 2$ the shown terms vanish and
$\mathcal{F}_2 = - 2 \omega^2$.)

\section{The $n$-suppressed terms}\label{sec:L2}

Following our plan to clarify step-by-step the basic RG scenario of section
\ref{basicRG}, we will discuss here the effects associated with the
$n$-suppressed terms, which were grouped in the $\mathcal{L}_2$ part of the
Cardy-transformed Lagrangian {\eqref{lagmain}}. As explained in section
\ref{sec:Sn}, these terms can be alternatively characterized as those which
break the $S_{n \rightarrow 0}$ symmetry of the $\mathcal{L}_0 +\mathcal{L}_1$
Lagrangian. This implies that the terms $n$-suppressed in the bare Lagrangian
remain $n$-suppressed along the RG flow. If we set $n = 0$, these terms vanish
in the bare Lagrangian and are not regenerated in the RG flow. Still, it is
instructive to analyze what would happen if we worked at a tiny but nonzero
$n$. Note that $\mathcal{L}_2$ contains several relevant terms so that, while
$n$-suppressed, they grow in the IR. One such term is the operator $\vf^2$,
which comes from the operator $\sigma_2 \ni n \vf^2$ (while the $n = 0$ part
of $\sigma_2$ goes into $\mathcal{L}_0$). What would be the role of these
terms for the IR behavior of the theory?

For concreteness let us just focus on this very operator $\vf^2$, the
discussion being similar for any other relevant part of $\mathcal{L}_2$ such
as $\varphi^4$ or $(\partial \varphi)^2$. We thus consider $\mathcal{L}_0$
action perturbed by
\begin{equation}
  \label{On} g_2 \, \Lambda_{\mathrm{UV}}^4 \int d^d x \hspace{0.17em}
  \vf^2,
\end{equation}
with $\Lambda_{\mathrm{UV}}$ the UV cutoff energy scale, and $g_2$ a
dimensionless coupling. The power of $\Lambda_{\mathrm{UV}}$ is fixed by the
dimension of the perturbing operator, $d - 4$ in the case at hand.

We are considering the situation when in absence of the perturbation the
$\mathcal{L}_0$ part of the action flows to an IR fixed point. When we add the
perturbation the coupling $g_2$ starts growing. Since $g_2$ starts at
order-$n$ at the UV scale, it reaches order-1 values at the scale
\begin{equation}
  \Lambda_{\mathrm{IR}} \sim n^{1 / 4} \Lambda_{\mathrm{UV}} \hspace{0.17em} .
\end{equation}
At that point we can no longer treat it as a small perturbation.

\begin{figure}[h]
	\centering \includegraphics[width=250pt]{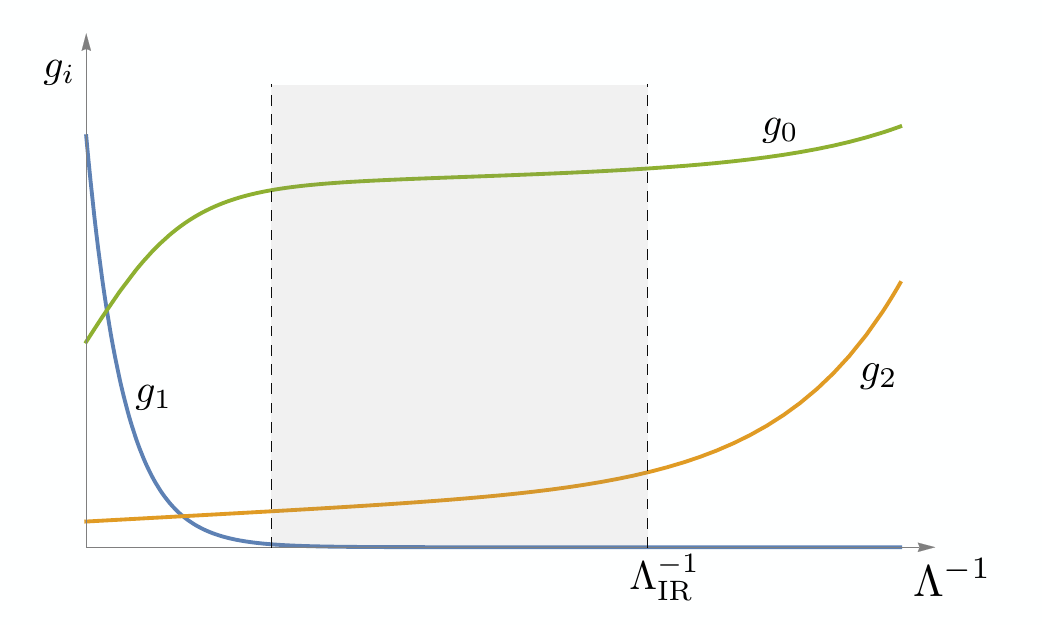}
  \caption{\label{OnRG}Schematic RG flow (from short to long distances)
  including $n$-suppressed terms. The flow has three parts (left to right):
  (1) the transitory part where the couplings $g_0$ of $\mathcal{L}_0$ flows
  to a fixed point, while the irrelevant couplings $g_1$ of $\mathcal{L}_1$ go
  to zero; (2) the shaded part where the flow stays close to the
  $\mathcal{L}_0$ fixed point; (3) the part where the relevant $\mathcal{L}_2$
  terms finally grow to overcome suppression by $n$, and the flow deviates
  from the fixed point. }
\end{figure}

The conclusions from this discussion is that, first of all, the $n = 0$ fixed
point is unstable with respect to turning on nonzero $n$. For $n$ tiny but
nonzero, the RG trajectory stays for a long time near the fixed point before
finally deviating. Thus, for a very small $n$, we expect that in a range of
distances the theory will be approximately described by the $n = 0$ fixed
point and will have an approximate scale invariance, and this range will
become longer and longer as $n \to 0$. However, no matter how small $n$ is,
the trajectory eventually deviates (see Fig.~\ref{OnRG}).

We do not know what happens with the small $n$ trajectory afterwards -- it may
flow to a gapped phase or to another fixed point. Note that even if the
trajectory flows to a fixed point, such a fixed point would have no
significance for the disordered physics, not being continuously connected to
the $n = 0$ fixed point.\footnote{\label{bond}Incidentally, this is radically
different from what happens in the \tmtextit{bond-disordered} Ising model,
where the fixed point is believed to exist for any $n$ so that we can compute
CFT data as a function of $n$ and perform the $n \rightarrow 0$ limit at the
CFT level (see section 8.3 of {\cite{Cardy-book}}, and
{\cite{Komargodski:2016auf}}).} One example of such a fixed point for nonzero
$n$ can be found in the work of Br{\'e}zin and De Dominicis
{\cite{Brezin-1998}}. Their fixed point has couplings scaling as inverse
powers of $n$, a clear feature of being disconnected from $n = 0$. For the
above reasons we will not consider their fixed point any further in the main
text, although we provide more details in App.~\ref{BDD}.

To summarize, when $n \to 0$, the range in which the flow is close to the
$\mathcal{L}_0$ fixed point becomes infinitely large. This clarifies in which sense
the $n$-suppressed terms can be safely discarded in the limit $n \to 0$.

\section{RG flow in the $\mathcal{L}_0 +\mathcal{L}_1$ theory}\label{L0L1}

At this point we are left with studying the RG flow in the theory consisting
of $\mathcal{L}_0 +\mathcal{L}_1$ part of the Cardy Lagrangian, working in the
strict $n = 0$ limit. As discussed in section \ref{RGbasic}, the
$\mathcal{L}_0$ theory by itself flows to an IR fixed point
equivalent\footnote{As stressed several times this equivalence holds only in
the sector of operators invariant under $O (n - 2)$ rotations of $\chi_i$'s.}
to the SUSY fixed point of $\mathcal{L}_{\tmop{SUSY}}$. The key remaining
question is whether this RG flow is stable under $\mathcal{L}_1$
perturbations.

In the original definition, $\mathcal{L}_1$ included all terms irrelevant in
$d$ just below 6, coming from the replicated bare Lagrangian with the quartic
potential. This was not completely general: the bare Lagrangian can be
expected to contain any possible $\mathbb{Z}_2$-even $S_n$ singlets, and even
if not initially, included such terms will be generated under RG flow
{\cite{Brezin-1998}}. From now on we will extend the definition of the bare
Lagrangian to include all $\mathbb{Z}_2$-even $S_n$ singlets. E.g., at the
quartic level we should consider all terms given in {\eqref{gens4}} (while
only the first of these five singlets was included so far). It is easy to see
that with the new definition we do not get any additional relevant terms in $d
= 6 - \eps$. So all new terms end up in $\mathcal{L}_1$.

Now that we have the full bare Lagrangian, we should ask: can it be that some
perturbations, while irrelevant near 6d, become relevant for smaller
$d$?\footnote{\label{DIO}Sometimes in high-energy physics one calls
``dangerously irrelevant'' operators which are irrelevant at the UV fixed
point but become relevant at the IR fixed point. We will refrain from this
usage of the term, which is different from statistical physics, where
dangerously irrelevant operator is a property of a single fixed point, not of
an RG flow (it is an irrelevant operator whose perturbation effect on the fixed
point is non-analytic in the coupling {\cite{Amit}}, a typical example being
the $\varphi^4$ operator around free massless scalar fixed point in $d > 4$
dimensions). } If this happens, the $\mathcal{L}_0$ fixed point will not be
reached for those $d$. The RG flow will be instead deviated to another fixed
point, which does not have SUSY if the new relevant interaction is
SUSY-breaking (something to be checked).

We wish to explore this mechanism for the loss of Parisi-Sourlas SUSY. The
problem is well defined, at least in perturbation theory: we need to consider
$\mathcal{L}_1$ perturbations one by one, and see which of them get anomalous
dimensions of sign and size likely to render them relevant. We are interested
in stability with respect to $S_n$ singlet perturbations, because only such
perturbations are present in the microscopic replicated Lagrangian (i.e.~before the Cardy transform).

\subsection{Leader and followers: quartic term}\label{renH}

In section \ref{sec:Sn} we saw that after the Cardy transform, a generic $S_n$
singlet is a sum of the leader (the lowest scaling-dimension part) and the
followers (higher scaling dimension parts). The quadratic terms in the
replicated Lagrangian do not have any followers, while the quartic term has
both the leader and the followers, see {\eqref{s4}}.

It is instructive to consider first the RG flow of the $\mathcal{L}_0
+\mathcal{L}_1$ theory truncated just to the quartic perturbation
{\eqref{s4}}. We start the RG flow at an energy scale $\Lambda$ with the
Lagrangian in the Cardy basis (in equations below we use the notation $\chi_i^k
\equiv \sum_i' \chi_i^k$)
\begin{equation}
  \partial \varphi \partial \omega + \frac{1}{2} (\partial \chi_i)^2 -
  \frac{H}{2} \omega^2 + \frac{m^2}{2} \left( 2 \vf \omega + \chi_i^2 \right)
  + \frac{\lambda}{4!} \left[ \left( 4 \omega \vf^3 + 6 \chi_i^2  \vf^2
  \right) + 4 \varphi \chi_i^3 + \ldots \right],
\end{equation}
where {$\ldots$} stands for the other $\sigma_4$ followers visible in
{\eqref{s4}}. Performing the integrating-out step down to the energy scale
$\Lambda' = \Lambda / b$ (but not yet any field rescaling), we will find an
effective Lagrangian
\begin{equation}
  Z_1 \left[ \partial \varphi \partial \omega + \frac{1}{2} (\partial
  \chi_i)^2 \right] - Z_2 \frac{H}{2} \omega^2 + (m')^2 \left( \vf \omega +
  \frac{1}{2} \chi_i^2 \right) + \frac{\lambda'}{4!} \left[ \left( 4 \omega
  \vf^3 + 6 \chi_i^2  \vf^2 \right) + 4 \varphi \chi_i^3 + \ldots \right] .
  \label{Z1Z2}
\end{equation}
Crucially, $S_n$ invariance guarantees that the kinetic terms $\partial
\varphi \partial \omega + \frac{1}{2} (\partial \chi_i)^2$, the mass terms $2
\vf \omega + \chi_i^2$, and the whole quartic interaction renormalize by
overall rescaling, since the form of these terms is fixed uniquely by
transforming $\sigma_{2 (\mu) (\mu)}$, $\sigma_2$ and $\sigma_4$ to the Cardy
basis. We now perform field rescaling
\begin{eqnarray}
  \varphi (x) & \rightarrow & Z_1^{- 1 / 2} b^{- \Delta_{\varphi}^0} \varphi
  (x / b), \nonumber\\
  \chi_i (x) & \rightarrow & Z_1^{- 1 / 2} b^{- \Delta_{\chi}^0} \chi_i (x /
  b), \\
  \omega (x) & \rightarrow & Z_1^{- 1 / 2} b^{- \Delta_{\omega}^0} \omega (x /
  b), \nonumber
\end{eqnarray}
where $\Delta_{\varphi}^0, \Delta_{\chi_i}^0, \Delta_{\omega}^0$ are the
Gaussian fixed point dimensions {\eqref{dimensions}}. After rescaling, the
fields again have momenta up to $\Lambda$ while the Lagrangian becomes:
\begin{eqnarray}
  &  & \partial \varphi \partial \omega + \frac{1}{2} (\partial \chi_i)^2 -
  \frac{Z_2}{Z_1} \frac{H}{2} \omega^2 + \frac{(m')^2 b^2}{Z_1} \left( \vf
  \omega + \frac{1}{2} \chi_i^2 \right) \nonumber\\
  &  & + \frac{\lambda' b^{\varepsilon}}{4! Z_1^2} \left[ \left( 4 \omega
  \vf^3 + 6 \chi_i^2  \vf^2 \right) + \frac{1}{b} 4 \varphi \chi_i^3 +
  \frac{1}{b^2} \left( \text{} \chi_i^4 - 6 \varphi \omega \chi_i^2 \right) -
  \frac{1}{b^3} 2 \omega \text{} \chi_i^3 + \frac{1}{b^4} \left( \frac{3}{2}
  \omega^2 \text{} \chi_i^2 + \varphi \omega^3 \right) \right] . 
  \label{sigma4rescaled} \label{}
\end{eqnarray}
We see that in general $Z_2 \neq Z_1$ and $H$ will be renormalized. However as
discussed in section \ref{RGL0} we can expect that this effect is transient
and disappears in deep infrared so that $H$ flows to a constant. Here we are
focusing on the behavior of the $\sigma_4$ followers, this time written in
full. We see that their coefficients rescale with an additional positive
integer power of $b$ compared to that of the leader. But, apart from this
additional rescaling, the relative coefficients stay fixed because determined
by $S_n$ invariance. This explain our choice for the leader-follower
terminology.

After many RG steps the coefficients of the followers will flow to zero, and
we approach the fixed point of the $\mathcal{L}_0$ theory. It is not so
surprising that the follower coefficients flow to zero as these operators are
irrelevant. What is more surprising is that the coefficients of these
irrelevant terms go to zero in a prescribed fashion. This feature of the
$\mathcal{L}_0 +\mathcal{L}_1$ RG flow is dictated by $S_n$ invariance.

We can rephrase the above conclusions as follows. Consider the perturbation
$(\delta \lambda) \sigma_4$ on top of the $\mathcal{L}_0$ fixed point,
splitting it into the leader and the followers:
\begin{equation}
  (\delta \lambda) \sigma_4 = \delta \lambda \left[ \left( 4 \omega \vf^3 + 6
  \chi_i^2  \vf^2 \right) + 4 \varphi \chi_i^3 + \ldots \right] .
\end{equation}
At a lower scale $\Lambda / b$ the perturbation will become
\begin{equation}
  \delta \lambda (b) \left[ \left( 4 \omega \vf^3 + 6 \chi_i^2  \vf^2 \right)
  + \frac{1}{b} 4 \varphi \chi_i^3 + \ldots \right] .
\end{equation}
We see here two effects. First, the coefficients of the followers are
suppressed compared to that of the leader by integer powers of $b$. Second,
assuming that the fixed point is reached, $\delta \lambda (b)$ flows to zero
(which is the same as $\lambda$ flowing to a constant). Let us introduce the
RG eigenvalue $y$ for $\delta \lambda$:
\begin{equation}
  \frac{d}{d \log b} \delta \lambda = - y \delta \lambda,
\end{equation}
where $y$ must be positive for $\delta \lambda$ to flow to zero.

The simplest way to compute $y$ is to go to deep IR. There the coefficients of
the followers are tiny and can be neglected. We are therefore reduced to the
problem of computing the anomalous dimension of the leader as a perturbation
of the $\mathcal{L}_0$ fixed point. This recipe is a key simplification: it
would have been much more awkward to compute anomalous dimension if we had to
keep track of both the leader and the followers.

For the quartic coupling case at hand, $y$ is related to the anomalous
dimension of $\left( 4 \omega \vf^3 + 6 \chi_i^2  \vf^2 \right)$ perturbing
the $\mathcal{L}_0$ fixed point. This being a susy-writable operator, its
anomalous dimension is the same as that of $\left( 4 \omega \vf^3 + 3 \psi
\bar{\psi}  \vf^2 \right)$ perturbing the $\mathcal{L}_{\tmop{SUSY}}$ fixed
point. In turn, by dimensional reduction, this is the same as the anomalous
dimension of $\hat{\phi}^4$ at the Wilson-Fisher fixed point in $d - 2$
dimensions (see section \ref{susy-irr} below). The latter operator is
irrelevant since the Wilson-Fisher fixed point has only one relevant
$\mathbb{Z}_2$ even singlet $(\hat{\phi}^2)$, hence indeed $y > 0$.

\subsection{Leader and followers: general case}

We will now generalize the quartic coupling perturbation considered in the
previous section to any other singlet perturbation $g\mathcal{O}$ inside the
$\mathcal{L}_0 +\mathcal{L}_1$ flow. Near the $\mathcal{L}_0$ fixed point,
this perturbation takes the form
\begin{equation}
  g (b) \left[ \mathcal{O}_L + \frac{1}{b} \mathcal{O}_{F 1} + \frac{1}{b^2}
  \mathcal{O}_{F 2} + \ldots + \right], \qquad \frac{d}{d \log b} g = -
  y_{\mathcal{O}} g, \label{LeaderFollowersRG}
\end{equation}
where $\mathcal{O}_L$ is the leader, while $\mathcal{O}_{F 1}, \mathcal{O}_{F
2}, \ldots$ are the followers. If the leader coefficient flows to zero
($y_{\mathcal{O}} > 0$), the follower coefficients flow to zero as well, and
faster. The RG eigenvalue $y_{\mathcal{O}}$ can be computed as
\begin{equation}
  y_{\mathcal{O}} = \Delta (\mathcal{O}_L) - d,
\end{equation}
where $\Delta (\mathcal{O}_L)$ is the scaling dimension of $\mathcal{O}_L$ as
a perturbation of the $\mathcal{L}_0$ fixed point. A very convenient feature
is that the followers do not enter into the latter computation.

We are thus converging on a well-defined problem of quantum field theory. We
have to classify all perturbations of the $\mathcal{L}_0$ fixed point which
can be realized as leaders of $\mathbb{Z}_2$-even $S_n$ singlets, and compute
their anomalous dimensions. If one of these becomes relevant, stability of the
$\mathcal{L}_0$ fixed point is lost. This program will be realized in section
\ref{leaders} and \ref{anomdim} below.

\subsection{Followers as individual $\mathcal{L}_0$
perturbations}\label{followers}

The reader may find somewhat puzzling the feature of the above discussion that
followers completely ``go for the ride''. In other words, we are not supposed
to consider followers as individual perturbations of the $\mathcal{L}_0$ fixed
point. Let us give a few more explanations concerning this fact. We are
studying stability of the $\mathcal{L}_0$ fixed point in the IR, by adding to
it infinitesimal $S_n$ singlet perturbations and seeing if they grow or decay.
In this setup, perturbing the $\mathcal{L}_0$ fixed point by a follower alone
would not be consistent: the follower always accompanies a leader, whose
coefficient is enhanced by the RG flow with respect to that of the follower.
That is why the correct procedure is to perturb infinitesimally by the leader,
while the follower perturbation then is ``doubly infinitesimal'' in IR, and
can be neglected.

But what if we nevertheless perturb the $\mathcal{L}_0$ fixed point by a
follower alone and compute the anomalous dimension of such a perturbation?
What would be the physical meaning of such a computation? The answer is
instructive. In addition to $S_n$ singlet perturbations, the $\mathcal{L}_0
+\mathcal{L}_1$ RG flow possesses perturbations breaking $S_n$ invariance.
Were we to perturb the $\mathcal{L}_0$ fixed point by a follower alone, we
would be computing dimensions of such $S_n $-breaking perturbations. These
perturbations are not important for the problem of $S_n$-invariant RG
stability studied in this paper, but they do exist.

To convince ourselves in the reality of $S_n$-breaking perturbations, we
found useful the following toy model. Consider the $S_n$-invariant RG flow
with initial conditions corresponding to the quadratic part of the
$\mathcal{L}_0$ Lagrangian perturbed by 5 quartic singlets without derivatives
from Eq.~{\eqref{gens4}}:
\begin{equation}
  \label{toy} \left[ \partial \varphi \partial \omega + \frac{1}{2} \partial
  \chi_i \partial \chi_i - \frac{H}{2} \omega^2 \right] + h_1 \sigma_4 + h_2
  \sigma_2^2 + h_3 \sigma_1 \sigma_3 + h_4 \sigma_1^2 \sigma_2 + h_5
  \sigma_1^4 .
\end{equation}
When we transform these singlets into the Cardy basis, we get a total of 11
monomials. We then \ consider a more general RG flow introducing 11
independent couplings for each of these monomials:
\begin{equation}
  \begin{array}{l}
    \label{Toy} \left[ \partial \varphi \partial \omega + \frac{1}{2} \partial
    \chi_i \partial \chi_i - \frac{H}{2} \omega^2 \right] + 6 g_1 \varphi^2
    \chi_i^2 + 4 g_2 \varphi^3 \omega + 4 g_3 \varphi \chi_i^3 + g_4
    \chi_i^4\\
    \hspace{4em} + g_5 \varphi \omega \chi_i^2 + g_6 \omega \chi_i^3 + g_7
    \omega^2 \chi_i^2 + g_8 \varphi \omega^3 + g_9 \chi_i^4 + g_{10} \varphi
    \omega \chi_i^2 + g_{11} \omega^4 .
  \end{array}
\end{equation}
When these 11 couplings are set to particular linear combinations of 5
$h_i$'s, we are back to the $S_n$-invariant flow {\eqref{toy}}, while when we
relax this condition, we get an $S_n$-breaking RG flow. In this setup we can
do renormalization and see how these couplings evolve when we approach the IR
fixed point. These computations are carried out in Appendix \ref{sec:Toy}, and
they give a concrete illustration and a confirmation of the picture developed
above.

\section{Classification of leaders}\label{leaders}

As the first step of the program set in section \ref{L0L1}, let us classify
the $\mathbb{Z}_2$-even $S_n$ singlet leader operators. Of course, the total
number of leaders is infinite. We will carry out a detailed classification for
leaders up to scaling dimension $12$ in $d = 6$, and we will make some
comments about operators of arbitrarily high dimensions. This will be
sufficient for our goal of understanding the loss of stability of the
$\mathcal{L}_0$ fixed point.

We will pay close attention to symmetries. Symmetries control mixing of
operators under RG evolution, importantly for the next section where we
compute anomalous dimensions. We know that the $\mathcal{L}_0$ fixed point has
Parisi-Sourlas supersymmetry upon replacing $\chi$ bilinears by $\psi$
bilinears. Some leaders (the susy-writable ones) can thus be located inside
SUSY multiplets. Their anomalous dimensions can then be determined easily, by
reusing known Wilson-Fisher results. This method is not available for leaders
which are not susy-writable, whose anomalous dimensions will be computed
independently starting from the $\mathcal{L}_0$ Lagrangian.

\subsection{General remarks}

We are interested in classifying the scalar leader operators up to classical
dimension $\Delta_{\max} = 12$ in $d = 6$. A general singlet operator is
constructed, in the replicated basis, as a product
\begin{equation}
  \mathcal{O}= A_{k_1} \ldots A_{k_p}, \label{Ogen}
\end{equation}
where each $A_k$ is either $\sigma_k$ or one of its dressings by derivatives,
Eq.~{\eqref{skder}}. The classical scaling dimension of the leader will be
\begin{equation}
  \Delta (\mathcal{O}_L) = N_{\phi} + 2 p + N_{\tmop{der}}, \label{OLdim}
\end{equation}
where $N_{\phi} = k_1 + \ldots + k_p$ is the total power of $\phi$ in
$\mathcal{O}$ (an even number for the considered $\mathbb{Z}_2$-even fields),
and $N_{\tmop{der}}$ is the total number of derivatives (also even, since
indices are contracted to get a scalar). The $N_{\phi} + 2 p$ in
{\eqref{OLdim}} is obtained when we replace in each $A_{k_i}$ one $\phi$ by
$\omega$ and the rest by $\varphi$, as in the first term in Eq.
{\eqref{ACardy}}. Linear combinations of operators {\eqref{Ogen}} may have
leaders of higher dimensions than {\eqref{OLdim}} if the leading terms cancel.

So we need to consider all possible products {\eqref{Ogen}} such that
$N_{\phi} + 2 p + N_{\tmop{der}} \leqslant \Delta_{\max}$, do the Cardy transform,
and separate the leaders. Let us show how this works for the case $N_{\phi} =
4$, $N_{\tmop{der}} = 0$. The basis of singlets is given in Eq.
{\eqref{gens4}}. Performing the Cardy transform we find:
\begin{eqnarray}
  \sigma_4 & = & \left[ 4 \omega \vf^3 + 6 \chi_i^2  \vf^2 \right]_{\Delta =
  6} + \ldots, \nonumber\\
  \sigma_1 \sigma_3 & = & [3 \varphi^2 \omega^2 + 3 \varphi \omega \chi_i^2
  ]_{\Delta = 8} + \ldots, \nonumber\\
  \sigma_2^2 & = & [4 \varphi^2 \omega^2 + 4 \varphi \omega \chi_i^2 +
  (\chi_i^2)^2]_{\Delta = 8},  \label{BDlist}\\
  \sigma_1^2 \sigma_2 & = & [2 \varphi \omega^3 + \omega^2 \chi_i^2]_{\Delta =
  10}, \nonumber\\
  \sigma_1^4 & = & [\omega^4]_{\Delta = 12} . \nonumber
\end{eqnarray}
Here are below we will continue to omit $\sum\nolimits'$: $\chi_i^k \equiv \sum\nolimits'
\chi_i^k$, $(\chi_i^2)^2 \equiv \left( \sum\nolimits' \chi_i^2 \right)^2$, etc.

Recall that the operators involving $\chi_i$'s only in $O (n - 2)$ symmetric
combinations, like $\chi_i^2$, are called susy-writable. Their correlators can
be computed in the SUSY theory $\mathcal{L}_{\tmop{SUSY}}$ replacing $\chi$
bilinears by $\psi$ bilinears: $\chi_i^2 \rightarrow 2 \psi \bar{\psi}$, etc.
The full rules are given in Appendix \ref{chi-psi}.

We will use the name ``susy-writable'' only for $O (n - 2)$ invariant
operators which do not vanish upon the SUSY substitution of $\chi$'s by
$\psi$'s. Operators which do vanish, because of the Grassmann nature of the
$\psi$ and $\bar{\psi}$, will be called ``susy-null''. The simplest example is
$(\chi_i^2)^2$, which maps to $(2 \psi \bar{\psi})^2 \equiv 0$. Although one
might think that susy-null operators do not have any physical effect, this is
not quite true because they may have non-null followers (see section
\ref{SUSY-null} below). The susy-null operators will not mix with
susy-writable nor with non-susy-writable operators under RG, which is another
reason to put them into a separate category.

Now, in {\eqref{BDlist}}, $\sigma_1 \sigma_3$ and $\sigma_2^2$ have the same
susy-writable part of their leader, up to a constant factor We thus can
perform a linear transformation to exhibit a singlet with a purely susy-null
leader:
\begin{equation}
  \sigma_2^2 - \frac{4}{3} \sigma_1 \sigma_3 = \frac{1}{6} \mathcal{F}_4 =
  [(\chi_i^2)^2]_{\Delta = 8} - \frac{4}{3} [\omega \chi_i^3]_{\Delta = 9} +
  \ldots, \label{SNullEx}
\end{equation}
where we also exhibited the non-susy-writable follower, coming from $\sigma_1
\sigma_3$. Interestingly, this special linear combination turns out
proportional to the Feldman operator $\mathcal{F}_4$, see Eqs. {\eqref{Fk}},
{\eqref{FkCardy}}.

This completes classification of leaders with $N_{\phi} = 4$, $N_{\tmop{der}}
= 0$ (see Table \ref{Nf4Nd0}). We stress that the leader type (susy-writable,
non-susy-writable or susy-null) is determined based on the expression for the
leader, not for the followers.

\begin{table}[h]
	\centering
    \begin{tabular}{@{}lll@{}}
      \toprule
      Singlet & Leader($+ 1^{\tmop{st}}$ follower if susy-null) & Leader
      type\\
      \midrule
      $\sigma_4$ & $\left[ 4 \omega \vf^3 + 6 \vf \chi_i^2  \right]_{\Delta =
      6}$ & susy-writable\\
      $\sigma_1 \sigma_3$ & $[3 \varphi^2 \omega^2 + 3 \varphi \omega
      \chi_i^2]_{\Delta = 8}$ & susy-writable\\
      $\frac{1}{6} \mathcal{F}_4 = \sigma_2^2 - \frac{4}{3} \sigma_1 \sigma_3$
      & $[(\chi_i^2)^2]_{\Delta = 8} - \frac{4}{3} [\omega \chi_i^3]_{\Delta =
      9}$ & susy-null\\
      $\sigma_1^2 \sigma_2$ & $[2 \varphi \omega^3 + \omega^2
      \chi_i^2]_{\Delta = 10}$ & susy-writable\\
      $\sigma_1^4$ & $[\omega^4]_{\Delta = 12}$ & susy-writable\\
      \bottomrule
    \end{tabular}
  \caption{\label{Nf4Nd0}Leaders with $N_{\phi} = 4$, $N_{\tmop{der}} = 0$.}
\end{table}

The described procedure can be analogously carried out for any $N_{\phi}$ and
$N_{\tmop{der}}$ (see Appendix \ref{class}). When classifying leaders
containing derivatives, we separate total derivatives since those do not
affect RG stability, and also do not mix with other operators of the same
classical dimensions. We will next highlight conceptual aspects of this
classification, separately for each leader type.

\subsection{Non-susy-writable leaders}\label{sec:nonsusywritableleaders}

We start with the non-susy-writable leaders. These operators break the
accidental $O (n - 2)$ symmetry of the $\mathcal{L}_0$ Lagrangian to the $S_{n
- 1}$ symmetry permuting the $\chi_i$ fields.\footnote{Note the subgroup
relation $S_{n - 1} \subset O (n - 2)$, familiar for integer $n.$ E.g.~$S_4
\subset O (3)$ acts by permuting the vertices of the tetrahedron centered at
the origin of $\mathbb{R}^3$.}

One might think that non-susy-writable leaders should be more
numerous than susy-writable ones because of their smaller symmetry. However
this turns out not to be true. The point is that while there are many
non-susy-writable operators, most of them end up being followers rather than
leaders. We have seen this already in Eq.~{\eqref{s4}}, where $\varphi
\chi_i^3$, $\chi_i^4$ and $\omega \chi_i^3$ are all followers. Systematic
enumeration (Appendix \ref{class}) finds only one non-susy-writable leader up
to $\Delta = 12$, which comes from the Feldman operator $\mathcal{F}_6$:
\begin{equation}
  \left( - \frac{1}{20} \mathcal{F}_6 \right)_L = [(\chi_i^3)^2 - \frac{3}{2}
  (\chi_i^2) (\chi_i^4)]_{\Delta = 12} . \label{F6leader}
\end{equation}
At higher $\Delta$, non-susy-writable leaders could be constructed e.g. from
the singlets
\begin{equation}
  \sum_{i, j = 1}^n (\phi_i - \phi_j)^6 P (\phi_i, \phi_j), \label{NSW}
\end{equation}
with $P (\phi_i, \phi_j)$ an arbitrary polynomial. In particular, for $P
(\phi_i, \phi_j) = (\phi_i - \phi_j)^{k - 6}$ these would be the higher
Feldman operators $\mathcal{F}_k$ whose non-susy-writable leaders are given in
Eq.~{\eqref{FkCardy}}. Still more non-susy-writable leaders can be obtained by
dressing singlets {\eqref{NSW}} with derivatives, or multiplying them by other
singlets. We will not attempt here a full classification.

\subsection{Susy-writable leaders}\label{sec:susywritableleaders}

Looking at Table \ref{Nf4Nd0} and Appendix \ref{class}, we see that most
leaders up to $\Delta \leqslant 12$ are susy-writable. It
would be somewhat tedious to have to compute the anomalous dimensions of all
these operators. Fortunately this turns out unnecessary because general
arguments (section \ref{susy-irr}) will establish that most of them are
guaranteed to be irrelevant. But before we come to that, let us have a
general discussion of this class of operators.

We will refer to susy-writable leaders transformed to SUSY
fields as ``susy-written''. Consider first the following question: what
distinguishes susy-written leaders from all other operators of the SUSY
theory? As one may expect, this has a neat answer based on symmetry, which is
as follows: \tmtextit{The susy-written leaders correspond to
supertranslation-invariant Sp(2)-invariant operators.} In other words,
supertranslations {\eqref{suptrans}} and $\tmop{Sp} (2)$ take the role of
$S_n$ in fixing linear combinations corresponding to leaders.

Let's explain how this comes about. The Sp(2) invariance acting on $\psi,
\bar{\psi}$ is manifest in the rule {\eqref{appB-bilinears}}. As an example of
supertranslation invariance, consider susy-writable leaders in Table
\ref{Nf4Nd0}. Transforming to SUSY fields we get $\varphi^3 \omega + 3
\varphi^2 \psi \bar{\psi}$, $\varphi^2 \omega^2 + 2 \varphi \omega \psi
\bar{\psi}$, $\varphi \omega^3 + \omega^2 \psi \bar{\psi}$, $\omega^4$. Indeed
these are all invariant under $\tmd  \vf = - \vareps  \psib, \tmd \psi =
\vareps \omega, \tmd  \psib = \tmd \omega = 0$, and only for these relative
coefficients. More generally, susy-writable leaders appear from terms in the
first line of Eq.~{\eqref{ACardy}}, and it is easy to check that these become
supertranslation-invariant upon passing to SUSY fields. This statement remains
true also in presence of derivatives. It would be interesting
to give a formal general proof, although we have tested this property so
extensively that we are absolutely sure in its validity.

As any $\mathcal{L}_{\tmop{SUSY}}$ operator, any susy-written leader can be
expressed in terms of superfield $\Phi$ given in {\eqref{supefield}} and its
(super)derivatives. E.g.
\begin{equation}
  \varphi^3 \omega + 3 \varphi^2 \psi \bar{\psi} = \Phi^3 \Phi_{, \theta
  \bar{\theta}} + 3 \Phi^2 \Phi_{, \bar{\theta}} \Phi_{, \theta} |_{\theta =
  \bar{\theta} = 0} \nobracket .
\end{equation}
Let us think in terms of superprimaries, i.e.~composite operators
$\mathcal{O}$ built out of the superfield $\Phi$ which transform simply under
the (super)conformal symmetry of the SUSY fixed point {\cite{paper1}}.
Superprimaries have well-defined anomalous dimensions at the SUSY fixed point,
equal to those of primaries in the Wilson-Fisher fixed point in $\hat{d} = d -
2$ dimensions {\cite{paper1}}. Identifying susy-written leaders with
components of superprimaries, we will easily determine their anomalous
dimensions. A general superprimary is expanded in components as
({\cite{paper1}}, Eq.~(3.20))
\begin{equation}
  \mathcal{O}^{(a)} (x, \theta, \bar{\theta}) =\mathcal{O}^{(a)}_0 (x) +
  \theta \mathcal{O}^{(a)}_{\theta} (x) + \bar{\theta}
  \mathcal{O}^{(a)}_{\bar{\theta}} (x) + \theta \bar{\theta} \tmmathbf{}
  \mathcal{O}^{(a)}_{\theta \bar{\theta}} (x),
\end{equation}
where $(a)$ is a collection of $\tmop{OSp} (d| 2)$ indices if superprimary
transforms in a nontrivial representation. Leaders $\Upsilon$ will be found in
the component $\mathcal{O}^{(a)}_{\theta \bar{\theta}} \equiv D_{\bar{\theta}}
D_{\theta} \mathcal{O}^{(a)}$ which is supertranslation invariant and has
scaling dimension $\Delta_{\mathcal{O}} + 2$. The indices $(a)$, if present,
have to be contracted to get a scalar leader. So we will have $\Upsilon =
t_{(a)} \mathcal{O}^{(a)}_{\theta \bar{\theta}}$ where $t_{(a)}$ is an
$\tmop{SO} (d) \times \tmop{Sp} (2)$ invariant tensor, or simply $\Upsilon
=\mathcal{O}_{\theta \bar{\theta}}$ if $\mathcal{O}$ is a scalar superprimary.
Total derivative leaders would correspond to $x$-derivatives of superprimary
components; they do not affect RG flow.

In general, $\mathcal{O}^{(a)}$ will transform under $\tmop{OSp} (d| 2)$ as a
traceless tensor with mixed graded symmetry represented by a Young tableau
{\cite{paper1}}. Because of the tracelessness condition, the $\tmop{SO} (d)
\times \tmop{Sp} (2)$ invariant tensor $t_{(a)}$ above can be chosen as a
product of $\varepsilon_{p q}$'s where $p, q \in \theta, \bar{\theta}$ run
over the Grassmann directions.\footnote{E.g.~$\mathcal{T}^{\mu \mu} \propto
\mathcal{T}^{\theta \bar{\theta}}$ for spin two representation and does not
have to be considered separately.} In other words, all indices $(a)$ will be
pairwise assigned to $\theta \bar{\theta}$. Since graded symmetry means
antisymmetry for Grassmann directions, we may conclude that the only Young
tableaux giving rise to nonzero $\tmop{Sp} (2)$ invariant components are those
of shape $(2, 2, \ldots, 2)$ (i.e.~2 boxes in each row).
This observation is very important, as it
radically reduces the number of representations we need to examine. The
representations with more than 2 rows do not occur below dimension 12, and we
will not discuss them except for a few comments below.

In summary, all needed susy-written scalar leaders are the highest components
$\mathcal{O}^{(a)}_{\theta \bar{\theta}}$ of superprimaries in the scalar
$\mathcal{S}$, spin-two $\mathcal{J}^{a b}$, or box $\mathcal{B}^{a b, c d}$
representations of $\tmop{OSp} (d| 2)$, where the graded symmetric pairs of
indices $(a b)$ and $(c d)$ have to be set to $\theta \bar{\theta}$, namely:
\begin{equation}
  \mathcal{S}_{\theta \bar{\theta} \tmmathbf{}}, \qquad \mathcal{J}^{\theta
  \bar{\theta}}_{\theta \bar{\theta} \tmmathbf{}}, \qquad \mathcal{B}^{\theta
  \bar{\theta}, \theta \bar{\theta}}_{\theta \bar{\theta} \tmmathbf{}} .
\end{equation}
Let us now discuss superrotations {\eqref{suprot}}. For a leader $\Upsilon
=\mathcal{O}_{\theta \bar{\theta}}$ where $\mathcal{O}$ is a scalar
superprimary, superrotation transformation generalizes that of $\omega$ in
{\eqref{suprot1}}:
\begin{equation}
  \delta \Upsilon = - \varepsilon_{\mu \theta} \partial^{\mu}
  \mathcal{O}_{\bar{\theta}} - \varepsilon_{\mu \thetab} \partial^{\mu}
  \mathcal{O}_{\theta} .
\end{equation}
This only produces total derivatives, and so $\int d^d x\, \Upsilon$ will be
preserving superrotations (and thus full SUSY). On the other hand, leaders
$\Upsilon$ built out of $\tmop{SO} (d) \times \tmop{Sp} (2)$-invariant
components $\mathcal{O}^{(a)}_{\theta \bar{\theta}}$ of a tensor superprimary
(like $\mathcal{J}^{a b}$ or $\mathcal{B}^{a b, c d}$) will superrotate to
other components (in addition to the total derivative terms). For such
leaders, $\int d^d x \,\Upsilon$ will break superrotations (and thus not
preserve full SUSY).

Coming back to the problem of identifying susy-written leaders with
superprimary components, we can go through the list of superprimaries, and see
what the corresponding leaders are. We are to classify superprimaries of the
Gaussian part of $\mathcal{L}_{\tmop{SUSY}}$, making use of the SUSY equation
of motion $D^2 \Phi = 0$.

In the sector with two superfields, the lowest two superprimaries are $\Phi^2
$ and the super stress tensor $\mathcal{T}^{a b},$see Eq.~(C.4) of
{\cite{paper1}}. Superconservation fixes the dimension of $\mathcal{T}^{a b}$
at $d - 2$ for any $d$, while anomalous dimensions of $\Phi^2 $ will be the
same as for the Wilson-Fisher operator $\hat{\phi}^2$. The supertranslation
invariant components are $(H = 2)$,
\begin{eqnarray}
  (\Phi^2)_{\theta \bar{\theta}} & = & 2 \varphi \omega + 2 \psi \bar{\psi},
  \nonumber\\
  \mathcal{T}^{\mu \mu}_{\theta \bar{\theta}} & = & 2\mathcal{T}^{\theta
  \bar{\theta}}_{\theta \bar{\theta}} = - \partial \varphi \partial \omega -
  \partial \psi \partial \bar{\psi} + 4 \omega^2 .  \label{stress}
\end{eqnarray}
The first one is the SUSY mass term, while the second is a particular linear
combinations of $\partial \varphi \partial \omega + \partial \psi \partial
\bar{\psi}$ and $\omega^2$. Another linear combination with a well-defined
anomalous dimension sits in the total derivative
\begin{equation}
  \partial^2 (\Phi^2)_{\theta \bar{\theta}} = 4 (\partial \varphi \partial
  \omega + \partial \psi \partial \bar{\psi} - \omega^2), \label{totdero2}
\end{equation}
where one uses the Gaussian EOM $\partial^2 \varphi = - 2 \omega$, $\partial^2
(\omega, \psi, \bar{\psi}) = 0$.

Higher spin $l \geqslant 4$ superprimaries built out of two superfields, e.g.
the spin-4 $\mathcal{J}^{a b c d}$, are graded symmetric-traceless tensors.
They have Young tableaux with $l$ boxes in one row. \ As discussed above, such
Young tableau do not give rise to supertranslation- and $\tmop{Sp}
(2)$-invariant scalars, as the corresponding components vanish by
graded-symmetric tracelessness (too many $\theta$'s, e.g. $\mathcal{J}^{\theta
\bar{\theta} \theta \bar{\theta} } = 0$).

Let us carry out a similar exercise in the sector with four superfields. Two
low-dimension superprimaries are $\Phi^4$ and $\Phi^2 \mathcal{T}^{a b}$ of 6d
scaling dimension 4 and 6 respectively. They give rise to supertranslation
invariant components of dimension 6 and 8:
\begin{eqnarray}
  (\Phi^4)_{\theta \bar{\theta}} & = & 4 \varphi^3 \omega + 12 \varphi^2 \psi
  \bar{\psi}, \nonumber\\
  (\Phi^2 \mathcal{T}^{\mu \mu})_{\theta \bar{\theta}} & = & 6 \varphi^2
  \omega^2 + 12 \varphi \omega \psi \bar{\psi} \\
  &  & - \varphi^2 \partial \psi \partial \bar{\psi} - 2 \varphi \partial
  \varphi (\partial \psi \bar{\psi} + \psi \partial \bar{\psi}) - (\partial
  \varphi)^2 \psi \bar{\psi} - \varphi \omega (\partial \varphi)^2 - \varphi^2
  \partial \varphi \partial \omega . \nonumber
\end{eqnarray}
The first one is the SUSY quartic interaction. The second one is recognized as
a linear combination of the dimension 8 leaders $(\sigma_1 \sigma_3)_L = 3
\varphi^2 \omega^2 + 3 \varphi \omega \chi_i^2$ and $(\sigma_{4 (\mu)
(\mu)})_L = (\partial \chi_i)^2 \varphi_{\nosymbol}^2 + \ldots$ (see Tables
\ref{Nf4Nd0}, \ref{Nf4Nd2}). Another linear combination corresponds to
$\partial^2 (\Phi^4)_{\theta \bar{\theta}}$.

To extend this story to higher $\Delta$, it is useful to take into account
that Parisi-Sourlas superprimaries in $6$ dimensions are in correspondence
with the free massless scalar primaries in $4$ dimensions. The latter can be
counted using conformal characters
{\cite{Barabanschikov:2005ri}}.\footnote{For systematic applications to
Wilson-Fisher see {\cite{Liendo:2017wsn}} and {\cite{Meneses:2018xpu}},
appendix A. Note that this method only determines the number of primaries of
each spin for every dimension. To find their explicit expressions in terms of
the fundamental field one would have to use other techniques, such as directly
imposing the primary condition $[K_{\mu}, O (0)] = 0$.} This gives the number
of primaries and their spin for each dimension. Denoting 4d primaries as
$\Delta_{j_1, j_2}$ where $\Delta$ is the scaling dimension and $j_1, j_2 \in
\mathbb{Z}/ 2$ label the $SO(4)$ representation,\footnote{Here $j_1$ and $j_2$ are the quantum numbers that label the two $SU(2)$ in $SO(4)=SU(2)\times SU(2)$. Representations $(j_1,j_2)$ can be simply related to Young tableaux $(l_1,l_2)$, i.e. $l_1$ boxes in the first row and $l_2$ boxes in the second row. E.g. spin $l$ representations are obtained by setting $j_1=j_2=l/2$. More generically, mixed symmetric Young tableaux $(l_1,l_2)$ are related to representations $(j_1,j_2)\oplus(j_2,j_1)$ where $j_1=(l_1+l_2)/2$ and $j_2=(l_1-l_2)/2$.
} up to $\Delta = 10$ we find the
following counting:\footnote{We go up to $\Delta = 10$ because the leader will
sit in the $\mathcal{O}^{(a)}_{\theta \bar{\theta}}$ component and have
dimension 2 higher, and we are classifying leaders up to $\Delta = 12$.}
\begin{eqnarray}
  4 \tmop{fields} : &  & 4_{0, 0}, 6_{1, 1}, 7_{3 / 2, 3 / 2}, \nonumber\\
  &  & 2 \times 8_{2, 2}, 8_{2, 0 \oplus 0, 2}, 8_{1, 1}, 8_{0, 0}, 
  \label{class4d}\\
  &  & 9_{5 / 2, 5 / 2}, 9_{5 / 2, 3 / 2 \oplus 3 / 2, 5 / 2}, 9_{5 / 2, 1 /
  2 \oplus 1 / 2, 5 / 2}, 9_{3 / 2, 1 / 2 \oplus 1 / 2, 3 / 2}, \nonumber\\
  &  & 3 \times 10_{3, 3}, 10_{3, 2 \oplus 2, 3}, 2 \times \overline{} 10_{3,
  1 \oplus 1, 3}, \overline{} 2 \times 10_{2, 2}, 10_{2, 1 \oplus 1, 2}, 2
  \times 10_{1, 1}, 10_{0, 0}, \nonumber\\
  6 \tmop{fields} : &  & 6_{0, 0}, 8_{1, 1}, 9_{3 / 2, 3 / 2}, 2 \times 10_{2,
  2}, 10_{2, 0 \oplus 0, 2}, 10_{1, 1}, 10_{0, 0}, \nonumber\\
  8 \tmop{fields} : &  & 8_{0, 0}, 10_{1, 1} . \nonumber
\end{eqnarray}
The only representations from this list giving rise to $\tmop{SO} (d) \times
\tmop{Sp} (2)$ invariant components in 6d (which are not total derivatives)
are scalars ($j_1, j_2 = 0$), rank-2 tensors ($j_1, j_2 = 1, 1$) and
mixed-symmetry 4-index tensor corresponding to the $(2, 2)$ ``box'' Young
tableau ($j_1, j_2 = 2, 0 \oplus 0, 2$). Mixed symmetry tensors of shape $(2,
2, \ldots)$ with more than two rows are not realized in 4d, although they may
exist in 6d. 6d tensors with such symmetry are examples of representations
which project to zero under dimensional reduction {\cite{paper1}} in physical
dimension. However when we go to $d = 4 - \eps$ dimensions, such
representations reappear as ``evanescent operators''
{\cite{Hogervorst:2014rta,Hogervorst:2015akt}}. It is possible to study
evanescent operators in the $\varepsilon$-expansion
{\cite{Hogervorst:2015akt}}, but since their classical dimension is rather
high (the lowest scalar evanescent has dimension 15 in 4d), we will not
consider them in this work as already mentioned above. We will however
consider the ``box'' tensors in full seriousness.

\subsection{Susy-null leaders}\label{SUSY-null}

Susy-null operators are closely related to susy-writable operators. Like
susy-writable operators, susy-null operators are $O (n - 2)$ invariant and can
be mapped to the $\psi$-formulation using the map described in appendix
\ref{chi-psi}. The special feature of these operators compared to
susy-writables is that in the $\psi$-formulation they exactly vanish. The
simplest instance of this class of operator is $(\chi^2_i)^2 $ mapped to
$(\psi \bar{\psi})^2$ which clearly vanishes because of anticommutation of
$\psi$.

These operators are evidently null in the susy theory, namely any correlation
function of a susy-null operator $\mathcal{O}_{\text{null}}$ with any other
operator $\mathcal{O}_{\tmop{SUSY}}^i$ will vanish:
\begin{equation}
  \left\langle \mathcal{O}_{\text{null}} \mathcal{O}_{\tmop{SUSY}}^1 \ldots
  \mathcal{O}_{\tmop{SUSY}}^k \right\rangle = 0 .
\end{equation}
The property above of course holds also when $\mathcal{O}_{\tmop{SUSY}}^i$ is
itself a susy-null operator, in particular the 2-point function
$\mathcal{O}_{\text{null}}$ must vanish,
\begin{equation} \left\langle \mathcal{O}_{\text{null}} \mathcal{O}_{\text{null}}
   \right\rangle = 0 . \end{equation}
One may be tempted to discard these operators, however this conclusion is too
quick. Indeed in the $\mathcal{L}_0$ theory we can also consider
non-susy-writable operators for which the vanishing condition does not hold,
\begin{equation} \left\langle \mathcal{O}_{\text{null}} \mathcal{O}^1_{\text{non-susy}}
   \ldots \mathcal{O}^k_{\text{non-susy}} \right\rangle \neq 0 . \end{equation}
A simple instance of this is the 5-point function $\langle (\chi^2_i)^2
\chi_{i_1} \chi_{i_2} \chi_{i_3} \chi_{i_4}  \rangle \neq 0$, as one can
easily verify in free theory by Wick contractions. In this sense susy-null
operators are physical operators of the $\mathcal{L}_0$ theory.

Because of its special structure, this class of operators satisfies very
strict selection rules for mixing under RG. Namely, susy-null operators can
only mix with other susy-null operators. Indeed, as the susy-writable
operators, they cannot mix with non-susy-writables since the latter are
invariant under a smaller symmetry group: $S_{n - 1}$ symmetry instead of the
accidental $O (n - 2)$. Also they cannot acquire admixtures of susy-writable
operators (which are not null). This must be the case, otherwise we would find
that a null operator in the SUSY theory (which should be set to zero) would
mix non-trivially with a non-vanishing operator. Notice however that the
mixing can occur in the opposite direction: non-susy writable and
susy-writable operators can acquire admixtures of susy-nulls. Schematically,
we have the following triangular mixing:
\begin{eqnarray}
  \text{susy-null} & \leftrightarrow & \text{susy-null} \nonumber\\
  \text{susy-writable} & \rightarrow & \mathrm{} \text{susy-writable}, \quad
  \text{susy-null} \\
  \text{non-susy-writable} & \rightarrow & \text{non-susy-writable}, \quad
  \text{susy-writable}, \quad \text{susy-null} . \nonumber
\end{eqnarray}
More formally this block-triangular structure holds for the matrix $Z$
relating the bare and renormalized operators, see Eq.~{\eqref{BD}}. This in
particular implies that renormalized susy-writable and non-susy-writable
operators with a well-defined anomalous dimension may contain a susy-null
piece. On the other hand, all renormalized susy-null operators will always
stay susy-null.

Now that the definition of susy-null operators is set, let us comment on which
are the possible susy-null leaders with dimensions up to $12$ in $d = 6$.
Systematic enumeration in Appendix \ref{class} produced a few instances of
susy-null leaders. The first one is the unique susy-null operator at
dimensions 8: this is the Feldman $\mathcal{F}_4$ leader which can be written
as $(\chi_i^2)^2 $. Another susy-null leader is found at dimension 10:
$\varphi^2 (\chi_i^2)^2$. At dimension $12$ there are three susy-null leaders
built out of 6 fields, which can mix among themselves $\varphi \omega
(\chi_i^2)^2$, $(\chi_i^2)^3$, $\partial_{\mu} \varphi \partial^{\mu} \varphi
(\chi_i^2)^2$. Finally, also at dimension 12, there is a unique susy-null
leader built of 8 fields $\varphi^4 (\chi_i^2)^2$. In the next section we will
go though this list and compute all their anomalous dimensions.

\subsection{Fixed point destabilization}\label{destab}

{An $S_n$-singlet perturbation can destabilize the IR fixed point, when its leader becomes relevant.
This criterion is obvious for the susy-writable and non-susy-writable leaders. The same criterion applies also for the susy-null leaders.
Note that the susy-null operators by themselves do not affect the correlation functions of all operators in the susy-writable sector.
However when the coefficient of a susy-null leader grows and becomes $O(1)$ (as it may happen when such a leader is relevant), it will enter and modify RG evolution equations of other perturbations. E.g., non-susy-writable leaders which were irrelevant, may become relevant in presence of such large susy-null perturbations.

There is a small loophole, because in principle it may happen that the offending susy-null coupling flows to a nearby fixed point and never becomes $O(1)$. Whether this happens of not, depends on the higher-order terms in the beta-function, and on the sign of the initial value of the RG evolving susy-null coupling. If such a fixed point does occur, SUSY observables will be unaffected.\footnote{While the SUSY observables are unaffected, it may be possible to see a presence of a susy-null coupling
	in more complicated correlation functions involving
	non-susy-writable operators. E.g.~the correlation function $\langle \chi_2
	(x_1) \chi_2 (x_2) \chi_2 (x_3) \chi_2 (x_4) \rangle$ would be affected if the
	leaders $(\chi_i^2)^2$ or $\varphi^2 (\chi_i^2)^2$ become relevant. This
	correlation can be mapped to the replica variables and back to the random
	field formulation as follows (namely we substitute
		$\chi_2 = \phi_2 - \frac{1}{n - 1} (\phi_3 + \ldots + \phi_n)$, expand the
		correlator for generic $n$, translate replicated correlators to the random
		field correlators using {\eqref{form2}}, and finally take the limit $n
		\rightarrow 0$ for the coefficients),
	\begin{eqnarray}
		\langle \chi_2 (x_1) \chi_2 (x_2) \chi_2 (x_3) \chi_2 (x_4) \rangle & = & 14
		\overline{\langle \phi (x_1) \phi (x_2) \phi (x_3) \phi (x_4) \rangle}
		\nonumber\\
		&  & - 10 (\overline{\langle \phi (x_1) \phi (x_2) \rangle  \langle \phi
			(x_3) \phi (x_4) \rangle} + 2 \tmop{perms}) \nonumber\\
		&  & - 14 (\overline{\langle \phi (x_3) \rangle  \langle \phi (x_1) \phi
			(x_2) \phi (x_4) \rangle} + 3 \tmop{perms})  \label{kkkk}\\
		&  & + 24 (\overline{\langle \phi (x_1) \phi (x_2) \rangle \langle \phi
			(x_3) \rangle \langle \phi (x_4) \rangle} + 5 \tmop{perms}) \nonumber\\
		&  & - 72 \overline{\langle \phi (x_1) \rangle \langle \phi (x_2) \rangle
			\langle \phi (x_3) \rangle \langle \phi (x_4) \rangle} . \nonumber
	\end{eqnarray}
	It may be possible to consider the r.h.s. in a simulation and see if it
	deviates from the result of the l.h.s predicted by using $\mathcal{L}_0$. If
	so, one can check if the correct result is obtained by perturbing
	$\mathcal{L}_0$ with a relevant susy-null operator. Admittedly, this is more a
	question of principle than a concretely realizable proposal, since simulating
	4-point functions is a very hard task. In any case a deviation in
	{\eqref{kkkk}} from the $\mathcal{L}_0$ prediction does not count as a
	violation of SUSY, since this observable was not protected by SUSY in the
	first place.} 
This possibility looks somewhat exotic. In this paper we will be conservative, and will count relevant susy-null leaders as potentially destabilizing perturbations.\footnote{In the preliminary report given in \cite{zoom}, as well as in the earlier version of the paper, we used a different criterion for when a susy-null leader destabilizes the fixed point, which required looking at the first follower, and requiring that follower be relevant. For this to happen, the susy-null leader dimension has to be below $d-1$. We now think that  conclusion, based on a linearized analysis, was not correct.}

}

%
%
%
%
\section{Anomalous dimensions}\label{anomdim}

In the previous section we classified the leaders of $\mathbb{Z}_2$-even $S_n$
singlets up to 6d scaling dimension 12. We will now discuss their anomalous
dimensions. Some anomalous dimensions will be computed at two loops, and some
at one loop. We will consider separately the three classes of leaders
(susy-null, susy-writable, and non-susy-writable). We will identify in each
class at least one perturbation which becomes less irrelevant as $d$ is
lowered. The next section will discuss the critical dimension $d_c$ where
these candidate perturbations may cross the relevance threshold.

We will start in section \ref{susy-irr} with the susy-writable leaders, the
most numerous class. Their anomalous dimensions can be determined, as
discussed above, by writing them as components of supermultiplets whose
dimensions are known from dimensional reduction to Wilson-Fisher theory. This
strategy is not available for the susy-null and non-susy-writable leaders,
whose anomalous dimensions have to be computed from scratch (sections
\ref{s-null-dim}, \ref{sec:dimensions_non-susy-writable}).

As we explained in sections \ref{L0L1}, the anomalous dimension computation is
greatly simplified by the fact that close to the IR fixed point all follower
operators can be dropped. We are therefore led to consider the anomalous
dimensions of leader operators in the theory defined by the Gaussian
$\mathcal{L}_0$ Lagrangian perturbed by the interaction $4 \omega \vf^3 + 6
\chi_i^2  \vf^2$, at its IR fixed point in $d = 6 - \varepsilon$ dimensions.
We use dimensional regularization.

Consistency of our method to compute anomalous dimensions by restricting to
the leader operators requires that leaders and followers do not mix. For
susy-writable leaders this is guaranteed explicitly by their supertranslation
invariance when transformed to the SUSY fields (section
\ref{sec:susywritableleaders}). From the point of view of the $\mathcal{L}_0$
Lagrangian the absence of leader-follower mixing may appear puzzling, since
these operators may have the same classical dimensions and the same number of
fields. However, our extensive checks confirm the absence of this mixing in
all cases we looked at. It would be interesting to find a formal proof, based
on selection rules following from $S_n$ invariance, and for all three classes
of leaders (see section \ref{sym-meaning}).

\subsection{Susy-writable leaders}\label{susy-irr}

The general remarks in section \ref{sec:susywritableleaders} give many handles
on the susy-writable leaders. We will now discuss their IR scaling dimension.
We have one low-dimension susy-writable leader: the SUSY mass term
$(\Phi^2)_{\theta \bar{\theta}} = 2 \varphi \omega + 2 \psi \bar{\psi}$. This
operator is relevant in any $d$, its anomalous dimension being the same as for
the Wilson-Fisher operator $\hat{\phi}^2$ in $\hat{d} = d - 2$ dimensions. The
coefficient of this operator is finetuned to reach the SUSY fixed
point.\footnote{For a non-finetuned coefficient the RG flow with SUSY initial
conditions would end up in a SUSY massive phase. The $\mathcal{L}_0
+\mathcal{L}_1$ flow is then also expected to end up in a massive phase, which
however is not going to be equivalent to the SUSY one, because of the residual
$\mathcal{L}_1$ effects, which will not have time to decay completely to
zero.}

Are there any other susy-writable $\mathbb{Z}_2$-even leaders which are
relevant? As we explained, apart from total $x$-derivatives, susy-writable
leaders $\Upsilon$, when transformed to the SUSY fields, are $\tmop{SO} (d)
\times \tmop{Sp} (2)$ invariant components $\mathcal{O}^{(a)}_{\theta
\bar{\theta} \tmmathbf{}}$ of superprimaries $\mathcal{O}^{(a)}$. Their
scaling dimension are thus
\begin{equation}
  \Delta_{\Upsilon} = \Delta_{\mathcal{O}} + 2 = \Delta_{\hat{O}} + 2,
  \label{YO}
\end{equation}
where we used that the scaling dimension of $\mathcal{O}$ equals that of the
Wilson-Fisher primary $\hat{O}$ to which $\mathcal{O}$ projects under
dimensional reduction {\cite{paper1}} (see also appendix 
\ref{app:susy_writables_an_dim} for a few one-loop examples). We have seen
above the example $\hat{O} = \hat{\phi}^2$ for $\mathcal{O}= \Phi^2$. By Eq.
{\eqref{YO}}, $\Upsilon$ is relevant $d$ dimensions if and only if $\hat{O}$
is relevant in $\hat{d}$ dimensions:
\begin{equation}
  \Delta_{\Upsilon} < d \quad \Longleftrightarrow \quad \Delta_{\hat{O}} <
  \hat{d} = d - 2 .
\end{equation}
In addition, as mentioned in section \ref{sec:susywritableleaders}, we are
interested in operators $\hat{O}$ which are either scalars, spin-2 tensors, or
$2 \times 2$ ``box'' Young tableau mixed symmetry tensors, since otherwise
$\mathcal{O}$ will not have $\tmop{SO} (d) \times \tmop{Sp} (2)$ invariant
components.\footnote{As mentioned in section \ref{sec:susywritableleaders},
Young tableau of shape $(2, 2, \ldots)$ could also be important but we will
neglect them since they have high classical dimension. They are harder to
study since they project to zero in 4d.} Let us then discuss what is known
about the spectrum of such $\mathbb{Z}_2$-even operators at the Wilson-Fisher
(WF) fixed point.

For any $\hat{d}$ the WF fixed point has one $\mathbb{Z}_2$-even relevant
scalar, $\hat{\phi}^2$, connected to the relevant SUSY mass term by the above
argument. All other $\mathbb{Z}_2$-even scalars are irrelevant, which
corresponds to the fact that the Ising phase transition is reached by tuning
one $\mathbb{Z}_2$ even parameter (temperature).

In the spin-2 sector the lowest operator is the stress tensor, of dimension
exactly $\hat{d}$. All other spin-2 operators are irrelevant in 4d and
expected to stay irrelevant in $\hat{d} < 4$, by two arguments. First, we have
the unitarity bound $\Delta \geqslant \hat{d}$ for any spin-2 primary in a
unitary CFT. This argument is rigorous for integer $\hat{d}$ but it has a caveat
for intermediate $\hat{d}$. In fact, the Wilson-Fisher theory in $\hat{d} = 4 -
\varepsilon$ is known to be not quite unitary because of the evanescent
operators {\cite{Hogervorst:2014rta,Hogervorst:2015akt}} mentioned in section
\ref{sec:susywritableleaders}. However the violations of unitarity appear
secluded at high dimension where all evanescent operators belong, and the
unitarity bound for low-lying operators seems safe even in non-integer
dimensions. The second argument does not rely on unitarity but on the
observation that to pass from irrelevant to relevant a spin-2 operator would
have to cross the stress tensor, and level crossing is believed unlikely in an
interacting non-integrable theory.

Finally, let us discuss the ``box'' tensors. The unitarity bound for these
tensors is relatively weak:\footnote{Put $\{ h_i \} = (2, 2, 0, \ldots)$ in
Eq.~(2.41) in {\cite{Minwalla:1997ka}}. This also agrees with
{\cite{Minwalla:1997ka}}, (2.45) using ``box''$= (2, 0) \oplus (0, 2)$ in
4d.}
\begin{equation}
  \Delta_{\tmop{box}} \geqslant \hat{d} - 1,
\end{equation}
which unfortunately does not guarantee irrelevance (even modulo caveats about
the lack of unitarity in non-integer $d$). So we have to enter into the
details. The lowest box tensor is $8_{2, 0 \oplus 0, 2}$ in 4d (see Eq.
{\eqref{class4d}}). This operator has 4 fields and 4 derivatives and an
expression in terms of fields of the form\footnote{The relative coefficient
between the two terms can be found by imposing the primary condition
$[K_{\mu}, \mathcal{O} (0)] = 0$, or by requiring zero two-point functions
with the lower primaries $\hat{\phi}^2$ and $\hat{\phi}^2 \hat{T}_{\mu \nu}$.}
\begin{equation}
  \left( \hat{\phi}_{, \mu \nu} \hat{\phi}_{, \rho \sigma} \hat{\phi}^2 -
  \frac{2 \hat{d}}{\hat{d} - 2} \hat{\phi}_{, \mu} \hat{\phi}_{, \nu}
  \hat{\phi}_{, \rho \sigma} \hat{\phi} \right)^Y,
\end{equation}
where $()^Y$ means that we should apply the box Young symmetrizer and subtract
traces. The IR scaling dimension of this operator in $\hat{d} = 4 -
\varepsilon$ is given by
\begin{equation}
  \Delta_{\hat{B}} = (8 - 2 \varepsilon)_{\tmop{class}} + \left( \frac{7}{9}
  \varepsilon \right)_{\text{1-loop}} + O (\varepsilon^2) = 8 - \frac{11}{9}
  \varepsilon + O (\varepsilon^2), \label{Bhat}
\end{equation}
where the one-loop correction is from {\cite{Kehrein:1994ff}}, Table 4 (line
``(2,0),(0,2)'', $n = 4$). Unfortunately we are not aware of a two-loop
computation.

So by Eq.~{\eqref{YO}}, the dimension of the leader $\mathcal{B}$ in $d = 6 -
\varepsilon$ is two units higher than {\eqref{Bhat}}:
\begin{equation}
  \Delta_{\mathcal{B}} = 10 - \frac{11}{9} \varepsilon + O (\varepsilon^2) .
  \label{Bcal}
\end{equation}
In Appendix \ref{app:susy_writables_an_dim} we write the form of the box
operator in Cardy variables, and perform an independent computation of its
one-loop anomalous dimension. This agrees with the Wilson-Fisher computation,
providing a further interesting check of dimensional reduction.

By Eq.~{\eqref{Bcal}}, the leader $\mathcal{B}$ is becoming less irrelevant
as the dimension is lowered, but only very slowly so. So in section \ref{Who}
it will not be our prime candidate to destabilize the SUSY fixed point.

\subsection{Susy-null leaders}\label{s-null-dim}

Here we will summarize computations of anomalous dimensions of susy-null
leaders (see Appendix \ref{app:dim_susy_null} for details).

\begin{figure}[h]
	\centering \includegraphics[width=200pt]{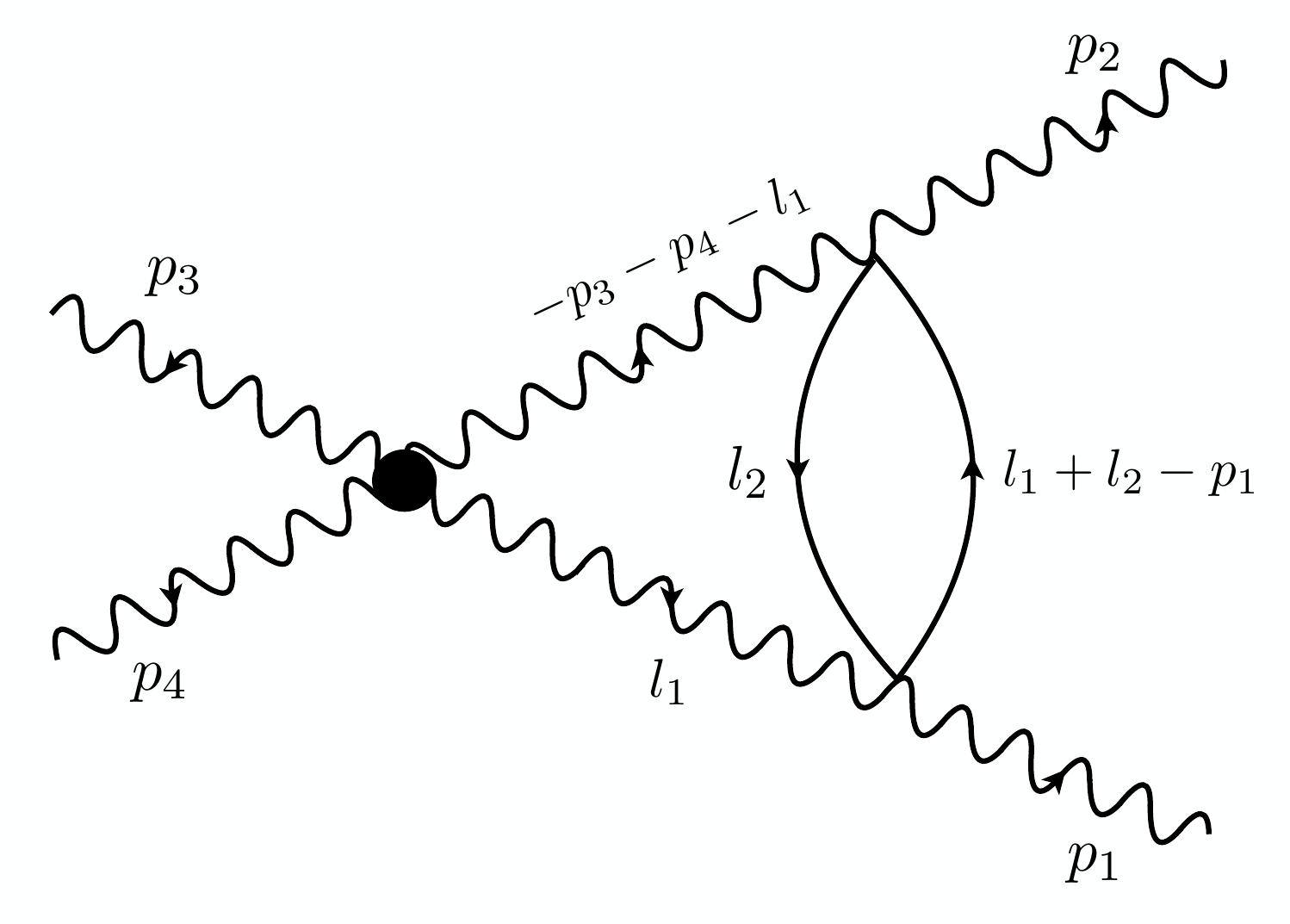}
  \caption{\label{chi22diag}Two-loop correction to the correlator $\langle
  (\chi_i^2)^2 (p = 0) \chi_j (p_1) \chi_k (p_2) \chi_l (p_3) \chi_m (p_4)
  \rangle$. The black dot indicates the bare composite operator $(\chi^2)^2$.
  We also see two $\vf^2 \chi^2$ vertices. The conventions for the propagators are explained in appendix \ref{App:prop}.
  }
\end{figure}

The susy-null operator with the smallest UV dimension is $(\chi_i^2)^2 .$ It
is the leader operator of the singlet combination $\sigma_3 \sigma_1 +
\frac{3}{4} \sigma_2^2$, which is also the Feldman operator $\mathcal{F}_4$,
see Eq.~{\eqref{SNullEx}}. Its anomalous dimension receives no one-loop
contribution, while the two-loop correction is given by the diagram in Fig.
\ref{chi22diag}. Using standard techniques (appendix \ref{chi22sec}) we obtain
the IR scaling dimension
\begin{equation}
  \Delta_{(\chi_i^2)^2} = 8 - 2 \varepsilon - \frac{8}{27} \varepsilon^2 + O
  (\varepsilon^3) . \label{dimchi4}
\end{equation}
Going higher in the UV dimension, we encounter  the susy-null leader $\varphi^2
(\chi_i^2)^2$ with classical dimension $10 - 3 \varepsilon$. Since it is the only
susy-null leader at this dimension, it does not mix with any other operator.
It receives a positive one-loop anomalous dimension equal to $3 \varepsilon$, as
discussed in appendix \ref{Delta10n6Null}. We have not evaluated its two-loop
anomalous dimension.

At classical dimension $12 - 3 \varepsilon$, we have three susy-null leaders made
of six fields, which mix with one another in a nontrivial way: $\varphi \omega
(\chi_i^2)^2, (\chi_i^2)^3$ and $(\partial_{\mu} \varphi)^2 (\chi_i^2)^2$. In
appendix \ref{Delta12n6Null} we compute their anomalous dimension matrix at
one loop. As explained there, the matrix is not completely diagonalizable,
rather it can be brought to the Jordan form (see also appendix \ref{ope}) with
eigenvalues 0 and $(11 / 9) \varepsilon$, the latter associated to a rank-two
Jordan block. In the CFT context, the Jordan block structure of the mixing
matrix signals the presence of a logarithmic multiplet, and is a symptom that
we are dealing with a logarithmic CFT {\cite{Hogervorst:2016itc}}. This fact
is not very surprising since our theory arises from the $n \rightarrow 0$
limit of $S_n$-symmetric replica Lagrangian (see {\cite{Cardy:2013rqg}}).

At classical dimension $12 - 4 \varepsilon$ there is a single composite of eight
fields, $\varphi^4 (\chi_i^2)^2$. This also receives a positive one-loop
anomalous dimension equal to $(22 / 3) \varepsilon$, as shown in appendix
\ref{Delta12n8Null}.

We summarize these results in Table \ref{null ops}. As discussed in section \ref{destab}, {we count relevant susy-null leaders as possible sources of the SUSY fixed point destabilization. The leader $(\chi_i^2)^2$ has the smallest dimension and is one candidate which may cause such
a destabilization, once it becomes relevant (see section \ref{Who}).}

\begin{table}[h]\centering
  \begin{tabular}{@{}lcc@{}}
    \toprule
    \quad Leaders $\mathcal{O}_{\text{null}}$\quad & Full $S_n$-singlet
    perturbation \ensuremath{\mathcal{O}}\quad &  IR dimension:
    $\Delta_{\mathcal{O}_{\text{null}}}$\\
    \midrule
    $(\chi_i^2)^2 $ & $\sigma_3 \sigma_1 + \frac{3}{4} \sigma_2^2$ &  $8 - 2
    \varepsilon - \frac{8}{27} \varepsilon^2 + O (\varepsilon^3)$\\
    $\varphi^2 (\chi_i^2)^2$ & $\sigma_2 \sigma_4 - \frac{8}{5} \sigma_1
    \sigma_5$ & $10 + O (\varepsilon^2)$\\
    $\varphi \omega (\chi_i^2)^2$ & $\sigma_1 \sigma_2 \sigma_3 - \frac{3}{2}
    \sigma_1^2 \sigma_4$ & \multirow{3}{*}{$ \left.\vphantom{\displaystyle\int\limits_0^\infty}\right\} \begin{array}{c}
      12 - 3 \e + O \left( \e^2 \right)\\
      12 - \frac{16}{9} \e + O \left( \e^2 \right)
    \end{array}$}\\
    $(\chi_i^2)^3$ & $\sigma_2^3 - 2 \sigma_1 \sigma_2 \sigma_3 + \sigma_1^2
    \sigma_4$ & \\
    $\partial_{\mu} \varphi \partial^{\mu} \varphi (\chi_i^2)^2$ & $\sigma_{3
    (\mu)}^2 - \frac{4}{3} \sigma_{2 (\mu)} \sigma_{4 (\mu)} + \frac{1}{3}
    \sigma_{1 (\mu)} \sigma_{5 (\mu)}$ & \\
    $\varphi^4 (\chi_i^2)^2$ & $\sigma_2 \sigma_6 - \frac{12}{7} \sigma_1
    \sigma_7$ & $12 + \frac{10}{3} \e + O \left( \e^2 \right)$\\
    \bottomrule
  \end{tabular}
  \caption{\label{null ops}Summary of anomalous dimension computations for all
  susy-null leaders with $\Delta_{\tmop{UV}} \leqslant 12 + O \left( \e
  \right)$. For the leaders $\varphi \omega (\chi_i^2)^2$, $\varphi \omega
  (\chi_i^2)^2$ and $\partial_{\mu} \varphi \partial^{\mu} \varphi
  (\chi_i^2)^2$ we show the two scaling dimensions arising after mixing (the
  second one being associated to a rank-two logarithmic multiplet).}
\end{table}

\subsection{Non-susy-writable leaders}\label{sec:dimensions_non-susy-writable}

In section \ref{sec:nonsusywritableleaders}, we have identified only one
non-susy writable leader up to $\Delta= 12$ in $d = 6$. It is the leader
$(\mathcal{F}_6)_L$ of the Feldman operator $\mathcal{F}_6$. Given its
expression {\eqref{F6leader}} in Cardy fields, its anomalous dimension is studied
via the 6-point correlation function $\langle (\mathcal{F}_6)_L (p) \chi_i
(p_1) \chi_j (p_2) \chi_k (p_3) \chi_l (p_4) \chi_m (p_5) \chi_n (p_6)
\rangle$. Its leading anomalous dimension appears at two loops, from the first
diagram in Fig.~\ref{Feldman6corr} (see Appendix \ref{app:dimFk} for details),
and it is negative. The two-loop corrected dimension IR dimension of
$(\mathcal{F}_6)_L$ is given by:
\begin{equation}
  \Delta_{(\mathcal{F}_6)_L} = 12 - 3 \varepsilon - \frac{7}{9} \e^2 + O
  (\varepsilon^3) . \label{DF6L}
\end{equation}
This leader is becoming less irrelevant as $d$ gets smaller, and is another
candidate which might destabilize the SUSY RG flow, as we discuss in section
\ref{Who}.

In Appendix \ref{app:dimFk} we also considered anomalous dimensions of higher
Feldman leaders $(\mathcal{F}_k)_L$, finding at two loops
\begin{equation}
  \Delta_{(\mathcal{F}_k)_L} = 2 k - \frac{k}{2} \e - \frac{k (3 k - 4)}{108}
  \e^2 + O (\varepsilon^3) . \label{FkLgen}
\end{equation}
This confirms the original result of Feldman {\cite{Feldman}}. It should be
noted that Ref. {\cite{Feldman}} used the ``old'' formalism for computing
anomalous dimensions, working in the replicated basis with propagator
{\eqref{eq:G}}. The agreement shows that the ``old'' formalism is not wrong,
if one is careful. We believe however that our new formalism (working in the
Cardy basis in the vicinity of the Gaussian fixed point, distinguishing
leaders and followers, classifying leaders by their symmetry) is more
systematic, hence less error prone.

While we confirm Feldman's result {\eqref{FkLgen}} for the anomalous
dimension, we disagree with his conclusion that this implies instability of
the SUSY fixed point for an arbitrary small $\varepsilon$; see section
\ref{Who}.

\section{Scenarios for the loss of SUSY}\label{Who}

In the previous sections we carried out the program of classifying the leaders
of $S_n$ singlet perturbations, and we described many anomalous dimension
computations. This is a vast body of knowledge about the spectrum of
potentially destabilizing perturbations. On the basis of this
information we will now discuss possible scenarios for how SUSY may be lost
below a critical dimension $d_c$.

The lowest leaders in each of the three classes have dimensions
{\eqref{Bcal}}, {\eqref{dimchi4}}, {\eqref{DF6L}}. In this section we will use
them truncated to the known terms:
\begin{eqnarray}
  \text{susy-writable:} &  & \Delta_{\mathcal{B}} = 10 - \frac{11}{9}
  \varepsilon, \nonumber\\
  \text{susy-null:} &  & \Delta_{(\chi_i^2)^2} = 8 - 2 \varepsilon - \frac{8}{27}
  \varepsilon^2,  \label{knownDelta}\\
  \text{non-susy-writable:} &  & \Delta_{(\mathcal{F}_6)_L} = 12 - 3 \varepsilon
  - \frac{7}{9} \e^2 . \nonumber
\end{eqnarray}
{In Fig.~\ref{FigDeltas} we plot these scaling dimensions 
as a function of $d$ in the range of interest $3 \leqslant d
\leqslant 6$. In the same plot we show the marginality threshold line $\Delta
= d$.

\begin{figure}[h]\centering \includegraphics[width=400pt]{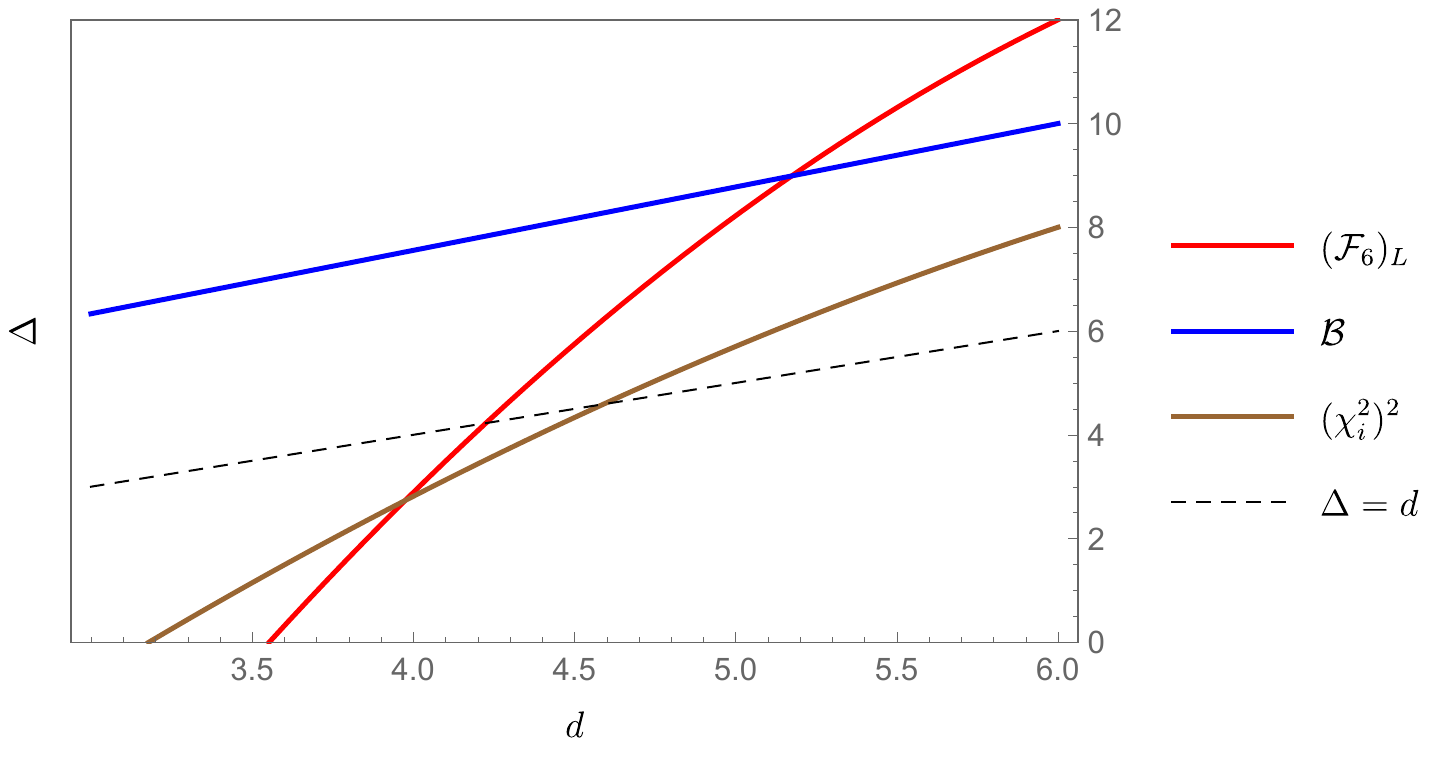}
  \caption{\label{FigDeltas}Scaling dimensions of the three lowest leaders
  (one per each class) as a function of the space dimension $d$, per Eqs.
  {\eqref{knownDelta}}.
}
\end{figure}

The immediate observation is that $\mathcal{B}$ does not become relevant in
this range of $d$, while $(\mathcal{F}_6)_L$ and 
$(\chi_i^2)^2$ do so at:
\begin{eqnarray*}
  \Delta_{(\chi_i^2)^2} = d &  & \quad \tmop{at} \quad d \approx 4.6,\\
  \Delta_{(\mathcal{F}_6)_L} = d &  & \quad \tmop{at} \quad d \approx 4.2.
\end{eqnarray*}
Taking this at face value, SUSY may be lost between $d=4$ and 5 because of these two perturbations.
 
The first uncertainty in this result is due to the shortness of available
perturbative series, which hopefully will be improved in the future by
higher-loop computations. E.g., using instead a $\textrm{Pad\'e}_{[1,1]}$ rational approximant for the conformal dimensions, we find that $\Delta_{(\chi_i^2)^2}$ crosses marginality at $d \approx 4.7$, while $ \Delta_{(\mathcal{F}_6)_L}$ at $d \approx 4.5$. This provides a rough idea of this uncertainty.\footnote{We thank \'Edouard Br\'ezin and Nicolas Sourlas for suggesting a Pad\'e check.}

Another uncertainty is associated with the fact that the coupling of the susy-null $(\chi_i^2)^2$, even if relevant, may flow to a nearby fixed point (``small loophole'' from section \ref{destab}), in which case SUSY may be preserved. But even if this happens, we still have the non-susy-writable leader $(\mathcal{F}_6)_L$, which will destroy SUSY at a nearby dimension.
}

Additional uncertainty may be due to the nonperturbative mixing with higher-dimension operators, as we will now
discuss.

By nonperturbative mixing we mean the following phenomenon. In perturbation
theory, mixing happens between operators of the same symmetry, with additional
selection rules that they should have the same number of fields and the same
scaling dimension in $d = 6$. Beyond perturbation theory, symmetry remains the
only selection rule. Let then $\Delta_1 (d)$ and $\Delta_2 (d)$ be dimensions
of two operators computed in perturbation theory, and suppose these two curves
intersect at some $d < 6$ (``level crossing''). If these two operators have
different symmetry, e.g. belong to different leader classes like in Fig.
\ref{FigDeltas}, then the level crossing is allowed also beyond perturbation
theory. On the other hand, if the two operators have the same symmetry, then
we should not believe the level crossing literally, as it will be modified by
nonperturbative mixing effects.

\begin{figure}[h]\centering \includegraphics[height=100pt]{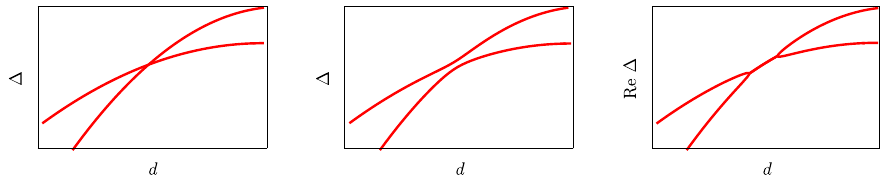}
  \caption[]{\label{FigCross}\tmtextit{Left:} Level crossing for two operator
  dimensions $\Delta_1 (d)$, $\Delta_2 (d)$ of the same symmetry in
  perturbation theory. \tmtextit{Center:} Crossing resolved via level
  repulsion (for norm of the same sign). \tmtextit{Right:} Crossing resolved
  via going to the complex plane (for norm of the opposite sign). These plots
  were obtained by diagonalizing the matrix $\left(\begin{array}{cc}
    \Delta_1 (d) & p\\
    p & \Delta_2 (d)
  \end{array}\right)$, where $p$ is a parameter characterizing the
  nonperturbative mixing strength, taken real or purely imaginary for norms
  of equal/opposite sign. \ }
\end{figure}

Normally, level crossing will be resolved via level repulsion (Fig.~\ref{FigCross}, center). Our theory being non-unitary, level crossing may also
be resolved via operator dimensions becoming complex conjugate (Fig.~\ref{FigCross}, right). Which of the two resolutions is realized depends on
the sign of the norm of the crossing operators (which is the same as the sign
of their two-point functions). Operators whose norm has the same sign will
repel (as is always the case in unitary theories), while for the norm of
opposite sign, dimension will go into the complex plane. If the mixing
operators have zero norm, as in the susy-null case, both options are possible.

After these general comments, let us see which of the curves in Fig.
\ref{FigDeltas} can be affected by nonperturbative mixing.

The susy-writable leader $\mathcal{B}$ could in principle mix
nonperturbatively with other susy-writable leaders coming from the box
representations, the first of which is the $10_{2, 0 \oplus 0, 2}$ in Eq.
{\eqref{class4d}}. The scaling dimension of the corresponding leader
$\mathcal{B}'$ is:\footnote{The one-loop correction is from
{\cite{Kehrein:1994ff}}, Table 4 (line ``(2,0),(0,2)'', $n = 6$). The number
in the table needs to be multipled by $\varepsilon / 3$.}
\begin{equation} \Delta_{\mathcal{B}'} = (12 - 3 \varepsilon)_{\tmop{class}} + \left(
   \frac{10}{3} \varepsilon \right)_{\text{1-loop}} + O (\varepsilon^2) = 12 +
   \frac{\varepsilon}{3} + O (\varepsilon^2) . \end{equation}
At $O (\varepsilon)$ the scaling dimension curves do not intersect, and we do
not consider this case any further.

\begin{figure}[h]
	\centering \includegraphics[width=280pt]{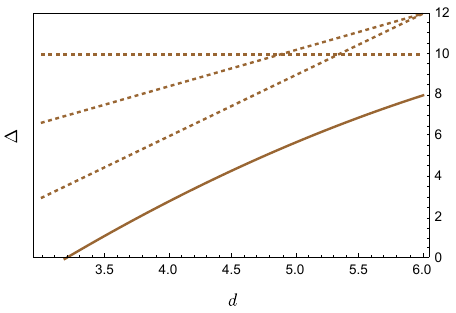}
  \caption{\label{FigNull}Scaling dimensions of susy-null leaders from Table
  \ref{null ops} (known terms).}
\end{figure}

The susy-null leader $(\chi_i^2)^2$ could in principle mix nonperturbatively
with other susy-null leaders shown in Table \ref{null ops}. We plot the
perturbative predictions for their scaling dimensions (to known order) in Fig.
\ref{FigNull}. We see that while the higher curves intersect among themselves,
they do not reach the lower $(\chi_i^2)^2$ curve.

\begin{figure}[h]\centering \includegraphics[width=300pt]{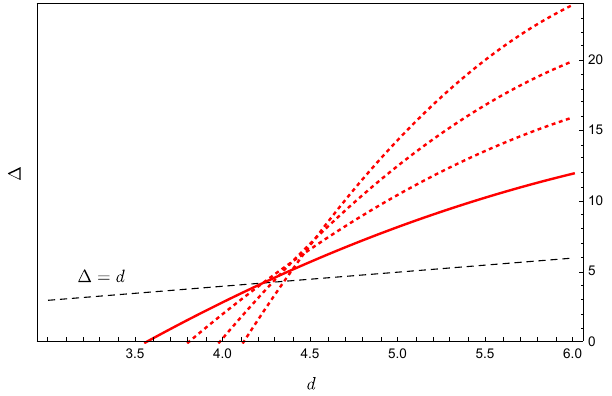}
  \caption{\label{FigFk}Perturbative scaling dimensions of the leaders of the
  first four Feldman operators.}
\end{figure}

Finally, the non-susy-writable sector $(\mathcal{F}_6)_L$ may mix
nonperturbatively with any of the higher non-susy-writable leaders. We know a
part of this higher spectrum, namely the two-loop dimensions of the higher
Feldman operators, Eq.~{\eqref{FkLgen}}. In Fig.~\ref{FigFk} we plot the
scaling dimension of the first four $(\mathcal{F}_k)_L$. This time we do see
level crossings. In fact, since the 2-loop correction is negative and grows
with $k$, it looks like the dimensions of $(\mathcal{F}_k)_L$ intersect the
dimensions of all lower $(\mathcal{F}_{k'})_L$. At least for the first few
Feldman operators, these crossings all appear to happen slightly above $d =
4$, close to the point where $(\mathcal{F}_6)_L$ becomes relevant.

\begin{figure}[h]\centering \includegraphics[width=250pt]{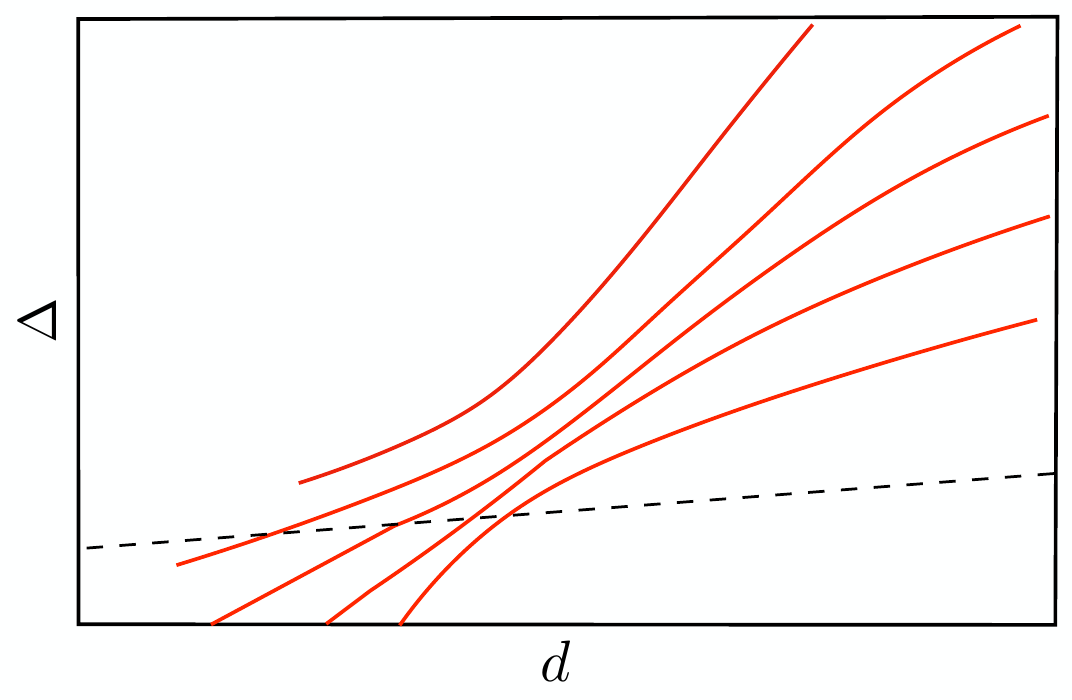}  
  \caption{\label{mixF1}Schematic dimension curves for Feldman operators after
  taking nonperturbative mixing into account. In making this figure we
  assumed the simplest scenario that all Feldman operators have the norm of
  the same sign and hence repel each other rather than go into the complex
  plane. It would be interesting to verify the norm sign in the future. For
  this one would have to determine the eigenperturbations and compute the sign
  of the two-point functions. This plot also does not take into account that
  there exist other non-susy-writable operators, with which higher Feldman
  operators are expected to mix beyond two-loop order, see section
  \ref{app:dimFk}.}
\end{figure}

This multitude of crossings deserves a discussion. The nonperturbative mixing
will likely repel the Feldman operators. As a result the lower Feldman
operator will become relevant at a slightly higher $d_c$ than without taking
mixing into account, while the higher Feldman operators will become relevant
at a slightly  lower $d$, if at all. SUSY will be valid for $d > d_c$. The
final picture may perhaps resemble that of Fig.~\ref{mixF1}. This deserves
further study, but we emphasize that any future discussion of this problem
should take nonperturbative mixing into account. It should also be remembered
that Feldman operators are but a small subset of all the non-susy-writable
operators which can be expected to take part in this mixing, see section
\ref{app:dimFk}.

Finally we would like to discuss what would happen if one naively extrapolated
Fig.~\ref{FigFk} to extremely high $k$. The two-loop result Eq.
{\eqref{FkLgen}} would seem to predict that $(\mathcal{F}_k)_L$ becomes
relevant for $\varepsilon_c \sim \sqrt{72 / k}$ which goes to zero for $k
\rightarrow \infty$. This was the argument advanced by Feldman
{\cite{Feldman}}, who thus concluded that arbitrarily close to $d = 6$ there
will be some sufficiently high $(\mathcal{F}_k)_L$ which is already relevant.
Hence, he argued, SUSY will be not present at any $d < 6$. We do not trust
this argument for two reasons. First, because it ignores nonperturbative
mixing discussed above. Second, because to have $\varepsilon_c$ small we would
have to consider very large $k$ indeed. E.g.~for $k = 50$ we still have
$\varepsilon_c = 1.01$ from the two-loop result. On the other hand the
coefficients of the perturbative series grow with $k$, as already visible in
{\eqref{FkLgen}}. Therefore the two-loop prediction at very large $k$ will be
trustworthy only in a tiny range of $\varepsilon$, and cannot by itself be
used to find where $(\mathcal{F}_k)_L$ becomes relevant and if this happens at
all.\footnote{See e.g. {\cite{Badel:2019oxl}} for a recent related discussion
of anomalous dimensions of composite operators made of many fields in the
Wilson-Fisher $4 - \varepsilon$ expansion. In this regard it is also
instructive to recall that in the $2 + \varepsilon$ expansion for the $O (N)$
vector model, operators with a large number of gradients $s$ acquire negative
anomalous dimensions at one {\cite{Wegner1990}} and two loops
{\cite{Castilla:1993zz,Castilla:1996qn}}. If one takes these terms too
literally, infinitely many such operators (namely all operators with $s \gg N
/ \varepsilon$) seem to become relevant, contrary to the usual expectation
that the Wilson-Fisher fixed point should have exactly one relevant $O
(N)$-singlet direction for any $2 < d < 4$. The general conclusion seems to be
that this paradox is due to applying $2 + \varepsilon$ expansion results
outside of their regime of validity
{\cite{Brezin-Hikami,Derkachov:1997qv,Derkachov:1997gc}}. } For these reasons,
we do not think it likely that infinitely many Feldman operators will cross or
approach the relevance threshold.

{To summarize, our computations suggest that Parisi-Sourlas SUSY will be lost
below a critical dimension $d_c \approx 4.2$ - $4.7$.}
Around this dimension, SUSY-breaking perturbations from two different symmetry
sectors (susy-null and non-susy-writable leaders) become relevant. In
particular, for integer dimension 5 all perturbations are irrelevant and
Parisi-Sourlas SUSY is expected to be present. For integer dimension 4, one
perturbation with a susy-null leader and at least one non-susy-writable
perturbations are relevant, and the RG flow is directed away from the SUSY
fixed point. The phase transition in the 4d RFIM is therefore not expected to
be supersymmetric.

\section{Discussion }\label{sec:conclusions}

In this paper we laid out a comprehensive framework to study RG stability of
the Parisi-Sourlas supersymmetric fixed point describing the phase transition
in the Random Field Ising Model. The key ingredients of our approach are:
\begin{itemizedot}
  \item We used the Cardy-transformed basis of fields $\varphi, \omega,
  \chi_i$, in which the RG flow looks manifestly as a Gaussian fixed point
  perturbed by a weakly-relevant interaction near the upper critical dimension
  6.
  
  \item We decomposed $S_n$-invariant perturbations (in the Cardy basis) into
  the leaders and followers. Scaling dimension of the leader then determines
  whether the whole perturbation is relevant.
  
  \item We classified the leaders into three classes by their symmetry
  (susy-writable, susy-null, and non-susy-writable).
  
  \item Susy-writable leaders were additionally classified as belonging to
  superprimary multiplets transforming in particular $\tmop{OSp} (d | 2
  \nobracket)$ representations.
  
  \item We enumerated all leaders up to 6d dimension $\Delta = 12$, and
  computed their perturbative anomalous dimensions (at one or two loops).
\end{itemizedot}
On the basis of the above, we identified two $S_n$-invariant perturbations
which become relevant at a critical dimension $d_c$ slightly above 4. In the
replicated field basis $\phi_i$, these perturbations correspond to
$\mathcal{F}_4$ and $\mathcal{F}_6$, where $\mathcal{F}_k = \sum_{i, j = 1}^n
(\phi_i - \phi_j)^k$ is the series of operators considered long ago by Feldman
{\cite{Feldman}}. Although looking similar in the replicated basis, in our
picture the two perturbations belong to two different classes: {$\mathcal{F}_4$
has a susy-null leader $(\chi_i^2)^2$, while $\mathcal{F}_6$ has a
non-susy-writable leader.}

The above conclusions were based on perturbative calculations of anomalous
dimensions around 6d, and on considering the lowest leaders in each class. In
the non-susy-writable class, perturbative calculations indicate level crossing
between $\mathcal{F}_6$ and the higher Feldman operators $\mathcal{F}_k$, $k >
6$. We discussed how this level crossing is expected to be resolved by
nonperturbative mixing effects, pushing slightly up the critical dimension
$d_c$ at which $\mathcal{F}_6$ will become relevant.

To summarize, the main features of our scenario for the loss of SUSY are:
\begin{enumeratenumeric}
  \item SUSY fixed point exists for any $3 < d \leqslant 6$.
  
  \item SUSY fixed point is stable for $d > d_c$ and unstable for $d < d_c$,
  where {$d_c \approx 4.2$ - $4.7$}.
\end{enumeratenumeric}
Our scenario is therefore different from the loss of SUSY in any $d < 6$
advocated for various reasons in some works such as
{\cite{ParisiLH,Parisi1992,Brezin-1998,Feldman}}.\footnote{See App. \ref{NPE}
concerning {\cite{ParisiLH,Parisi1992}}, App. \ref{BDD} concerning
{\cite{Brezin-1998}}, and section \ref{Who} concerning {\cite{Feldman}}.} It
is also different from the disappearance of the SUSY fixed point via fixed
point annihilation, found in Functional Renormalization Group (FRG) studies
{\cite{Tarjus1,TarjusIV,FRGother1,FRGother3,Balog:2019kxo}} at $d_c \approx
5.1$ (see App. \ref{FRG-RFIM}). For them, SUSY is absent for $d < d_c$ because
there is no longer any SUSY fixed point, while for us the SUSY fixed point
exists for any $d$, it just becomes unstable for $d < d_c$. [The value of
$d_c$ is also different but this is less important.] We do not see any signs
of fixed point annihilation in our picture. For us, this would require that a
SUSY-preserving operator crosses the relevance threshold. In the language of
section \ref{susy-irr}, this would be a scalar $\mathbb{Z}_2$-even
susy-writable leader, and as discussed there all such operators remain
irrelevant for all $d$.

Our work suggests two kinds of open problems: to explore our method further,
and to check our conclusions with other techniques. We discuss these in turn.

\subsection{Exploring our method further}

\subsubsection{Symmetry meaning of leaders in the Cardy
basis}\label{sym-meaning}

One aspect of our construction which deserves further thinking is the symmetry
understanding of leaders. Leaders were introduced as the lowest-dimension
components of $S_n$-singlet operators. We then argued (and checked
extensively) that leaders only mix with leaders under RG. If so there must be
a symmetry reason for this fact contained in the $\mathcal{L}_0$ Lagrangian
(without appealing to the original $S_n$ invariant Lagrangian), and it would
be interesting to identify such a reason. For susy-writable leaders we gave a
criterion (section \ref{sec:susywritableleaders}) that they become
supertranslation invariant when written in the SUSY field basis, and it would
be interesting to prove this more rigorously.

Another hint for such a symmetry may come from an interesting property which all leader operators have: their correlation functions vanish. One easy argument for this comes from considering correlation functions of $S_n$-singlets written in terms of the original replicated fields $\phi_i$ (namely without writing them in Cardy basis). It is indeed easy to prove that such correlation functions are proportional to $n$ and thus vanish in the $n \to 0$ limit. [This is related to the fact that the partition function of replicated theory is exactly 1 in the $n \to 0$ limit.] As an example let us consider $\langle \sigma_{k_1} \dots  \sigma_{k_m} \rangle$ in the free replicated theory (namely \eqref{Sr} with $V=0$). The result of this correlator can be written as a product of propagators \eqref{eq:G}, thus it is proportional to the trace of a product of the matrices $\mathds{1}$ and ${ \mathbf{M}}$ of \eqref{eq:G}. Since $\tr (\mathds{1} \cdots \mathds{1} \mathbf{M} \cdots \mathbf{M}) =O(n^a) $ with $a\geq 1$, the result must vanish as $n \to 0$. This argument generalizes straightforwardly for more generic $S_n$ singlets\footnote{This argument works unless we consider $S_n$-singlets rescaled by powers of $n$. In this paper we do not consider such rescaled operators since they would diverge in the $n\to 0$ limit. One may argue that e.g. $\sum_{i=1}^n \phi^k_i$ is also vanishing for $n\to 0$, and thus that dividing by $n$ is an allowed operation. However in Cardy variables  $\sum_{i=1}^n \phi^k_i$ is non-vanishing and for this reason in this formalism the rescaled operators are not allowed.} and it is also easy to see that when $V\neq0$, at any order in perturbation theory, the same property must hold, since the potential is also an $S_n$-singlet by construction. We thus conclude that correlation functions of $S_n$-singlets vanish. Since the leaders are lowest dimensional pieces of $S_n$-singlets, their correlators must vanish too.

But can we understand this vanishing of correlators directly in terms of leaders, without appealing to $S_n$-singlets from which the leaders originated? For susy-writable leaders we can, since they are the highest component of a supermultiplet, so, by supersymmetry, their correlation function must be zero (this is the same arguments that predicts $\langle \omega(x) \omega(0)\rangle=0$). Also correlation functions of susy-null leaders must vanish, since the operator are null (similarly this  happens for mixed correlation functions of  susy-null operators and susy-writable ones). A less trivial case is when non-susy-writable leaders are inserted in the correlation function. In this case SUSY arguments don't apply, and one has to find a new strategy to prove the statement.


\subsubsection{Higher-loop computations}

Perturbative computations of anomalous dimensions played an important role in
our considerations. In this work we relied on one- and two-loop predictions.
In comparison, 6- or 7-loop results are available nowadays for the leading
critical exponents in the Wilson-Fisher $4 - \varepsilon$ expansion
{\cite{Kompaniets:2017yct,Schnetz:2016fhy}}. It would be interesting to
compute higher loop anomalous dimensions of composite operators responsible
for destabilizing the SUSY fixed point. Resumming these series could lead to
improved determinations of $d_c$.

\subsubsection{Branched polymers and the Lee-Yang universality
class}\label{BP-LY}

One can consider the same problem as we studied in this paper, but replacing
the quartic potential $\phi^4$ with the imaginary cubic potential $i \phi^3$ \cite{papershort, paperphi3}.
As mentioned in section \ref{sec:intro} and reviewed in App. \ref{sec:LY},
this ``Random Field Lee-Yang Model'', with upper critical dimension 8,
describes the phase transition in the system of random polymers. Dimensional
reduction from $d$ to $d - 2$ dimensions in this case is verified to high
precision by Monte Carlo simulations.\footnote{As discussed in App.~\ref{sec:LY}, there is also a rigorous proof for dimensional reduction of
branched polymers {\cite{zbMATH02068689}}. Done for a model of branched
polymers preserving SUSY at the microscopic level, that proof does not shed
light on the stability of the SUSY fixed point with respect to SUSY-breaking
perturbations.} Therefore, we do not expect to find instability phenomena
that we found here for the RFIM case. In other words, leader anomalous
dimensions in the $i \phi^3$ theory should be such that they will not cross the
relevance threshold. It would be very interesting to verify this explicitly.
Note that both $\mathbb{Z}_2$ invariant and $\mathbb{Z}_2$ breaking leaders
should be considered in this computation, since $\mathbb{Z}_2$ is not a
symmetry of the Lee-Yang model.

\subsection{Checking our conclusions with other techniques}

\subsubsection{Numerical simulations}\label{numsimideas}

A feature of our scenario is that the Parisi-Sourlas SUSY fixed point exists
in any $d > 3$ (although it becomes unstable for $d < d_c$). In principle,
this could be tested in numerical lattice simulations. State-of-the-art
simulations study the RFIM phase transition at zero temperature, by tuning the
disorder distribution. In practice one considers a one-parameter family of
distributions with a fixed shape and varying overall strength (i.e.~dispersion). E.g.~Ref. {\cite{Picco1}} did this in $d = 4$ for the Gaussian
and Poisson disorders. In both cases they found identical, non-SUSY, critical
exponents (as reviewed in App. \ref{d=34}). If the SUSY fixed point exists in
$d = 4$, it should be possible to observe it by tuning within a family of
disorder distributions depending on a larger number of parameters (as many as
there are relevant perturbations at the SUSY fixed point). As discussed in
section \ref{Who}, it looks like at least two more perturbations in addition
to the SUSY mass become relevant in $d = 4$ ($\mathcal{F}_4$ and
$\mathcal{F}_6$). If that is the case, one would have to consider a
3-parameter family of distributions to find a SUSY fixed point in a lattice
simulation---a daunting task. But perhaps the higher-loop terms slightly
modify the behavior of anomalous dimensions on $d$, and only one extra
perturbation becomes relevant in $d = 4$, so that a 2-parameter family would
suffice. See App. \ref{app:tuning} for a schematic discussion of this
possibility.

Another question is whether one can measure some of the observables considered in this paper through a lattice simulation, e.g. in $d=5$, where the SUSY fixed point is expected to be reached. 
In particular it would be interesting to compute the scaling dimensions of the leaders of the $S_n$-singlet operators $\mathcal{F}_4$ and $\mathcal{F}_6$ and check that these are indeed irrelevant in $d=5$. 
Let us explain how this can be done.
Any correlation function of any operator in Cardy variables can be mapped to a correlator of the random field theory by first rewriting it in terms of the replicated fields $\phi_i$ following Eqs.~{\eqref{eq:Cardy1}},{\eqref{eq:Cardy2}}
and then by using the recipe in Eqs.~{\eqref{eq:Sr1}},{\eqref{form2}}. An example of such computation is done in (\ref{Gtmo}-\ref{conndisc}) and \eqref{kkkk}, but it can be easily generalized to any kind of operator, in particular to the leaders of $\mathcal{F}_4$ and $\mathcal{F}_6$.
It is however important to mention that correlation functions of both $S_n$-singlets and of their leaders are vanishing as we explain in section \ref{sym-meaning}. Hence one cannot extract the dimension of a leader by computing its two point function, since this would be zero. 
One should consider instead more complicated correlation functions involving non-singlet operators. E.g.~$\Delta_{(\mathcal{F}_6)_L}$ can be studied with the help of the 3-point function $\langle (\mathcal{F}_6)_L(x_1) \varphi(x_2) \varphi(x_3) \rangle$. On the other hand, the susy-null dimension $\Delta_{(\mathcal{F}_4)_L}$ may be accessible via through a 3-point function $\langle (\mathcal{F}_4)_L(x_1) \calO(x_2) \calO(x_3) \rangle$, where $\calO$ has to be non-singlet and in addition non-susy-writable, otherwise the result will vanish.\footnote{E.g. $\calO=(\chi_2)^2$ gives a nonzero 3-point function. On the other hand $\calO=\chi_2$ seems to give a vanishing 3-point function because the two point function $\langle \chi_2 \chi_2\rangle$ can be expressed in the susy variables as $\langle \psi \bar\psi\rangle$ times a prefactor, see \eqref{chi-chi}.}
These kind of observables would be very interesting to study. Even if higher point functions are significantly more complicated to compute in a lattice simulation, we hope that progress can be done in this direction. E.g. the same scaling dimensions $\Delta_{(\mathcal{F}_6)_L}$ and $\Delta_{(\mathcal{F}_4)_L}$ can be extracted from finite-volume corrections to scaling of the two point functions $\langle \varphi \varphi\rangle$ and $\langle \calO \calO\rangle$.

\subsubsection{Conformal bootstrap}\label{CB}

In recent years, the conformal bootstrap has emerged as a powerful method to
study nonperturbative CFTs in any dimension (see the review
{\cite{Poland:2018epd}}). Most of its successes have been for unitary CFTs,
which can be analyzed rigorously due to the unitarity bounds for the operator
dimensions and reality constraints on the OPE coefficients, as was first shown
in {\cite{Rattazzi:2008pe}}. These rigorous methods do not directly apply to
the Parisi-Sourlas SUSY fixed point which is non-unitary, as visible e.g. in
$\Delta_{\varphi}$ below the scalar unitarity bound, and in the violation of
spin-statistics by spinless fermions.

Scaling dimensions of susy-writable leaders could be obtained by looking at
the corresponding primaries in the Wilson-Fisher fixed point in $\hat{d} = d -
2$ dimensions. This is complicated by the fact that in integer $d$ some
interesting representations project to zero, while for non-integer $d$ the
Wilson-Fisher theory is also non-unitary {\cite{Hogervorst:2015akt}} (as
already mentioned in section \ref{susy-irr}). For us, the most interesting
susy-writable leader is the box operator $\mathcal{B}$, whose dimension equals
that of the Wilson-Fisher box primary $\hat{B}$. The number of components ($O
(\hat{d})$ irrep dimension) of $\hat{B}$ is given by
(see Eq.~\eqref{dim(2,2)}) $\dim (\hat{B}) = \frac{(\hat{d} + 2)
(\hat{d} + 1) \hat{d} (\hat{d} - 3)}{12} .$ We have \ $\dim (\hat{B}) = 10$
for $\hat{d} = 4$ consistently with box=$(2, 0) \oplus (0, 2)$ in that
dimension. Note that $\dim (\hat{B})$ vanishes for $\hat{d} = 3$, which means
that $\hat{B}$ does not exist in 3d, i.e.~$\mathcal{B}$ projects to zero under
dimensional reduction from $d = 5$ to 3. For $3 < \hat{d} < 4$, $\dim
(\hat{B})$ is positive. Perhaps in this range of $\hat{d}$ some trustworthy,
albeit non-rigorous, information about the scaling dimension of $\hat{B}$ can
be provided by the standard numerical bootstrap (applying it e.g. to the 4-point functions of spin-one operators or stress tensors).\footnote{While being still technically challenging to set up the conformal bootstrap for such operators in arbitrary dimensions, we might see developments in this direction in the near future. E.g. $U(1)$ currents and stress tensors were recently considered in $d=3$ \cite{Dymarsky:2017xzb, Reehorst:2019pzi, Dymarsky:2017yzx}.}

Apart from the lack of positivity, there are other complications in applying
the conformal bootstrap method to the RFIM. First, the fixed point of interest
exists only for $n = 0$, thus precluding the analytic continuation of CFT data
in $n$ (section \ref{sec:L2}). Second, the CFT is expected to contain
logarithmic multiplets (section \ref{s-null-dim}), for which logarithmic
conformal blocks need to be used instead of the usual conformal blocks
{\cite{Hogervorst:2016itc}}.

At present, the only bootstrap algorithm not relying on positivity and thus
applicable also to non-unitary CFTs is the one proposed by Gliozzi
{\cite{Gliozzi:2013ysa}}.\footnote{Its notable further investigations are
{\cite{Gliozzi:2014jsa,Gliozzi:2015qsa}}. See {\cite{Poland:2018epd}}, Section
IV.E for a review and more references.} In its present incarnation this
algorithm is not rigorous, as well as less systematic than the standard
numerical bootstrap algorithms. In spite of the lack of rigor and the
above-mentioned complications, it would be interesting to try to apply
Gliozzi's algorithm to the RFIM fixed point. This was attempted by Hikami
{\cite{Hikami:2018mrf}}, who found support for the loss of dimensional
reduction below $d_c \approx 5$. As we explain in App. \ref{Hikami}, more work
in this direction is needed to verify the robustness of Hikami's results and
to extract what they say about the mechanism for the loss of dimensional
reduction.

\subsubsection{Functional renormalization group}

Previous functional renormalization group (FRG) studies of the RFIM phase
transition, such {\cite{Tarjus1,TarjusIV,FRGother1,FRGother3,Balog:2019kxo}}
(see App. \ref{FRG-RFIM}) used the $S_n$-symmetric formalism. In principle, it
should also be possible to apply the FRG method in the Cardy field basis. For
that one would have to package an infinite series of singlet operators in a
general function. E.g., one could generalize Feldman operators to an
$S_n$-invariant interaction parametrized by a general function $R$:
\begin{equation} \sum_{i, j = 1}^n R (\phi_i - \phi_j), \end{equation}
which in the Cardy basis becomes
\begin{equation} 2 \sum_{i = 2}^n R (\omega - \chi_i) + \sum_{i, j = 2}^n R (\chi_i -
   \chi_j) . \end{equation}
[Separating it into a leader plus followers may not be necessary in a
nonperturbative framework such as FRG.] One could then derive an RG flow
equation for the function $R$, leading to the anomalous dimension predictions
for the Feldman operators. Note that this RG equation will be different from
that in the FRG studies of the interface disorder (section \ref{interface}),
because of the presence of the quartic coupling in our problem. Also unlike in
section \ref{interface}, the function $R$ for us does not have to satisfy any
conditions at infinity. It would be interesting to carry out this exercise, as
a way to confirm our expectations from section \ref{Who} about the mixing and
level repulsion of these operators.

\section*{Acknowledgements}

We are grateful to Connor Behan, Lorenzo Di Pietro and Johan Henriksson for discussions of
nonperturbative mixing. We also thank Edouard Br\'ezin, John Cardy, Marco Meineri, Giorgio Parisi, Leonardo Rastelli, Aninda Sinha, Nicolas Sourlas, Gilles Tarjus and Kay Wiese for various useful comments and discussions. 

This work was partially supported by the Simons Foundation grant 488655
(Simons Collaboration on the Nonperturbative Bootstrap) and by Mitsubishi
Heavy Industries (MHI-ENS Chair).

\appendix\section{History and prior work}\label{sec:history}

\subsection{Early work}\label{early}

That critical exponents of random-field $\phi^4$ theory in $d = 6 -
\varepsilon$ agree with Wilson-Fisher in $d = 4 - \varepsilon$ was first
proved by Aharony, Imry and Ma in 1976 {\cite{Aharony:1976jx}} (see also Young
{\cite{Young1977}}). They picked out a subclass of perturbative diagrams most
divergent in IR, and showed them identical to those of the Wilson-Fisher
theory with coupling $\lambda' = \lambda H$, for $(d - 2)$-dimensional
external momenta.

Parisi and Sourlas {\cite{Parisi:1979ka}} rephrased this calculation in terms
of supersymmetry. The diagrams in question being tree-level in $\phi$ (before
integrating over the random field), their sum can be done by solving the
classical equation of motion in external field $h$. This can be then neatly
interpreted as a path integral with insertion $\delta (- \Delta \phi + V'
(\phi) - h) \det (- \Delta + V'' (\phi))$. Introducing a Lagrange multiplier
field $\omega$ for the $\delta$-function and a pair of anticommuting scalars
to represent the determinant, one lands on a supersymmetric action.
Dimensional reduction is then argued in perturbation theory using
superpropagators and the identity $\int d^{d - 2} x\, F (Y_i x, x^2) = \int d^d
x d \theta d \bar{\theta}\, F (Y_i x, x^2 + \bar{\theta} \theta)$ where $Y_i$
are $d - 2$ dimensional vectors. The position space argument is thus easier than the momentum space one.

\subsection{No SUSY and no dimensional reduction in $d = 3, 4$ }\label{d=34}

The Parisi-Sourlas conjecture fails in $d = 3$: the $d = 3$ RFIM is ordered at
low temperatures (for weak disorder) by the Imry-Ma criterion
{\cite{ImryMapaper}}, while $d = 1$ Ising is of course disordered at all
temperatures and does not even have a phase transition. The ordering of 3d
RFIM was also proved rigorously by Imbrie {\cite{imbrieLCD2,ImbrieLCD}} ($T =
0$) and by Bricmont and Kupiainen {\cite{PhysRevLett.59.1829,BK}} (small $T$).

The transition is believed continuous in 3d. Numerical simulations can be
done at $T = 0$ varying the disorder strength to reach the transition. One
uses fast algorithms to find exact ground states for a collection of disorder
samples, and then performs disorder average. In this setup, a large-scale
study on cubic lattices with size up to $L = 192$ and with $10^7$ disorder
samples was performed by Fytas and Martin-Mayor {\cite{Fytas3}}. They found a
continuous transition with exponents $\nu \approx 1.4 (1)$, $\eta \approx
0.515 (1)$. Defining different exponents for the connected and disconnected
propagators as
\begin{equation} \overline{\partial \langle S_x \rangle / \partial h_y} \sim 1 / r^{d - 2 +
   \eta}, \quad \overline{\langle S_x \rangle \langle S_y \rangle} \sim 1 /
   r^{d - 4 + \bar{\eta}} \end{equation}
they find $\bar{\eta} \approx 2 \eta$. SUSY is ruled out as it would predict
$\bar{\eta} = \eta$. Correction to scaling exponent is \ $\omega = 0.52 (11)$.

The $d = 4$ RFIM and $d = 2$ Ising both have a phase transition but exponents
do not agree. The $d = 4$ RFIM exponents were measured precisely in
{\cite{Picco1}}, with results $\nu \approx 0.872 (6)$, $\eta \approx 0.1930
(13)$, \ $2 \eta - \bar{\eta} \approx 0.032 (2)$, $\omega \approx 1.3 (1)$. In
particular $\bar{\eta} \neq \eta$ and SUSY is ruled out.

\subsection{SUSY in $d = 5 ?$}\label{d=5}

The $d = 5$ RFIM study {\cite{Picco2}} found $\nu = 0.626 (18)$, $\eta = 0.055
(15)$, $2 \eta - \bar{\eta} = 0.058 (8)$, $\omega = 0.66 (15)$. Within error
bars, this is largely consistent with both SUSY $\bar{\eta} = \eta$ and with
the 3d Ising exponents ($\nu = 0.629971 (4)$, $\eta = 0.036298 (2)$, $\omega =
0.82966 (9)$ from the conformal bootstrap
{\cite{Kos:2016ysd,Simmons-Duffin:2016wlq}}).

Further evidence for SUSY and dimensional reduction in 5d RFIM was presented
in {\cite{Picco3}}, which simulated elongated hypercube geometries with $d -
2$ dimensions fixed at $L$ and 2 remaining ones at $R L$. For $R \rightarrow
\infty$ and $L$ fixed, SUSY imposes relations between connected and
disconnected propagators, and these relations were found to be satisfied in $d
= 5$ (working at $R = 5$). E.g.~three independent finite-volume correlation
lengths (which scale with $L$) were found equal at a percent level: for the
connected and disconnected propagators in 5d RFIM and for the 3d Ising.

\subsection{Branched polymers and the Lee-Yang universality
class}\label{sec:LY}

Another interesting case of dimensional reduction occurs for the statistics of
branched polymers which can be modeled  as connected clusters of $N$ points on
a lattice (``lattice animals''). Their number $P (N)$ and average size $R$
scales as $P (N) \sim N^{- \theta} \lambda^N$, $R \sim N^{\nu}$ where $\theta,
\nu$ are critical exponents and $\lambda$ is a lattice-dependent non-universal
constant. Lubensky and Isaacson {\cite{PhysRevA.20.2130}} proposed a
field-theoretic description for extracting these exponents from a scalar
theory in $n \rightarrow 0$ limit, similar to de Gennes's description of
self-avoiding walks but with extra cubic vertices breaking $O (n)$ symmetry to
$S_n$. Parisi and Sourlas {\cite{PhysRevLett.46.871}} then interpreted this as
the replica method limit for the random field model {\eqref{LGrfim}} with the
cubic potential $V (\phi) = m^2 \phi^2 + i g \phi^3$. Via supersymmetry,
critical exponents should be the same as for the Lee-Yang universality class
in $d - 2$ dimensions. The lattice animals exponents are known from precise
Monte Carlo simulations {\cite{LatticeAnimals}} for all $d < d_{uc} =
8$ and indeed they agree with the Lee-Yang exponents (known exactly or
approximately depending on $d$).

Brydges and Imbrie {\cite{zbMATH02068689}} found a model of branched polymers
which has manifest supersymmetry at the microscopic level (see also review
{\cite{Cardy-branched}}). In their model branched polymers are represented as
a gas of particles in $d$ dimensions with weight $\prod_{i \sim j} Q (r^2_{i
j}) \prod_{i \nsim j} P (r_{i j}^2)$ where $Q$ keeps neighboring nodes at a
distance $r \sim a$ apart, while $P$ repels all other nodes. Supersymmetry is
present if $Q (x) = d P (x) / d x$. In this case, the model can be written as
a classical gas of particles in $\mathbb{R}^{d | 2 \nobracket}$ interacting
with repulsive potential $V$, $e^{- V} = P$. Dimensional reduction follows, to
a classical repulsive gas in $\mathbb{R}^{d - 2}$. The latter model has a
critical point at negative fugacity which is one of two famous microscopic
realizations of the Lee-Yang universality class
{\cite{Repulsive-core,Park-Fisher}} (the other being the Ising model in
imaginary magnetic field {\cite{Fisher:1978pf}}). Brydges and Imbrie's result
is limited to their model of polymers and to the finetuned case of $Q = P'$.
It does not explain why supersymmetry should emerge in a generic model of
branched polymers (i.e.~why supersymmetry breaking deformations are
irrelevant). This explanation may come from repeating the analysis of our
paper for the cubic potential (see section \ref{BP-LY}).

\subsection{Zero-temperature fixed points}\label{T=0}

Long-distance behavior of disordered systems is often described in terms of
``zero-temperature fixed points''. Some features of these fixed points appeared
to us rather unusual, and we will attempt here a review for non-expert
audience, according to our own incomplete understanding. We found particularly useful
the original references {\cite{Bray1985,Fisher86}}, and section 8.5 of
{\cite{Cardy-book}}.

To set the notation, recall that the usual ``fixed point'' is a system with an
action $S (\{ \phi \}, \{ \lambda \})$ depending on a collection of fields $\{
\phi \}$ and couplings $\{ \lambda \}$, which remains invariant under an RG
transformation corresponding to the RG change of scale $x' = b x$, $b > 1$, in
the sense that all couplings are invariant: $\{ \lambda' \} = \{ \lambda \}$.
To exhibit this invariance one may have to reparametrize the fields after
performing RG transformation (e.g. rescale them). One of the RG invariant
couplings may be taken to be temperature itself. E.g., the fixed point
describing the ferromagnetic phase transition in the usual, non-disordered
Ising model has a fixed nearest-neighbor coupling $J$ which can be identified
also with the inverse temperature, as well as infinitely many other couplings
corresponding to the next-to-nearest and other possible $\mathbb{Z}_2$
invariant couplings, which may be neglected in an approximate treatment.

In the same notation, by a ``zero-temperature fixed point'' one means a system
whose action is written with an explicit $T^{- 1}$ factor, $\frac{1}{T} S (\{
\phi \}, \{ \lambda \})$, and whose behavior under RG amounts to the change
$T' = b^{- \theta} T$ with $\theta > 0$ a critical exponent, while all the
other couplings $\{ \lambda \}$ remain invariant. So $T \rightarrow 0$ as one
iterates RG. Two well-known examples are the low-temperature fixed points
describing the ordered phase of the Ising model, as well as of the $O (N)$
model with $N > 2$, $d \geqslant 3$. Both of these cases lead to simple
long-range behavior (gapped for Ising, Gaussian with massless Goldstones for
$O (N)$).

\begin{figure}[h]\centering \includegraphics[width=250pt]{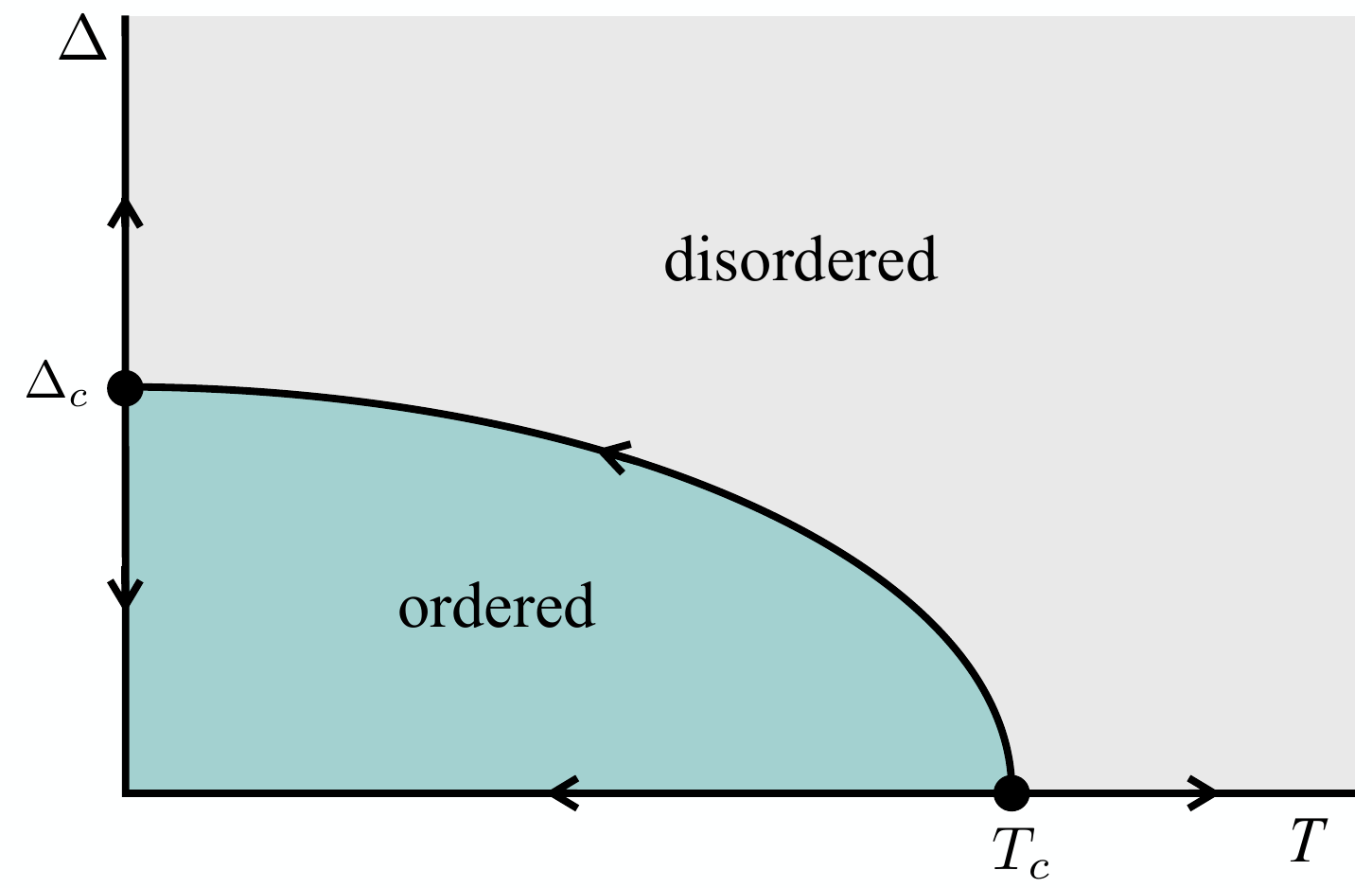}
  \caption{Phase diagram of the RFIM for $d > 2$. \label{TDelta}}
\end{figure}

What is unusual is that disordered phase transition are often described by
non-Gaussian zero-temperature fixed points. RFIM is one example. Rewriting the
Hamiltonian {\eqref{RFIM}} as $\frac{1}{T} \Bigl[ - \sum_{\langle i j \rangle}
s_i s_j + \sum_i h_i s_i \Bigr]$, $\overline{h_i^2 } = \Delta^2$, the $d > 2$
phase diagram in the space $(T, \Delta)$ is shown in Fig.~\ref{TDelta}. It
contains the usual non-disordered fixed point at $T = T_c$ which is unstable
under arbitrarily small disorder (Harris criterion) and flows to the
disordered fixed point at $T = 0$, $\Delta = \Delta_c$. This fixed point is a
zero temperature fixed point according to the above definition. The critical
exponent $\theta$ is known to be 2 in $d = 6 - \varepsilon$ to all orders,
while $\theta = 1 + \varepsilon / 2$ for $d = 2 + \varepsilon$
{\cite{Bray1985}}.

Physically, this means that the thermal fluctuations are negligible to those
due to disorder at the phase transition. This is very useful in numerical
simulations of the RFIM phase transition as it can be done at zero temperature
with $\Delta = \Delta_c$. One picks a large ensemble of disorder
representatives $\{ h_i \}$, for each of these one computes the ground state
(the configuration of spins with minimal energy $- \sum_{\langle i j \rangle}
s_i s_j + \sum_i h_i s_i$). This can be done in polynomial time with the
push-relabel algorithm {\cite{push-relabel}}, the solution is generically
unique, and the critical slow-down is not too bad
{\cite{Middleton,MiddletonFisher}}. These zero-temperature algorithms are used
in modern large scale numerical simulations
{\cite{Fytas3,Picco1,Picco2,Picco3}}.

Let us discuss how the zero-temperature fixed point concept is reconciled
with perturbation theory in $d = 6 - \varepsilon$ and Parisi-Sourlas SUSY
({\cite{Fisher86}} and {\cite{Cardy-book}}, section 8.5). Consider the
replicated action {\eqref{Sr}}. Restoring the temperature dependence, the
Lagrangian takes the form $ \frac{1}{T} \sum_{i = 1}^n \frac{1}{2}
(\partial_{\mu} \phi_i)^2 + \lambda_0 \phi^4_i - \frac{\tmH_0}{2 T^2} \left(
\sum_{i = 1}^n \phi_i \right)^2$ where $\overline{h (x) h (x')} = H_0 \delta
(x - x')$. Rescaling the fields we get the action as in {\eqref{Sr}} with
$\lambda = T \lambda_0$, $H = H_0 / T$. If we use the classical scalar
dimension in 6d, then an RG step with rescaling factor $b > 1$ will give
$\lambda' = b^{- 2} \lambda$, $H' = b^2 H$. We see that this corresponds to
$\lambda_0, H_0$ being invariant, while $T' = T / b^2$, so we flow to a
zero-temperature fixed point $(\theta = 2)$.\footnote{Note that in the replica
formalism zero-temperature fixed points are described with disorder-averaged
terms having higher order in $1 / T$.}

Let us next consider $d < 6$ and connect with the Cardy-transformed
description. In the basic scenario of section \ref{basicRG}, the IR fixed
point is described by the Lagrangian {\eqref{L0}} with $H$ and the quartic
$\lambda$ both RG-invariant, provided that the fields $f \in \{ \varphi',
\chi_i', \omega' \}$ are rescaled by $f' (x) = b^{\Delta_f} f (b x)$ where
$\Delta_{\chi} = \Delta_{\varphi} + 1$, $\Delta_{\omega} = \Delta_{\varphi} +
2$. If on the other hand we do the rescaling using $\Delta_{\chi}$ instead of
$\Delta_f$ for all three fields, the Lagrangian will retain its form with the
unit-normalized kinetic terms, but $H, \lambda$ will change as $H' = b^2 H$,
$\lambda' = \lambda / b^2$, which as we have seen above is equivalent to a
zero-temperature fixed point. We conclude that $\theta = 2$ in any $d < 6$
where Parisi-Sourlas SUSY holds.

We thus see that the zero-temperature fixed point can manifest itself in many
guises. In lattice simulations we can just set $T = 0$. On the other hand, in
the standard replicated description {\eqref{Sr}} we are forced to keep $T \neq
0$ even though it flows to zero (a coupling which flows to zero but cannot be
simply dropped is ``dangerously irrelevant'', footnote \ref{DIO}).
Alternatively, we are forced to keep the quartic coupling $\lambda_0$ in the
action although it is irrelevant and flows to zero, because it combines with
another coupling $H_0$ which flows to infinity. Finally, in the Cardy field
basis the zero-temperature fixed point looks like the usual non-zero
temperature one: nothing flows to zero or infinity provided that we use the
correct dimensions of fields in the SUSY multiplet. The latter property is of
course the chief reason why we used the Cardy basis throughout this paper.

But what if we decide, for the sake of the argument, to forego RG and set $T =
0$ directly in the Landau-Ginzburg description {\eqref{LGrfim}}? This would
mean that we have to find, for each random $h (x)$, the field $\phi_h$ solving
the classical EOM
\begin{equation}
  - \partial^2 \phi_h + V' (\phi_h) + h = 0, \label{classEOM}
\end{equation}
and having the minimal energy. This gives in fact the fastest way to
``derive'' Parisi-Sourlas SUSY. We represent the resulting path integral as
\begin{equation}
  \int \mathcal{D} \phi \,\delta (\phi - \phi_h) \mathcal{P} (h) \mathcal{D}h =
  \int \mathcal{D} \phi\, \mathcal{P} (h) \mathcal{D}h \,\delta (- \partial^2 \phi
  + V' (\phi) + h) \det (- \partial^2 + V'' (\phi)) .
\end{equation}
where we reexpressed $\delta (\phi - \phi_h)$ assuming the solution of
{\eqref{classEOM}} is unique. We can further rewrite this as
\begin{equation}
  \int \mathcal{D} \varphi \mathcal{D} \omega \mathcal{D} \psi \mathcal{D}
  \bar{\psi} \mathcal{P} (h) \mathcal{D}h\, e^{\int \omega (- \partial^2 \phi +
  V' (\phi) + h) + \int \psi (- \partial^2 + V'' (\varphi)) \bar{\psi}},
  \label{PS-path}
\end{equation}
and upon averaging over Gaussian $h$ we land on the SUSY Lagrangian. This was
in fact the derivation used originally by Parisi and Sourlas
{\cite{Parisi:1979ka}} except that they argued the localization to solutions
of {\eqref{classEOM}} based on the structure of most important terms in
perturbation theory, not on the fact that the fixed point is at zero
temperature. In retrospect, their argument proves that fixed point
\tmtextit{is} at zero temperature in perturbation theory {\cite{Fisher86}}.

By convexity arguments, Eq.~{\eqref{classEOM}} has a unique solution for any
$h$ as long as $m^2 \geqslant 0$. At the same time, for $m^2 < 0$ there are
multiple solutions for some $h$. This is discussed in Parisi's Les Houches
1982 lectures {\cite{ParisiLH}} (see also {\cite{Villain,Lancaster}}). The
bare mass at the phase transition is expected to be negative, and so multiple
solutions are always present at the short-distance scale. In presence of
multiple solutions, Parisi-Sourlas path integral {\eqref{PS-path}} can be
reduced to the sum over all solutions weighted by the determinant. This
deviates from what one originally wanted: just the minimal energy solution.
One can try to relate this deviation to nonperturbative effects, which may
lead to exponentially small $e^{- C / \varepsilon}$ deviations from
dimensional reduction already in $d = 6 - \varepsilon$
{\cite{ParisiLH,Parisi1992}}.\footnote{We thank Giorgio Parisi for a
discussion.} It appears that the constant $C$ has never been computed, which
makes the reality of these corrections somewhat nebulous. In fact, multiple
solutions may be a short-distance phenomenon, their effect renormalized away
when one flows to long distances. To see whether this scenario is realized,
one may wish to study the whole RG flow leading to the zero-temperature fixed
point and see if the RG fixed point is stable (rather than set $T = 0$ from
the start). That is exactly what we did in the main text of the paper.

\subsection{Interface disorder}\label{interface}

Another famous occurrence of disorder is in statistical physics of interfaces (see e.g.~lecture notes by Kay Wiese {\cite{Wiese-notes}} and by Leon
	Balents {\cite{Balents}}). Some aspects of this problem are analogous to RFIM, although there are also differences because the symmetry is not the same.

An interface is a scalar function $u (x)$, $x \in \mathbb{R}^d$, described by an
action
\begin{equation}
  \frac{1}{T} \int d^d x \left[ \frac{1}{2} (\partial u)^2 + V (x, u (x))
  \right], \label{Sinterface}
\end{equation}
where $V$ is a random potential. [E.g.~one may imagine the interface between
two phases of $d + 1$ dimensional Ising model at $T < T_c$, the disorder
potential $V$ coming from impurities.] We kept $T$ explicit in the action.
The potential is taken
Gaussian random with
\begin{equation}
  \overline{V (x, u)} = 0, \quad \overline{V (x, u) V (x', u')} = \delta^d (x
  - x') R (u - u') .
\end{equation}
As a consequence of the shift symmetry $u \rightarrow u + \tmop{const}$, the
variance depends on the difference $R (u - u')$, assumed an even function. One
is interested in computing various correlators of $u (x)$, e.g.
\begin{equation}
  \overline{(\langle u (x) \rangle - \langle u (y) \rangle)^2} \sim | x - y
  |^{2 \zeta} \label{roughness}
\end{equation}
at long distances, where $\zeta$ is called the roughness exponent .

Replica method then leads to the Lagrangian:
\begin{equation}
  \frac{1}{T} \sum_i \frac{1}{2} (\partial u_i)^2 - \frac{1}{2 T^2} \sum_{i j}
  R (u_i (x) - u_j (x)) . \label{ReplInt}
\end{equation}
Compared to the RFIM replicated Lagrangian {\eqref{Sr}} (with $\phi_i
\leftrightarrow u_i$), this Lagrangian lacks the individual potential $V
(u_i)$ which is forbidden by the shift symmetry. The term $\left( \sum_i u_i
\right)^2$ is reproduced for the quadratic $R (u)$, because $\sum_{i j} (u_i -
u_j)^2 = - 2 \sum_{i j} u_i u_j = - 2 \left( \sum u_i \right)^2$ in the $n
\rightarrow 0$ limit. For $R (u) = u^k$, the perturbation in {\eqref{ReplInt}}
is nothing but the Feldman operator $\mathcal{F}_k$ which played such an
important role in our paper. Applying the Cardy transform $u_1 = \varphi +
\omega / 2$, $u_i = \varphi - \omega / 2 + \chi_i$ $\left( i = 2, \ldots, n,
\sum \chi_i = 0 \right)$ gives, in the $n \rightarrow 0$ limit, the Lagrangian
\begin{eqnarray}
  &  & \frac{1}{T} \left[ \partial \varphi \partial \omega + \frac{1}{2} \sum
  (\partial \chi_i)^2 \right] + \frac{R'' (0)}{2 T^2} \omega^2 - \nonumber\\
  &  & \quad - \frac{R^{(4)} (0)}{8 T^2} \left[ \left( \sum \chi_i^2
  \right)^2 - \frac{4}{3} \omega \sum \chi_i^3 \right] + \ldots 
  \label{CardyInt}
\end{eqnarray}
The leading Lagrangian (first line of
{\eqref{CardyInt}}) is a free SUSY theory. The second line is the Feldman $\mathcal{F}_4$ \eqref{SNullEx} in the Cardy basis, with a susy-null leader $\left( \sum \chi_i^2
\right)^2$. In our paper we had to calculate the scaling dimension of this perturbation working in the $6-\eps$ expansion, but in the interface problem the SUSY fixed point is Gaussian (because the shift symmetry forbids the quartic interaction), and we can just read off the leader scaling dimension as $\Delta=2d-4$, which becomes relevant for $d<4$. Thus, according to the discussion in section \ref{destab}, we expect that at $d<4$ the Gaussian SUSY fixed point may be destabilized.

Furthermore, the problem at hand provides a nice illustration of how to deal with the ``small loophole'' mentioned in section \ref{destab}, namely that we have to consider the nonlinear terms in the beta-function to see whether the susy-null coupling actually grows to become $O(1)$, or perhaps flows to a nearby fixed point, in which case the SUSY would not really be broken. Denoting by $g=- \frac{R^{(4)} (0)}{8 T^2}$ the $\left( \sum \chi_i^2
\right)^2$ coupling, the beta-function at marginality (i.e.~in 4d) has the form:
\begin{equation}
	\beta_g=C g^2\,, \label{betag0}
\end{equation}
where $C$ is a positive order 1 constant. It turns out that the initial conditions for the RG evolution in the interface problem are always $g(\Lambda_{\rm UV})<0$ at the microscopic scale \cite{Wiese-notes}.\footnote{We thank Kay Wiese for emphasizing this point to us.} With such an initial condition and with $C>0$ in \eqref{betag0}, the RG evolution leads to a blowup of the coupling $g$ in the IR (similarly to how the QCD gauge coupling formally blows up at $\Lambda_{\rm QCD}$). We conclude that in the interface problem the ``small loophole'' does not realize, and once the $\left( \sum \chi_i^2 \right)^2$ passes the marginality threshold, the Gaussian SUSY fixed point is truly destabilized.

It should be said that we do not know if the same will happen when $\left( \sum \chi_i^2 \right)^2$ crosses the marginality threshold in our RFIM problem, because we have not calculated the sign of the beta-function (which may be different in our case due to quantum corrections) and we do not know if the initial sign of the coupling will be the same in our problem as for the interface.

Going back to the interface problem, while the SUSY fixed point instability can be easily understood using our language, there remains the question where the flow goes. For the interface, this question has been studied using FRG, by writing the RG equation for the whole function $R(u)$ \cite{DanFisher} rather than performing the expansion as in \eqref{CardyInt}.\footnote{Note that, physically, the function $R (u)$ must go to
	zero at large $u$. E.g.~we might have $R (u) \sim e^{- u^2}$ at the UV cutoff
	scale for random-bond type of disorder in the underlying Ising model which
	hosts the interface, while for random-field type disorder we have instead $R''
	(u) \rightarrow 0$ at large $u$ {\cite{Wiese-notes}}. The requirement that $R(u)$ go to zero implies nontrivial correlations between the coefficients of its Taylor series. E.g.~if we set all expansion
	coefficients with $k \geqslant k_0$ to zero, the resulting $R (u)$ is a
	polynomial, growing at infinity, which is not allowed by physical
	constraints.}

Consider the replica action {\eqref{ReplInt}} with
$n = 0$ and some UV cutoff scale $\Lambda$. It is convenient to rescale $R
= \Lambda^{\varepsilon} \bar{R}$. Then, the one-loop RG equation for $R
(u)$ in $d = 4 - \varepsilon$ has the form
\begin{equation} - \frac{d \bar R}{d \log \Lambda} = \varepsilon \bar R (u) + \tilde C \left[  \frac{1}{2}
   [\bar R'' (u)]^2 - \bar R'' (u) \bar R'' (0) \right], 
\end{equation}
where $\tilde C$ is an order 1 positive dimensionless coefficient (it is proportional to $C$ in \eqref{betag0}). 
Starting the RG evolution with an analytic
function $R (t)$ having a physically acceptable behavior at infinity, as well as having $R^{(4)}(0)>0$ (which is always the case for the random interface), one finds that the function $\Delta(u)=- R'' (u)$ develops a
cusp at $u = 0$, i.e.~$\Delta (u) = c_0 + c_1 | u | + \ldots$ at small $t$. This cusp appears at the scale where $R^{(4)}(0)\to \infty$, which we found above based on \eqref{betag0}.

Once the cusp forms, one would be tempted to declare that the problem became strongly coupled and no further analytic progress is possible. That is what happens in perturbative QCD, where below $\Lambda_{\rm QCD}$, when the gauge coupling becomes strong, one has to resort to lattice Monte Carlo simulations or descriptions in terms of an effective action using very different degrees of freedom from those in the UV, and whose parameters cannot be determined from the first principles.

It is somewhat surprising however, that the literature on the disordered interfaces claims to be able to go beyond the cusp formation point (see \cite{Wiese-notes}).
One looks for a self-similar solution of the
RG equation in the form
\begin{equation}
  R (u) = \Lambda^{- 4 \zeta} K (\Lambda^{\zeta} u) \label{RK}
\end{equation}
which gives a fixed point equation for $K (t)$
\begin{equation}
  0 = (\varepsilon - 4 \zeta) K (t) + \zeta t K (t) + C \left[  \frac{1}{2}
  [K'' (t)]^2 - K'' (t) K'' (0) \right] \label{eqK}
\end{equation}
This equation turns out to have nontrivial solutions with ``cuspy'' $K'' (t)$,
for an appropriately chosen value of $\zeta$.

With $R$ on the self-similar trajectory {\eqref{RK}}, redefining the variable
$u (x) = \Lambda^{- \zeta} \bar{u} (\Lambda y)$, effective action in terms of
the $\bar{u}$ field (which has momenta of order 1) takes the same form as
{\eqref{ReplInt}} but with a rescaled temperature
\begin{equation} T \rightarrow \alpha T, \quad \alpha = \Lambda^{d + 2 \zeta - 2} < 1 \end{equation}
So the effective temperature goes to zero at long distances, while $\zeta$ in
{\eqref{RK}} may be identified with the roughening exponent, giving values in agreement with Monte Carlo simulations.

\subsection{Functional renormalization group studies of RFIM}\label{FRG-RFIM}

Tarjus, Tissier and collaborators (Tarjus et al in what follows) applied the
functional renormalization group (FRG) method to the RFIM phase transition
{\cite{Tarjus1,TarjusIV,FRGother1,FRGother3,Balog:2019kxo}}. Here we will
attempt to review some aspects of their work in spite of the fact that we
understand it only partially, and compare to our approach.\footnote{We thank
Gilles Tarjus for a discussion.}

Tarjus et al ({\cite{Tarjus1}}, Section 2.C) allow for unequal sources for
different replicas, which they contrast with the conventional replica
approaches using, they say, equal sources. We tend to disagree that this
difference is so crucial. The usual replica formalism allows to describe all
experimentally observable correlators, see Eq.~{\eqref{form2}}. The
Cardy-transformed Lagrangian used in our work is equivalent to the usual
replica Lagrangian, as long as one does not drop any terms without due RG
justification. In this paper we talked about fields and correlators, which is
of course equivalent to introducing sources and differentiating with respect
to them, although we did not find it necessary to stress this standard part of
QFT dictionary explicitly.

In the rest of this appendix we will comment on Ref.
{\cite{TarjusIII,TarjusIV}}, devoted to the loss of Parisi-Sourlas SUSY. As
far as we can see, this FRG analysis is applied to an action arrived at by not
fully justified assumptions. Namely, one first derives the Parisi-Sourlas
Lagrangian using the original Parisi-Sourlas argument. Then one observes that
the obtained result is wrong due to multiple solutions to the classical
equation of motion. It is then proposed to fix this via an auxiliary parameter
$\beta$ providing a Boltzmann suppression for the extra contributions, see
{\cite{TarjusIII}}, Eq.~(28).\footnote{This is somewhat similar in spirit to
the modification of mean-field theory considered in {\cite{Lancaster}}.} While
such a direct modification of the Parisi-Sourlas Lagrangian may appear
physically reasonable, it does not seem a first-principle derivation from the
replicated action. This should be contrasted with our approach, where the
terms $\mathcal{L}_1, \mathcal{L}_2$ modifying the Parisi-Sourlas Lagrangian
came from an explicit Cardy transform.

This line of reasoning leads Tarjus et al to a theory (see
{\cite{TarjusIII}}, Eq.~(42)) with $N$ superfields\footnote{Whose number is
denoted by $n$ in Ref {\cite{TarjusIII}}.} $\Phi_a (x, \theta, \bar{\theta})$,
$a = 1, \ldots, N$ and the action ($\int_{\theta} \assign \int d \bar{\theta}
d \theta (1 + \beta \theta \bar{\theta})$
\begin{equation}
  \sum_{a = 1}^N \int d^d x \int_{\theta} \left[ \frac{1}{2} (\partial_{\mu}
  \Phi_a)^2 + V (\Phi_a) \right] - \frac{H}{2} \sum_{a, b = 1}^N \int d^d x
  \int_{\theta_1} \int_{\theta_2} \Phi_a (x, \theta_1, \overline{\theta_1})
  \Phi_b (x, \theta_2, \overline{\theta_2}) . \label{Taction}
\end{equation}
As mentioned it is not clear to us if this action is correct in the first
place. Nevertheless, for the sake of comparison, let us try to get a similar
action from our point of view. We will not fully succeed, but we will learn
some interesting lessons along the way. Take our replicated action
{\eqref{Sr}} and replace $n \rightarrow n N$ for a fixed integer $N$. The
limits $n \rightarrow 0$ and $n N \rightarrow 0$ being equivalent, imagine
that we have $N$ groups of $n \rightarrow 0$ fields, and apply the Cardy transform
in each group separately. We will get a Lagrangian for fields $\varphi_a,
\chi_{a, i}, \omega_a,$ $a = 1 \ldots N,$ $i = 2 \ldots n$. The kinetic term
in the $n \rightarrow 0$ limit is
\begin{equation}
  \sum_{a = 1}^N \left\{ \del \varphi_a  \del \omega_a + \frac{1}{2}  \sum\nolimits'
  (\del \chi_{a, i})^2 \right\} - \frac{\tmH}{2}  \left( \sum_{a = 1}^N
  \omega_a \right)^2 . \label{kinN}
\end{equation}
Assume now for a second that the fields $\varphi_a, \chi_{a, i}, \omega_a$ for
each $a$ have the same scaling dimensions as those of $\varphi, \omega,
\chi_i$ given in {\eqref{dimensions}} [this is not quite correct, see below].
Then, dropping the interaction terms irrelevant in $d = 6 - \varepsilon$ with
these scaling dimension assignments, we would get the interaction Lagrangian
\begin{equation}
  \sum_{a = 1}^N \left\{ V' (\varphi_a) \omega_a + \frac{1}{2} V'' (\varphi_a)
  \sum\nolimits' \chi_{a, i}^2 \right\} . \label{intN}
\end{equation}
If we now introduce $N$ supermultiplets $\Phi_a = (\varphi_a, \psi_a,
\bar{\psi}_a, \omega_a)$, and replace the $\chi_a$-bilinears by $\psi_a
\bar{\psi}_a$ ones, the sum of {\eqref{kinN}} and {\eqref{intN}} maps on the
Lagrangian
\begin{equation}
  \sum_{a = 1}^N \left\{ \del \varphi_a  \del \omega_a + \partial \psi_a
  \partial \bar{\psi}_a + V' (\varphi_a) \omega_a + V'' (\varphi_a) \psi_a
  \bar{\psi}_a \right\} - \frac{\tmH}{2}  \left( \sum_{a = 1}^N \omega_a
  \right)^2,
\end{equation}
which can be also written in terms of superfields as ($A |_{\theta
\bar{\theta}} \nobracket : = \int d \bar{\theta} d \theta A$)
\begin{eqnarray}
  &  & \sum_{a = 1}^N \left[ \frac{1}{2} \partial_{\mu} \Phi_a \partial_{\mu}
  \Phi_a + V (\Phi) \right]_{\theta \bar{\theta}} - \frac{\tmH}{2}  \left(
  \sum_{a = 1}^N \Phi_a |_{\theta \bar{\theta}} \nobracket \right)^2 . 
\end{eqnarray}
This would correspond to the $\beta = 0$ case of the Tarjus et al action
{\eqref{Taction}}. It has $N$ independent supertranslation invariances, one
for each supermultiplet.

However, it was incorrect to assign to $\varphi_a, \chi_{a, i}, \omega_a$ for
each $a$ the same scaling dimensions as for $\varphi, \chi_i, \omega$. We had
to diagonalize the kinetic part before assigning scaling dimensions, and
{\eqref{kinN}} is not fully diagonalized, since $\omega_a$ appear coupled
through $\left( \sum \omega_a \right)^2$. For a better treatment, we have to
introduce fields $\omega_0 = \sum \omega_a$, $\varphi_0 = \frac{1}{N} \sum
\varphi_a$, $\tilde{\omega}_a = \omega_a - \frac{1}{N} \omega_0$,
$\tilde{\varphi}_a = \varphi_a - \varphi_0$ ($\sum \tilde{\varphi}_a = \sum
\tilde{\omega}_a = 0$) in terms of which the kinetic Lagrangian takes the form
\begin{equation}
  \del \varphi_0  \del \omega_0 - \frac{\tmH}{2} \omega_0^2 + \sum_{a = 1}^N
  \left\{ \del \tilde{\varphi}_a  \del \tilde{\omega}_a + \frac{1}{2}  \sum\nolimits'
  (\del \chi_{a, i})^2 \right\} .
\end{equation}
Therefore, the pair of fields $\varphi_0, \omega_0$ has scaling dimensions
like $\varphi, \omega$ while all the other fields ($\chi_{a, i}$,
$\tilde{\varphi}_a, \tilde{\omega}_a$) should be assigned scaling dimensions
$d / 2 - 1$. With the new dimension assignments, more terms in the interaction
Lagrangian {\eqref{intN}} become irrelevant and should be dropped (while no
previously dropped terms became relevant). The only remaining terms are:
\begin{equation}
  V' (\varphi_0) \omega_0 + V'' (\varphi_0) \left\{ \sum_{a = 1}^N
  \tilde{\varphi}_a \tilde{\omega}_a + \frac{1}{2} \sum\nolimits' \chi_{a, i}^2
  \right\} .
\end{equation}
Now, the total number of fields $\tilde{\varphi}_a, \tilde{\omega}_a, \chi_{a,
i}$ is $n N - 2 \rightarrow - 2$ in the $n \rightarrow 0$ limit, and their
effect can be reproduced \ by a single pair of fermions $\psi_0,
\bar{\psi}_0$. We then end up with a theory containing one supermultiplet
$(\varphi_0, \psi_0, \bar{\psi}_0, \omega_0)$, not $N$ supermultiplets like
Tarjus et al.

This discussion suggests that splitting $n \rightarrow 0$ fields into $N$
groups does not add new effects when using the Cardy transform, provided that one
correctly identifies scaling dimensions. As we mentioned several times in this
paper, the Cardy transform is just a change of the field basis, and all bases
should be equivalent no matter how one slices and dices the fields, as long as
we do not drop any terms without justification.

We next discuss the terms in the Tarjus et al action {\eqref{Taction}} which
appear for nonzero $\beta$. Focusing on $N = 1$, the full Lagrangian in
superfield components takes the form:\footnote{As observed in
{\cite{TarjusIII}} this action retains a supertranslation invariance even for
nonzero $\beta$, corresponding to Killing vectors of the curved superspace
with a constant-curvature metric $d \bar{\theta} d \theta (1 + \beta \theta
\bar{\theta})$.}
\begin{equation}
  \mathcal{L}_{\tmop{SUSY}} + \beta \left[ \frac{1}{2} (\partial \varphi)^2 +
  V (\varphi) \right] - \frac{H}{2} (\beta \omega \varphi + \beta^2 \varphi^2
  \nobracket) \nobracket . \label{betane0}
\end{equation}
Recall that in our picture the Parisi-Sourlas Lagrangian follows from the
replicated action in the $n \rightarrow 0$ limit, and the underlying $S_n$
invariance has to be respected. It therefore appears to us worrisome that the
extra terms in {\eqref{betane0}} are not $S_n$ singlets according to our
classification. The $S_n$ invariance is thus explicitly broken by these
$\beta$-terms. For $N > 1$ action {\eqref{Taction}} preserves only $S_N$
invariance, while we would insist that the full $S_{n N}$ has to be preserved
at the microscopic level.

After we reviewed all the worries that we have about action {\eqref{Taction}},
let us describe the results that Tarjus et al derive from this action. In
{\cite{TarjusIV}}, Section III, they describe loss of SUSY in terms of its
``spontaneous breaking''. Since their action {\eqref{Taction}} is not fully
supersymmetric in the first place (superrotation invariance being broken by
the $\beta \neq 0$ terms), it is not clear to us why it is legitimate to talk
about spontaneous breaking (in the standard high-energy physics terminolgy the
situation at hand would be called explicit SUSY breaking). Recall also that
the usual spontaneous SUSY breaking is associated with the appearance of
massless particles (goldstinos), which does not appear consistent with the
RFIM phenomenology.

When Tarjus et al derive the FRG equations they set $\beta = \infty$, in
which limit the equations involve only the bosonic components of the
superfields, a property they call ``Grassmannian ultralocality''. They claim
they have an argument that in this limit supersymmetry is restored, which
seems to us rather counterintuitive, but we should admit that our
understanding of this part of their work is very limited. Working with the FRG
for the second cumulant, they give further arguments relating the loss
dimensional reduction to an appearance of a ``cusp'' in this quantity. Finally
by solving the FRG numerically, they do find a cusp below $d_c \approx 5.1$
{\cite{TarjusIV}}. While in {\cite{TarjusIV}} they attribute the loss of
dimensional reduction to ``spontaneous SUSY breaking'' (see our misgivings
above), in subsequent work {\cite{Baczyk,Balog:2019kxo}} they ascribe it
instead to an annihilation of two SUSY fixed points.

Note that in more familiar situations where two fixed point of the same
symmetry annihilate, they disappear into the complex plane (e.g. in the 2d
Potts model where a stable and an unstable fixed points annihilate for $Q =
4$, for $Q > 4$ there is no fixed point and the transition is weakly first
order, see {\cite{Gorbenko:2018ncu,Gorbenko:2018dtm}} for a recent
discussion). This is not what is found in the FRG work where two SUSY fixed
points existing at $d > d_c$ are claimed to annihilate and yet a third,
non-SUSY, fixed point emerges at $d < d_c$. It would be interesting to
understand what makes their scenario consistent, from the point of view of the
quantum numbers of the operator crossing the marginality bound (which should
be a full SUSY singlet if the annihilating fixed points are both SUSY, and it
is then unclear how it can give rise to a non-SUSY fixed point at $d < d_c$).

This concludes our outline of the FRG RFIM studies by Tarjus et al. In the
next section we will try to make contact with their work {\cite{Tarjus_2016}},
where cuspy interactions were discussed in the context of perturbative
expansion in $d = 6 - \varepsilon$.

\subsection{Comments on the perturbative ``cusp operators''}\label{cusp}

In sections \ref{interface} and \ref{FRG-RFIM} we saw that ``cusp''
interactions appear to play a role in the nonperturbative descriptions of
disordered fixed points. We are happy to admit that this might well be the
case in the FRG context, not being experts in that technique. We feel more
confident however in perturbative QFT aspects, and here we wish to comment on
Ref. {\cite{Tarjus_2016}}, which considered a cusp operator in the context of
a perturbative expansion around a Gaussian fixed point in $d = 6 -
\varepsilon$ dimension. Working with the usual replicated RFIM action
{\eqref{Sr}}, this ``cusp'' operator was defined in {\cite{Tarjus_2016}} in
terms of the replica fields as (see their Eq.~(5))
\begin{equation}
  \mathcal{C}= \sum_{i, j = 1}^n \phi_i \phi_j | \phi_i - \phi_j | .
  \label{cuspy}
\end{equation}
Before we discuss how they deal with this operator, let us consider a simpler
case of a single free massless scalar $\phi$ (in any $d > 2$). In this
Gaussian theory, the full spectrum of perturbations $\mathcal{O}_i$ with
well-defined scaling dimensions is given by normal ordered products of $\phi$
and its derivatives. Operators involving non-analytic functions of $\phi$,
such as $| \phi |$, are not in that list. Now one can ask, what if one does
consider a correlation function involving $| \phi |$? E.g.~one can imagine
measuring $\langle \nobracket | \phi | (x) | \phi | (y) \rangle$ in a Monte
Carlo simulation in a lattice-discretized $(\partial \phi)^2$ theory (or with
another UV cutoff). What would be a behavior of this correlation function at
large distances where the theory is scale invariant? The answer to this
follows from the fact that the operator $| \phi |$ will have an expansion
\begin{equation}
  | \varphi | (x) = \sum_{i = 1}^{\infty} a_i \mathcal{O}_i (x),
\end{equation}
where $\mathcal{O}_i$ are the above operators with well-defined scaling
dimension. In addition, since $| \phi |$ is $\mathbb{Z}_2$-even, only
$\mathbb{Z}_2$-even operators will appear on the r.h.s. We will have
$\mathcal{O}_1 = 1$ (unit operator), $\mathcal{O}_2 =\, :\! \phi^2\! :$, etc. So
subtracting the constant, the leading behavior of the connected two-point
function of the operator $| \phi |$ will be the same as for $\phi^2$ (see
section \ref{2ptabs} below for a proof via an explicit computation). This just
illustrates that we do not enlarge the spectrum of scaling dimensions by
considering non-polynomial operators. This should not be surprising: e.g. when
we study the spectrum of perturbations of the Wilson-Fisher fixed points, we
always consider only polynomial interactions and compute their anomalous
dimensions. If we had to consider non-polynomial operators, there would be
many more anomalous dimensions to compute, and there is no evidence that this
is necessary, from theory, experiment, or simulations.

For similar reason, we believe that operator {\eqref{cuspy}} does not exist as
a scale-invariant perturbation of the RFIM replicated Gaussian fixed point in
$d = 6 - \varepsilon$ dimensions. At long distances, this operator should be
expandable in polynomial operators considered by Br{\'e}zin and De Dominicis
{\cite{Brezin-1998}}, Feldman {\cite{Feldman}}, and other operators that we
studied in our work.

On the other hand, Ref. {\cite{Tarjus_2016}} does consider operator
{\eqref{cuspy}} as an independent perturbation of classical scaling dimension
$\Delta_{\mathcal{C}}^0 = d + 1 - \varepsilon / 2 = 7 - 3 \varepsilon / 2$
(see e.g. their Eq.~(9)), without explaining in detail how they arrived to
this dimension (no correlator which would correspond to such a scaling is
exhibited). Given that we do not understand the origin of this classical
dimension, and in fact oppose the very existence of $\mathcal{C}$ as a
scale-invariant perturbation, we will not enter into the discussion of how
Ref. {\cite{Tarjus_2016}} computes the anomalous dimension of $\mathcal{C}$.

Finally we would like to show a property of the Feldman operators
$\mathcal{F}_k$, which might have some positive connections to the work of
{\cite{Tarjus_2016}}. The requirement used in {\cite{Tarjus_2016}} to fix the
form of $\mathcal{C}$ is that the second cumulant of the partition function
perturbed by $\mathcal{C}$ should behave as the absolute value of $| \phi_a -
\phi_b |$ in the limit $\phi_b \to \phi_a$, namely $\frac{\delta}{\delta
\phi_a (x)}  \frac{\delta}{\delta \phi_b (y)} \int d^d z\,\mathcal{C} (z) =
2 \delta (x - y) | \phi_a - \phi_b | (1 + O (\phi_a - \phi_b))$. As we explained
above, we think that the perturbation $\mathcal{C}$ should not be considered.
On the other hand in our work we presented some perturbations which we think
could destabilize the IR SUSY fixed point, the most dangerous candidates being
the Feldman operators $\mathcal{F}_k$. These operators affect the second
cumulant as follows,
\begin{equation}
  \frac{\delta}{\delta \phi_a (x)}  \frac{\delta}{\delta \phi_b (y)} \int d^d
  z\,\mathcal{F}_k (z) = - 2 k (k - 1) \delta (x - y)  (\phi_a - \phi_b)^{k - 2}
  (1 + O (\phi_a - \phi_b)) .
\end{equation}
Of course the behavior above is very different from the one of $\mathcal{C}$
since it is analytic. However the fact that all $\mathcal{F}_k$ behave as
positive powers of $(\phi_a - \phi_b)$ ---in contrast with other operators of
the replicated theory which would scale like a constant, e.g. $\sum_i
\phi_i^k$--- is a tantalizing observation. A similar observation is that the
absolute value $| \phi_a - \phi_b |$ can be expanded (using the regularization
described in section \ref{2ptabs} below) in terms of operators $\mathcal{F}_k$.
It would be interesting to see if there exists a connection between the
alleged cuspy behavior of the susy-broken IR fixed point and the Feldman
perturbations $\mathcal{F}_k$.

\subsubsection{Two-point function of $| \phi |$}\label{2ptabs}

To convince the reader that our picture is indeed correct, in the following we
perform an explicit computation of the two point function of $| \phi |$ in the
free massless scalar theory of $\phi$ regulated with a UV cutoff $\Lambda$ in
momentum space.\footnote{We do not know any way to make sense of this operator
in a theory without UV cutoff.} We will show that this can be expanded in an
infinite sum of two-point functions of $\mathbb{Z}_2$-even operators. Naively,
to do this computation one may wish to expand $| \phi |$ in terms of monomials
$\phi^k$. Of course this is not possible since the absolute value is not an
analytic function and it does not admit a power expansion. We will use an
alternative definition of $| \phi | = \phi \hspace{0.27em} \text{sign} (\phi)$
representing $\tmop{sign} (\phi)$ as a limit of an analytic function
\begin{equation}
  \label{signregular} \text{sign} (x) = \lim_{\e \to 0} f \left(
  \frac{x}{\varepsilon} \right) \hspace{0.27em}, \qquad f (x) = \frac{2}{\pi} 
  \text{Si} (\pi x) \hspace{0.17em},
\end{equation}
where ${\rm Si}\,(x) \equiv \int_0^x \frac{d y}{y} \sin y$ is an entire function known
as ``sine integral'' (see below for why we choose this particular regulator).
We therefore Taylor expand the function $f$ as follows
\begin{equation}
  f (x) = \sum_{n = 0}^{\infty} a_n x^n \hspace{0.17em}, \qquad a_{2 n + 1} =
  \frac{2 (- \pi^2)^n}{(2 n + 1)^2  (2 n) !} \hspace{0.17em}, \quad a_{2 n} =
  0 \hspace{0.17em} .
\end{equation}
Using this expansion we rewrite the two point function of $| \phi |$ as
follows:
\begin{equation}
  \label{twoPTabs} \langle | \phi (x) || \phi (y) | \rangle = \lim_{\varepsilon
  \to 0}  \sum_{m, n = 0}^{\infty} \frac{a_n a_m}{\varepsilon^{n + m}} \langle
  \phi^{n + 1} (x) \phi^{m + 1} (y) \rangle .
\end{equation}
Notice that the operators $\phi^{n + 1}$ and $\phi^{m + 1}$ in
{\eqref{twoPTabs}} are not normal ordered. It is then convenient to rewrite
the correlator $\langle \phi^{n + 1} (x) \phi^{m + 1} (y) \rangle$ as a sum of
normal ordered correlation functions,
\begin{equation}
  \langle \phi^{n + 1} (x) \phi^{m + 1} (y) \rangle = \sum_{M = 0, 2, 4,
  \ldots} \binom{n + 1}{M} \binom{m + 1}{M} \langle \phi^{n + 1 - M} (x)
  \rangle \langle \phi^{m + 1 - M} (y) \rangle \langle \NO{\phi^M (x)} \NO{\phi^M
  (y)}\rangle \hspace{0.17em}, 
\end{equation}
where $M$ must be even since both $n + 1$ and $m + 1$ are even. Here the
correlation functions of $k$ operators inserted at the same point can be
computed as $\langle \phi^k (x) \rangle = (k - 1) !!G (0)^{k / 2}$, where the
double factorial $(k - 1)$!! is the combinatorial factor which counts the
number of pairings in $k$ elements, while $G (0)$ is the two-point function at
coincident points $G (0) \equiv \langle \phi (x) \phi (x) \rangle = \int_{|k|
< \Lambda} \frac{d^d k}{(2 \pi)^d}  |k|^{- 2} = \tmop{const} \times \Lambda^{d -
2}$. Combining these results we find that the sums over $n$ and $m$ factorize
and can be easily performed,
\begin{equation}
  S_M (X) \equiv \sum_{n = 0}^{\infty} (n - M) \text{!!} \binom{n + 1}{M} a_n
  X^n = X \hspace{0.17em} 2^{\frac{M}{2} + 1}  \hspace{0.27em} \frac{_2 F_2
  \left( \frac{1}{2}, 2 ; \frac{3}{2}, 2 - \frac{M}{2} ; - \frac{\pi^2 X^2}{2}
  \right)}{\Gamma \left( 2 - \frac{M}{2} \right) \Gamma (M + 1)}
  \hspace{0.17em},
\end{equation}
where the expansion parameter $X$ takes the form $X = \frac{\sqrt{G
(0)}}{\varepsilon}$. In particular we are interested in taking the limit of
$\varepsilon \to 0$ which corresponds to sending $X$ to infinity,
\begin{equation}
  S_M^{\infty} \equiv \lim_{X \to \infty} S_M (X) = \frac{2^{\frac{M -
  1}{2}}}{M! \Gamma \left( \frac{3 - M}{2} \right)} .
\end{equation}
Putting everything together we thus find
\begin{equation}
  \label{twoPTabsRESULT} \langle | \phi (x) || \phi (y) | \rangle = \sum_{M =
  0, 2, 4, \ldots} G (0)^{1 - M} (S_M^{\infty})^2 \langle \NO{\phi^M (x)}\NO{\phi^M (y)}\rangle .
\end{equation}
The result {\eqref{twoPTabsRESULT}} confirms our expectation: the two-point
function of $| \phi |$ can be written as sum of two-point functions of
$\mathbb{Z}_2$-even operators. In particular at large distances (for large $|x
- y|$) {\eqref{twoPTabsRESULT}} behaves as the two-point function of the
identity plus the one of $\phi^2$. One can eliminate the contribution of the
identity by defining a normal ordered version of $| \phi |$ with vanishing
one-point function. The resulting normal ordered operator at large distances
would thus behave as $\phi^2$.

To clarify all steps of our computation, we would like to comment on the
choice of the function $f (x)$ in {\eqref{signregular}}. There are many
possible analytic functions which tend to the sign function in some limit. We
chose to use {\eqref{signregular}} because of its excellent convergence
properties. Indeed it is not enough to consider a function $f (x)$ with a
uniformly convergent expansion. For our computation we must also require that
it is possible to commute the path integral with the series expansion. It is
easy to see that this is a more restrictive requirement. E.g.~for the
one-point function of $| \phi |$, after commuting the path integral with the
sum we find a new series of the form $\sum_{n = 0}^{\infty} (n) !!a_n x^n$,
where the new coefficients $(n) !!a_n$ grow much faster than $a_n$. If one
chooses functions $f$ with weaker convergence it may happen that the
path-integral and the series cannot be swapped or, in other words, that the
integrated expansion diverges.\footnote{The above is not the only way to
perform this computation. E.g., denoting $a = \phi (x)$, $b = \phi
(y)$, one can integrate out all the space-points in the path integral except
for $x, y$, and obtain the probability distribution density $P (a, b)$. Since
the theory is Gaussian, this is given by a Gaussian distribution $P (a, b)
\propto \exp (- u (a^2 + b^2) - 2 v a b)$, and the coefficients $u, v$ can be
fixed uniquely by requiring that the two point functions $\langle \phi (x)
\phi (x) \rangle$ and $\langle \phi (x) \phi (y) \rangle$ are correctly
reproduced. Then the two-point function $\langle | \phi (x) || \phi (y) |
\rangle$ can be computed as $\int d a\, d b\, | a | | b | P (a, b)$. This gives
the same result as {\eqref{twoPTabsRESULT}}. See also \cite{Fradkin:1963vva} for how to deal with non-polynomial operators in field theory (we thank Giorgio Parisi for mentioning this early work, whose focus is on the UV).} 

We hope that this explicit computation clarifies that it is sufficient to
consider perturbations around free theory of the polynomial form.

\subsection{Comparison to the work of Br{\'e}zin and De Dominicis}\label{BDD}

In this paper we used the observation of Br{\'e}zin and De Dominicis
{\cite{Brezin-1998}} concerning the need to consider additional $S_n$
invariant interactions in the effective Lagrangian. On the other hand, we
disagree with Ref. {\cite{Brezin-1998}} on how to interpret the instability
of $n = 0$ fixed point with respect to turning on nonzero $n$, and in
particular about the role of the additional fixed point identified in \
{\cite{Brezin-1998}}. In this appendix we will review this disagreement in
more detail.

Ref. {\cite{Brezin-1998}} worked in the ``old'' formalism (replicated field
basis with propagator {\eqref{eq:G}}). The quadratic part of the replicated
Lagrangian was perturbed by a general linear combination of the $S_n$-singlet
perturbations with 4 fields given in Eq.~{\eqref{gens4}}: $u_1 \sigma_4 + u_2
\sigma_1 \sigma_3 + u_3 \sigma_2^2 + u_4 \sigma_1^2 \sigma_2 + u_5
\sigma^4_1$. All of the couplings $u_i$ were assigned in $d = 6 - \varepsilon$
the same scaling dimension $\varepsilon$ as $u_1$, as is visible from RG
equations (3.2) in {\cite{Brezin-1998}}, which all have the form $\beta_{u_i}
= - \varepsilon u_i + H C_{i j k} u_j u_k$ with dimensionless $C_{i j k} .$ In
our scheme the couplings $u_{2,} u_3, u_4, u_5$ would have dimensions
$\varepsilon - 2$, $\varepsilon - 2$, $\varepsilon - 4$, $\varepsilon - 6$,
looking at the scaling dimension of the leader of the corresponding
interaction. Up to this difference, RG equations (3.2) in {\cite{Brezin-1998}}
bear some similarity with the Wilsonian RG equations {\eqref{betah}} in
Appendix \ref{sec:Toy} below. Eqs. {\eqref{betah}} were derived for $n = 0$,
but (3.2) in {\cite{Brezin-1998}} contain some terms proportional to powers of
$n$. Ref. {\cite{Brezin-1998}} observed that the fixed point $u_1 = u_{1
\ast}$, $u_2 = u_3 = \ldots = u_5 = 0$ is unstable with respect to the
inclusion of these $n \neq 0$ terms, and identified another fixed point for
nonzero $n$ where the couplings scale singularly as
\begin{equation}
  u_1 = O (1), \quad u_2, u_3 = O (1 / n), \quad u_4 = O (1 / n^2), \quad u_5
  = O (1 / n^3) \label{BDfp} .
\end{equation}
Because of the mentioned mismatch in the scaling dimensions of $u_i$, we are
not sure to agree with the details of this computation, although we do
completely agree with the conclusion that the $n = 0$ fixed point should be
unstable with respect to turning on $n \neq 0$ (see section \ref{sec:L2}). We
disagree however with the interpretation of this instability. As discussed in
section \ref{sec:L2}, even though the $n = 0$ fixed point is unstable, the
approximately scale-invariant regime becomes longer and longer as $n$ gets smaller and smaller. The Br{\'e}zin-De Dominicis fixed
point {\eqref{BDfp}} at nonzero $n$ is pushed to longer and longer distances
as $n$ gets smaller, and cannot describe the RFIM phase transition. It is, as
we said in section \ref{sec:L2}, disconnected from the $n = 0$ physics. This
is why in the main text we did not at all consider this fixed point.

In particular, we do not believe that one can shed light on the Parisi-Sourlas
conjecture by considering the properties of the Br{\'e}zin-De Dominicis fixed
point. (That is what Ref. {\cite{Brezin-1998}} tried to do. They observed that
their fixed point {\eqref{BDfp}} is unstable, and argued that this instability
may lead to the violation of the Parisi-Sourlas conjecture for any $d < 6$.)

Ref. {\cite{Brezin-1998}} is also sometimes cited (e.g. in
{\cite{Cardy:2013rqg}}) for the fact that RG in RFIM is singular as $n
\rightarrow 0$. Some loop effects singular in $n$ are indeed mentioned in
Section 2 of {\cite{Brezin-1998}}. We are puzzled by that section: e.g. in a
Wilsonian RG scheme with a UV and an IR cutoff we do not see any singularity
in their Eq.~(2.6) for $d = 6$. Independently of what these ``singularities''
might mean, they do not trickle down to their $d = 6 - \varepsilon$ RG
equations (\cite{Brezin-1998}, Section 3), which are completely smooth in
the limit $n \rightarrow 0$, in agreement with our discussion in section
\ref{sec:Sn}.

\subsection{Conformal bootstrap approach to dimensional
reduction}\label{Hikami}

In section \ref{CB} we described prospects for applying the conformal
bootstrap approach to study the RFIM fixed point. The only prior work in this
direction is by Hikami {\cite{Hikami:2018mrf}}. We will now briefly describe
how we understand the computations reported in that paper.
We only comment on the part of
{\cite{Hikami:2018mrf}} which concerns the RFIM, leaving aside the branched
polymer case also treated there.

In our language, Ref. {\cite{Hikami:2018mrf}} studies the 4pt function of
$\chi_i$'s which is called there ``$\phi$'' and we will use the same notation
in this appendix. This identification is visible from {\cite{Hikami:2018mrf}},
Eq.~(27). This 4pt function is considered in the strict $n \rightarrow 0$
limit\footnote{This should correspond to $- 2$ linearly independent $\chi$'s.
Since the crossing equations are not written explicitly in
{\cite{Hikami:2018mrf}}, it is impossible to verify the exact number of fields
used in the computations. If Ref. {\cite{Hikami:2018mrf}} used 0-component
$\chi$, it is a mistake.} and at the fixed point, assuming conformal
invariance. The spectrum of exchanged CFT operators in the OPE $\phi \times
\phi$ is limited to 3 scalar operators of dimension $\Delta_1,
\Delta_{\varepsilon}, \Delta_{\varepsilon'}$ and one spin-4 operator whose
dimension is called $Q$. (The operator of dimension $\Delta_1$ is
non-susy-writable in our language.) Determinant method of Gliozzi is applied
to solve crossing approximately and determine the scaling dimensions of these
operators as a function of the spatial dimension $4 \leqslant d \leqslant 6$
(\cite{Hikami:2018mrf}, Table 4). The so determined scaling dimensions of
$\phi$ and of the energy operator $\varepsilon$ are seen to satisfy the
dimensional reduction predictions reasonably well for $d > 5$, while for $d <
5$ larger deviations are observed.

In each of the figures 7-13, Ref. {\cite{Hikami:2018mrf}} varies
$\Delta_{\phi}$ and $\Delta_{\varepsilon}$ to find the intersection points
(approximate solution of crossing) while parameters $\Delta_1$ and $Q$ are
fixed to particular values. It is not clearly reported how those values are
arrived at, and how the predictions would change if different values were
chosen. It is also not clear why a spin-4 operator is included in this study
but not spin-2 operators. One may also question the accuracy of truncation,
because at $d = 6$, $\Delta_1 = 4.3$ in Table 4, deviating significantly from
the Gaussian prediction $\Delta_1 = 4$.

Ref. {\cite{Hikami:2018mrf}} does not investigate the mechanism by which
dimensional reduction is lost at the critical dimension $d_c$. On theoretical
grounds, we know that this loss is associated with the loss of SUSY, which
should happen because some operator becomes relevant. The operator becoming
relevant may be either a SUSY singlet, in which case SUSY would be lost via
fixed point annihilation, as in FRG studies cited in section
\ref{sec:conclusions}. Or, as we found in our work, SUSY fixed point may
become unstable because a SUSY-breaking leader operator becomes relevant. In
the former case there should be two fixed points above $d_c$: the two SUSY
fixed points which annihilate. In the later case there should be two fixed
points below $d_c$: the unstable SUSY one, and the stable non-SUSY.

Focusing on just one 4pt function, Ref. {\cite{Hikami:2018mrf}} is not
sensitive to finer aspects of Parisi-Sourlas supersymmetry apart from
predictions for operator dimensions from dimensional reduction. It reports
only one CFT for any $d$, which appears incompatible with either scenario. No
operator is reported to become relevant at $d_c$. (A SUSY-singlet should
appear in the OPE $\phi \times \phi$ and hence be visible in this study.)

In our opinion, while the observations of {\cite{Hikami:2018mrf}} are
suggestive, a much more careful study is needed to verify that they are
physical and are not instead due to truncation effects. This study should
confirm explicitly the existence of SUSY above $d_c$ and its absence below
$d_c$ (beyond dimensional reduction operator dimensions), and clarify the
mechanism by which SUSY is lost.

\subsection{Other approaches}\label{NPE}

Without trying to judge the merit, we will briefly mention two other
theoretical ideas about the RFIM transition.

It has been proposed to connect the loss of dimensional reduction to
``formation of bound state of replicas''.\footnote{We are not sure, but
perhaps one can think of this mechanism as due to fluctuations with large
values of fields which render the fixed point unstable, in spite of stability
with respect to small fluctuations. This would be somewhat similar to
instabilities in fixed points of scalar theories with unstable potentials
(like cubic with a real coupling or quartic with a negative coupling, see e.g.
{\cite{Giombi:2019upv}}).} This scenario was discussed e.g. in
{\cite{Brezin2001}}, Section 5, where references to prior work can be found.
In Ref. {\cite{PSbound}}, numerical simulations of the RFIM in $d = 3$ seemed
to provide support for bound states of replicas. Note that as mentioned
several times, we do not expect a SUSY fixed point in $d = 3$. It would be
interesting to know if the non-self-averaging phenomena observed in
{\cite{PSbound}} persist in $d = 4$ and $d = 5$, and whether they are present
in modern high-statistics simulations {\cite{Fytas3,Picco1,Picco2}} which do
not comment on this issue.

Recently, Ref. {\cite{Parisi2019}} proposed to expand the RFIM around an exact
solution on the ``Bethe lattice'' (an infinite tree without loops with all
vertices equivalent and having coordination number $2 d$, like for the cubic
lattice in $d$ dimension). While this approach is very different from the
traditional one, their calculations were consistent with dimensional reduction
in $d$ close to $6$.

Finally, RFIM critical exponents can be studied using the high temperature expansion. Ref.~\cite{HighT} thus obtained $\gamma=1.13(3),1.45(5),2.1(2)$ in $d=5,4,3$, which using $\gamma=\nu(2-\eta)$ is in the ballpark of the more recent accurate Monte Carlo results cited in sections \ref{d=34},\ref{d=5}. 

\section{Toy model for the $\Lcal_0 + \Lcal_1$ RG flow}\label{sec:Toy}

In this section we develop a very concrete toy model for the $\Lcal_0 +
\Lcal_1$ RG flow, mentioned in section \ref{followers}. It is important to
stress that the aim of this section is not to study all interesting operators
which may have an important role in destabilizing the RG. Here we want only to
show a computation which clarifies some features of the RG (e.g. the role of
$S_n$ symmetry, leaders, followers, etc.).

We consider a setup where the Gaussian piece of the $\Lcal_0$ Lagrangian is
perturbed by $5$ $S_n$-singlet operators, chosen to be all the perturbations
which contain $4$ fields and no derivatives. Since the perturbations are free
of derivatives, they can be written as products of the $\sigma_i$ fields. So
we get Eq.~{\eqref{toy}}, where each $S_n$-singlet multiplies a coupling
$h_i$. It is instructive write this Lagrangian in the Cardy basis. When $n = 0$,
the $5$ $S_n$ singlets are written as linear combinations of $11$ fields,
\begin{equation}
  \label{Snoprphi4} \begin{array}{cl}
    \sigma_4 & = 6 \varphi^2 \chi^2 + 4 \varphi^3 \omega + 4 \varphi \chi_i^3
    + (\chi_i^4 - 6 \varphi \omega \chi_i^2) - 2 \omega \chi_i^3 +
    (\frac{3}{2} \omega^2 \chi_i^2 + \varphi \omega^3),\\
    \sigma_2^2 & = 4 \varphi \omega \chi_i^2 + 4 \varphi^2 \omega^2 +
    (\chi_i^2)^2,\\
    \sigma_1 \sigma_3 & = 3 \varphi \omega \chi_i^2 + 3 \varphi^2 \omega^2 +
    \omega \chi_i^3 - \frac{3}{2} \omega^2 \chi_i^2 + \frac{\omega^4}{4},\\
    \sigma_1^2 \sigma_2 & = \omega^2 \chi_i^2 + 2 \varphi \omega^3,\\
    \sigma_1^4 & = \omega^4 .
  \end{array}
\end{equation}
We are therefore led to write a Gaussian action perturbed with eleven
independent couplings $g_i$, Eq.~{\eqref{Toy}}. This Lagrangian exactly
matches equation (\ref{toy}) when the couplings $g_i$ satisfy the following
$S_n$-invariance condition obtained by substituting (\ref{Snoprphi4}) in
(\ref{toy})
\begin{equation}
  \label{Sn} \begin{array}{lll}
    g_1 = g_2 = g_3 = g_4 = h_1, & g_5 = - 6 h_1 + 4 h_2 + 3 h_3, & g_6 = h_3
    - 2 h_1,\\
    g_7 = \frac{3}{2}  (h_1 - h_3) + h_4, & g_8 = h_1 + 2 h_4, & g_9 = 4 h_2 +
    3 h_3,\\
    g_{10} = h_2, & g_{11} = \frac{h_3}{4} + h_5 \hspace{0.17em} . & 
  \end{array}
\end{equation}
In the nest subsections we want to investigate how the couplings $g_i$ evolve
under RG when the $S_n$-condition (\ref{Sn}) are or are not imposed in the UV.

\subsection{Integrating out}

Let us start by considering $11$ independent couplings $g_i$ which are all
small perturbation of the same order. We work in $d = 6 - \e$ dimensions in a
theory with a momentum-space cutoff $\Lambda$. We want to compute how the
couplings change as we integrate out degrees of freedom from $\Lambda' <
\Lambda$ to $\Lambda$. The resulting couplings $\tilde{g}_i$, at order $O
(g_i^2)$, take the form
\begin{equation}
  \label{giprime} \begin{array}{l}
    \tilde{g}_1 = g_1 + 12 (2 g_1 + g_2) g_1 I\\
    \tilde{g}_2 = g_2 + 36 g_2^2 I\\
    \tilde{g}_3 = g_3 + 36 g_1 g_3 I\\
    \tilde{g}_4 = g_4 + 36 g_3^2 I\\
    \tilde{g}_5 = g_5 + 12 (2 g_1 g_5 + g_2 g_5 + 2 g_1 g_9) I\\
    \tilde{g}_6 = g_6 + 12 g_3 g_5 I\\
    \tilde{g}_7 = g_7 + (g_5^2 + g_9 g_5 + 18 g_1 g_8) I\\
    \tilde{g}_8 = g_8 + 4 (g_9^2 + 9 g_2 g_8) I\\
    \tilde{g}_9 = g_9 + 60 g_2 g_9 I\\
    \tilde{g}_{10} = g_{10} + 6 (6 g_3^2 + g_1 g_5) I\\
    \tilde{g}_{11} = g_{11} + 3 g_8 g_9 I,
  \end{array}
\end{equation}

Here, for simplicity, we consider only the contributions given by the one-loop
integral $I$ which depends logarithmically on the ratio $b \equiv \Lambda /
\Lambda'$ of the cutoff scales:\footnote{These are indeed the only
contributions which would survive in other schemes, like dimensional
regularization. For completeness we also performed a computation which takes
into account all the one-loop integrals  --- also the ones which scale as
powers of the cutoff ---  and we found that, at leading order in $\e$, the IR
fixed point does not change. Since the result is unchanged, but all the
intermediate step are more complicated, we decided to present this simpler
setup.}
\begin{equation}
  I = \frac{H}{2}  \int_{\Lambda'}^{\Lambda} \frac{d^d k}{(2 \pi)^d} 
  \frac{1}{(k^2)^2  (p - k)^2} = \frac{1}{2}  \frac{H}{(4 \pi)^3} \log b + O
  (\e) .
\end{equation}
From (\ref{giprime}) we can explicitly test if $S_n$ symmetry is preserved by
integrating-out. Namely we want to check that, when (\ref{Sn}) is satisfied by
the bare couplings, the renormalized couplings $\tilde{g}_i$ satisfy the same
condition (\ref{Sn}) where all couplings are tilded, namely
\begin{equation}
  \begin{array}{lll}
    \tilde{g}_1 = \tilde{g}_2 = \tilde{g}_3 = \tilde{g}_4 = \tilde{h}_1, &
    \tilde{g}_5 = - 6 \tilde{h}_1 + 4 \tilde{h}_2 + 3 \tilde{h}_3, &
    \tilde{g}_6 = \tilde{h}_3 - 2 \tilde{h}_1,\\
    \tilde{g}_7 = \frac{3}{2}  (\tilde{h}_1 - \tilde{h}_3) + \tilde{h}_4, &
    \tilde{g}_8 = \tilde{h}_1 + 2 \tilde{h}_4, & \tilde{g}_9 = 4 \tilde{h}_2 +
    3 \tilde{h}_3,\\
    \tilde{g}_{10} = \tilde{h}_2, & \tilde{g}_{11} = \frac{\tilde{h}_3}{4} +
    \tilde{h}_5 \hspace{0.17em}, & 
  \end{array} \label{Sntilde}
\end{equation}
where $\tilde{h}_i$ define the new values of the couplings $h_i$ after
integrating out. This amounts to check that, after imposing (\ref{Sn}) and
(\ref{Sntilde}), the eleven equations (\ref{giprime}) for the couplings $g_i$
reduce to only five equations for the renormalization of the couplings $h_i$.
For example let us consider what happens to the first four equations of
(\ref{giprime}) which only involve couplings $g_1, \ldots, g_4$ which are all
set to $h_1$ by (\ref{Sn}) (and similarly for their tilded companions). For
$S_n$ to be respected it is crucial that these four equation reduce to the
same one in terms of $h_1$. It is easy to see that this indeed happens giving
rise to $\tilde{g}_1 = \tilde{g}_2 = \tilde{g}_3 = \tilde{g}_4 = \tilde{h}_1 =
h_1 + 36 h_1^2 I$. By applying the same logic to the other equations we obtain
the wanted $5$ renormalization equations for $h_i$,
\begin{equation}
  \label{hitilde} \begin{array}{l}
    \tilde{h}_1 = h_1 + 36 h_1^2 I\\
    \tilde{h}_2 = h_2 + (24 h_1 h_2 + 18 h_1 h_3) I\\
    \tilde{h}_3 = h_3 + (48 h_1 h_2 + 36 h_1 h_3) I\\
    \tilde{h}_4 = h_4 + (32 h_2^2 + 48 h_3 h_2 + 18 h_3^2 + 36 h_1 h_4) I\\
    \tilde{h}_5 = h_5 + (24 h_2 h_4 + 18 h_3 h_4) I
  \end{array} \hspace{0.17em} .
\end{equation}
In other words, when $S_n$ symmetry is present in the UV, the renormalization
of the $11$ couplings $\tilde{g}_i$ can be computed using (\ref{Sntilde}) and
the renormalization of only $5$ $S_n$-symmetric couplings (\ref{hitilde}).

Computationally, this is a non-trivial check of $S_n$ symmetry, although
conceptually this should not be surprising. In fact we knew that $S_n$
invariance was actually present in the initial action, even if it was hidden
by the use of Cardy variables. One could have been worried that $S_n$ would be
spoiled by dropping the $n$-suppressed terms. This computation exemplifies
that even when $n$ is set to zero, $S_n$ symmetry continues to exist and plays
an important role.

\subsection{Rescaling}

To complete the RG step, and get a Lagrangian of the same form as the initial
one but with new couplings $g_i (b)$, we need to rescale the cutoff to its
original value. This amounts to rescale the couplings $\tilde{g}_i$ defined at
$\Lambda'$ by a factor $b^{d - \D_0}$ where $b \equiv \Lambda / \Lambda'$ and
the power is dictated by the classical dimensions $\D_0$ of the fields,
\begin{equation}
  \begin{array}{llll}
    g_1 (b) = \tilde{g}_1 b^{\varepsilon} \hspace{0.17em}, & g_2 (b) =
    \tilde{g}_2 b^{\varepsilon} \hspace{0.17em}, & g_3 (b) = \tilde{g}_3
    b^{\varepsilon - 1} \hspace{0.17em}, & g_4 (b) = \tilde{g}_4 b^{\varepsilon - 2}
    \hspace{0.17em},\\
    g_5 (b) = \tilde{g}_5 b^{\varepsilon - 2} \hspace{0.17em}, & g_6 (b) =
    \tilde{g}_6 b^{\varepsilon - 3} \hspace{0.17em}, & g_7 (b) = \tilde{g}_7
    b^{\varepsilon - 4} \hspace{0.17em}, & g_8 (b) = \tilde{g}_8 b^{\varepsilon - 4}
    \hspace{0.17em},\\
    g_9 (b) = \tilde{g}_9 b^{\varepsilon - 2} \hspace{0.17em}, & g_{10} (b) =
    \tilde{g}_{10} b^{\varepsilon - 2} \hspace{0.17em}, & g_{11} (b) =
    \tilde{g}_{11} b^{\varepsilon - 6} \hspace{0.17em} & 
  \end{array} \label{rescalinggi}
\end{equation}
By rescaling the couplings, equations (\ref{Sntilde}) get rescaled as follows
\begin{equation}
  \label{gib} \begin{array}{l}
    g_1 (b) = g_2 (b) = g_3 (b) b = g_4 (b) b^2 = \tilde{h}_1 b^{\varepsilon}\\
    g_5 (b) = - (6 \tilde{h}_1 - 4 \tilde{h}_2 - 3 \tilde{h}_3) b^{\varepsilon -
    2}\\
    g_6 (b) = - (2 \tilde{h}_1 - \tilde{h}_3) b^{\varepsilon - 3}\\
    g_7 (b) = \frac{1}{2}  (3 \tilde{h}_1 - 3 \tilde{h}_3 + 2 \tilde{h}_4)
    b^{\varepsilon - 4}\\
    g_8 (b) = (\tilde{h}_1 + 2 \tilde{h}_4) b^{\varepsilon - 4}\\
    g_9 (b) = (4 \tilde{h}_2 + 3 \tilde{h}_3) b^{\varepsilon - 2}\\
    g_{10} (b) = \tilde{h}_2 b^{\varepsilon - 2}\\
    g_{11} (b) = \frac{1}{4}  (\tilde{h}_3 + 4 \tilde{h}_5) b^{\varepsilon - 6}
  \end{array} \hspace{0.17em} .
\end{equation}
So while before rescaling $S_n$ symmetry sets certain couplings equal to each
other, after rescaling it relates them by powers of the rescaling factor $b$.
This is due to the fact that fields $\varphi, \omega, \chi_i$ related by the
$S_n$ symmetry have different classical scaling dimensions (contrary to the
usual situation that fields forming a multiplet under a symmetry have the same
dimensions). We say that ``$S_n$ symmetry does not commute with rescaling''.
This makes $S_n$ symmetry less manifest, since some couplings which start in
the UV with the same value, may evolve differently. However it is important to
stress that $S_n$ symmetry is still present (so it is not broken), and
constrains the RG at all scales as it is clear from the relations (\ref{gib}).

By setting all $\tilde{h}_{i > 1} = 0$ in (\ref{gib}), we can recover exactly
formula (\ref{sigma4rescaled}) for the form of $\sigma_4$ after an RG step.
Similarly, by keeping only one non-zero $\tilde{h}_i$, we can obtain how the
other $S_n$ singlets rescale after one RG step. The result is as follows
\begin{equation}
  \begin{array}{ccl}
    h_1 \sigma_4 & \to & h_1 (b)  (6 \varphi^2 \chi^2 + 4 \varphi^3 \omega + 4
    \frac{\varphi \chi_i^3}{b} + \frac{\chi_i^4 - 6 \varphi \omega
    \chi_i^2}{b^2} - 2 \frac{\omega \chi_i^3}{b^3} + \frac{\frac{3}{2}
    \omega^2 \chi_i^2 + \varphi \omega^3}{b^4}) \hspace{0.17em},\\
    h_2 \sigma_2^2 & \to & h_2 (b)  (4 \varphi \omega \chi_i^2 + 4 \varphi^2
    \omega^2 + (\chi_i^2)^2) \hspace{0.17em},\\
    h_3 \sigma_1 \sigma_3 & \to & h_3 (b)  (3 \varphi \omega \chi_i^2 + 3
    \varphi^2 \omega^2 + \frac{\omega \chi_i^3}{b} - \frac{3}{2} 
    \frac{\omega^2 \chi_i^2}{b^2} + \frac{\omega^4}{4 b^4}) \hspace{0.17em},\\
    h_4 \sigma_1^2 \sigma_2 & \to & h_4 (b)  (\omega^2 \chi_i^2 + 2 \varphi
    \omega^3) \hspace{0.17em},\\
    h_5 \sigma_1^4 & \to & h_5 (b) \omega^4,
  \end{array} \label{RGLeaderFollowerQuartic}
\end{equation}
where we introduced a natural definition for the rescaled couplings $h_i (b)$,
\begin{equation}
  \label{hib} h_1 (b) \equiv \tilde{h}_1 b^{\varepsilon} \hspace{0.17em}, \quad
  h_2 (b) \equiv \tilde{h}_2 b^{\varepsilon - 2} \hspace{0.17em}, \quad h_3 (b)
  \equiv \tilde{h}_3 b^{\varepsilon - 2} \hspace{0.17em}, \quad h_4 (b) \equiv
  \tilde{h}_4 b^{\varepsilon - 4} \hspace{0.17em}, \quad h_5 (b) \equiv
  \tilde{h}_5 b^{\varepsilon - 6} \hspace{0.17em} .
\end{equation}
Expression (\ref{RGLeaderFollowerQuartic}) is an explicit example of formula
(\ref{LeaderFollowersRG}) of the main text.

\subsection{Beta functions and fixed point}

After integrating-out (\ref{giprime}) and rescaling (\ref{rescalinggi}) we can
finally define the beta functions for the eleven couplings by $\beta_{g_i}
\equiv \frac{d}{d \log b} g_i (b)$. This gives
\begin{equation}
  \label{betag} \begin{array}{l}
    \beta_{g_1} = - g_1 \varepsilon + 12 g_1  (2 g_1 + g_2)  \hspace{0.27em} J\\
    \beta_{g_2} = - g_2 \varepsilon + 36 g_2^2  \hspace{0.27em} J\\
    \beta_{g_3} = g_3  (1 - \varepsilon) + 36 g_1 g_3  \hspace{0.27em} J\\
    \beta_{g_4} = g_4  (2 - \varepsilon) + 36 g_3^2  \hspace{0.27em} J\\
    \beta_{g_5} = g_5  (2 - \varepsilon) + 12 (g_2 g_5 + 2 g_1  (g_5 + g_9)) 
    \hspace{0.27em} J\\
    \beta_{g_6} = g_6  (3 - \varepsilon) + 12 g_3 g_5  \hspace{0.27em} J\\
    \beta_{g_7} = g_7  (4 - \varepsilon) + (g_5^2 + g_9 g_5 + 18 g_1 g_8) 
    \hspace{0.27em} J\\
    \beta_{g_8} = g_8  (4 - \varepsilon) + 4 (g_9^2 + 9 g_2 g_8)  \hspace{0.27em}
    J\\
    \beta_{g_9} = g_9  (2 - \varepsilon) + 60 g_2 g_9  \hspace{0.27em} J\\
    \beta_{g_{10}} = g_{10}  (2 - \varepsilon) + 6 (6 g_3^2 + g_1 g_5) 
    \hspace{0.27em} J\\
    \beta_{g_{11}} = g_{11}  (6 - \varepsilon) + 3 g_8 g_9  \hspace{0.27em} J
  \end{array} \hspace{0.17em},
\end{equation}
where $J \equiv \frac{H}{2 (4 \pi)^3}$. We are interested in fixed points
which can be reached from the $S_n$ invariant initial conditions {\eqref{Sn}}.
In particular any such fixed point will have $g_1 = g_2$, as is clear from
(\ref{gib}). Imposing this condition, we find a single non-trivial fixed
point:
\begin{equation}
  g_1^{\star} = g_2^{\star} = \frac{\e}{36 J} \hspace{0.17em}, \hspace{0.17em}
  \qquad g_{i > 2}^{\star} = 0 \hspace{0.17em} .
\end{equation}
Since all couplings $g_{i > 2}^{\star}$ vanish, this fixed point is the same
as that of $\Lcal_0$ (equivalent to $\Lcal_{\tmop{SUSY}}$). We conclude that
every computation done close to the IR fixed point of Lagrangian (\ref{Toy})
with $S_n$ symmetric initial conditions (\ref{Sn}) can be equivalently done in
a much simpler setup where the Gaussian piece of $\Lcal_0$ is perturbed by the
single susy-writable operator $6 \varphi^2 \chi_i^2 + 4 \varphi^3 \omega$.

Finally we show the $\tmb$-functions for the couplings $h_i (b)$ of
(\ref{hib}):
\begin{equation}
  \begin{array}{l}
    \beta_{h_1} = - h_1 \varepsilon + 36 h_1^2 J\\
    \beta_{h_2} = h_2  (2 - \varepsilon) + 24 h_1 h_2 J + 18 h_1 h_3 J\\
    \beta_{h_3} = h_3  (2 - \varepsilon) + 48 h_1 h_2 J + 36 h_1 h_3 J\\
    \beta_{h_4} = h_4  (4 - \varepsilon) + 32 h_2^2 J + 48 h_3 h_2 J + 18 h_3^2 J
    + 36 h_1 h_4 J\\
    \beta_{h_5} = h_5  (6 - \varepsilon) + 24 h_2 h_4 J + 18 h_3 h_4 J
  \end{array} \hspace{0.17em} . \label{betah}
\end{equation}
The fixed point of the $h_i$ flow is given by $h_1^{\star} = \frac{\e}{36
\hspace{0.27em} J}$ and $h_{i > 1}^{\star} = 0$.

\subsection{Perturbations around the IR fixed point}

Next we want to study the perturbations around the fixed point.

We first linearize the RG flow of $g_i$ around the fixed point and study the
11 eigenvectors $v_a$ and eigenvalues $\lambda_a$ of the matrix $M_{ij} =
\partial_{g_j} \tmb_{g_i} |_{g^{\star}} \nobracket$. The eigenvectors $v_a$
define the IR perturbations $\Ocal_a$ in operator space, while the eigenvalues
define the correspondent $1$-loop scaling dimension as $\D_a = d + \lambda_a$.
Here it is necessary to make a disclaimer. Our toy model does not include
operators with derivatives, which can mix with the operators of (\ref{Toy})
(e.g. the operator $\varphi^2 \omega^2$ may mix with $\varphi \omega
\partial^{\m} \varphi \partial_{\m} \varphi$ of the same scaling dimension,
which was not included in the toy Lagrangian (\ref{Toy})). In cases affected
by such mixings, we do not expect that our toy model computation will obtain
the correct renormalized operators nor their correct anomalous dimensions. (On
the contrary in the serious calculations in section \ref{anomdim} and App.
\ref{an-details} we were careful to take all possible mixings into account.)
So in practice the dimensions $\D_a$ reported below should be only considered
as a component of a mixing matrix, which encodes how the operator $\Ocal_a$
renormalizes itself. Only when $\Ocal_a$ does not mix with any other operators
outside of (\ref{Toy}), then we should expect that $\D_a$ defines its correct
conformal dimension at $1$-loop. In the end of the section we will come back
to this point. With this in mind, the result of the diagonalization of
$M_{ij}$ is summarized in table \ref{Tab:Toy11}.

\begin{table*}[h]
  \begin{center}
    \begin{tabular}{@{}cc@{}}
      \toprule
      $\D_a$ ($1$-loop) & $\Ocal_a$\\
      \midrule
      $6$ & $6 \varphi^2 \chi_i^2 + 4 \varphi^3 \omega$\\
      $8 - \frac{\varepsilon}{3}$ & $(\chi_i^2)^2 + 10 \varphi \omega \chi_i^2 +
      10 \varphi^2 \omega^2$\\
      $8 - 2 \varepsilon$ & $(\chi_i^2)^2$\\
      $10 - \varepsilon$ & $\omega^2 \chi_i^2 + 2 \varphi \omega^3$\\
      $12 - 2 \varepsilon$ & $\omega^4$\\
      \midrule
      $6 - \frac{\varepsilon}{3}$ & $\varphi^2 \chi_i^2$\\
      $7 - \varepsilon$ & $\varphi \chi_i^3$\\
      $8 - 2 \varepsilon$ & $\chi_i^4$\\
      $8 - \varepsilon$ & $\varphi \omega \chi_i^2 + 6 (\chi_i^2)^2$\\
      $9 - 2 \varepsilon$ & $\omega \chi_i^3$\\
      $10 - 2 \varepsilon$ & $\omega^2 \chi_i^2$\\
      \bottomrule
    \end{tabular} 
  \end{center}
  \caption{\label{Tab:Toy11}Toy model: all the 11 IR perturbations. The first
  $5$ are $S_n$-preserving, the last $6$ are $S_n$-breaking. $\D_a$ is the
  $1$-loop mixing matrix element which encodes how $\Ocal_a$ renormalizes
  itself.}
\end{table*}

This table lists 11 linear combinations of perturbations of the IR fixed point
by quartic operators without derivatives, which have well-defined anomalous
dimensions (in our toy model). What is their relation with the $S_n$ symmetry?
We know that the $S_n$-preserving directions form a $5$-dimensional subspace
$U$ of the $11$-dimensional space $V_{11}$ of couplings. The complementary
directions should be classified as $S_n$-breaking. As the RG flow progresses,
$U$ changes, ``rotating'' inside $V_{11}$ in accordance with (\ref{gib}).
However, the number of $S_n$-preserving (and of $S_n$-breaking) directions is
preserved along the RG flow. At the IR fixed point $U$ reaches a final form
$U_{\tmop{IR}}$, defining the $5$ different $S_n$-preserving IR perturbations.
Any flow starting in the subspace $U$ in the UV will approach the IR fixed
point along a linear combination of these 5 directions. From this argument we
expect that $U_{\tmop{IR}}$ has a basis of operators with well-defined IR
anomalous dimensions. Operators with well-defined IR dimensions which are not
in $U_{\tmop{IR}}$ will span a complementary space denoted by
$\bar{U}_{\tmop{IR}}$. So we have $V_{11} = U_{\tmop{IR}} \oplus
\bar{U}_{\tmop{IR}}$, where $U_{\tmop{IR}}$ is a 5-dimensional subspace of
$S_n$-preserving IR perturbations, and $\bar{U}_{\tmop{IR}}$ is a
complementary $6$-dimensional subspace of $S_n$-breaking IR perturbations.

So, which directions are which? We claim that the $S_n$-preserving IR
perturbations are the first $5$ entries of the table \ref{Tab:Toy11}. To see
this, we repeat the diagonalization exercise for the $\beta$-functions
{\eqref{betah}} associated to the $S_n$ couplings $h_i$. When we diagonalize
$\partial_{h_j} \beta_{h_i} |_{h^{\star}} \nobracket$ we get 5 eigenvalues and
eigenvectors given in table \ref{Tab:Toy5} (where we give $S_n$ singlets to
which the 5 eigencouplings couple).

\begin{table*}[h]
  \begin{center}
    \begin{tabular}{@{}cc@{}}
      \toprule
      $\D_a$ ($1$-loop) & $\Ocal_a$\\
      \midrule
      $6$ & $\sigma_4$\\
      $8 - \frac{\varepsilon}{3}$ & $\sigma_2^2 + 2 \sigma_1 \sigma_3$\\
      $8 - 2 \varepsilon$ & $\sigma_2^2 - \frac{4}{3} \sigma_1 \sigma_3$\\
      $10 - \varepsilon$ & $\sigma_1^2 \sigma_2$\\
      $12 - 2 \varepsilon$ & $\sigma_1^4$\\
      \bottomrule
    \end{tabular} 
  \end{center}
  \caption{\label{Tab:Toy5}Toy model: the $5$ $S_n$-preserving IR
  perturbations coming from the $\beta$ functions {\eqref{betah}} linearized
  near the fixed point. $\D_a$ is the $1$-loop mixing matrix element which
  encodes how $\Ocal_a$ renormalizes itself.}
\end{table*}

The leader pieces of the singlets in table \ref{Tab:Toy5} exactly match the
first 5 operators in table \ref{Tab:Toy11}:
\begin{eqnarray}
  (\sigma_4)_L & = & 6 \varphi^2 \chi_i^2 + 4 \varphi^3 \omega
  \hspace{0.17em}, \nonumber\\
  (\sigma_2^2 + 2 \sigma_1 \sigma_3)_L & = & (\chi_i^2)^2 + 10 \varphi \omega
  \chi_i^2 + 10 \varphi^2 \omega^2 \hspace{0.17em}, \nonumber\\
  (\sigma_2^2 - \frac{4}{3} \sigma_1 \sigma_3)_L & = & (\chi_i^2)^2
  \hspace{0.17em}, \\
  (\sigma_1^2 \sigma_2)_L & = & \omega^2 \chi_i^2 + 2 \varphi \omega^3
  \hspace{0.17em}, \nonumber\\
  (\sigma_1^4)_L & = & \omega^4 \hspace{0.17em}, \nonumber
\end{eqnarray}
and the values of $\D_a$ in both tables \ref{Tab:Toy5} also agree. This proves
the claim that the first 5 operators in table \ref{Tab:Toy11} \ are $S_n$
invariant directions (and hence, by exclusion, the last $6$ directions are
$S_n$-breaking).

Let us now return to the problem of understanding which $\D_a$ of tables
\ref{Tab:Toy11} and \ref{Tab:Toy5} correspond to the actual $1$-loop
dimensions of the respective operator $\Ocal_a$. As we said above, this
happens when $\Ocal_a$ does not mix with operators containing derivatives,
since those operators were not considered in (\ref{Toy}). Let us consider this
question for the $S_n$-invariant directions. Are there additional $S_n$
singlets producing leaders with the same number of fields, with the same
classical dimensions and the same symmetry properties (recall that
susy-writable, susy-null, and non-susy-writable leaders do not mix with each
other\footnote{More precisely there is only triangular mixing, which does not
affect scaling dimensions.})? Fortunately this exercise is already done in
appendix \ref{class}, where the classification of all quartic operators with
dimensions $\D \leq 12$ is given. For our purpose it is enough to consider
table \ref{Nf4Nd2}. There, we see one leader $(\chi_{i \mu}^2)
\varphi_{\nosymbol}^2 + \ldots$ involving two derivatives, susy-writable and
of classical dimensions $8$ (at $d = 6$), which can mix with $(\sigma_2^2 + 2
\sigma_1 \sigma_3)_L$ (the second line of tables \ref{Tab:Toy11} and
\ref{Tab:Toy5}). There are also three susy-writable operators with dimensions
$10$, that can mix with $(\sigma_1^2 \sigma_2)_L$. Finally, there are $2$
susy-writable operators of dimensions $12$ which can mix with
$(\sigma_1^4)_L$. The value of $\D_a$ for these three operators therefore
should not be confused with their scaling dimension. On the other hand, there
are no operators which can mix with $(\sigma_4)_L$ and the (susy-null)
$(\sigma_2^2 - \frac{4}{3} \sigma_1 \sigma_3)_L$, thus their dimension is
indeed given by $\D_a$. We can easily check that the result is correct: the
anomalous dimension of $(\sigma_4)_L$ is the well-known one of $\hat{\phi}^4$
of the Wilson-Fisher fixed point, while $(\chi_i^2)^2$ gets no one-loop
correction (see App. \ref{chi22sec}).

These computations represent a nice toy model to better understand our RG
setup, where $S_n$ symmetry does not commute with rescaling. The final tables
\ref{Tab:Toy11} and \ref{Tab:Toy5} show that by diagonalizing the possible IR
perturbations we get some directions which are $S_n$-preserving while others
are $S_n$-breaking. From table \ref{Tab:Toy11} we see that the
$S_n$-preserving IR perturbations are captured by the leaders of the
correspondent $S_n$-singlets of table \ref{Tab:Toy5}. Moreover table
\ref{Tab:Toy11} exhibits other eigenperturbations, which are linear
combinations of the followers and which correspond to $S_n$-breaking
directions, in agreement with the interpretation given in section
\ref{followers}. \ It is also important to notice that when two $S_n$-singlets
have leaders of the same classical dimension (e.g. $\sigma_2^2$ and $\sigma_1
\sigma_3$), eigenperturbations are their particular linear combinations, which
sometimes can be determined by looking at the leader type (e.g. $(\sigma_2^2 -
\frac{4}{3} \sigma_1 \sigma_3)_L = (\chi_i^2)^2$ is the only susy-null linear
combination at this dimension, so must be an eigenperturbation). These
observations illustrate the general algorithm proposed in sections \ref{L0L1}
and \ref{leaders} to organize the spectrum of all the $S_n$-preserving IR
perturbations. Hopefully this discussion convinces the reader that the
proposed organization principle is indeed correct.

\section{Correspondence between correlators of $\chi_i$ and $\psi,
\bar{\psi}$}\label{chi-psi}

Consider first the Gaussian theory of $n - 1$ $\chi_i$'s subject to the
constraint $\sum_{i = 2}^n \chi_i = 0$ and with the action $S [\chi] = -
\frac{1}{2} \int d^d x \chi_i \partial^2 \chi_i$ (sum over repeated $i$'s
implicit here and elsewhere in this section, unless noted otherwise) and the
Gaussian theory of Grassmann fields $\psi, \bar{\psi}$ with the action $S
[\psi, \bar{\psi}] = - \int d^d x\, \psi \partial^2 \bar{\psi}$. We can compute
correlators from the partition functions coupled to sources:\footnote{To do
the first path integral it is convenient to represent the constraint $\sum_{i =
2}^n \chi_i = 0$ by a Lagrange multiplier.}
\begin{eqnarray}
  &  & Z_{\chi} [J_i] = \int \mathcal{D} \chi_i\, e^{- S [\chi] + \int J_i
  \chi_i} =\mathcal{N}_{\chi} (\det \partial^2)^{- \frac{n - 2}{2}} \exp
  \left( \frac{1}{2} \int K_{i j} J_i (\partial^2)^{- 1} J_j \right),
  \nonumber\\
  &  & Z_{\psi} [J, \bar{J}] = \int \mathcal{D} \psi\, \mathcal{D} \bar{\psi}\,
  e^{- S [\psi, \bar{\psi}] + \int \psi \bar{J} + J \bar{\psi}}
  =\mathcal{N}_{\psi} (\det \partial^2) \exp \left( \int J (\partial^2)^{- 1}
  \bar{J} \right), 
\end{eqnarray}
where $K_{i j} = \delta_{i j} - \frac{1}{n - 1} \Pi_{i j}$ is the matrix
appearing in {\eqref{propsL0}}. From here we get the $\chi \chi$ and $\psi
\bar{\psi}$ propagators shown in {\eqref{propsL0}} and {\eqref{befCut}}.

As mentioned in section \ref{PSSUSY}, the $\mathcal{L}_0$ theory defined in
terms of $\chi$ field contains more operators than its $\psi, \bar{\psi}$
counterpart $\mathcal{L}_{\tmop{SUSY}}$. E.g.~operators of the form $\sum\nolimits'
\chi^n_i$ do not have any correspondent due to the Grassmann nature or $\psi,
\bar{\psi}$. Let us show that observable of the $\chi$-formulation which
involve $\text{O} (n - 2)$ singlets, can be recovered from the
$\psi$-formulation. To this end we compute the path integrals with sources for
bilinear operators inserted at different points
\begin{eqnarray}
  Z_{\chi} [A (x, y)] = \int \mathcal{D} \chi_i \,e^{- S [\chi] - \frac{1}{2}
  \int \chi_i (x) \chi_i (y) A (x, y)} =\mathcal{N}_{\chi}  [\det (-
  \partial^2 + A)]^{- \frac{n - 2}{2}} \quad, &  &  \nonumber\\
  Z_{\psi} [A (x, y)] = \int \mathcal{D} \psi \mathcal{D} \bar{\psi}\, e^{- S
  [\psi, \bar{\psi}] - \frac{1}{2} \int (\psi (x) \bar{\psi} (y) + \psi (y)
  \bar{\psi} (x)) A (x, y) } =\mathcal{N}_{\psi} \det (- \partial^2 + A) . & 
  & 
\end{eqnarray}
Here $A (x, y) = A (y, x)$ is a symmetric function. We consider the Gaussian
actions $S [\chi]$, $S [\psi, \bar{\psi}]$ but it is easy to introduce the
coupling to $\varphi$ via $\partial^2 \rightarrow \partial^2 + V'' (\varphi)$.
We see that the results coincide in the limit $n \rightarrow 0$, discarding
the overall normalization which cancels in the computation of any correlator.
By taking derivatives in $A (x, y)$ it is straightforward to see that all
correlation functions of the bilocal operators $\mathcal{O}_{\chi} (x, y)
\equiv \chi_i (x) \chi_i (y)$ and $\mathcal{O}_{\psi} (x, y) \equiv \psi (x)
\bar{\psi} (y) + \psi (y) \bar{\psi} (x)$ exactly match. For example
\begin{eqnarray}
  \langle \psi (x_1) \bar{\psi} (x_2 \rangle + \langle \nobracket \psi (x_2)
  \bar{\psi} (x_1) \rangle & = & \langle \chi_i (x_1) \chi_i (x_2) \rangle, \\
  \langle (\nobracket \psi (x_1) \bar{\psi} (x_2) + \psi (x_2) \bar{\psi}
  (x_1) (\nobracket \psi (x_3) \bar{\psi} (x_4) + \psi (x_4) \bar{\psi} (x_3)
  \rangle & = & \langle \chi_i (x_1) \chi_i (x_2) \chi_j (x_3) \chi_j (x_4)
  \rangle . \nonumber
\end{eqnarray}
We therefore obtain an equivalence map between bilocal operators of the two
theories $\mathcal{O}_{\chi} (x, y) \longleftrightarrow \mathcal{O}_{\psi} (x,
y)$.

As a next step, we pass from bilocal to local operators. To this end we
differentiate an arbitrary number of times in $x$ and in $y$ and take a limit
as $y \rightarrow x$. This way we obtain that any two bilinear local operators
of this form are equivalent between the two theories:
\begin{equation}
  (\partial^{(\alpha)} \chi_i ) (\partial^{(\beta)} \chi_i)
  \longleftrightarrow (\partial^{(\alpha)} \psi) (\partial^{(\beta)}
  \bar{\psi}) + (\partial^{(\beta)} \psi) (\partial^{(\alpha)} \bar{\psi}),
  \label{appB-bilinears}
\end{equation}
where $(\alpha)$, $(\beta)$ are arbitrary collections of indices. E.g.
(denoting derivatives $\partial_{\mu}$ as $()_{\mu}$ etc)
\begin{eqnarray}
  \chi_i \chi_{i, \mu}  & \longleftrightarrow & \psi \bar{\psi}_{\mu} +
  \psi_{\mu} \bar{\psi}, \nonumber\\
  \chi_i \chi_{i, \mu \nu} & \longleftrightarrow & \psi \bar{\psi}_{\mu \nu} +
  \psi_{\mu \nu} \bar{\psi}, \nonumber\\
  \chi_{i, \sigma} \chi_{i, \rho \mu \nu}  & \longleftrightarrow &
  \psi_{\sigma} \bar{\psi}_{\rho \mu \nu} + \psi_{\rho \mu \nu}
  \bar{\psi}_{\sigma}, 
\end{eqnarray}
Finally we can extend this correspondence to products of bilinear operators,
e.g.
\begin{equation}
  \chi_i \chi_{i, \mu \nu} \chi_j \chi_{j, \rho \sigma} \longrightarrow [\psi
  \bar{\psi}_{\mu \nu} + \psi_{\mu \nu} \bar{\psi}] [\psi \bar{\psi}_{\rho
  \sigma} + \psi_{\rho \sigma} \bar{\psi}] .
\end{equation}
However one should be careful of ambiguities which may arise at this level.
E.g.~we have
\begin{eqnarray}
  \chi_i \chi_{i, \mu} \chi_j \chi_{j, \mu} & \longrightarrow & [\psi
  \bar{\psi}_{\mu} + \psi_{\mu} \bar{\psi}] [\psi \bar{\psi}_{\mu} +
  \psi_{\mu} \bar{\psi}] = 2 \psi \bar{\psi} \psi_{\mu} \bar{\psi}_{\mu}, 
  \label{ambpsi4}\\
  \frac{1}{2} \chi_i \chi_i \chi_{j, \mu} \chi_{j, \mu} & \longrightarrow & 2
  \psi \bar{\psi} \psi_{\mu} \bar{\psi}_{\mu}, \nonumber
\end{eqnarray}
i.e.~two different $\chi$ operators map to the same $\psi$ operator, meaning
that their difference is susy-null.

This gives us a dictionary to map operators of the two theories. It is
important to stress that a large part of operators of the $\chi$-theory is
left out from the dictionary. Indeed only $O (n - 2)$ singlets with all
indices repeated twice like $\chi_i \chi_{i, \mu \nu}$ have a nice
interpretation. $S_{n - 1}$ singlet operators with three or more $\chi_i$
carrying the same index cannot be represented in terms of the fermionic
variables. In the opposite direction, also a sector of the fermionic theory
cannot be represented in terms of the $\chi$-theory. Indeed only the
$\tmop{Sp} (2)$ bilinear singlets have meaning in the $\chi$-theory, therefore
operators of the form $\psi$, $\psi \psi_{\mu \nu}$, and so on, do not have a
representative. Moreover also in $\tmop{Sp} (2)$ singlet sector only operators
with derivatives acting symmetrically on $\psi$ and $\bar{\psi}$ (see
{\eqref{appB-bilinears}}) make sense in the $\chi$-theory.

One could have hoped that the correspondence can be extended also to $\chi$
correlators with indices non-contracted, at the expense of introducing
tensorial coefficients. Sometimes this can be done, but not in full
generality. E.g.~this fails for general 4-point functions, as one cannot find
tensorial coefficients $T^{(I)}_{i j k l}$, $I = 1, 2, 3$, making the two
sides of the following equation agree (already in free theory):
\begin{eqnarray}
  \langle \chi_i (1) \chi_j (2) \chi_k (3) \chi_l (4) \rangle & \neq &
  T^{(1)}_{i j k l} \langle \psi (1) \psi (2) \bar{\psi} (3) \bar{\psi} (4)
  \rangle \nonumber\\
  &  & + T^{(2)}_{i j k l} \langle \psi (1) \psi (3) \bar{\psi} (2)
  \bar{\psi} (4) \rangle + T^{(3)}_{i j k l} \langle \psi (1) \psi (4)
  \bar{\psi} (2) \bar{\psi} (3) \rangle . 
\end{eqnarray}
\section{Tables of leaders up to $\Delta = 12$}\label{class}

\subsection{$N_{\phi} = 2$}

The $N_{\phi} = 2$ singlets are $\sigma_2$, $\sigma_1^2$, or derivative
dressings thereof. These singlets are particularly simple for two reasons.
First, they have well defined classical dimension when expressed in the Cardy
basis (see section \ref{singletsCardy}). This means they give pure leaders (no
followers). Second, they involve at most two powers of $\chi_i$, and so are
susy-writable.

The $N_{\phi} = 2$ singlets without derivatives are given in Table
\ref{Nf2Nd0}. We will not write explicitly the $N_{\phi} = 2$ singlets with
derivatives. One familiar such singlet is the kinetic term $\sigma_{2 (\mu)
(\mu)} = \left[ 2 \partial \omega \partial \vf + (\partial \chi_i)^2 
\right]_{\Delta = 6}$.

\begin{table}[h]\centering
    \begin{tabular}{@{}lll@{}}
      \toprule
      Singlet & Leader & Leader type\\
      \midrule
      $\sigma_2$ & $\left[ 2 \omega \vf + \chi_i^2  \right]_{\Delta = 4}$ &
      susy-writable\\
      $\sigma_1^2$ & $[\omega^2]_{\Delta = 6}$ & susy-writable\\
      \bottomrule
    \end{tabular}
  \caption{\label{Nf2Nd0}Scalar $\mathbb{Z}_2$-even leaders with $N_{\phi} =
  2$, $N_{\tmop{der}} = 0$.}
\end{table}

\subsection{$N_{\phi} = 4$}

The $N_{\phi} = 4$ leaders without derivatives were given in Table
\ref{Nf4Nd0}.

Table \ref{Nf4Nd2} lists scalar $N_{\phi} = 4$ leaders with $N_{\tmop{der}} =
2$ derivatives and dimension $\Delta \leqslant 12$. (The Greek indices on
$\varphi, \chi_i, \omega$ denote partial derivatives: $\varphi_{\mu} =
\partial_{\mu} \varphi$, etc.) They arise from 7 singlets:
\begin{equation}
  \sigma_{4 (\mu) (\mu)}, \quad \sigma_{1 (\mu)} \sigma_{3 (\mu)}, \quad
  \sigma_1 \sigma_{3 (\mu) (\mu)}, \quad \sigma_{2 (\mu)}^2, \quad \sigma_2
  \sigma_{2 (\mu) (\mu)}, \quad \sigma_2 \sigma_{1 (\mu)}^2, \quad \sigma_1^2
  \sigma_{2 (\mu) (\mu)} . \label{Nd2list}
\end{equation}
When classifying singlets involving derivatives, we make use of the equations
of motion (EOM) of the Gaussian part of the $\mathcal{L}_0$ Lagrangian
(working in normalization $H = 2$):
\begin{equation}
  \partial^2 \varphi = - 2 \omega, \qquad \partial^2 \omega = \partial^2
  \chi_i = 0,
\end{equation}
which can be written equivalently as
\begin{equation}
  \partial^2 \phi_i = - 2 \omega = - 2 \sigma_1 .
\end{equation}
This equation means that we never have to consider, in the replicated basis,
the singlets {\eqref{skder}} involving $\partial^2 \phi_i$, such as $\sigma_{k
(\mu \mu)}$. This explains their absence in {\eqref{Nd2list}}.\footnote{This
use of EOM is analogous to using the EOM when classifying fields at the
Wilson-Fisher fixed point
{\cite{Kehrein:1992fn,Kehrein:1994ff,Kehrein:1995ia}},
{\cite{Hogervorst:2015akt}}. In the interacting theory, EOM get modified with
a non-linear term appearing in the r.h.s.: $\partial^2 \phi_i = - H \sigma_1 -
\frac{\lambda}{3!} \phi_i^3$. We can still classify fields modulo EOM because
fields involving EOM only have correlators at coincident points. Such fields
correspond to redundant operators {\cite{Wegner}} and their scaling dimensions
do not influence RG stability of the theory; they also do not mix with the
non-redundant fields. So we can write any field involving $\partial^2 \phi_i$
as a redundant operator (which we drop from consideration) plus a field which
does not involve $\partial^2 \phi_i$.}

One particular linear combination of singlets {\eqref{Nd2list}} is equivalent,
modulo EOM, to the total derivative $\partial^2 \mathcal{F}_4$ and has a
susy-null leader $\partial^2 [(\chi_i^2)^2]$. Indeed, applying Eq.
{\eqref{appB-bilinears}} to the expression in Table \ref{Nf4Nd2} we get zero:
\begin{equation}
  (\chi_i \chi_{i \mu})^2 + \frac{1}{2} (\chi_i^2) (\chi_{i \mu}^2)
  \rightarrow (\psi \partial_{\mu} \bar{\psi} + \partial_{\mu} \psi
  \bar{\psi})^2 + 2 \psi \bar{\psi} \partial_{\mu} \psi \partial_{\mu}
  \bar{\psi} = 0 .
\end{equation}
\begin{table}[h]\centering
    \begin{tabular}{@{}lll@{}}
      \toprule
      Singlet & Leader & Leader type\\
      \midrule
      $\sigma_{4 (\mu) (\mu)}$ & $\nobracket [(\chi_{i \mu}^2)
      \varphi_{\nosymbol}^2 + 4 (\chi_i \chi_{i \mu}) \varphi_{\nosymbol}
      \varphi_{\mu} + (\chi_i^2) \varphi_{\mu}^2 + 2 \varphi_{\nosymbol}
      \varphi_{\mu}^2 \omega_{\nosymbol} + 2 \varphi_{\nosymbol}^2
      \varphi_{\mu} \omega_{\mu}]_{\Delta = 8} \nobracket$ & susy-writable\\
      $\sigma_{1 (\mu)} \sigma_{3 (\mu)}$ & $\nobracket [2 (\chi_i \chi_{i
      \mu}) \varphi_{\nosymbol} \omega_{\mu} + (\chi_i^2) \varphi_{\mu}
      \omega_{\mu} + 2 \varphi_{\nosymbol} \varphi_{\mu} \omega_{\nosymbol}
      \omega_{\mu} + \varphi_{\nosymbol}^2 \omega_{\mu}^2]_{\Delta = 10}
      \nobracket$ & susy-writable\\
      $\sigma_1 \sigma_{3 (\mu) (\mu)}$ & $[(\chi_{i \mu}^2)
      \varphi_{\nosymbol} \omega_{\nosymbol} + 2 (\chi_i \chi_{i \mu})
      \varphi_{\mu} \omega_{\nosymbol} + \varphi_{\mu}^2 \omega_{\nosymbol}^2
      + 2 \varphi_{\nosymbol} \varphi_{\mu} \omega_{\nosymbol}
      \omega_{\mu}]_{\Delta = 10}$ & susy-writable\\
      $\frac{1}{48} \partial^2 \mathcal{F}_4$ & $\left. [(\chi_i \chi_{i
      \mu})^2 + \frac{1}{2} (\chi_i^2) (\chi_{i \mu}^2)]_{\Delta = 10}
      \right.$ & susy-null\\
      $\sigma_2 \sigma_{2 (\mu) (\mu)}$ & $[(\chi_i^2) (\chi_{j \mu}^2) + 2
      (\chi_{i \mu}^2) \varphi_{\nosymbol} \omega_{\nosymbol} + 2 (\chi_i^2)
      \varphi_{\mu} \omega_{\mu} + 4 \varphi_{\nosymbol} \varphi_{\mu}
      \omega_{\nosymbol} \omega_{\mu}]_{\Delta = 10}$ & susy-writable\\
      $\sigma_2 \sigma_{1 (\mu)}^2$ & $[(\chi_i^2) \omega_{\mu}^2 + 2
      \varphi_{\nosymbol} \omega_{\nosymbol} \omega_{\mu}^2]_{\Delta = 12}$ &
      susy-writable\\
      $\sigma_1^2 \sigma_{2 (\mu) (\mu)}$ & $[(\chi_{i \mu}^2)
      \omega_{\nosymbol}^2 + 2 \varphi_{\mu} \omega_{\nosymbol}^2
      \omega_{\mu}]_{\Delta = 12}$ & susy-writable\\
      \bottomrule
    \end{tabular}
  \caption{\label{Nf4Nd2}Scalar $\mathbb{Z}_2$-even leaders with $N_{\phi} =
  4$, $N_{\tmop{der}} = 2$. $\frac{1}{48} \partial^2 \mathcal{F}_4 = \sigma_{2
  (\mu)}^2 - \sigma_{1 (\mu)} \sigma_{3 (\mu)} + \frac{1}{2} \sigma_2
  \sigma_{2 (\mu) (\mu)} - \sigma_1 \sigma_{3 (\mu) (\mu)}$.}
\end{table}

The other linear combination produce susy-writable leaders. Some of these,
such as $(\sigma_{4 (\mu) (\mu)})_L$ and one linear combination of $\sigma_{1
(\mu)} \sigma_{3 (\mu)}$ and $\sigma_1 \sigma_{3 (\mu) (\mu)}$, are total
derivatives (modulo EOM) of the $N_{\phi} = 4$ leaders without derivatives
given in Table \ref{Nf4Nd0}. Others are new primaries. We will not carry out
the separation.

Now let us move to scalar $N_{\phi} = 4$ leaders with $N_{\tmop{der}} = 4$
derivatives and dimension $\Delta \leqslant 12$. Without giving full
expressions, the following 12 singlets:
\begin{eqnarray}
  &  & \sigma_{4 (\mu) (\mu) (\nu) (\nu)}, \sigma_{4 (\mu) (\nu) (\mu \nu)},
  \sigma_{4 (\mu \nu) (\mu \nu)}, \nonumber\\
  &  & \sigma_1 \sigma_{3 (\mu) (\nu) (\mu \nobracket \nobracket \nu)},
  \sigma_1 \sigma_{3 (\mu \nobracket \nobracket \nu) (\mu \nobracket
  \nobracket \nu)}, \sigma_{1 (\nu)} \sigma_{3 (\mu) (\mu) (\nu)}, \sigma_{1
  (\nu)} \sigma_{3 (\mu) (\mu \nobracket \nobracket \nu)}, \sigma_{1 (\mu
  \nu)} \sigma_{3 (\mu) (\nu)}, \sigma_{1 (\mu \nu)} \sigma_{3 (\mu \nobracket
  \nobracket \nu)}, \nonumber\\
  &  & \sigma_2 \sigma_{2 (\mu \nu) (\mu \nu)}, \sigma_{2 (\mu \nu)}
  \sigma_{2 (\mu) (\nu)}, \sigma_{2 (\nu)} \sigma_{2 (\mu) (\mu \nu)} 
\end{eqnarray}
give rise to manifestly susy-writable leaders (i.e.~either at most quadratic
in $\chi$'s, or with quartic in $\chi$ terms none of which vanish upon $\chi
\rightarrow \psi$ substitution).

Three more singlets $\sigma^2_{2 (\mu \nu)}$, $\sigma^2_{2 (\mu) (\nu)},
\sigma_{2 (\mu) (\mu)} \sigma_{2 (\nu) (\nu)}$ give rise to leaders containing
at least some quartic in $\chi$ terms vanishing upon $\chi \rightarrow
\psi$ substitution. An equivalent basis of leaders (modulo EOM) is obtained by
replacing these three singlets by the following total derivatives
combinations:
\begin{equation}
  \partial_{\mu} \partial_{\nu} [\sigma_2 \sigma_{2 (\mu \nu)}],
  \partial_{\mu} \partial_{\nu} [\sigma_2 \sigma_{2 (\mu) (\nu)}],
  (\partial^2)^2 \mathcal{F}_4 .
\end{equation}
It can be verified that $\sigma_2 \sigma_{2 (\mu \nu)}$ and $\sigma_2
\sigma_{2 (\mu) (\nu)}$ have susy-writable leaders. To summarize, at the
$N_{\phi} = 4$, $N_{\tmop{der}} = 4$ level with $\Delta \leqslant 2$ we have
only one susy-null leader, and it is the total derivative of
$(\mathcal{F}_4)_L$.

Finally, all scalar leaders with $N_{\phi} = 4$, $N_{\tmop{der}} = 6$ and
dimension $\Delta \leqslant 12$ originate from dressing $\sigma_4$ with
derivatives. They are all susy-writable.

\subsection{$N_{\phi} = 6$}

Leaders with $N_{\phi} = 6$ and $N_{\tmop{der}} = 0$ with $\Delta \leqslant
12$ arise from 7 singlets:
\begin{equation}
  \sigma_6, \quad \sigma_1 \sigma_5, \quad \sigma_1^2 \sigma_4, \quad \sigma_2
  \sigma_4, \quad \sigma_3^2, \quad \sigma_1 \sigma_2 \sigma_3, \quad
  \sigma_2^3 . \label{Nf6Nd0list}
\end{equation}
Three of them are susy-writable. We also identify one susy-null linear
combination of dimension 10 and two SUSY-nulls of dimension 12. Finally, there
is a non-susy-writable leader of dimension 12, corresponding to the Feldman
$\mathcal{F}_6$ operator. See Table \ref{Nf6Nd0}.

\begin{table}[h]\centering
    \begin{tabular}{@{}lll@{}}
      \toprule
      Singlet & Leader (+ First follower if susy-null) & Leader type\\
      \midrule
      $\sigma_6$ & $\nobracket [15 (\chi_i^2) \varphi_{\nosymbol}^4 + 6
      \varphi_{\nosymbol}^5 \omega_{\nosymbol}]_{\Delta = 8} \nobracket$ &
      susy-writable\\
      $\sigma_1 \sigma_5$ & $\nobracket [10 (\chi_i^2) \varphi_{\nosymbol}^3
      \omega_{\nosymbol} + 5 \varphi_{\nosymbol}^4
      \omega_{\nosymbol}^2]_{\Delta = 10} \nobracket$ & susy-writable\\
      $\sigma_1^2 \sigma_4$ & $[6 (\chi_i^2) \varphi_{\nosymbol}^2
      \omega_{\nosymbol}^2 + 4 \varphi_{\nosymbol}^3
      \omega_{\nosymbol}^3]_{\Delta = 12}$ & susy-writable\\
      $\sigma_2 \sigma_4 - \frac{8}{5} \sigma_1 \sigma_5$ & $\nobracket
      \nobracket [6 (\chi_i^2)^2 \varphi_{\nosymbol}^2]_{\Delta = 10} + [4
      (\chi_i^2) (\chi_i^3) \varphi_{\nosymbol} - 8 (\chi_i^3)
      \varphi_{\nosymbol}^2 \omega_{\nosymbol}]_{\Delta = 11}$ & susy-null\\
      $\sigma_1 \sigma_2 \sigma_3 - \frac{3}{2} \sigma_1^2 \sigma_4$ & $[3
      (\chi_i^2)^2 \varphi_{\nosymbol} \omega_{\nosymbol}]_{\Delta = 12} +
      [(\chi_i^2) (\chi_i^3) \omega_{\nosymbol} - 4 (\chi_i^3)
      \varphi_{\nosymbol} \omega_{\nosymbol}^2]_{\Delta = 13}$ & susy-null\\
      $\sigma_2^3 - 2 \sigma_1 \sigma_2 \sigma_3 + \sigma_1^2 \sigma_4$ &
      $[(\chi_i^2)^3]_{\Delta = 12} - [2 (\chi_i^2) (\chi_i^3)
      \omega_{\nosymbol}]_{\Delta = 13} $ & susy-null\\
      $- \frac{1}{20} \mathcal{F}_6$ & $\left. [(\chi_i^3)^2 - \frac{3}{2}
      (\chi_i^2) (\chi_i^4)]_{\Delta = 12} \right.$ & non-susy-writable\\
      \bottomrule
    \end{tabular}
  \caption{\label{Nf6Nd0}Scalar $\mathbb{Z}_2$-even leaders with $N_{\phi} =
  6$, $N_{\tmop{der}} = 0$. $- \frac{1}{20} \mathcal{F}_6 = \sigma_3^2 -
  \frac{3}{2} \sigma_2 \sigma_4 + \frac{3}{5} \sigma_1 \sigma_5$.}
\end{table}

Moving to the $N_{\phi} = 6$, $N_{\tmop{der}} = 2$ case, the following 5
singlets give rise to scalar susy-writable leaders with $\Delta \leqslant 12$:
\begin{equation}
  \sigma_{6 (\mu) (\mu)}, \sigma_{1 (\mu)} \sigma_{5 (\mu)}, \sigma_1
  \sigma_{5 (\mu) (\mu)}, \sigma_{2 (\mu) (\mu)} \sigma_4, \sigma_3 \sigma_{3
  (\mu) (\mu)} .
\end{equation}
Three more singlets give $\Delta \leqslant 12$ leaders with some quartic in
$\chi$ pieces which vanish upon $\chi \rightarrow \psi$:
\begin{eqnarray}
  (\sigma_{2 (\mu)} \sigma_{4 (\mu)})_L & = & \left[ 3 (\chi_i \chi_{i \mu})^2
  \varphi_{\nosymbol}^2 + 3 (\chi_i^2) (\chi_i \chi_{i \mu})
  \varphi_{\nosymbol} \varphi_{\mu} + \text{susy-writable} \right]_{\Delta =
  12}, \nonumber\\
  (\sigma_2 \sigma_{4 (\mu) (\mu)})_L & = & \left[ \nobracket 4 (\chi_i^2)
  (\chi_i \chi_{i \mu}) \varphi_{\nosymbol} \varphi_{\mu} + (\chi_i^2)^2
  \varphi_{\mu}^2 \nobracket + \text{susy-writable} \right]_{\Delta = 12}, \\
  (\sigma_{3 (\mu)}^2)_L & = & \left[ 4 (\chi_i \chi_{i \mu})^2
  \varphi_{\nosymbol}^2 + 4 (\chi_i^2) (\chi_i \chi_{i \mu})
  \varphi_{\nosymbol} \varphi_{\mu} + (\chi_i^2)^2 \varphi_{\mu}^2 +
  \text{susy-writable} \right]_{\Delta = 12} . \nonumber
\end{eqnarray}
We can form two total derivative combinations including these singlets:
\begin{eqnarray}
  \frac{1}{12} \partial_{\mu} (\sigma_4 \sigma_{2 \nobracket (\mu)
  \nobracket}) & \rightarrowlim^L &  (\chi_i \chi_{i \mu})^2
  \varphi_{\nosymbol}^2 + (\chi_i^2)  (\chi_i \chi_{i \mu})
  \varphi_{\nosymbol} \varphi_{\mu} + \text{susy-writable}, \\
  \frac{1}{12} \partial^2 (\sigma_4 \sigma_{2 \nobracket \nobracket}) &
  \rightarrowlim^L & 4 (\chi_i \chi_{i \mu})^2 \varphi_{\nosymbol}^2 + 8
  (\chi_i^2)  (\chi_i \chi_{i \mu}) \varphi_{\nosymbol} \varphi_{\mu} +
  (\chi_i^2)^2 \varphi_{\mu}^2 - 2 (\chi_i^2)^2 \varphi_{\nosymbol}
  \omega_{\nosymbol} + \text{susy-writable} . \nonumber
\end{eqnarray}
A third one $\partial_{\mu} (\sigma_2 \sigma_{4 \nobracket (\mu) \nobracket})$
is linearly dependent with these two at the susy-null level. One can also
check that $\frac{1}{3} \partial_{\mu} (\sigma_3 \sigma_{3 \nobracket (\mu)
\nobracket}) = \frac{1}{18} \partial^2 (\sigma_3^2)$ has the same susy-null
part as $\frac{1}{12} \partial^2 (\sigma_4 \sigma_{2 \nobracket \nobracket})$.
Taking these into account, there remains exactly one susy-null singlet scalar
at this level which is not a total derivative, whose explicit expression is
\begin{equation}
  \sigma_{3 (\mu)}^2 - \frac{4}{3} \sigma_{2 (\mu)} \sigma_{4 (\mu)} +
  \frac{1}{3} \sigma_{1 (\mu)} \sigma_{5 (\mu)} = [(\chi_i^2)^2
  \varphi_{\mu}^2]_{\Delta = 12} + [\tmop{follower}]_{\Delta = 13} + \ldots
  \label{Nf6Nd2}
\end{equation}
Finally, at $N_{\phi} = 6$, $N_{\tmop{der}} = 4$ with $\Delta \leqslant 12$ we
find only susy-writable leaders, obtained from $\sigma_6$ dressed with
derivatives.

\subsection{$N_{\phi} = 8, 10$}

Leaders with $N_{\phi} = 8$ and $N_{\tmop{der}} = 0$ with $\Delta \leqslant
12$ arise from 5 singlets:
\begin{equation}
  \sigma_8, \quad \sigma_1 \sigma_7, \quad \sigma_2 \sigma_6, \quad \sigma_3
  \sigma_5, \quad \sigma_4^2,
\end{equation}
but there are only three independent leaders with $\Delta \leqslant 12$: two
susy-writable and one susy-null (Table \ref{Nf8Nd0}). This is because three
linear combinations $\sigma_2 \sigma_6 - \frac{12}{7} \sigma_1 \sigma_7$,
$\sigma_3 \sigma_5 - \frac{15}{7} \sigma_1 \sigma_7$ and $\sigma_4^2 -
\frac{16}{7} \sigma_1 \sigma_7$ all have the same $\Delta = 12$ leading term
$(\chi_i^2)^2 \varphi_{\nosymbol}^4$. Taking further differences we could
cancel this leading term and exhibit further leaders of higher dimensions. We
will not do it here since we are interested only in $\Delta \leqslant 12$.

\begin{table}[h]\centering
    \begin{tabular}{@{}lll@{}}
      \toprule
      Singlet & Leader (+ First follower if susy-null) & Leader type\\
      \midrule
      $\sigma_8$ & $\nobracket [28 (\chi_i^2) \varphi_{\nosymbol}^6 + 8
      \varphi_{\nosymbol}^7 \omega_{\nosymbol}]_{\Delta = 10} \nobracket$ &
      susy-writable\\
      $\sigma_1 \sigma_7$ & $\nobracket [21 (\chi_i^2) \varphi_{\nosymbol}^5
      \omega_{\nosymbol} + 7 \varphi_{\nosymbol}^6
      \omega_{\nosymbol}^2]_{\Delta = 12} \nobracket$ & susy-writable\\
      $\sigma_2 \sigma_6 - \frac{12}{7} \sigma_1 \sigma_7$ & $[15 (\chi_i^2)^2
      \varphi_{\nosymbol}^4]_{\Delta = 12} + [20 (\chi_i^2) (\chi_i^3)
      \varphi_{\nosymbol}^3 - 20 (\chi_i^3) \varphi_{\nosymbol}^4
      \omega_{\nosymbol}]_{\Delta = 13}$ & susy-null\\
      \bottomrule
    \end{tabular}
  \caption{\label{Nf8Nd0}Scalar $\mathbb{Z}_2$-even leaders with $N_{\phi} =
  8$, $N_{\tmop{der}} = 0$.}
\end{table}

The only scalar leader with $N_{\phi} = 8$, $N_{\tmop{der}} = 2$ and $\Delta
\leqslant 12$ comes from $\sigma_{8 (\mu) (\mu)}$, which is equivalent to the
total derivative $\partial^2 \sigma_8$ (modulo EOM and $\sigma_1 \sigma_7$).

There is only one $\Delta \leqslant 12$ leader with $N_{\phi} = 10$, and it is
susy-writable:
\begin{equation}
  \sigma_{10} = [45 (\chi_i^2) \varphi^8 + 10 \varphi^9 \omega]_{\Delta = 12}
  + \ldots
\end{equation}
\section{Free $\mathcal{L}_0$ propagators}\label{props}\label{App:prop}

Propagators follow from {\eqref{L0}}, putting $V (\vf) = 0$. The $\vf$-$\vf$
and $\omega$-$\vf$ propagators are obtained by diagonalizing the quadratic
terms containing $\vf$ and $\omega$:
\begin{equation}
  \label{freeprop} G_{\varphi \varphi} (q) = \frac{H}{q^4}, \qquad G_{\omega
  \vf} (q) = \hspace{0.27em} \frac{1}{q^2} .
\end{equation}
The $\chi$-$\chi$ is obtained from the $\chi$ kinetic term, taking into
account the constraint $\sum_{i = 2}^n \chi_i = 0$; it is given by:
\begin{equation}
  \label{prop-1} G_{\chi_i \chi_j} (q) = \frac{K_{i j}}{q^2}, \qquad K_{ij}
  \equiv \delta_{ij} - \frac{1}{n - 1} \Pi_{ij},
\end{equation}
where $\Pi_{ij} = 1$ for all $i, j = 2, \ldots, n$. This computation is done
either by realizing the constraint by a Lagrange multiplier, or equivalently
by eliminating one of the fields via the constraint, and inverting the
quadratic term for the remaining independent fields. E.g.~by eliminating
$\chi_2$ we get the quadratic action $\frac{1}{2}  \sum_{i, j = 3}^n
(\partial_{\m} \chi_i)  (\delta_{ij} + \Pi_{ij})  (\partial^{\m} \chi_j)$
which gives the above propagator.

\begin{figure}[h]\centering \includegraphics[width=400pt]{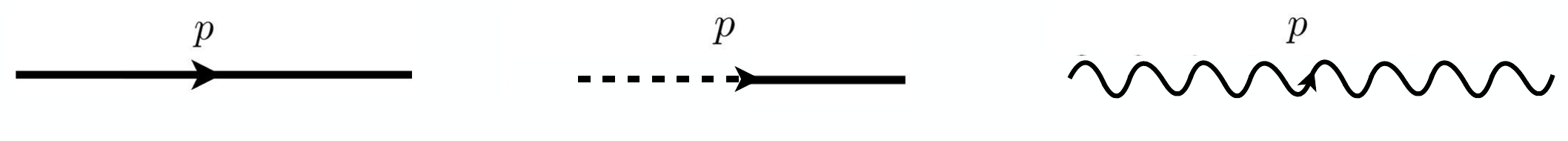}
  \
  \caption{\label{propagators}(From left to right) Propagators $G_{\vf \vf}
  (p)$ [solid line], $G_{\omega \vf} (p)$ [dotted half connects to $\omega$,
  solid to $\varphi$] and $G_{\chi_i \chi_j} (p)$ [wavy line, indices $i, j$
  understood].}
\end{figure}

In position space the propagators read
\begin{equation}
  \label{2ptfree} G_{\varphi \varphi} (x) = \frac{H A_d}{2 (d - 4)} 
  \frac{1}{(x^2)^{\frac{d}{2} - 2}}, \quad G_{\omega \vf} (x) =
  \frac{A_d}{(x^2)^{\frac{d}{2} - 1}}, \quad G_{\chi_i \chi_j} (x) = A_d 
  \frac{K_{i j}}{(x^2)^{\frac{d}{2} - 1}},
\end{equation}
where $A_d = \frac{\Gamma (\frac{d}{2} - 1) 2^{d - 2}}{(4
\pi)^{\frac{d}{2}}}$. It is easy to check that $\partial^2 G_{\varphi \varphi}
= - H G_{\omega \varphi}$ consistently with the equation of motion $\partial^2
\varphi = - H \omega$. When drawing Feynman diagrams, propagators are denoted
as in Fig.~\ref{propagators}.

The matrix $K_{ij}$ satisfies some useful relations, which are easy to check:
\begin{eqnarray}
  &K^T = K, \qquad K^2 = K, \qquad \tmop{tr} K = n - 2, \qquad \sum_{i = 2}^n
  K_{ij} = 0, \qquad \sum_{i, j = 2}^n K_{ij} K_{ij} = n - 2,  & 
  \label{properties}\\
 & \sum_{i = 2}^n K_{ij} \chi_i = \chi_j, \qquad \sum_{i, j = 2}^n K_{ij}
  \chi_i^m \chi_j = \sum_{i = 2}^n \chi_i^{m + 1} .  &  \nonumber
\end{eqnarray}
The last two relations follow using $\sum_{i = 2}^n \chi_i = 0$.

Our RG calculations will only involve the $\chi_i$ fields of the
$\mathcal{L}_0$ theory. Some calculations could be equivalently performed
using the $\mathcal{L}_{\tmop{SUSY}}$ theory. For completeness we give the
corresponding propagators obtained by setting $V = 0$ in {\eqref{LSUSY1}}. The
$\varphi$-$\varphi$ and $\varphi$-$\omega$ propagators are the same as for
$\mathcal{L}_0$, while the $\psib$-$\psi$ one is
\begin{equation}
  \label{prop} G_{\psib \psi} (x) = G_{\vf \omega} (x) \hspace{0.17em},
\end{equation}
in agreement with the general relation {\eqref{Ward}} for SUSY 2pt functions,
be that free or interacting. All individual propagators can be extracted from
the superfield propagator
\begin{equation}
  \label{supprop} G_{\Phi \Phi} (x, \theta) = \frac{A_d}{d - 4}  \frac{1}{(x^2
  - (4 / H) \theta \bar{\theta})^{\frac{d - 4}{2}}} .
\end{equation}
\section{RG at one loop}\label{ope}

In this section we will discuss how to set up RG computations of beta
functions and anomalous dimensions. To keep technical details to a minimum, we
discuss renormalization here at one loop, and then in Appendix \ref{2loop} at
two loops. At one loop there is no wavefunction renormalization, and we can
think in terms of the free theory $\mathcal{L}^{(0)}$ defined by setting $V =
0$ in {\eqref{L0}}, in $d = 6 - \e$ dimensions and perturbed by scalar
operator $\mathcal{V}= \left( 4 \omega \vf^3 + 6 \chi_i^2  \vf^2 \right)$ of
dimension $\D^0_{\mathcal{V}} = 6 - 2 \e$:

\begin{equation}
  \mathcal{L}=\mathcal{L}^{(0)} + \mu^{\varepsilon} \frac{\lambda}{4!}
  \mathcal{V}. \label{Hpert}
\end{equation}

Here $\mu$ is an arbitrary mass scale. We use dimensional regularization as the
regulator. We start by discussing renormalization of local operators in the
theory $\mathcal{L}$, and later use the same principles for renormalization of
$\lambda$. Renormalized operators $\mathcal{O}_i$ are related to the bare
operators $\mathcal{O}_i^B$ built from bare fields ($B$ stands for ``bare''),
via a mixing (renormalization) matrix $Z_{i j}$ as follows:
\begin{equation}
  \mathcal{O}^B_i = Z_{i j} \mathcal{O}_j . \label{Zij}
\end{equation}
While correlators of bare operators have poles in $\varepsilon$, renormalized
operators \ $\mathcal{O}_j$ are defined so that their correlation functions do
not have such poles.

The matrix $Z_{i j}$ admits an expansion in powers of $1 / \varepsilon$ and
$\lambda$, which at one loop as we are interested here takes the form
\begin{equation}
  Z_{i j} = \tmd_{i j} + \frac{\lambda}{\varepsilon} z_{i j} + \ldots
  \hspace{0.17em} . \label{ZijForm}
\end{equation}
It can be shown that the matrix $z_{i j}$ has a simple block diagonal form,
where each block corresponds to operators with equal number of fields and the
same classical dimension, \ $\D^0_i = \D^0_j$, where $\D_i^0$ is the bare
dimension of $\mathcal{O}_i$.

The anomalous dimension matrix is defined in terms of the $Z$ matrix as
follows:
\begin{equation}
  \begin{array}{ll}
    \Gamma (\lambda) & \equiv \: Z^{- 1} . \frac{d}{d \log \mu} Z \,
  \end{array} . \hspace{0.17em}  \label{anomdimmat}
\end{equation}
Then, by using \ $\frac{d}{d \log \mu} Z = \frac{d Z}{d \lambda}  \frac{d
\lambda}{d \log \mu}$ and $\beta_{\lambda} \equiv \frac{\partial
\lambda}{\partial \log \mu} = - \e \lambda + O (\lambda^2)$ (see below for the
discussion of the beta-function), we obtain a simple formula relating $\G_{i
j}$ and $z_{i j}$:
\begin{equation}
  \G_{i j} (\lambda) = - \lambda z_{i j} + O (\lambda^2) . \label{ztoGamma}
\end{equation}
To compute the anomalous dimensions of operators at the fixed point, we should
evaluate the anomalous dimension matrix at the fixed point coupling: $\G
\equiv \G (\lambda_{\ast})$.

Diagonalizing $\G$ one obtains the set of renormalized operators with well
defined anomalous dimensions. Namely given an eigenvector $e^{(m)}$ such that
$\sum_i e^{(m)}_i \G_{i j} = \g_m e^{(m)}_i$, one obtains the renormalized
operators $\mathcal{O}^R_m = \left( 1 + \frac{\g_m}{\e} \right) \sum_j
e^{(m)}_j \mathcal{O}^B_j$ with anomalous dimension $\g_m$.

Sometimes it will happen for us that $\G$ is not fully diagonalizable, i.e.~it has fewer eigenvalues than its size. In this case we can still bring $\G$ to
a Jordan normal form, and define generalized eigenvectors $e^{(1)}_i, \ldots
., e^{(r)}_i$ associated to each eigenvalue $\g$, where $r$ is the rank of the
corresponding Jordan block, which satisfy the Jordan chain property $\sum_i
e^{(k)}_i \left( \G_{i j} - \g 1 \right) = (1 - \delta_{k 1}) e^{(k - 1)}_i$.
The associated renormalized operators $\mathcal{O}^R_k = \left( 1 +
\frac{\g}{\e} \right) \sum_j e_j^{(k)} \mathcal{O}^B_j$ \ ($k = 1, \ldots,
r$) define a logarithmic multiplet which satisfies the property
\begin{equation}
  D \left(\begin{array}{c}
    \mathcal{O}^R_r\\
    \mathcal{O}^R_{r - 1}\\
    \vdots\\
    \mathcal{O}^R_1
  \end{array}\right) = \, \left(\begin{array}{cccc}
    \Delta & 1 &  & \\
    & \Delta & 1 & \\
    &  & \ddots & \ddots\\
    &  &  & 
  \end{array}\right) \left(\begin{array}{c}
    \mathcal{O}^R_r\\
    \mathcal{O}^R_{r - 1}\\
    \vdots\\
    \mathcal{O}^R_1
  \end{array}\right), \label{Jordan}
\end{equation}
where $D$ is the dilatation operator and $\Delta = \Delta^0 + \gamma$ (where
$\Delta^0$ is the classical dimension of the operators which undergo mixing).
The presence of logarithmic multiplets signals that we are working in a
logarithmic CFT. This is somewhat expected since we are studying a theory with
$S_n$ symmetry in the limit $n \rightarrow 0$ {\cite{Cardy:2013rqg}}.

\subsection{OPE method }\label{OPEmethod}

Another simplification which arises at one loop is that the RG functions can
be quickly computed using the OPE method, which we will review here. This is
not obligatory and the same results can be obtained with Feynman diagrams. The
OPE method saves a lot of time especially in situations when one has to
disentangle mixing of many operators. Our presentation of the OPE method
mimics {\cite{Hogervorst:2015akt}}. A classic reference for the OPE method is
{\cite{cardy_1996}}, Chapter 5 (although it uses a real-space cutoff, not
dim.reg. like us).

We consider a correlation functions $\langle \mathcal{O}^B_i (0) \ldots
\rangle$ of a bare scalar operator $\mathcal{O}^B_i$ with an arbitrary number
{$\ldots$} of other operators. The leading order correction to this
correlator can be computed by expanding the functional integral (associated to
the Lagrangian {\eqref{Hpert}}) at first order, and is given by
\begin{equation}
  - \frac{\lambda \mu^{\e}}{4!} \int d^d x \langle \mathcal{V} (x)
  \mathcal{O}^B_i (0) \ldots \rangle, \label{leading}
\end{equation}
with $d = 6 - \varepsilon$. To renormalize $\mathcal{O}_i$, we need to understand
the $1 / \varepsilon$ pole of this expression, associated with the $x \rightarrow
0$ part of the integration region. This is easy to do using the OPE between
the operators $\mathcal{O}^B_i$ and the interaction. The OPE takes the form
\begin{equation}
  \mathcal{V} (x) \times \mathcal{O}^B_i (0) = \sum_j C_{i, j}  | x |^{-
  \D^0_{\mathcal{V}} + \Delta_i^0 - \Delta_j^0} \mathcal{O}^B_j (0) 
  \label{OPEpartial}
\end{equation}
(this form is adequate for the case when the operator $\mathcal{O}^B_i$ does
not contain derivatives, see below for the general case). This needs to be
integrated for $x$ near 0, say over $| x | \leqslant 1$, the upper limit being
arbitrary. For $\D^0_{\mathcal{V}} = 6 - 2 \varepsilon$ as we are considering,
the integral $\int_{| x | \nobracket \leqslant 1 \nobracket} d^d x\, | x |^{-
\D^0_{\mathcal{V}} + \Delta_i^0 - \Delta_j^0}$ gives a pole in $\varepsilon$ as
long as $\Delta_i^0 - \Delta_j^0 = O (\varepsilon)$. It can be shown similarly to
{\cite{Hogervorst:2015akt}} that a selection rule guarantees that all such
operators with nonzero OPE coefficients $C_{i, j}$ have $\Delta_i^0 =
\Delta_j^0$.

In our case the OPE matrix $C_{i, j}$ also has selection rules between the
three operator classes:
\begin{eqnarray}
  \mathcal{V} \times \text{susy-null} & = & \text{susy-null}, \nonumber\\
  \mathcal{V} \times \text{susy-writable} & = & \text{susy-writable} +
  \text{susy-null},  \label{selRule}\\
  \mathcal{V} \times \text{non-susy-writable} & = & \text{non-susy-writable +
  susy-writable} + \text{susy-null} . \nonumber
\end{eqnarray}
I.e. OPE of $\mathcal{V}$ with a susy-null operator produces only susy-null
operators in the r.h.s., etc. These rules follow by the symmetry
considerations as in the mixing discussion in section \ref{SUSY-null}.

Going back to Eq.~{\eqref{OPEpartial}}, it implies that any correlator
$\langle \mathcal{O}^B_i (0) \ldots \rangle$ has poles in $\varepsilon$
proportional to $\lambda \langle \mathcal{O}^B_j (0) \ldots \rangle$. The
renormalized operators are defined by correcting $\mathcal{O}^B_i$ to cancel
these poles. We see that this can be achieved by
\begin{equation}
  \mathcal{O}_i =\mathcal{O}^B_i + \varepsilon^{- 1} \frac{\lambda}{4!} \sum_j
  C_{i, j} S_d \mathcal{O}^B_j, \label{OiOiB}
\end{equation}
where $S_d = \frac{2 \pi^{d / 2}}{\Gamma (d / 2)}$ denotes the area of the
unit sphere in $d$ dimensions. Inverting this relation, we obtain the matrix
$z_{i, j}$ {\eqref{ZijForm}} as: $z_{i j} = - \frac{\lambda}{4!} C_{i, j}
S_d$.

By Eq.~{\eqref{selRule}}, this matrix (and hence $Z$
itself) has a block-triangular structure among the three operator classes
(susy-null, susy-writable, and non-susy-writable, in this order):
\begin{equation}
  Z = \left(\begin{array}{ccc}
    \ast & 0 & 0\\
    \ast & \ast & 0\\
    \ast & \ast & \ast
  \end{array}\right) . \label{BD}
\end{equation}

Symmetry considerations from section \ref{SUSY-null} show that this structure
should hold generally, at any number of loops. Furthermore, the same
block-triangular structure in inherited by the anomalous dimension matrix
{\eqref{anomdimmat}}.

The above explains the general idea, up to the need to generalize Eq.
{\eqref{OPEpartial}} a bit when considering operators containing derivatives.
The more general expression used in our computations is
\begin{equation}
  \mathcal{V} (x) \times \mathcal{O}_i^B (0) \quad \sim \hspace{1.4em} \sum_k
  C_{i, k} \frac{x^{\mu_1} \ldots x^{\mu_{\ell}}}{| x |^{\D^0_{\mathcal{V}} +
  \ell_k}}  (\mathcal{T}_k)_{\mu_1 \ldots \mu_{\ell_k}} (0) \hspace{0.17em},
  \label{opesimp}
\end{equation}
where $(\mathcal{T}_k)_{\mu_1 \ldots \mu_{\ell_j}}$ is a tensor of rank
$\ell_k$ (not necessarily traceless) and of dimension $\D^0_k = \D^0_i$. The
OPE coefficient function is now a tensorial function of scaling $-
\D^0_{\mathcal{V}}$, and it will give an $\varepsilon$ pole when integrated near
$x = 0$, projecting the operator $\mathcal{T}_k$ on its scalar component (see
below).

Let us give an example of how {\eqref{opesimp}} works. If we consider the OPE
of $\mathcal{O}_i^B = \vf \omega$ we find 3 possible operators $\mathcal{T}_k$
with the same dimension of $\mathcal{O}_i^B$: the two scalars $\vf \omega$,
$\chi_i^2$ and the rank-2 tensor operator $\partial^{\mu_1} \partial^{\mu_2}
\vf^2$. The scalars are the remaining operators after we take two Wick
contractions of fields in $\mathcal{O}_i $ and in $\mathcal{V}$,
\begin{equation}
  \left( 4 \omega \vf^3 + 6 \chi_i^2  \vf^2 \right) (x) \times \vf \omega (0)
  \sim 6 \times 2 \langle \vf \vf \rangle_0 \langle \vf \omega \rangle_0
  \chi_i^2 (0) \: + 4 \times 6 \langle \vf \vf \rangle_0 \langle \vf \omega
  \rangle_0  \vf \omega (0) \: + \ldots \label{opesimpEg} 
\end{equation}
Here the factor 2 in the first term and 6 in the second, are the combinatorial
factors which count the possible Wick contractions. We use the notation
$\langle \ldots \rangle_0$ to denote the 2-point functions of
$\mathcal{L}^{(0)}$. Since it is free, we have simply $\langle \vf \omega
\rangle_0 = \langle \vf (x) \omega (0) \rangle_0 = \langle \vf (0) \omega (x)
\rangle_0 = G_{\omega \vf} (x)$ and $\langle \vf \vf \rangle_0 = G_{\vf \vf}
(x)$. The 2-tensor operator arises from the following OPE:
\begin{equation}
  \begin{array}{lll}
    \left( 4 \omega \vf^3 + 6 \chi_i^2  \vf^2 \right) (x) \times \vf \omega
    (0) & \sim & 4 \times 3 \langle \vf \omega \rangle_0 \langle \vf \omega
    \rangle_0  \vf^2 (x) \: + \ldots\\
    & = & 4 \times 3 \langle \vf \omega \rangle_0 \langle \vf \omega
    \rangle_0  \frac{x^{\mu_1} x^{\mu_2}}{2} \partial_{\mu_2} \partial_{\mu_2}
    \vf^2 (0) + \ldots
  \end{array} \label{OPEeg1}
\end{equation}
In this case the tensor structure is obtained by Taylor expanding the
remaining field $\vf^2$ of $\mathcal{V} (x)$, keeping only the second order
term since it gives a field of the same dimensions as $\vf \omega$ (others do
not produce poles in $\varepsilon$ when integrated near $x = 0$).

When integrating {\eqref{opesimp}} over $x$, we have to deal with tensorial
integrals. E.g.~after using {\eqref{OPEeg1}} we are led to an integral of the
form $\int_{| x | \leqslant 1} d^d x\, | x |^{- \D^0_{\mathcal{V}} - 2} x^{\mu_1} x^{\mu_2}$, whose pole part is easily seen equal to  $\varepsilon^{-
1} S_d  \tmd^{\mu_1 \mu_2} / d$ by rotation invariance. For general $\ell$ we
have:
\begin{equation}
  \int d \Omega\, \hat{x}^{\mu_1} \ldots \hat{x}^{\mu_{\ell}} = \,
  P_{\ell}^{(d)}  \tmd^{(\nobracket \mu_1 \mu_2} \ldots \tmd^{\nobracket
  \mu_{\ell - 1} \mu_{\ell})} S_d, \label{sphereintegralsl}
\end{equation}
where $P_{2 \ell}^{(d)} = \frac{(2 \ell - 1) ! \nobracket !}{2^{\ell} (d /
2)_{\ell}} $, $P_{2 \ell + 1}^{(d)} = 0$ and the brackets imply symmetrization
of the indices.

After performing these integrals we are left with a product of Kronecker
deltas contracted with the tensor operators, i.e.~scalars. E.g.~the operator
$\partial_{\mu_2} \partial_{\mu_2} \vf^2 (0)$ of {\eqref{OPEeg1}} after
integration is contracted with \ $\tmd^{\mu_1 \mu_2}$ and becomes equal to
$\partial^2 \vf^2 = 2 \partial \vf \partial \vf + 2 \vf \partial^2 \vf = 2
\partial \vf \partial \vf - 2 H \vf \omega$. More generally we want to write
the contraction of $(\mathcal{T}_k)_{\mu_1 \ldots \mu_{\ell}}$ with $\tmd^{\mu
\nu}$'s in {\eqref{sphereintegralsl}} in terms of the scalar operators
$\mathcal{O}_i^B$ that we want to study. We write these as
\begin{equation}
  \tmd^{(\nobracket \mu_1 \mu_2} \ldots \tmd^{\nobracket \mu_{\ell - 1}
  \mu_{\ell})} (\mathcal{T}_k)_{\mu_1 \ldots \mu_{\ell}} = \sum_j n^{(k)}_j
  \mathcal{O}_j^B (0), \label{contractionindices}
\end{equation}
where the above formula can be read as a definition for the coefficients
$n^{(k)}_j$.

Putting these ingredients together, we rewrite the integral of the $k$-th
operator in {\eqref{opesimp}} \ (up to the overall factor $- \frac{\lambda
\mu^{\e}}{4!} C_{i, j}$) as follows,
\begin{equation}
  \int d^d x\, \frac{x^{\mu_1} \ldots x^{\mu_{\ell}}}{| x |^{\D^0_{\mathcal{V}}
  + \ell}} \langle (\mathcal{T}_k)_{\mu_1 \ldots \mu_{\ell}} (0) \ldots
  \rangle = \frac{1}{\e} P_{\ell_k}^{(d)} S_d  \sum_j n^{(k)}_j \langle
  \mathcal{O}_j^B (0) \ldots \rangle .
\end{equation}
This shows how one should generalize {\eqref{OiOiB}}. From here we obtain a
final formula for $z_{i j}$, and therefore for the 1-loop anomalous dimensions
matrix,
\begin{equation}
  \begin{array}{ll}
    \Gamma_{i j} (\lambda) & = \: \frac{\lambda}{4!} S_d \sum_k
    P_{\ell_k}^{(d)} \, C_{i, k} n^{(k)}_j . \,
  \end{array} \label{anommatres}
\end{equation}
Given the block-triangular structure {\eqref{BD}} with
respect the three operator classes, only diagonal blocks of this matrix matter
for the purposes of computing the anomalous dimensions. The off-diagonal
blocks do influence the eigenperturbations (e.g. scaling susy-writable
operators will have an admixture of susy-nulls), but do not modify the
eigenvalues.

Furthermore, while using equation {\eqref{anommatres}} we will encounter
operators that will be related to one another by addition of a total
derivative. Perturbing the action by an integral of a total derivative of
course has no effect on physics since the integral vanishes. This can be also
expressed by saying that under RG, total derivatives may generate only total
derivatives. In the OPE formalism this manifests itself as the fact that when
we take the OPE of $\mathcal{V}$ with a total derivative operator, only total
derivatives occur in the r.h.s.\footnote{This is obvious, by differentiating
the OPE of the parent operator.} These observations can be used to simplify
the computations of anomalous dimensions, as follows. We will say that two
operators $\mathcal{O}$ and $\mathcal{O}'$ belong to the same 
equivalence class $\{ \mathcal{O} \}$ if they are proportional up to adding a
total derivative, i.e.~$\mathcal{O}' = \alpha \mathcal{O}+ \partial^{\mu}
\tilde{\mathcal{O}}_{\mu}$ for some constant $\alpha$ and operator
$\mathcal{O}_{\mu}$. For each $\{ \mathcal{O}_i \}$ we will choose a single
representative element $\mathcal{O}_i$ while all other elements can be written
in terms of $\mathcal{O}_i$ by adding a suitable total derivative. Eg. in the
equivalence class $\{ \omega^2 \}$ we have an operator $\mathcal{O}=
\omega^2$, and also another operator $\mathcal{O}' = \partial^{\mu} \vf
\partial_{\mu} \omega = H \omega^2 + \partial^{\mu} \left( \omega
\partial_{\mu} \vf \right)$, which we can see can be written in terms of
$\mathcal{O}$ by using equation of motion and adding a total derivative.

Replacing operators by their equivalence classes (picking one representative
in each class), we eliminate total derivative operators from the problem and
get a smaller eigenvalue system to solve, which however gives rise to the same
anomalous dimension for the non-total-derivative operators as the full system.
In some cases one may be interested to recover in each equivalence class a
primary, i.e.~an operator which has good scaling behavior under RG, including
the total derivative part, and has zero 2-point functions with other
primaries. The problem of finding such a scaling operator is harder, and to
solve it one has to work with the full operators, not with the equivalence
classes, i.e.~to diagonalize the full matrix $\G$.

\subsection{Beta function}

One can also compute beta function of the coupling $\lambda$ with the OPE
approach. Once again Feynman diagrams will give the same result. Consider the
interaction in {\eqref{Hpert}} with $\mu^{\varepsilon} \frac{\lambda}{4!}$
replaced by $\frac{\lambda_B}{4!}$. We should find a relation between the bare
and renormalized couplings, $\lambda_B$ and $\lambda$ of the
form,\footnote{Strictly speaking we should treat the two vertices $\vf^3
\omega$ and $\vf^2 \chi^2$ differently. If their coupling constants are
$\lambda_B^{(1)}$ and $\lambda_B^{(2)}$ respectively, we should write
$\lambda_B^{(1)} = Z_{\lambda}^{(1)} \lambda^{(1)}$ and $\lambda_B^{(2)} =
Z_{\lambda}^{(2)} \lambda^{(2)}$. However as commented later, when
$\lambda_B^{(1)} = \lambda_B^{(2)}$ it turns out $Z_{\lambda}^{(1)} =
Z_{\lambda}^{(2)}$. \ \ }
\begin{equation}
  \lambda_B = \mu^{\e} Z_{\lambda} \lambda \label{Zlam},
\end{equation}
which removes poles in $\varepsilon$ when everything is expressed in terms of
$\lambda$. Here at leading order $Z_{\lambda} = 1 + \lambda / \varepsilon$. Note
that the wavefunction renormalization of fields $\vf$, $\omega$ and $\chi$
starts at $O (\lambda^2)$, see the next section. Now take any correlator
$\langle A \rangle$ where $A$ is a product of some fields. Consider its first
and second order corrections, which are given below:
\begin{equation}
  - \frac{\lambda_B}{4!} \int d^d x \langle \mathcal{V} (x) A \rangle +
  \frac{1}{2} \left( \frac{\lambda_B}{4!} \right)^2 \int d^d x \int d^d y
  \langle \mathcal{V} (x + y) \mathcal{V} (x) A \rangle . \label{lamcorr}
\end{equation}
The second term will have an $\varepsilon$ pole due to the singularity as $y
\rightarrow 0$. We choose $Z_{\lambda}$ such that this pole is canceled by the
first term. For this we consider the OPE (we can do this for $x = 0$ by
translational invariance):
\begin{eqnarray}
  \vf^3 \omega (y) \times \vf^3 \omega (0) \; & \sim & 36 \langle \vf \omega
  \rangle_0 \langle \vf \vf \rangle_0  \vf^3 \omega (0) \hspace{0.17em} +
  \ldots 
\end{eqnarray}
The product $\langle \vf \omega \rangle_0 \langle \vf \vf \rangle_0$ gives a
pole in $\varepsilon$ when integrated over $y$ near 0. Plugging this into
{\eqref{lamcorr}} and demanding the cancellation we get
\begin{equation}
  Z_{\lambda} = 1 + \frac{3 \lambda H}{(4 \pi)^3 \e} \label{ZPhi4} .
\end{equation}
The beta function $\beta_{\lambda} \equiv \mu \frac{\partial \lambda}{\partial
\mu}$ can then be obtained using that the bare coupling does not depend on
$\mu$:
\begin{equation}
  0 = \frac{\partial \log \lambda_B }{\partial \log \mu} =
  \frac{\partial}{\partial \log \mu} \left[ \log \left( \mu^{\e} Z_{\lambda}
  \lambda \right) \right] = \e + \frac{3 H}{(4 \pi)^3 \e} \beta_{\lambda} +
  \frac{1}{\lambda} \beta_{\lambda}, \label{betaevl}
\end{equation}
from where
\begin{equation}
  \beta_{\lambda} = - \varepsilon \lambda+ \frac{3 H \lambda^2}{64 \pi^3} + O (\lambda^3)
  \hspace{0.17em} . \label{betaapp}
\end{equation}
This is the beta function {\eqref{betaSUSY}} of our theory. It gives a fixed
point at $\lambda_{\ast} = \frac{64 \pi^3 \varepsilon}{3 H} + O \left( \e^2 \right)
.$

We can also consider the OPEs that generate $\chi_i^2  \vf^2$. It will give
rise to the same $Z_{\lambda}$ and the same beta function. This can be seen a
consequence of the $O (n)$ symmetry in $n \rightarrow 0$ limit, or of the
hidden supersymmetry which becomes manifest in the form {\eqref{LSUSY1}} that
we get from {\eqref{L0}} once we substitute $\chi_i$ with $\psi, \psib$ using
{\eqref{fromchitopsi}}.

We could also discuss renormalization of the mass term $m^2 \left( \vf \omega
+ \frac{1}{2} \chi_i^2 \right)$ in the same language. This term renormalizes
as a whole for the same reasons. \ This matches nicely with the fact that to
reach the critical point (phase transition) we have to tune a single parameter
($m^2$).

\section{RG at two loops}\label{2loop}

After one-loop RG in App. \ref{ope}, here we set up the more general scheme
valid at any loop order. In practice we will go to the maximum of two loops in
some anomalous dimension computations for which the one-loop result vanishes.
The regulator will be the same as in App. \ref{ope} (dim. reg.). The OPE
method losing its simplicity beyond one loop, here we will be using instead
Feynman diagram to extract poles in $\varepsilon$. We are assuming the reader is
somewhat familiar with dimensionally regulated computations for the
Wilson-Fisher fixed point (see e.g. a nice review in
{\cite{Kleinert:2001ax}}), to which our case is rather similar.

We will present the setup for computations in terms of the fields $\chi_i$,
Although susy-writable computations can be done (and even become easier) in
terms of $\psi, \psib$, we need the general setup for computations of
anomalous dimensions of susy-null and non-susy-writable operators.

\subsection{Beta function}

We start with Lagrangian {\eqref{L0phi4}} with zero mass term and all
quantities set to their bare values:
\begin{equation}
  \Lcal_0 = \partial \varphi_B \partial \omega_B - \frac{H}{2} \omega_B^2 +
  \frac{1}{2} (\partial \chi_B)^2 + \frac{\lambda_B}{4!} \left( \left. 4
  \omega_B  \vf_B^3 + 6 \chi_B^2  \vf_B^2 \right) \right. . \label{lagL0}
\end{equation}
The bare quantities are related to the renormalized ones by:
\begin{equation}
  \vf_B = Z_{\vf} \vf, \hspace{0.5cm} \omega_B = Z_{\omega} \omega,
  \hspace{0.5cm} \chi_B = Z_{\chi} \chi, \hspace{0.5cm} \lambda_B =
  Z_{\lambda} \mu^{\e} \lambda . \label{counterterms}
\end{equation}
where $Z_i$ are renormalization constants, and $\mu$ is an arbitrary mass
scale. Correlators of renormalized fields $\vf, \omega, \chi$ have to be free
of poles in $\varepsilon$ when expanded in the renormalized coupling
dimensionless coupling $\lambda$. Compared to App. \ref{ope} we are adding
wavefunction renormalization constants $Z_{\varphi}, Z_{\omega}, Z_{\chi}$,
necessary beyond one loop.

Plugging {\eqref{counterterms}} into {\eqref{lagL0}} we get:
\begin{equation}
  \quad \Lcal_0 = Z_{\vf} Z_{\omega} \partial \varphi \partial \omega -
  Z_{\omega}^2  \frac{H}{2} \omega^2 + \frac{Z_{\chi}^2}{2} (\partial
  \chi_i)^2 + \frac{Z_{\lambda} \mu^{\e} \lambda}{4!} \left( 4 Z_{\vf}^3
  Z_{\omega} \omega \vf^3 + 6 Z_{\chi}^2 Z_{\vf}^2 \chi_i^2  \vf^2 \right) .
  \label{lagct}
\end{equation}

In the Minimal Subtraction (MS) scheme, the renormalization constant
$Z_{\lambda}$ takes the form:
\begin{equation}
  \quad Z_{\lambda} = 1 + \sum_{p \geqslant 1} \sum_{1 \leqslant q \leqslant
  p} z^{(p, q)}_{\lambda} \lambda^p \varepsilon^{- q} . \label{Zform}
\end{equation}
(compare to {\eqref{ZijForm}}). The quantities $Z_{\vf}$, $Z_{\omega}$,
$Z_{\chi}$ have similar expansions, except that the corresponding $z^{(1, 1)}$
vanishes (see App. \ref{wfrenorm}).

\begin{figure}[h]
 \centering \includegraphics[width=450pt]{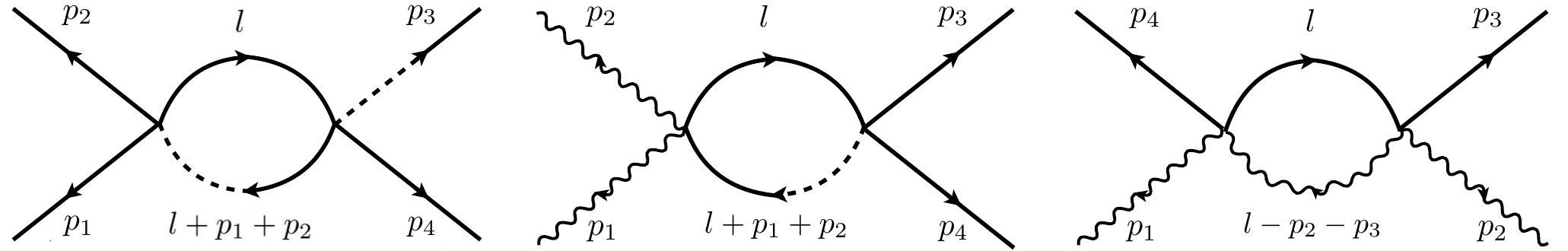}
  \caption{\label{betafnfig}The Feynman diagrams for the one-loop
  renormalization of $\lambda$. The first diagram corrects $\langle \vf (p_1)
  \vf (p_2) \vf (p_3) \vf (p_4) \rangle$, while the other two would arise for
  $\langle \chi_j (p_1) \chi_j (p_2) \vf (p_3) \vf (p_4) \rangle$.}
\end{figure}

Let us use this formalism to re-compute the one-loop beta function (two-loop
beta function is not needed in this paper). We will consider the
momentum-space 4-point function $\langle \vf (p_1) \vf (p_2) \vf (p_3) \vf
(p_4) \rangle$ and the condition that it should be free of poles in $\varepsilon$
will determine $z_{\lambda}^{(1, 1)}$. At tree level this 4-point function
involves a single $\omega \vf^3$ vertex. At one loop, we have the first
diagram in Fig \ref{betafnfig} (and two diagrams for the crossed channels).
Factoring out the trivial dependence on the external momenta coming from the
propagators on the external legs, we get the amputated 4-point function
\begin{equation}
  \langle \vf (p_1) \vf (p_2) \vf (p_3) \vf (p_4) \rangle_{\tmop{amp}} \, =
  Z_{\lambda} \mu^{\e} \lambda + \left( Z_{\lambda} \mu^{\e} \lambda \right)^2
  I_{\omega \vf^3}, \label{4ptamp}
\end{equation}
where we have set the wavefunction renormalization constants to 1, and
$I_{\omega \vf^3}$ is the one-loop integral
\begin{equation}
  I_{\omega \vf^3} = - \frac{H}{(2 \pi)^d} \int \frac{d^d l}{(l^2)^2  (p_1 +
  p_2 + l)^2} + \left( \text{t,u} \tmop{channels} \right),
\end{equation}
having an $\e^{- 1}$ pole which we need to cancel. This $\varepsilon$-pole is
extracted in the usual way (we need the UV $\varepsilon$-pole, and the external
momenta propagating through the loop serve as a IR regulator). Omitting these
standard details (see e.g. {\cite{Kleinert:2001ax}}), we get
\begin{equation}
  I_{\omega \vf^3} = - \frac{3 H}{(4 \pi)^3 \e} + O \left( \e^0 \right) .
\end{equation}

Note that while the finite piece of $I_{\omega \vf^3}$ has nontrivial
dependence of the external momenta, the pole is $p$-independent so we can
cancel it against the tree level contribution in {\eqref{4ptamp}}. This
determines
\begin{equation}
  Z_{\lambda} = 1 + \frac{3 \lambda H}{(4 \pi)^3 \e}, \label{Zl}
\end{equation}
the same result as in the previous section. Hence we get the same beta
function {\eqref{betaapp}}.

Alternatively we could consider $\langle \chi_j (p_1) \chi_j (p_2) \vf (p_3)
\vf (p_4) \rangle$. At one loop we have the two diagrams in Figure \ref{betafnfig}. Here too,
at one-loop the dependence on the external momenta and $i, j$ indices is
proportional to that in tree level, but now involving the other vertex
$\chi_i^2  \vf^2$. The amputated function once again gives the same correction
$z^{(1, 1)}_{\lambda}$. This can be seen as a consequence of the equivalence
of {\eqref{L0}} with the SUSY theory {\eqref{LSUSY1}}. We omit the details.

\subsection{Wavefunction renormalization}\label{wfrenorm}

\begin{figure}[h]\centering \includegraphics[width=300pt]{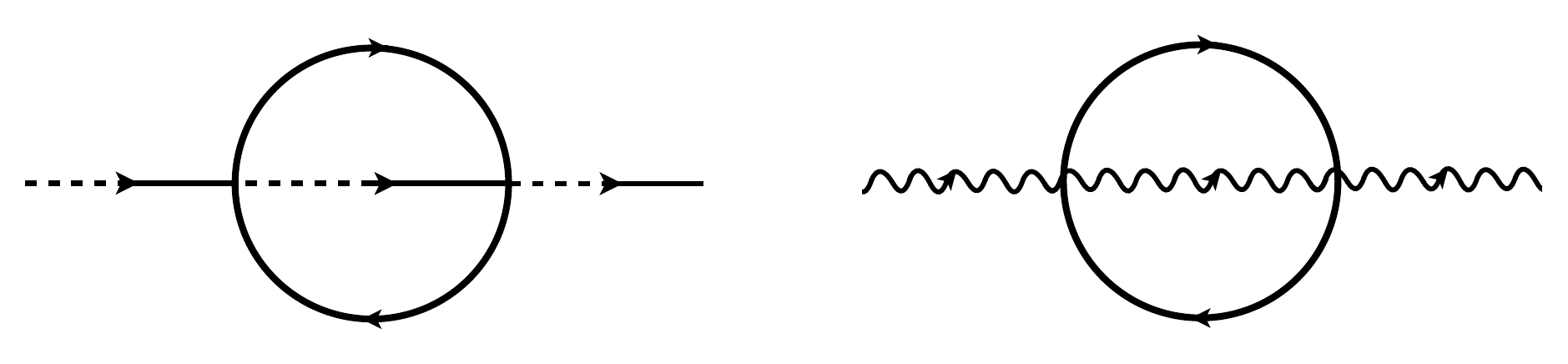}
  \caption{\label{phiomega} Two-loop corrections to $\langle \vf (p) \omega (-
  p) \rangle$ and $\langle \chi_i (p) \chi_j (- p) \rangle$ (labels $i, j$
  omitted).}
\end{figure}

\begin{figure}[h]\centering \includegraphics[width=300pt]{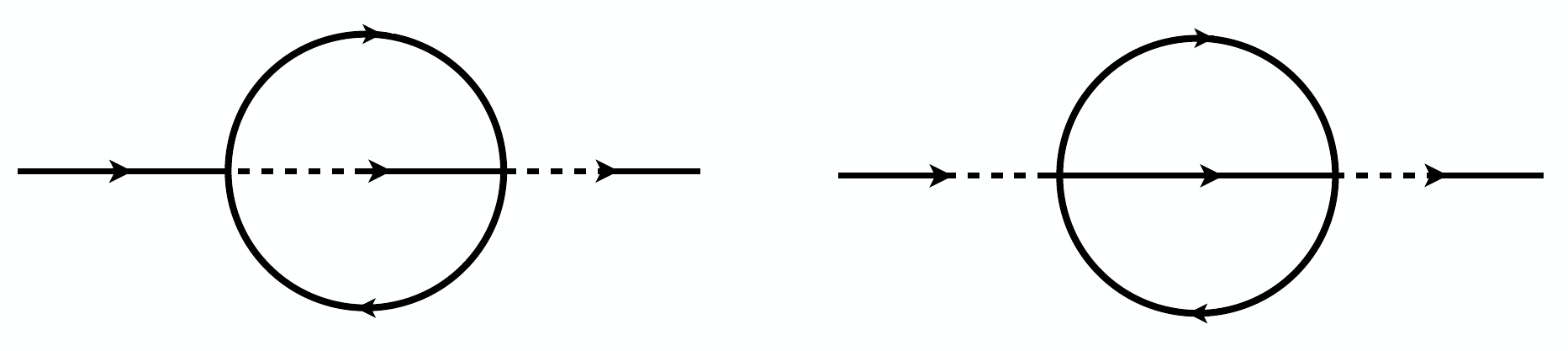}
  \caption{\label{phiphi}Two-loop corrections to $\langle \vf (p) \vf (- p)
  \rangle$.}
\end{figure}

Here we will calculate the wavefunction renormalization constants $Z_{\vf},
Z_{\omega}, Z_{\chi}$, which get the first correction at two loops (we will
need it in our two-loop anomalous dimension computations). These
renormalization constants are determined by requiring the 2-point functions
$\langle \vf (p) \omega (- p) \rangle$, $\langle \vf (p) \vf (- p) \rangle$
and $\langle \chi_i (p) \chi_j (- p) \rangle$ be free of $\varepsilon$ poles.
These 2-point functions receive two-loop corrections shown in Figures
\ref{phiomega} and \ref{phiphi}.

These loop integrals are similar to the usual Wilson-Fisher two-loop field
renormalization integral {\cite{Kleinert:2001ax}}. One integral common to all
corrections is:
\begin{equation}
  I_{\vf \omega} (p^2) = \frac{\lambda^2 H^2}{(2 \pi)^{2 d}} \int \int
  \frac{d^d l_1 d^d l_2}{(l_1^2)^2 (l_2^2)^2  (p + l_1 + l_2)^2} = \; -
  \frac{\lambda^2 H^2 p^2}{6 (4 \pi)^6  \e} + O \left( \e^0 \right) .
  \label{phiomegaint}
\end{equation}
For the correction to $\langle \vf (p) \vf (- p) \rangle$ we have an extra
integral:
\begin{equation}
  I_{\vf \vf} = \frac{\lambda^2 H^2}{(2 \pi)^{2 d}} \int \int \frac{d^d l_1
  d^d l_2}{(l_1^2)^2 (l_2^2)^2  ((p + l_1 + l_2)^2)^2} = \; - \frac{\lambda^2
  H^2}{2 (4 \pi)^6  \e} + O \left( \e^0 \right) .
\end{equation}
Putting in the appropriate symmetry factors, we obtain the following two-loop
corrected 2-point functions (where we keep $Z_{\vf}, Z_{\omega}, Z_{\chi}$ at
tree level but set them to one in the correction):
\begin{eqnarray}
  \left\langle \vf (p) \omega (- p) \right\rangle & = & \frac{1}{p^2} \left[
  \frac{1}{Z_{\vf} Z_{\omega}} - \frac{\lambda^2 H^2}{12 (4 \pi)^6  \e} + O
  \left( \lambda^2 \e^0 \right) \right], \nonumber\\
  \langle \chi_i (p) \chi_j (- p) \rangle & = & \frac{1}{p^2} \left[
  \frac{1}{Z_{\chi}^2} - \frac{\lambda^2 H^2}{12 (4 \pi)^6  \e} + O \left(
  \lambda^2 \e^0 \right) \right], \\
  \langle \vf (p) \vf (- p) \rangle & = & \frac{H}{(p^2)^2} \left[
  \frac{1}{Z_{\vf}^2} - \frac{\lambda^2 H^2}{12 (4 \pi)^6  \e} + O \left(
  \lambda^2 \e^0 \right) \right] . \nonumber
\end{eqnarray}
Requiring that $\varepsilon$ poles cancel determines:
\begin{equation}
  \hspace{2.4em} Z_{\vf} Z_{\omega} = Z_{\vf}^2 = Z_{\chi}^2 = \: 1 -
  \frac{\lambda^2 H^2}{12 (4 \pi)^6  \e} \: + O (\lambda^3) . \label{propcor}
\end{equation}
Thus we find $Z_{\vf} = Z_{\omega} = Z_{\chi}$. In particular $Z_{\vf} =
Z_{\omega}$ can be interpreted as the non-renormalization of $H$. Recall that
our theory is equivalent to the supersymmetric theory {\eqref{LSUSY1}}, and
$H$ is a parameter in the SUSY transformations. As commented in section
\ref{RGsusy}, dimensional regularization preserves full SUSY and hence $H$.
(In section \ref{renH} we instead discussed that in other regularization
schemes $H$ can renormalize, using a slightly different notation for
wavefunction renormalization constants, see Eq.~{\eqref{Z1Z2}}.)

Finally, from $Z_{\vf}, Z_{\omega}, Z_{\chi}$ by the usual definitions we
compute the anomalous dimensions of the fields:
\begin{equation}
  \g_{\vf} = \g_{\omega} = \g_{\chi} = \left[ \frac{\partial \log
  Z_{\vf}}{\partial \log \mu} \right]_{\lambda = \lambda_{\ast} ^{} } =
  \frac{\e^2}{108} + O \left( \e^3 \right),
\end{equation}

equal to $\gamma_{\hat{\phi}}$ at the usual Wilson-Fisher fixed point,
consistently with the dimensional reduction.

\section{Details of anomalous dimension computations}\label{an-details}

In this appendix we will show the details of the anomalous dimension
computations presented in section \ref{anomdim}. Throughout this section we
will denote by $\g_{\mathcal{O}} = \Delta_{\mathcal{O}} -
\Delta^0_{\mathcal{O}}$ the anomalous dimension for an operator $\mathcal{O}$,
where $\Delta_{\mathcal{O}}$ is the dimension of $\mathcal{O}$ at the fixed
point {\eqref{eq:l*}} and $\Delta_{\mathcal{O}}^0$ is its dimension at the
Gaussian (free theory) fixed point at $d = 6 - \e$.

Recall that the three operator classes have the block-diagonal mixing
structure shown in {\eqref{BD}}, which allows to compute anomalous dimensions
in each class separately (although the true scaling susy-writable operators
will have an admixture of susy-nulls, while non-susy-writables will have an
admixture of both susy-writables and susy-nulls). We will try to remind the
reader of that whenever a confusion might arise.

\subsection{Susy-writable operators }\label{app:susy_writables_an_dim}

As explained in section \ref{sec:susywritableleaders}, susy-writable operators
have well-defined anomalous dimensions equal to the ones of the Wilson-Fisher
(WF) in $d = 4 - \varepsilon$ dimensions. Here we give a few examples of such
operators and their anomalous dimensions. In this section we work at one loop
and the reported computations have been performed using the OPE formalism from
App. \ref{ope}.

In our computations we used the $\chi$-formulation, but in the discussions and
in the comparison with WF it is also convenient to use the superfields. The
superfield formulation by construction misses all contributions proportional
to susy-null operators, which are instead non-vanishing in the
$\chi$-formulation. As already mentioned several times, susy-null operators
cannot generate susy-writables under RG (while the opposite may happen), so
that their mixing matrix is triangular, which ensures that anomalous
dimensions of susy-writable operators can be computed by setting to zero
susy-null contributions. In the following we will use this shortcut. We will
be also able to recover the susy-null contributions by asking that the
complete operator is an eigenperturbation.

Let us start by considering the operators $(\Phi^2)_{\theta \bar{\theta}},$
$\mathcal{T}^{\mu \mu}_{\theta \bar{\theta}}$, $\partial^2 (\Phi^2)_{\theta
\bar{\theta}}$, $(\Phi^4)_{\theta \bar{\theta}}$, $(\Phi^2 \mathcal{T}^{\mu
\mu})_{\theta \bar{\theta}}$ discussed in the main text. As a first step we
rewrite all terms involving $\psi$, $\bar{\psi}$ using the $\chi$-formulation
(see appendix \ref{chi-psi}). It is then easy to check that the anomalous
dimensions of these operators are respectively equal to $\varepsilon / 3$,
$0$, $\varepsilon / 3$, $2 \varepsilon$, $(13 / 9) \varepsilon$, as expected
from the WF counterpart {\cite{Kehrein:1994ff}}.

We can further study if some susy-null contributions should be added. It is
easy to check that the first four operators above do not get modified by
susy-null terms. On the other hand the last operator is an eigenpertubation
only when we add to it a term proportional to $(\chi^2)^2$, namely
\begin{equation}
  (\Phi^2 \mathcal{T}^{\mu \mu})_{\theta \bar{\theta}} + \frac{15}{26} 
  (\chi^2)^2 .
\end{equation}
As we explain in the main text, the only susy-writable operator which may play
an important role in destabilizing the susy RG is the so called box superfield
$B^{a b, c d}$, which transforms in the $(2, 2)$ representation of $\tmop{OSp}
(d | 2 \nobracket)$ (recall that unitarity bounds for this representation are
too weak to ensure that the operator is irrelevant). Its WF counterpart is
defined in the main text as
\begin{equation}
  \hat{B}_{\mu \nu \comma \rho \sigma} = \left( \hat{\phi}_{, \mu \nu} \hat{\phi}_{,
  \rho \sigma} \hat{\phi}^2 - \frac{2 \hat{d}}{\hat{d} - 2} \hat{\phi}_{, \mu}
  \hat{\phi}_{, \nu} \hat{\phi}_{, \rho \sigma} \hat{\phi} \right)^Y .
  \label{BoxWF}
\end{equation}
The anomalous dimension of (\ref{BoxWF}) was computed in
{\cite{Kehrein:1994ff}} and it equals $(7 / 9) \varepsilon$. We would like to
reproduce this result by studying the superfield $\calB^{a b, cd}$. This is
a very non-trivial check that dimensional reduction works also for operators
in non-trivial $\tmop{OSp} (d | 2 \nobracket)$ representations. While doing
so, we will also show the explicit expression in components for this operator.

We mainly focus on $(\calB^{\theta \bar{\theta}, \theta \bar{\theta}})_{\theta
\bar{\theta}}$, since this component is bosonic, supertranslation invariant
(it is the highest component of a superfield) and it is a scalar with respect
to $\tmop{SO} (d)$. Because of these features, this operator is generated in
our RG flow and could destabilize it (if it becomes relevant).

First let us spell out the $\tmop{SO} (\hat{d})$ Young symmetrizer denoted by
$Y$ in (\ref{BoxWF}). This is a tensor with two sets of four indices, each set
transforming in the \ $\tmop{SO} (\hat{d})$ representation $(2, 2)$. By
contracting the indices $\mu \nu \rho \sigma$ inside the brackets of
(\ref{BoxWF}) with one set, we obtain an operator depending on the second set
of indices, which transforms properly in the $(2, 2)$ irrep. It is convenient
to represent the symmetrizer (see
{\cite{Costa:2014rya,Costa:2016hju,Lauria:2018klo}}) by contracting its eight
indices with auxiliary $\mathbb{R}^{\hat{d}}$ vectors--- the symmetric indices
in each row are contracted with the same vector. The first set is contracted
with $X_1$ and $X_2$, while the second set with $Z_1$ and $Z_2 .$ The result
is expressed in the following polynomial form:
\begin{align}
  & \!\!\! \! \Pi_{2, 2} (X_1, X_2 ; Z_1, Z_2) = c_{2, 2} \bigl\{ - (X_1 \cdot
  X_1) (Z_1 \cdot Z_1)  (X_2 \cdot Z_2)^2 
  - 2 (X_1 \cdot X_2)^2  (Z_1 \cdot Z_2)^2 
   \nonumber\\
  &    
   + 2 (X_1 \cdot X_1) (Z_1 \cdot Z_2) (X_2 \cdot Z_1) (X_2 \cdot Z_2)
  + 2 (X_1 \cdot X_1) (X_2
  \cdot X_2)  (Z_1 \cdot Z_2)^2 \nonumber\\
  &    + 2 (X_1 \cdot X_2)^2 (Z_1 \cdot Z_1) (Z_2 \cdot Z_2) - 2 (X_1 \cdot
  X_1) (X_2 \cdot X_2) (Z_1 \cdot Z_1) (Z_2 \cdot Z_2) \nonumber\\
  &    + (\hat{d} - 1) (X_2 \cdot X_2) (Z_1 \cdot Z_1)  (X_1 \cdot Z_2)^2
  - 2 (\hat{d} - 1) (X_1 \cdot X_2) (Z_1 \cdot Z_1) (X_1 \cdot Z_2) (X_2 \cdot Z_2)
  \nonumber\\
  &    - 2 (\hat{d} - 1) (X_2 \cdot X_2) (Z_1 \cdot Z_2) (X_1 \cdot Z_1) (X_1
  \cdot Z_2) + 2 (\hat{d} - 1) (X_1 \cdot X_2) (Z_1 \cdot Z_2) (X_1 \cdot Z_2) (X_2
  \cdot Z_1) \nonumber\\
  &    + 2 (\hat{d} - 1) (X_1 \cdot X_2) (Z_1 \cdot Z_2) (X_1 \cdot Z_1) (X_2
  \cdot Z_2) + (\hat{d} - 1) (X_2 \cdot X_2) (Z_2 \cdot Z_2)  (X_1 \cdot Z_1)^2 \\
  &    - 2 (\hat{d} - 1) (X_1 \cdot X_2) (Z_2 \cdot Z_2) (X_1 \cdot Z_1) (X_2
  \cdot Z_1) + (\hat{d} - 1) (X_1 \cdot X_1) (Z_2 \cdot Z_2)  (X_2 \cdot Z_1)^2
  \nonumber\\
  &    - (\hat{d} - 2)  (\hat{d} - 1)  (X_1 \cdot Z_2)^2  (X_2 \cdot
  Z_1)^2 + 2 (\hat{d} - 2)  (\hat{d} - 1) (X_1 \cdot Z_1) (X_1 \cdot Z_2) (X_2
  \cdot Z_1) (X_2 \cdot Z_2) \nonumber\\
  &    - (\hat{d} - 2)  (\hat{d} - 1)  (X_1 \cdot Z_1)^2  (X_2 \cdot
  Z_2)^2 + \hat{d} (X_1 \cdot X_1) (Z_1 \cdot Z_1)  (X_2 \cdot Z_2)^2 \nonumber\\
  &     - 2 \hat{d} (X_1 \cdot X_1) (Z_1 \cdot Z_2) (X_2 \cdot Z_1)
  (X_2 \cdot Z_2) \bigr\}, \nonumber
\end{align}
where $c_{2, 2} = - \frac{1}{3 (- 2 + \hat{d}) (- 1 + \hat{d})}$ is a
normalization constant which ensures that the symmetrizer is idempotent. In
order to get back the indices it is then sufficient to take derivatives with
respect to the auxiliary vectors.\footnote{E.g.~we can compute the dimensions of the representation as the trace of the projector, which entails freeing all the indices and contracting them pairwise (up to a combinatorial factor):
\begin{equation}
\label{dim(2,2)}
\dim_{(2, 2)}=
\frac{1}{16}
\partial^\m_{X_1}
\partial^\m_{Z_1}
\partial^\n_{X_1}
\partial^\n_{Z_1}
\partial^\rho_{X_2}
\partial^\rho_{Z_2}
\partial^\s_{X_2} 
\partial^\s_{Z_2} 
\Pi_{2, 2} (X_1, X_2 ; Z_1, Z_2) = \frac{1}{12} (\hat{d}-3) \hat{d} (\hat{d}+1) (\hat{d}+2) \, ,
\end{equation}
This matches e.g. Eq.~(10.68) of \cite{Cvitanovic:2008zz}. We use this result in section \ref{CB}.
} 
As explained in {\cite{paper1}}, this
contracted form of the Young symmetrizer is also convenient since it trivially
generalizes to $\tmop{OSp} (d | 2 \nobracket)  \tmop{representations}$, by
considering vectors $X_i, Z_i$ in $\mathbb{R}^{d | 2 \nobracket}$ and the
scalar product $X \cdot Y = X^a g_{_{a b}} Y^b$, with the usual $\tmop{OSp} (d
| 2 \nobracket)$ metric $g_{a b}$. We stress that the dependence of the
symmetrizer on the parameter $\hat{d}$ must not be changed (indeed $\hat{d} =
d - 2$ is equal to the supertrace).

By some manipulations of this projector, we are able to obtain the final form
for the box superfield,
\begin{eqnarray}
  & (\calB^{\theta \bar{\theta}, \theta \bar{\theta}})_{\theta \bar{\theta}} = &
  \frac{1}{6} \bigl\{  - \varphi^2 \psi_{, \mu \nu}  \bar{\psi}_{, \mu
  \nu} + \psi \bar{\psi}  (- \varphi_{, \mu \nu}^2 - 20 \varphi_{, \mu}
  \omega_{, \mu} + 54 \omega^2) + 4 \varphi \varphi_{, \mu}  (\psi_{, \nu}
  \bar{\psi}_{, \mu \nu} + \psi_{, \mu \nu}  \bar{\psi}_{, \nu}) \nonumber\\
  &  & + 2 \varphi_{, \mu} \varphi_{, \nu}  (\psi \bar{\psi}_{, \mu \nu} +
  \psi_{, \mu \nu} \bar{\psi} ) - 2 \varphi \varphi_{, \mu \nu}  (\psi
  \bar{\psi}_{, \mu \nu} + \psi_{, \mu \nu}  \bar{\psi}) - 42 \varphi \omega
  \psi_{, \mu}  \bar{\psi}_{, \mu} \nonumber\\
  &  & + 2 \varphi \varphi_{, \mu \nu}  (\psi_{, \mu}  \bar{\psi}_{, \nu} +
  \psi_{, \nu}  \bar{\psi}_{, \mu}) + 4 \varphi_{, \mu} \varphi_{, \mu \nu}
  (\psi \bar{\psi}_{, \nu} + \psi_{, \nu} \bar{\psi} ) - \varphi \omega
  \varphi_{, \mu \nu}^2 + 54 \varphi \omega^3 \nonumber\\
  &  & - 12 \omega \varphi_{, \mu}  (\psi \bar{\psi}_{, \mu} + \psi_{, \mu}
  \bar{\psi}) + 10 \varphi \omega_{, \mu}  (\psi \bar{\psi}_{, \mu} + \psi_{,
  \mu}  \bar{\psi}) - 30 \psi \bar{\psi} \psi_{, \mu}  \bar{\psi}_{, \mu}
  \nonumber\\
  &  & + 2 \omega \varphi_{, \mu} \varphi_{, \nu} \varphi_{, \mu \nu} + 4
  \varphi \omega_{, \mu} \varphi_{, \nu} \varphi_{, \mu \nu} + \varphi (2
  \varphi_{, \mu} \varphi_{, \nu} - \varphi \varphi_{, \mu \nu}) \omega_{, \mu
  \nu} + 5 \varphi^2 \omega_{, \mu}^2 \nonumber\\
  &  &  - 6 \omega^2 \varphi_{, \mu}^2 - 32 \varphi \omega
  \varphi_{, \mu} \omega_{, \mu} \bigr\},  \label{Boxthetathetab}
\end{eqnarray}
where for short the expression is written for $d = \hat{d} + 2 = 6$. We
checked that this operator, upon substitution $\psi, \bar{\psi} \rightarrow
\chi$ has indeed anomalous dimension $(7 / 9) \varepsilon$ as expected.

Notice that (\ref{Boxthetathetab}) contains the term $\psi \bar{\psi} \psi_{,
\mu} \bar{\psi}_{, \mu}$ which can be mapped in two different ways (as
explained in App. \ref{chi-psi}, Eq.~{\eqref{ambpsi4}}) in terms of $\chi$.
Both choices are equally good, since their difference is proportional to the
susy-null operator $\partial^2  (\chi^2)^2$ (which can be set to zero for
anomalous dimensions computations). As usual we can also determine the
susy-null contribution by requiring that the full operator is an
eigenperturbation: this gives the above $(\calB^{\theta \bar{\theta}, \theta
\bar{\theta}})_{\theta \bar{\theta}}$ with $\psi \bar{\psi} \psi_{, \mu}
\bar{\psi}_{, \mu} \rightarrow \frac{1}{4} \chi_i \chi_i \chi_{j, \mu}
\chi_{j, \mu} + \frac{5}{84} \partial^2  (\chi^2)^2$.

As a final example we compute the anomalous dimensions of all susy-writable
(and one susy-null) operators at dimensions $\Delta = 10$ made of four fields.
One of such operators is (\ref{Boxthetathetab}), and we would like to check
that all of the others also have dimensions consistent from dimensional
reduction. In practice we consider a list of 32 monomials: the 22 summands of
(\ref{Boxthetathetab}) (terms of the form e.g. $\psi \bar{\psi}_{, \nu} +
\psi_{, \nu} \bar{\psi} $ are counted as one after the $\psi, \bar{\psi}
\rightarrow \chi$ map), the susy-null operator $\partial^2  (\chi^2)^2$
discussed above and the following 9 extra operators
\begin{eqnarray}
  \varphi_{, \mu}^2 \chi_{, \nu}^2, \quad \varphi_{, \mu} \chi_{, \mu}
  \varphi_{, \nu} \chi_{, \nu}, \quad \varphi^2 \varphi_{, \mu \nu \sigma}^2,
  \quad \varphi \varphi_{, \mu} \varphi_{, \nu \sigma} \varphi_{, \mu \nu
  \sigma}, \quad \varphi_{, \nu}^2 \varphi_{, \mu \sigma}^2 &  & \\
  \varphi \varphi_{, \mu \nu} \varphi_{, \mu \sigma} \varphi_{, \nu \sigma},
  \quad \varphi_{, \mu} \omega_{, \mu} \varphi_{, \nu}^2, \quad \varphi_{,
  \mu} \varphi_{, \nu} \varphi_{, \sigma} \varphi_{, \mu \nu \sigma}, \quad
  \varphi_{, \nu} \varphi_{, \sigma} \varphi_{, \mu \nu} \varphi_{, \mu
  \sigma} . &  &  \nonumber
\end{eqnarray}
This list of 32 monomials is closed upon renormalization, mixing in a
non-trivial way. Diagonalizing the resulting $32 \times 32$ mixing matrix
gives the following list of anomalous dimensions:
\begin{eqnarray}
 & & 2 \varepsilon, 2 \varepsilon, \frac{13 \varepsilon}{9}, \frac{13 \varepsilon}{9},
  \frac{13 \varepsilon}{9}, \frac{13 \varepsilon}{9}, \frac{13 \varepsilon}{9},
  \frac{19 \varepsilon}{15}, \frac{10 \varepsilon}{9}, \frac{10 \varepsilon}{9},
  \varepsilon, \varepsilon, \varepsilon, \varepsilon,  \label{32list}\\
&&  \frac{14 \varepsilon}{15}, \frac{8 \varepsilon}{9}, \frac{8 \varepsilon}{9},
  \frac{8 \varepsilon}{9}, \frac{8 \varepsilon}{9}, \frac{7 \varepsilon}{9}, \frac{7
  \varepsilon}{9}, \frac{7 \varepsilon}{9}, \frac{7 \varepsilon}{9},
  \frac{\varepsilon}{2}, \frac{\varepsilon}{2}, \frac{4 \varepsilon}{9},
  \frac{\varepsilon}{3}, \frac{2 \varepsilon}{9}, \frac{\varepsilon}{9}, 0, 0, 0.
  \nonumber
\end{eqnarray}
Let us see next how this list can be related to WF computations. First, we
would like to emphasize that many of the numbers in (\ref{32list}) are
associated to descendants. Indeed, we can redo the computation only using
equivalence classes of operators defined up to total derivatives (as discussed
in appendix \ref{ope}). The result is that only 8 of the above operators are
not descendants. The associated anomalous dimensions are $\frac{10
\varepsilon}{9}, \frac{14 \varepsilon}{15}, \frac{8 \varepsilon}{9}, \frac{7
\varepsilon}{9}, \frac{\varepsilon}{2}, \frac{\varepsilon}{3}, \frac{\varepsilon}{9}, 0,
0$. This also explains why in (\ref{32list}) there are many repeated anomalous
dimensions: they correspond to total derivatives of different components of
the superfield. To clarify this, let us focus on the anomalous dimension
$\frac{7 \varepsilon}{9}$ (the one of the box operator). This appears four times
in (\ref{32list}) and only one of these is not a total derivative ---this
indeed corresponds to the operator (\ref{Boxthetathetab}). The other three
occurrences are related to the following descendants (we also checked this
explicitly):
\begin{equation} \partial^2 (\calB^{\theta \bar{\theta}, \theta \bar{\theta}})_0, \quad
   \partial_{\mu} \partial_{\nu} (\calB^{\theta \bar{\theta}, \mu \nu})_0, \quad
   \partial_{\mu} (\calB^{\theta \bar{\theta}, \theta \mu})_{\bar{\theta}} . \end{equation}
E.g.~the first one, before applying $\partial^2$, takes the form
\begin{eqnarray}
  (\calB^{\theta \bar{\theta}, \theta \bar{\theta}})_0 & \rightarrow & -
  \frac{1}{6} \varphi^2 \varphi_{, \mu \nu}^2 + \frac{2}{3} \varphi \varphi_{,
  \mu} \varphi_{, \nu} \varphi_{, \mu \nu} - \frac{5}{6} \varphi^2 \chi_{i,
  \mu}^2 + \frac{10}{3} \varphi \chi_i \varphi_{\mu} \chi_{i, \mu} - 2 \varphi
  \omega \varphi_{, \mu}^2 \nonumber\\
  &  & + \frac{7 \varphi^2 \omega^2}{3} - \frac{20}{3} \varphi \chi_i^2
  \omega - \left[ \frac{10 (\chi_i^2)^2}{7} \right], 
\end{eqnarray}
where we also included, in square brackets, the susy-null contribution.

We also notice that different anomalous dimensions in
{\eqref{32list}} occur a different number of times. It is easy to see why this
happens. This depends on the superprimary $\tmop{Osp} (d | 2 \nobracket)$
representation, and on how many times we need to differentiate to get to the
needed dimension. E.g.~$2 \varepsilon$ in (\ref{32list}) is related to the
superprimary \ $\Phi^4$ in the following two combinations: $(\partial^2)^2
\Phi_{\theta \bar{\theta}}^4 $, $(\partial^2)^3 \Phi_0^4$. Here, since
$\Phi^4$ is a scalar, there is a single $\tmop{OSp}$ component to use (in
constrast with non-scalar operators). Also only total derivatives appear
because the superfield itself has too low bare dimension: $[\Phi_0^4] = 4$,
$[\Phi_{\theta \bar{\theta}}^4] = 6$. As a final example, the value $\frac{13
\varepsilon}{9}$ in (\ref{32list}) is related to five possible descendants of the
superfield $\Phi^2 \mathcal{T}^{a b}$: $(\partial^2)^2 (\Phi^2
\mathcal{T}^{\theta \bar{\theta}})_0$, $\partial^2 \partial_{\mu}
\partial_{\nu}^{} (\Phi^2 \mathcal{T}^{\mu \nu})_0$, $\partial^2
\partial_{\mu} (\Phi^2 \mathcal{T}^{\mu \bar{\theta}})_{\theta}$, $\partial^2
(\Phi^2 \mathcal{T}^{\theta \bar{\theta}})_{\theta \bar{\theta}}$,
$\partial_{\mu} \partial_{\nu}^{} (\Phi^2 \mathcal{T}^{\mu \nu})_{\theta
\bar{\theta}}$.

We do not present here a detailed explanation for all anomalous dimensions in
(\ref{32list}). However we stress that we checked that they are all in
agreement with available WF results: not only the anomalous dimensions match
but also the occurrences (of primaries and descendants) are the ones
expected.

Most of the checks are done by comparing with table 4 of
{\cite{Kehrein:1994ff}} (using the entries with $n = 4$ fields),\footnote{As
explained in {\cite{Kehrein:1994ff}}, for $\mathbb{Z}_2$ symmetry, the
anomalous dimensions are the numbers in the table times $\varepsilon / 3$. }
where all operators built out of $5$ or less derivatives are presented.
However the list (\ref{32list}) is also sensitive to operators with six
derivatives: indeed the lowest component of a superfield with four $\Phi$'s
and six derivatives has dimension $10$. Unfortunately such operators were not
fully classified in the WF literature, except for a scalar operator considered
in {\cite{Liendo:2017wsn}} with $\gamma = \varepsilon / 3$ (see equation
(3.12)), which indeed matches our computation.

We can therefore predict that in the WF spectrum of operators with four fields
and six derivatives, there are three operators with anomalous dimensions $0$,
$\varepsilon / 9$, $(14 / 15) \varepsilon$. We can further say something about
their possible $\tmop{SO} (\hat{d})$ representation. Indeed the operators in
our list must have an even number of indices which are set to $\theta$ and
$\bar{\theta}$ (otherwise the resulting operator would be fermionic or it
would not be an $\tmop{SO} (\hat{d})$ scalar). We thus conclude that these
three operators must transform either in the scalar, or spin two, or $(2, 2)$
representation of $\tmop{SO} (\hat{d})$.\footnote{We thank Johan Henriksson, who confirmed us in a private communication that there exist two WF operators with four fields and six derivatives in the spin-two representation of $\tmop{SO} (\hat{d})$ with anomalous dimensions equal to $\varepsilon / 9$ and $(14 / 15) \varepsilon$. On the other hand the operator with anomalous dimensions $0$ was not found in the scalar and spin-two sector, so it must transform in the box $(2, 2)$ representation of $\tmop{SO} (\hat{d})$.  }

Finally let us comment on the three operators with
$\gamma = 0$. Only one of them is a descendant: the susy-null operator
$\partial^2  (\chi^2)^2$. One of the other two operators is a primary, an
operator with six derivatives. The third operator in this group is actually
not an eigenvector, but a generalized eigenvector forming a logarithmic
multiplet together with the $\gamma = 0$ primary. This is a recurrent feature
of our non-unitary theory: mixing matrices may not be fully diagonalizable and
can be organized in Jordan blocks, as discussed around Eq.~{\eqref{Jordan}}.

\subsection{Susy-null leaders }\label{app:dim_susy_null}

We next consider all the susy-null leaders with 6d classical dimension up to
12 (one at dimension 8, one at 10, and four at 12). For each of them we
compute the one-loop anomalous dimension, or the two-loop one if the one-loop
result is trivial. We use the OPE method at one loop, and Feynman diagrams
whenever we have to go to two loops.

\subsubsection{$\Delta^0 = 8 + O (\varepsilon) $, $n = 4$}\label{chi22sec}

The lowest susy-null leader is $(\chi_i^2)^2$, of classical dimension
$\Delta^0 = 8 - 2 \varepsilon$ in $d = 6 - \varepsilon$ (see Table
\ref{Nf4Nd0}). It is easy to see that it does not receive any anomalous
dimension at one loop, so we proceed to study the 2-loop contribution.
Denoting by $((\chi_i^2)^2)^B$ the operator built of bare fields, and by
$(\chi_i^2)^2$ the renormalized operator whose correlators should be free of
poles in $\varepsilon$, they are related by
\begin{equation}
  ((\chi_i^2)^2)^B = Z \, (\chi_i^2)^2,
\end{equation}
To find the renormalization factor $Z$ we consider the correlator $\langle
(\chi_i^2)^2 (p = 0) \chi_j (p_1) \chi_k (p_2) \chi_l (p_3) \chi_m (p_4)
\rangle$. A nonzero two-loop diagram is shown in Figure \ref{chi22diag}.
Another two loop diagram (a double bubble diagram of the type shown in Figure
\ref{Feldman6corr}) vanishes because it is proportional to $n$, from
contractions of the $K_{i j}$ matrices in the $\chi$-$\chi$ propagator
{\eqref{prop-1}}. At two loops we must also consider the propagator computed
in section \ref{wfrenorm}.

Taking everything into account, the two-loop corrected amputated correlator
equals the tree-level one, times
\begin{equation}
  Z_{\,}^{- 1} + \left[ I_{(\chi_i^2)^2} + \left( \text{t,u} \tmop{channels}
  \right) \right] + \frac{1}{4} \sum_{i = 1}^4 I_{\vf \omega} (p_i^2) / p_i^2
  . \label{correlchi}
\end{equation}
Here $I_{\vf \omega}$ is the integral in {\eqref{phiomegaint}} which comes
from the external leg corrections. The $I_{(\chi_i^2)^2}$ comes from the loop
diagram of Figure \ref{chi22diag}. It is given by (see
{\cite{Kleinert:2001ax}} for the standard details):
\begin{equation}
  I_{(\chi_i^2)^2} = \frac{H^2 \lambda^2}{(2 \pi)^{2 d}} \int \frac{d^d l_1
  d^d l_2}{l_1^2  (l_2^2)^2  (l_1 + p_3 + p_4)^2  ((l_1 + l_2 - p_1)^2)^2} =
  \frac{H^2 \lambda^2}{2 (4 \pi)^6  \e} + O \left( \e^0 \right) .
  \label{loopchi22}
\end{equation}
The $Z_{\,}$ is obtained by demanding that $\e^{- 1}$ poles cancel:
\begin{equation}
  Z_{\,}^{- 1} = 1 - \frac{4}{3} \frac{H^2 \lambda^2}{(4 \pi)^6  \e} .
  \label{ZH21}
\end{equation}
From this we get the anomalous dimension of $(\chi_i^2)^2 $ as follows:
\begin{equation}
  \gamma_{(\chi_i^2)^2} = \mu \frac{\partial}{\partial \mu} \log Z = -
  \frac{8}{27} \e^2 . \label{gammachi4}
\end{equation}
\subsubsection{$\Delta^0 = 10 + O (\varepsilon)$, $n = 6$
}\label{Delta10n6Null}

The next susy-null leader is $\vf^2 (\chi_i^2)^2$ (Table \ref{Nf6Nd0}), of
bare dimension $\Delta^0 = 10 - 3 \varepsilon$. It gets a nonzero one-loop
anomalous dimension, which we compute by the OPE method (App. \ref{ope}). The
following OPEs are important:
\begin{eqnarray}
  \vf^2 (\chi_i^2)^2 (x) \times \chi_j^2  \vf^2 (0) & \sim & 32
  \langle \vf (x) \vf (0) \rangle_0 \langle \chi_i (x) \chi_j (0) \rangle_0 \,
  \vf^2 (\chi_k^2) \chi_i \chi_j (0) + \ldots\nonumber\\[5pt]
  \vf^2 (\chi_i^2)^2 (x) \times \omega \vf^3 (0) & \sim & 6 \langle \vf (x) \vf
  (0) \rangle_0 \langle \vf (x) \omega (0) \rangle_0\,  \vf^2 (\chi_i^2)^2 (0)
  + \ldots
\end{eqnarray}
From this and using the formula {\eqref{anommatres}} we get
\begin{equation}
  \gamma_{\vf^2 (\chi_i^2)^2} = \left[ \frac{9 H \lambda}{64 \pi^3}
  \right]_{\lambda = \lambda_{\ast} } = 3 \e + O \left( \e^2 \right) .
\end{equation}
\subsubsection{$\Delta^0 = 12 + O (\varepsilon)$, $n = 6$
}\label{Delta12n6Null}

At dimension $\Delta^0 = 12 - 3 \varepsilon$ we have three independent
susy-null leaders which are composites of 6 fields. Two of them are shown in
Table \ref{Nf6Nd0} and the third in Eq.~{\eqref{Nf6Nd2}}.

To compute their anomalous dimensions it is convenient to work with
equivalence classes of operators defined up to total derivatives (as discussed
in the end of appendix \ref{OPEmethod}). We can parametrize the equivalence
classes by the following three operators: $\mathcal{O}_1 = \vf \omega
(\chi^2)^2, \: \mathcal{O}_2 = (\chi^2)^3$ and\footnote{We add a factor $1 /
H$ to $\left( \partial \vf \right)^2 (\chi^2)^2$ so that the mixing matrix is
free of $H$. } $\mathcal{O}_3 = \frac{1}{H} \left( \partial \vf \right)^2
(\chi_i^2)^2$. E.g.~$\vf \partial \vf (\chi_i \partial \chi_i) (\chi_j^2)$ is
not considered as an independent operator since it can be written in terms of
$\mathcal{O}_1$ and $\mathcal{O}_2$ up to a total derivative, namely $\vf
\partial \vf (\chi_i \partial \chi_i) (\chi_j^2) = \frac{1}{4} \partial^{\mu}
\left[ \vf \partial_{\mu} \vf (\chi_j^2)^2 \right] + \frac{H}{4} \mathcal{O}_1
- \frac{H}{4} \mathcal{O}_2$.

To compute the mixing matrix we consider the following OPEs of the operators
$\mathcal{O}_i$ with the interaction $\mathcal{V}$
\begin{eqnarray}
  \vf \omega (\chi^2)^2 (0) \times \chi^2  \vf^2 (x) & \sim & 16 \langle
  \vf \vf \rangle_0 \langle \chi_i \chi_j \rangle_0 \: \vf \omega \chi^2
  \chi_i \chi_j + 2 \langle \vf \vf \rangle_0 \langle \vf \omega \rangle_0 
  (\chi^2)^3  \label{openull1}\\
  &  & + 8 x^{\mu} x^{\nu} \langle \vf \omega \rangle_0 \langle \chi_i \chi_j
  \rangle_0 \: \vf \chi^2 \chi_i \partial_{\mu} \partial_{\nu} \left( \vf
  \chi_j \right) + \ldots \nonumber\\[5pt]
  \vf \omega (\chi^2)^2 (0) \times \vf^3 \omega (x) & \sim & 6 \langle \vf \vf
  \rangle_0 \langle \vf \omega \rangle_0  \vf \omega (\chi^2)^2 + 3 x^{\mu}
  x^{\nu} \langle \vf \omega \rangle_0 \langle \vf \omega \rangle_0 \left(
  \partial_{\mu} \partial_{\nu} \vf^2 \right) (\chi^2)^2 + \ldots \\[5pt]
  (\chi^2)^3 (0) \times \chi^2  \vf^2 (x) & \sim & 24 \langle \chi_i \chi_j
  \rangle_0 \langle \chi_i \chi_j \rangle_0 \left( \partial_{\mu}
  \partial_{\nu} \vf^2 \right)  (\chi^2)^2  \nonumber\\
  &  & + 6 \langle \chi_i \chi_j \rangle_0 \langle \chi_i \chi_k \rangle_0
  \left( \partial_{\mu} \partial_{\nu} \vf^2 \right) \chi^2 \chi_j \chi_k +
  \ldots \\[5pt]
  \left( \partial \vf \right)^2 (\chi^2)^2 (0) \times \vf^3 \omega (x) & \sim
  & 6 x^{\mu} x^{\nu} \partial^{\sigma} \langle \vf \vf \rangle_0 \,
  \partial_{\sigma} \langle \vf \omega \rangle_0  \left( \partial_{\mu}
  \partial_{\nu} \vf^2 \right) (\chi^2)^2 \nonumber\\
  &  & + 6 \partial^{\sigma} \langle \vf \vf \rangle_0 \, \partial_{\sigma}
  \langle \vf \vf \rangle_0  \vf \omega (\chi^2)^2 + \ldots \\[5pt]
  \left( \partial \vf \right)^2 (\chi^2)^2 (0) \times \chi^2  \vf^2 (x) & \sim
  & - 32 x^{\nu} \langle \chi_i \chi_j \rangle_0 \partial^{\mu} \langle \vf
  \vf \rangle_0  \left( \partial_{\mu} \vf \right) \chi^2 \chi_i
  \partial_{\nu} \left( \vf \chi_j \right) \nonumber\\
  &  & + 2 \partial \langle \vf \vf \rangle_0 \partial \langle \vf \vf
  \rangle_0  (\chi^2)^3 + \ldots 
\end{eqnarray}
In the above OPEs we have expanded the r.h.s. in $x$ wherever needed and kept
only the terms that give $1 / \varepsilon$ pole. Following the discussion in
appendix \ref{OPEmethod}, we finally obtain the $3 \times 3$ anomalous
dimension matrix:
\begin{equation}
  \Gamma = \frac{\varepsilon}{18}  \left( \begin{array}{ccc}
    24 & - 12 & 2\\
    3 & 0 & 3\\
    - 2 & 12 & 20
  \end{array} \right) . \label{GammaEx}
\end{equation}
This matrix is not fully diagonalizable, but it admits a Jordan decomposition
(see around formula {\eqref{Jordan}}) with eigenvalues 0 and $(11 / 9)
\varepsilon$, to which there correspond true eigenvectors $e^{(1)} = \left\{ -
1, - \frac{11}{6}, 1 \right\}$ and $e^{(2)} = \{ - 1, 0, 1 \}$. There is also
a generalized eigenvector $e^{(3)} = \left\{ - \frac{99}{2 \varepsilon}, -
\frac{27}{4 \varepsilon}, 0 \right\}$ which forms a rank-2 Jordan block with
$e^{(2)}$, namely $\left( \Gamma_{i j} - \frac{11}{9} \e  \tmd_{i j} \right)
e_j^{(3)} = e_i^{(2)}$. Therefore, corresponding to the anomalous dimension
$\frac{11}{9} \e$, there is a $2 \times 2$ Jordan block and a logarithmic
multiplet $\{ \mathcal{O}_1^R, \mathcal{O}_2^R \}$, such that\footnote{We
rescaled the operator $\mathcal{O}_2^R$ so that it has order one coefficients,
and as a consequence the upper right corner of the dilatation matrix is
$\varepsilon$ and not 1. The $\varepsilon \rightarrow 0$ limit is smooth in
this form.}
\begin{equation}
  D \left(\begin{array}{c}
    \mathcal{O}_2^R\\
    \mathcal{O}_1^R
  \end{array}\right) = \left(\begin{array}{cc}
    12 - \frac{16 \varepsilon}{9} & \varepsilon\\
    0 & 12 - \frac{16 \varepsilon}{9}
  \end{array}\right) \left(\begin{array}{c}
    \mathcal{O}_2^R\\
    \mathcal{O}_1^R
  \end{array}\right),
\end{equation}
where the dilatation operator $D$ is acting on the renormalized operators
$\mathcal{O}_1^R \equiv -\mathcal{O}_1 +\mathcal{O}_3$ and $\mathcal{O}_2^R
\equiv - \frac{99}{2} \mathcal{O}_1 - \frac{27}{4} \mathcal{O}_2$. Here the
operators $\mathcal{O}_1^R$ and $\mathcal{O}_2^R$ should not be considered as
the actual renormalized operators, since they only parametrize the equivalence
classes of operators up to derivatives.

For the sake of completeness we also performed a much more general computation
which gives us the exact form of the renormalized operators. Computing the
anomalous dimensions matrix of the 49 possible operators which are built out
of 6 fields and have dimensions 12 in $d = 6,$ we obtained that
$\mathcal{O}_1^R$ was actually a correct eigenperturbation. On the other hand,
the operator $\mathcal{O}_2^R$ should have been corrected by some total
derivatives, namely by adding to it $\frac{231}{32}
\partial_{\mu}  (\varphi_{, \mu} \varphi \chi_i^2 \chi_j^2 ) - \frac{33}{16}
\partial_{\mu} (\varphi^2 \chi_i \chi_{i, \mu} \chi_j^2)$.

\subsubsection{$\Delta^0 = 12 + O (\varepsilon)$, $n =
8$}\label{Delta12n8Null}

With 8 fields and dimension $\Delta = 12 + O \left( \e \right)$ we have a
single susy-null leader, $\vf^4 (\chi^2)^2$ (See Table \ref{Nf8Nd0}). Its $O
\left( \e \right)$ anomalous dimension can be obtained from the following
OPEs:
\begin{eqnarray}
  \vf^4 (\chi^2)^2 (0) \times \chi^2  \vf^2 (x) & \sim & 64 \langle \vf \vf
  \rangle_0 \langle \chi_i \chi_j \rangle_0 \: \vf^4 \chi^2 \chi_i \chi_j\nonumber\\[5pt]
  \vf^4 (\chi^2)^2 (0) \times \vf^3 \omega (x) & \sim & 36 \langle \vf \vf
  \rangle_0 \langle \vf \omega \rangle_0 \vf^4 (\chi^2)^2 .
\end{eqnarray}
Following {\eqref{anommatres}} we get the anomalous dimension:
\begin{equation}
  \gamma_{\vf^4 (\chi^2)^2} = \frac{22 \e}{3} .
\end{equation}
\subsection{Non-susy-writable leaders}\label{app:dimFk}

We will now compute the anomalous dimensions for some specific non-susy
writable leaders. For $\D \leqslant 12$ the only non-susy writable leader
operator comes from the Feldman $\mathcal{F}_6$. We will consider this
operator first and then will generalize to higher Feldman operators
$\mathcal{F}_k$. This way we revisit the result of {\cite{Feldman}} that these
operators have negative leading anomalous dimensions.

As shown in the main text the leader for Feldman operators have the
structure:
\begin{eqnarray}
  (\mathcal{F}_k)_L & = & \sum_{l = 2}^{k - 2} (- 1)^l \binom{k}{l} \left(
  \sum\nolimits' \chi^l_i \right)  \left( \sum\nolimits' \chi^{k - l}_j \right) . 
  \label{FkCardy1}
\end{eqnarray}
In particular $(\mathcal{F}_6)_L = (\chi_i^3)^2 - \frac{3}{2} (\chi_i^2)
(\chi_i^4)$ up to a constant factor. We will see in a second that
$(\mathcal{F}_6)_L$ has no anomalous dimension at one loop, so we are setting
up a two-loop computation. Let $\mathcal{O}^B_i$ be all leader
operators of bare dimension 12 with which $\mathcal{O}^B_1 =
(\mathcal{F}_6)_L$ might mix, related to the renormalized operators by
$\mathcal{O}^B_i = Z_{i j} \mathcal{O}_j$. Since $(\mathcal{F}_6)_L$ is the
lowest non-susy-writable leader, all these other operators are susy-null or
susy-writable, hence the mixing matrix has the form (see {\eqref{BD}})
\begin{equation}
  Z = \left(\begin{array}{ccc}
    Z_{11} & \ast & \ldots\\
    0 & \ast & \ldots\\
    0 & \ast & \ldots\\
    \vdots & \vdots & \ddots
  \end{array}\right), \label{ZF}
\end{equation}
i.e.~in the first column only $Z_{11}$ is nonzero. Because of this, we only
need to know $Z_{11}$ to compute the anomalous dimension of
$(\mathcal{F}_6)_L$. We would need to know the potentially nonzero entries
marked by $\ast$ to compute the full eigenoperator, but we will not do this
here.

\begin{figure}[h]\centering \includegraphics[width=400pt]{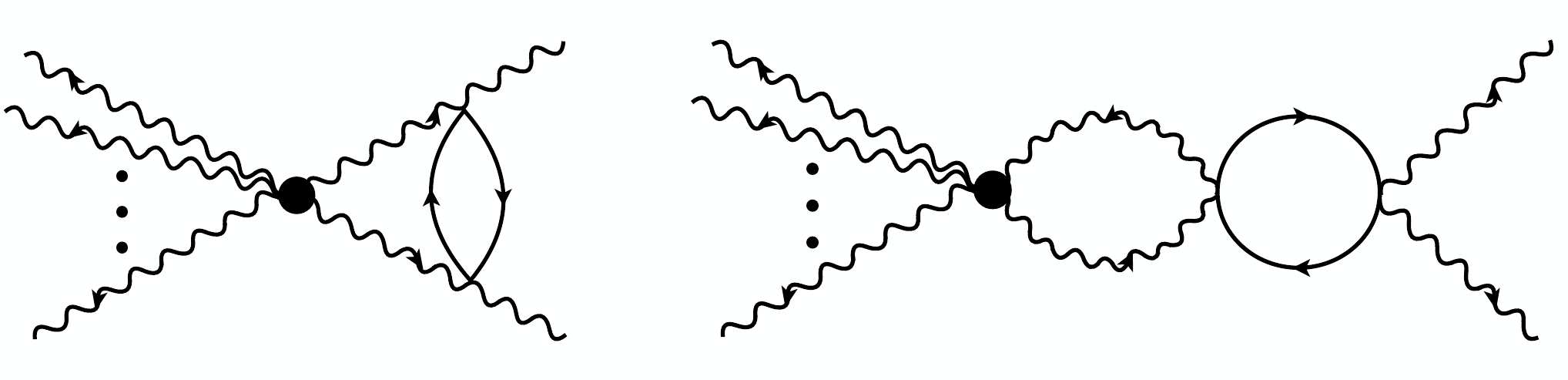}
  \caption{\label{Feldman6corr}Two-loop diagrams correcting the correlator
  $\langle (\mathcal{F}_k)_L (p = 0) \chi_{i_1} (p_1) \ldots \chi_{i_k} (p_k)
  \rangle$.}
\end{figure}

To compute $Z_{11}$ we consider the correlator $\langle (\mathcal{F}_6)_L (p
= 0) \chi_{i_1} (p_1) \ldots \chi_{i_6} (p_6) \rangle$. It is easy to see that
there is no one-loop diagram which contributes to this correlator. This
implies that $O (\lambda)$ correction to $Z_{11}$ vanishes, i.e.~as promised
$(\mathcal{F}_6)_L$ has no one-loop anomalous dimension. [On the other hand
some of the entries marked by $\ast$ in the first row are nonzero at one loop.
As a result the eigenoperator gets a susy-null admixture already at one loop.
See section \ref{sn-adm} for a discussion.] At two loops the correlator gets
contributions from the two diagrams in Fig.~\ref{Feldman6corr} (for $k = 6$),
both with a $1 / \varepsilon$ pole.

The first diagram's tensor structure in the $\chi_i$ indices is precisely
that of $(\mathcal{F}_6)_L$ itself (this is obvious because of $K^2 = K$, see
{\eqref{properties}}). Performing $K_{i j}$-tensor contractions, the second
diagram is instead proportional to the tree-level diagram with insertion of
the susy-null operator $(\chi_i^2)^3$. This is an example of a susy-null
admixture, a nonzero $\ast$ entry in the first row of {\eqref{ZF}}. So only
the first diagram contributes to $Z_{11}$.

That the second diagram does not contribute to $Z_{11}$ in fact holds for
general $k$, and we will need this fact below, so let us show it. Its tensor
structure splits into two groups of indices separated by the $\varphi$ loop.
The right tensor structure is just that of $\chi^2_i$. The left tensor
structure with its $\chi$-loop can be obtained by applying $K_{r s}
\frac{\delta}{\delta \chi_r} \frac{\delta}{\delta \chi_s}$ to the Feldman
leader expression. This calculation is greatly simplified using the equivalent
equation (see {\eqref{Fk0}})
\begin{eqnarray}
  (\mathcal{F}_k)_L & = & 2 \sum_{i = 2}^n (- \chi_i)^k + \sum_{i, j = 2}^n
  (\chi_i - \chi_j)^k,  \label{Fk1}
\end{eqnarray}
and it gives $2 k (k - 1) (\mathcal{F}_{k - 2})_L$ after a few lines of
algebra, i.e.~a multiple of the lower Feldman leader. Thus the total tensor
structure is that of $(\chi^2_i) (\mathcal{F}_{k - 2})_L$. For $k = 6$ this
reduces to $(\chi_i^2)^3$ as claimed, since $(\mathcal{F}_4)_L \propto
(\chi^2_i)^2$.

Let us next evaluate the first diagram in {\eqref{Feldman6corr}}, also for
general $k$. The loop integral is the same as $I_{(\chi_i^2)^2}$ in
{\eqref{loopchi22}}, the computation being similar to the one for the
$(\chi^2)^2$ in section \ref{chi22sec}. (Eq.~{\eqref{Z11}} below reduces to
{\eqref{ZH21}} for $k \rightarrow 4$.) Taking into account external leg
corrections and combinatorial factors, we get:
\begin{equation}
  \hspace{0.5cm} Z_{11}^{- 1} = 1 - \frac{k (k - 1)}{4} [I_{(\chi_i^2)^2}]_{1
  / \e} - \frac{k}{4} \left[ \frac{I_{\vf \omega} (p^2)}{p^2} \right]_{1 / \e}
  = 1 - \frac{k (3 k - 4)}{24} \frac{H^2 \lambda^2}{(4 \pi)^6  \e} .
  \label{Z11}
\end{equation}
Setting here $k = 6$ we get the anomalous dimension:
\begin{equation}
  \gamma_{(\mathcal{F}_6)_L} = - \frac{7}{9} \e^2 . \label{F6gamma}
\end{equation}
With a small extra input we can upgrade the above discussion and extract the
anomalous dimensions of $(\mathcal{F}_k)_L$ for general $k$. We just need to
check if one more element of the mixing matrix is zero. We have seen above
that $(\mathcal{F}_k)_L$ requires adding $(\chi^2_i) (\mathcal{F}_{k - 2})_L$
to get a finite operator. For $k = 6$ the latter operator was susy-null, hence
the inverse mixing was guaranteed not to happen. For $k > 6$ the operator
$(\chi^2_i) (\mathcal{F}_{k - 2})_L$ is non-susy-writable, so we need to see
if it mixes back to $(\mathcal{F}_k)_L$. It is however easy to see that this
does not happen. Using Eq.~{\eqref{FkCardy1}}, operator $(\chi^2_i)
(\mathcal{F}_{k - 2})_L$ can be expanded in monomials each of which is a
product of three $O (n - 2)$ singlets of the form $(\chi^2_{i_1})
(\chi_{i_2}^a) (\chi_{i_3}^b)$, where $a + b = k - 2$. If any of these
singlets is plugged into the second diagram in Fig.~\ref{Feldman6corr}, it
returns a monomial which is a product of four or three singlets (depending on
how the $\chi$-loop is contracted). Since $(\mathcal{F}_k)_L$ is made of
products of two singlets, its tensor structure cannot arise. This discussion
implies that the anomalous dimension of $(\mathcal{F}_k)_L$ for any $k$ is not
modified by mixing with $(\chi^2_i) (\mathcal{F}_{k - 2})_L$ and can be
computed from $Z_{11}$ given in {\eqref{Z11}}. We obtain:
\begin{equation}
  \gamma_{(\mathcal{F}_k)_L} = - \frac{k (3 k - 4)}{108} \e^2 .
  \label{gammaFkL}
\end{equation}
This nicely matches the result of {\cite{Feldman}}. In the future it would be
interesting to analyze other non-susy-writable leaders with the same classical
dimension as $(\mathcal{F}_k)_L$ for $k > 6$, one obvious example being
$\left( 2 \omega \vf + \chi_i^2 \right) (\mathcal{F}_{k - 2})_L$, to see if
any of these has the corrected dimension even lower than $(\mathcal{F}_k)_L .$

\subsubsection{Remark on admixture of susy-nulls}\label{sn-adm}

As mentioned above, the leader $(\mathcal{F}_6)_L$ experiences a susy-null
mixing already at one-loop level, which does not modify its anomalous
dimension (zero at one loop), but does modify the form of the eigenvector. In
fact the correct eigenvector at one loop is the linear combination:
\begin{equation}
  (\chi_i^3)^2 - \frac{3}{2} (\chi_i^2) (\chi_j^4) + \frac{3}{2} (\chi_i^2)^3,
  \label{fullF6}
\end{equation}
where the first two terms are (proportional to) $(\mathcal{F}_6)_L$, while the
last term is susy-null. The form of this one-loop eigenvector can be
determined e.g. using the OPE method (App. \ref{ope}).

Here we would like to point out that {\eqref{fullF6}} can also be determined
using group theory reasoning. Namely, we expect that leaders with well-defined
anomalous dimensions will transform in irreducible $O (n - 2)$
representations, which should correspond to symmetric \tmtextit{traceless}
tensors. Eq.~{\eqref{fullF6}} can be written as the contraction of 6 $\chi$'s
with the symmetric $6$-tensor
\begin{equation}
  T = \left( \delta_3 \otimes \delta_3 - \frac{3}{2} \delta_2 \otimes \delta_4
  + \frac{3}{2} \delta_2 \otimes \delta_2 \otimes
  \delta_2 \right)_{\tmop{sym}} . \label{TF6}
\end{equation}
Here $\tmop{sym}$ is the symmetrization, and $\delta_p$ denotes the rank
$k$-tensor whose only nonzero components are $(\delta_p)_{\underbrace{\scriptstyle i i
\ldots i}_p} = 1$ i.e.~when all $p$ indices coincide (the indices run from 2
to $n$). E.g.~$(\delta_2)_{i j} = \delta_{i j}$ is the Kronecker delta tensor,
while $\delta_1$ is the $(1, 1, \ldots)$ vector. The appropriate trace
taking into account the constraint $\sum\nolimits' \chi_i = 0$ is:\footnote{See
footnote \ref{Piij} for the definition of $\Pi_{i j}$; $\delta_{i j} + \Pi_{i
j}$ is the $n \rightarrow 0$ limit of $\delta_{i j} - \frac{1}{n - 1} \Pi_{i
j}$.}
\begin{equation}
  (\tmop{tr}' T)_{\ldots} = \sum_{i, j = 2}^n (\delta_{i j} + \Pi_{i j}) T_{i
  j \ldots} .
\end{equation}
It is then easy to work out (we define $\delta_0 = - 1$, a constant):
\begin{eqnarray}
  \tmop{tr}' \delta_p & = & 2 \delta_{p - 2}, \\
  \tmop{tr}'  (\delta_p \otimes \delta_q)_{\tmop{sym}} & = & [A_{p, q} 2
  \delta_{p - 2} \otimes \delta_q + A_{q, p} 2 \delta_p \otimes \delta_{q - 2}
  + (1 - A_{p, q} - A_{q, p}) (\delta_{p + q - 2} + \delta_{p - 1} \otimes
  \delta_{q - 1})]_{\tmop{sym}}, \nonumber
\end{eqnarray}
where $A_{p, q} = \binom{p + q - 2}{p - 2} \left/ \binom{p + q}{p} \right. =
\frac{p (p - 1)}{(p + q) (p + q - 1)}$, and similarly for higher tensor
products. Using these rules, one can check that the tensor {\eqref{TF6}} is
indeed traceless, while it would not have been traceless without the last
term.

\section{Remarks about tuning the disorder distribution}\label{app:tuning}

As discussed in section \ref{numsimideas}, one might be able to look for the
SUSY fixed point in numerical simulations of the RFIM, by tuning the disorder
distribution within a family depending on more than one parameter. Here we
discuss some ideas about what parameter to tune, to set the relevant operator
to 0, assuming for simplicity that a single perturbation has turned relevant,
the one corresponding to the susy-null leader $(\chi^2)^2$. This discussion is meant as schematic and non-rigorous.

Our starting point is the analysis of Br{\'e}zin-De Dominicis
({\cite{Brezin-1998}}, Section 1) who used the Hubbard-Stratonovich
identity to rewrite an Ising spin system in terms of a scalar field.
Introducing replicas and integrating out the disorder, they arrived at the
system of $n$ scalar fields on the lattice with the $S_n$-invariant potential
({\cite{Brezin-1998}}, Eq.~(1.9))
\begin{eqnarray}
  V & = & \frac{1}{2} (\tau_2 - 1)  \s_2 - \frac{\tau_2}{2}  \s_1^2 +
  \frac{1}{12}  (1 + 3 \tau_4 - 4 \tau_2)  \s_4 + \frac{1}{24}  (3 \tau_2^2 -
  \tau_4)  \s_1^4 \nonumber\\
  &  & + \frac{1}{8}  (\tau_2^2 - \tau_4)  \s_2^2 + \frac{1}{3}  (\tau_2 -
  \tau_4)  \s_1  \s_3 - \frac{1}{4}  (\tau_2^2 - \tau_4)  \s_1^2  \s_2 + O
  (\phi^6),  \label{Brdedom}
\end{eqnarray}
where $\sigma_k = \sum_{i = 1}^n \phi_i^k$ \ as in section \ref{Snsinglets},
and the quantities $\tau_p$ are defined as
\begin{equation}
  \tau_p = \int_{- \infty}^{\infty} dhP (h) (\cosh h)^n (\tanh h)^p
  \hspace{0.17em} \rightarrow \int_{- \infty}^{\infty} dh\,P (h) (\tanh h)^p 
  \qquad (n \rightarrow 0) .
\end{equation}
Now let us refer to the toy model RG analysis in App. \ref{sec:Toy}. We can
express the quartic part of the potential in terms of eigenperturbations given
in Table \ref{Tab:Toy5}: $\mathcal{O}_1 = \s_4$, $\mathcal{O}_2 = \s_1^4$,
$\mathcal{O}_3 = \s_1  \s_2^2$, $\mathcal{O}_4 = \s_2^2 + 2 \s_1  \s_3$ and
$\mathcal{O}_5 = \s_2^2 - \frac{4}{3}  \s_1  \s_3$: $V = \sum_{a = 1}^5 c_a
\mathcal{O}_a$. We are particularly interested in the coefficient of
$\mathcal{O}_5$, which comes out equal:
\begin{equation} 
c_5 = - \frac{1}{40}  (4 \tau_2 - 3 \tau_2^2 - \tau_4) . 
\end{equation}
Indeed, the operator $\mathcal{O}_5$ has the leader $(\chi^2)^2$ and we are
assuming that this direction is relevant, so we coefficient $c_5$ needs to be
tuned to reach the SUSY fixed point in the IR. The needed value of $c_5$ at
the UV scale depends on the microscopic details (it may be positive or
negative depending on the sign of the contributions that $c_5$ gets under RG
running). Since $c_5$ is a linear combination involving the second and fourth
moments of the disorder, one can imagine that the necessary tuning may be
obtained by adjusting the kurtosis of the distribution.\footnote{We thank
Giorgio Parisi for this remark. Recall that the kurtosis $K$ is defined, for
even distributions, as the normalized fourth moment $K = \frac{\overline{\,
h^4} }{\s^4} = \s^{- 4}  \int dh\,P (h) h^4$ where $\s = \left( \overline{\,
h^2} \right)^{1 / 2}$ is the standard deviation.} It should be stressed that
the $\mathcal{O}_a$'s are not exact nonperturbative eigenperturbations, and
so the tuning which we described should not be taken too literally. \\

\small
\bibliographystyle{utphys}
\bibliography{references}

\end{document}